\documentclass[structabstract]{aa}  
\usepackage{natbib}
\usepackage{graphicx}
\usepackage{txfonts}
\usepackage{color}

\begin{document}

\title{
Grain size limits derived from 3.6 $\mu$m and 4.5 $\mu$m coreshine
}

\author{
{J. Steinacker}\inst{1,2,3}
 \and
{M. Andersen}\inst{4}
 \and
{W.-F. Thi}\inst{5}
 \and
{R. Paladini}\inst{6}
 \and
{M. Juvela}\inst{7}
 \and
{A. Bacmann}\inst{1,2}
 \and
{V.-M. Pelkonen}\inst{7,8}
 \and
{L. Pagani}\inst{9}
 \and
{C. Lef{\`e}vre}\inst{9}
 \and
{Th. Henning}\inst{3}
 \and
{A. Noriega-Crespo}\inst{6,10}
       }
\institute{Univ. Grenoble Alpes, IPAG, F-38000 Grenoble, France\\
\email{stein@mpia.de}
\and
CNRS, IPAG, F-38000 Grenoble, France
\and
Max-Planck-Institut f\"ur Astronomie,
K\"onigstuhl 17, D-69117 Heidelberg, Germany
\and
Gemini Observatory, Casilla 603, La Serena, Chile
\and
Max-Planck-Institut für extraterrestrische Physik, Giessenbachstrasse 1, 85748 Garching,
Germany
\and
Infrared Processing and Analysis Center, California Institute of Technology, Pasadena, CA, 91125, USA
\and
Department of Physics, PO Box 64, University of Helsinki, 00014, Helsinki, Finland
\and
Finnish Centre for Astronomy with ESO, University of Turku, V\"{a}is\"{a}l\"{a}ntie 20, FI-21500 PIIKKI\"{O}, Finland
\and
LERMA \& UMR8112 du CNRS, Observatoire de Paris, 61, Av. de l'Observatoire, 75014 Paris, France
\and
Space Telescope Science Institute, Baltimore, MD, 21218, USA
     }
\date{Received, accepted}

  \abstract
{
Recently discovered scattered light from molecular cloud cores in the wavelength
range 3-5 $\mu$m 
(called "coreshine") 
seems to indicate the presence of grains with sizes above 0.5 $\mu$m.
}
{   
We aim to analyze 3.6 and 4.5 $\mu$m coreshine from molecular cloud cores
to probe the largest grains in the size distribution.
}
{
We analyzed dedicated deep Cycle 9 Spitzer 
IRAC observations in the 3.6 and 4.5 $\mu$m bands for a sample of 10 low-mass cores. 
We 
used a new modeling approach based on a combination of ratios of the two background- and 
foreground-subtracted surface brightnesses and observed limits of the optical depth.
The dust grains were modeled as ice-coated silicate and carbonaceous spheres.
We discuss the impact of local radiation fields with a spectral slope differing from what is
seen in the DIRBE allsky maps.
}
{
For the cores L260, ecc806, L1262, L1517A, L1512, and L1544, the model reproduces the data  
with maximum grain sizes around 0.9, 0.5, 0.65, 1.5, 0.6, and > 1.5 $\mu$m, respectively. 
The maximum coreshine intensities of L1506C, L1439, and L1498 in the individual
bands require smaller maximum grain sizes than derived from the observed distribution of band ratios. Additional isotropic local radiation fields with a spectral shape differing from the DIRBE map shape do not remove this discrepancy.
In the case of Rho Oph 9, we were unable to reliably disentangle the coreshine emission from background variations and the strong local PAH emission.
}
{
Considering surface brightness ratios in the 3.6 and 4.5 $\mu$m bands across a 
molecular cloud core
is an effective method of disentangling the complex interplay of structure and opacities 
when used in combination with observed limits of the optical depth.
}


\authorrunning{Steinacker et al.}
\titlerunning{Grain size limits derived from coreshine}
\maketitle

\section{Introduction}
Dust grains are present in most cosmic objects, and they add a rich spectrum 
of investigation methods to astrophysical research
\citep[see, e.g.,][]{2003ARA&A..41..241D,2010ARA&A..48...21H}.
Through the absorption, emission, and scattering of radiation they 
allow us to trace the local density and temperature structure of objects
\citep[for molecular cloud cores see, e.g.,][]{2012A&A...547A..11N,2013A&A...560A..41L,2013A&A...551A..98L},
and to reveal the action of young stellar objects (YSOs), which are enshrouded
at short wavelengths 
\citep[e.g.,][]{1990A&A...227..542H}.
In the interstellar medium (ISM), dense regions are predominantly cooled 
by dust radiation
\citep{2001ApJ...557..736G}. 
Chemical reactions on their surface contribute to the chemical processes 
in the gas
\citep{2005JPhCS...6...18H} with the available surface depending on the grain size distribution
and the fractal degree.
Dust grains are sensitive to magnetic forces because of their charge and 
composition
\citep{2007JQSRT.106..225L}.
In the planet formation process, small dust grains are the seeds for 
forming larger bodies in the accretion disks around YSOs 
\citep{2013prpl.conf1B012S}.

Important questions therefore include the origin of the grains, their 
composition, and size distribution. 
Dust grains are known to be produced and destroyed in supernovae events
\citep[e.g.,][]{2010MNRAS.403..474W, 2014ApJ...782L...2I}
and in the 
atmospheres of stars in their late evolutionary phases
\citep[e.g.,][]{2009MNRAS.394..831M}.
Their distribution into the ISM and their complex processing including 
destruction and growth is the subject of ongoing research
\citep{2011A&A...530A..44J, 2011ApJ...742....7A, 2014arXiv1407.8489Z}. 
Grain growth processes have been discussed mostly in the framework of 
accretion disks and planet formation. 
Cold and dense prestellar molecular cloud cores may also exist 
long enough to allow the collisional growth of grains (e.g.,
\citep{1993A&A...280..617O, 1994A&A...291..943O, 1994ApJ...430..713W,
2009A&A...502..845O,2011A&A...532A..43O}.
There is evidence that grains are ice-coated in the dense ISM
\citep{1983Natur.303..218W},
supporting the growth process by efficient sticking. However, it remains to
be investigated to what extent gas turbulence can provide the relative velocities
that are needed to coagulate grains as
assumed in these molecular cloud core models.

Concerning modeled grain size distributions,
\citet{1977ApJ...217..425M} 
were the first to propose a distribution for the diffuse ISM. 
They suggested a power-law distribution with a slope of -3.5 and with a size 
limit of 0.25 $\mu$m for non-graphite grains (abbreviated as "MRN distribution")
and emphasized that only weak 
constraints could be derived for the larger grains in the distribution.
In the following we denote size distributions as "MRN-type" when they have 
a power-law index of -3.5. When we refer to grain sizes, we will give the grain radius value.
Since then a large number of observations and models 
considering thermal emission and extinction of grains have suggested both varying size distributions and grains with radii beyond 0.25 $\mu$m in the denser parts of the ISM
\citep{
2003A&A...398..551S, 
2006MNRAS.373.1213K, 
2006A&A...451..961R, 
2008ApJ...684.1228S,
2009ApJ...690..496C, 
2009ApJ...699.1866C, 
2009ApJ...693L..81M,
2010A&A...520L...8P,
2010ApJ...713..959V,
2013A&A...559A.133Y}.

Scattering by dust grains is another process that is sensitive to the grain size.
\citet{1996A&A...309..570L}
modeled the 
1.25, 1.6, and 2.2 $\mu$m band (J, H, and Kn) scattered light images of the Thumbnail nebula 
and concluded that the limiting grain size exceeds 0.25 $\mu$m.
The strong decline of scattering efficiency of ISM grains beyond the Ks band caused people to 
assume that the cores become dark in scattered light in the mid-infrared (MIR).
It would take even larger grains of micron-size scale to elevate scattered light intensities
back to the surface brightness (SFB) measurable by MIR detectors.
Such radiation was indeed found inside the cloud L183 
and named {\it \emph{coreshine}} \citep{2010A&A...511A...9S}. This term distinguished it from {\it \emph{cloudshine}},
as 
\citet{2006ApJ...636L.105F}
named the 
near-infrared (NIR) scattered light from the core parts that are not shielded from NIR radiation. Beneficial for the detection of 
this radiation was the fact that extinction has a minimum in the wavelength range 3-5
$\mu$m, allowing interstellar radiation to penetrate deeper into the core.

\citet{2010A&A...511A...9S}
present the most complex spatial structure modeling that has been performed so far for
a molecular cloud core.
Based on {\it Spitzer}/IRAC data 
at 3.6, 4.5, and 8 $\mu$m, the 3D radiative transfer (RT) modeling of L183 required the assumption
of population of grains with sizes up to 1 $\mu$m.
They argue that the excess of SFB in the 4.5 $\mu$m band could not be 
fit by emission of transiently-heated grains 
due to the absence of a strong feature in this band.
Despite the strong central extinction ($A_V$ above 150)
and high Galactic latitude ($b=36.8^\circ$), the physical conditions
in the L183 clump are not intrinsically different from those seen in nearby cores. 
Correspondingly, the investigation data of {\it Spitzer} PI and 
Legacy survey data  of a sample of 110 cores led to the conclusion that about half of all cores show hints of
coreshine \citep{2010Sci...329.1622P}. 
It was shown that scattered light from cores is observable in a variety of star formation stages, 
from prestellar cores to disks around forming stellar objects, as well as for a wide range of 
morphologies, i.e. single or binary systems and in filamentary structures
\citep[see also][for a small subsample of cores with 3.6 $\mu$m emission]{2009ApJ...707..137S}.
In the same study, the RT analysis of model cores revealed that without background, the 
scattered light morphology
does not vary strongly with the position in the Galaxy aside 
from weak enhancement towards the Galactic center (GC). 
For more massive cores with 10 M$_\odot$, the morphology showed a stronger crescent 
of enhanced emission due to the combined action of extinction and multiple scattering.

Since then further studies have investigated the occurrence and conditions of 
coreshine detection, and this work is also devoted to scattered MIR radiation in molecular
clouds and the grain properties causing it.
\citet{2012A&A...541A.154P} 
find a drastically reduced occurrence of scattered light 
(3 detections plus 3 uncertain cases out of 24 objects) in the Gum/Vela region. 
It has been suggested that a connection
to the action of a supernova remnant blast wave affects the grain size population in this region. 
In general the analysis of coreshine in regions with an
enhanced interstellar radiation field (ISRF) like the Vela region is impeded
owing to the confusion with emission of stochastically-heated grains like PAHs (polycyclic aromatic hydrocarbons).

Observations in the bands at 2.2, 3.6, and 8 $\mu$m of the core L260 were modeled by 
\citet{2013A&A...559A..60A}.
Their simple core model was able to reproduce the observed SFB profiles that
assume an MRN-type size distribution with grains up to about 1 $\mu$m in size and
a locally enhanced incident radiation field (enhancement factor 1.7).

\citet{2014A&A...564A..96S}
have analyzed the coreshine seen from the low-density core L1506C and find that it requires the 
presence of grains with sizes exceeding the MRN distribution. 
Using grain growth models, they argued that L1506C must have passed through a period of 
higher density and stronger turbulence to grow the grains in situ.

The conditions of detecting scattered light from low-mass molecular cores at 3.6 $\mu$m have been
re-investigated in 
\citet{2014A&A...563A.106S} including background effects.
The limits derived from the RT equation indicate that extinction by the core 
prohibits detection in bright parts of the Galactic plane and especially near the GC. 
They show that the scattering phase function favors the detection of scattered light above and 
below the GC and, to some extent, near the Galactic anti-center. 
It was discussed that an enhanced radiation field may give rise to a coreshine signal even
in the presence of a strong background.

\citet{2014A&A...572A..20L} have investigated the key parameters for reproducing the general trend of SFBes and intensity ratios of both coreshine and near-infrared observations. 
Based on a careful determination of the background field, they find that for a sample of 72 sources,
starless cores show 3.6 $\mu$m/4.5 $\mu$m SFB ratios above 2, while cores with embedded sources can have lower values.
To constrain the dust properties, they used a rich grid of size-averaged dust models 
based on an extrapolation of standard spherical grains able to fit the observations in the diffuse medium. Moreover, the effects of fluffiness, ices, and a handful of classical grain size distributions were also tested.
As spatial density distribution, an inclined ellipsoid with a Plummer-like profile with two masses was 
assumed. Their density structure is well-suited to centrally-condensed cores like L1544, which show a central depression in their SFB pattern, while the vast majority of cores with coreshine have no depression.
They find that normal interstellar radiation field conditions are sufficient to explain coreshine with suitable grain models at all wavelengths for starless cores. According to their multiwavelength approach, the standard interstellar grains are not able to reproduce observations and only a few grain types meet the criteria set by the data. 
 
This paper investigates coreshine in a sample of 
ten cores. Coreshine has been detected for all cores except for Rho Oph 9 
where the emission could also be explained by expected PAH emission. 
We use deep warm {\it Spitzer} 
IRAC 3.6 and 4.5 $\mu$m band 
observations (Cycle 9 Coreshine Follow-up: Program ID 90109, PI R. Paladini, Paladini et al., in prep.)
\footnote{{\em \emph{"Cold"}} and {\em \emph{"Warm"}}
denote, respectively, the first part of the Spitzer mission - which lasted
from August 2003 to May 2009 - during which all the instruments, including
IRAC, were cooled to cryogenic temperatures, and the second part of the
mission - from May 2009 to present - which started at the end of cryogen
and in which only the channels 1 and 2 of the IRAC instrument are operating.}
.
The observations were motivated by the fact that silicate grains of a size around 1 $\mu$m
are expected to have their largest scattering-to-absorption efficiency in the 
wavelength range 3-5 $\mu$m and that the largest grains should have their
strongest impact in the 4.5 $\mu$m band.

In Sect.~2 we briefly summarize the data and their acquisition and processing.
The SFB ratio model is derived in Sect.~3, outlining the different RT transfer methods and optical depth effects. The modeling is described in Sect.~4, and
the maximum grain sizes are analyzed for each core assuming an MRN-type size distribution. In Sect.~5, we discuss the results and consider the influence of local isotropic radiation fields, and
summarize our findings in Sect.~6. The determination of the off-core SFB
is described in Appendix A, and in Appendix B we describe the opacities of the dust model.

\section{Data}\label{data}
\begin{figure*}[!ht]
\includegraphics[width=18cm]{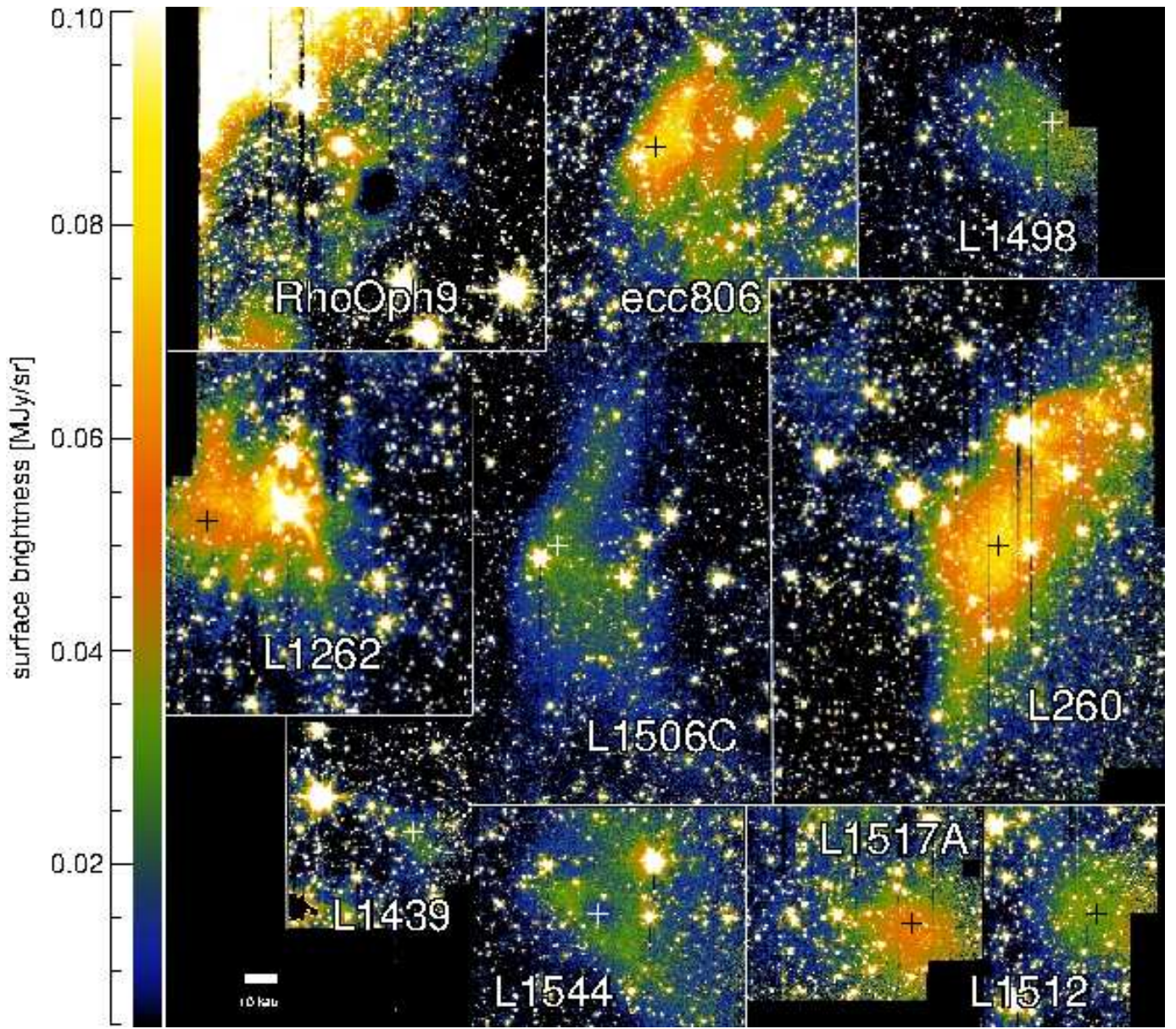}
\caption{
Off-subtracted surface brightness images at 3.6 $\mu$m for the
sources of the sample. The image sizes have been adjusted to show the 
sources at the same distance in order to visualize the source sizes. 
Most cores have a complex shape, and the diffuse signals differ with respect
to the peak brightness and point source contamination. 
In the case of Rho Oph 9, the image has not been off-subtracted and the bright emission in the upper left corner is caused by transiently-heated particles and not by coreshine.
The crosses indicate the approximate location of the core center
based on the corresponding 8 $\mu$m extinction maps or thermal emission maps in the
FIR/mm.
}
\label{all}
\end{figure*}
The 3.6 and 4.5 $\mu$m IRAC data used for this work were taken during 
cycle 9 of the {\it Spitzer} warm mission.
These observations targeted a set of ten Galactic cores partially selected from 
the
"Hunting for Coreshine" program (Program ID 80053, PI Paladini). 
In earlier data, nine cores show a clear indication of coreshine emission 
at 3.6 $\mu$m. Extinction at 8 $\mu$m can be found for all cores in the cold {\it Spitzer} data
except for PLCKECC G303.09-16.04 (we use the abbreviation ecc806 in this paper)
and L1506C. 
It is a core from the
Planck Early Release Cold Core Catalogue (ECC), which is part of the 
Planck Early Release Compact Source Catalog (ERCSC) \citep{2011A&A...536A...7P,2011A&A...536A..23P}.
We selected a varied sample of cores for this study, 
some with particularly strong coreshine (e.g., L260, ecc806, L1262, L1512), 
some also previously well-studied for coreshine emission (L260, L1506C) 
or otherwise (L1544, L1498). Some of these and other cores have particular features of interest: 
binary core L1262, L1517A having similar cores nearby, 
L1512 with a particularly simple shape, 
L1544 with a central coreshine depression, 
L1506C and L1498 with unusually low densities, 
L1439 near more evolved young stellar objects, 
and Rho Oph9 with an expected strong PAH contribution.
The original survey did not reach a comparable signal-to-noise (S/N) 
at both 3.6 and 4.5 $\mu$m, with the consequence that coreshine emission at 4.5 
$\mu$m was typically at the level or even below the instrumental noise. 
This situation was rectified in Cycle 9. These new observations were carried out 
in full array mode with 100 s integration. 
A detailed description of the source selection, observing strategy, and data processing will 
be provided in a forthcoming 
paper (Paladini et al., in prep.).
In this work we have used the IRAC Level 2 mosaics (or pbcd, post basic
calibration data),  generated  by the Spitzer Science Center (SSC)
pipeline and downloaded from the Spitzer Heritage Archive (SHA). These
images are affected by column-pulldown, an instrumental effect that
consists in a depression of the intensity registered by pixels along a
given array column, and which is caused by a bright source, i.e. a star or
a cosmic ray. The IRAC images are also not zody-subtracted, meaning that
the contribution from zodiacal light is not removed by individual frames
or from the final mosaics at any stage of the automatic processing.

Basic core properties gathered from the literature are given in Table \ref{table:1}.
The sources are
ordered by decreasing coreshine peak SFB (see also Table \ref{table:2}).
The estimates for the central column density have error bars (factors of a few) 
partly because the values are derived using a specific dust model. The 
ranges given in the Table are either specified in the 
cited work or based on assuming uncertainties by a factor of 2. Values without citation are based on
simple estimates and just serve as an indication.
The given masses and sizes (mean core diameter) are derived using different methods and also depend on the definitions of 
the core boundary. Moreover, it usually takes more than one parameter to
characterize the complex core structure. In some cases we have estimated the size from the {\it Spitzer}
map.
To overcome the variation in the optical depth due to the uncertainties, we will rely in the modeling on a maximum optical depth at 3.6 $\mu$m (2.2 $\mu$m
in the case of L260) that is given in the last column. Its choice and role in the modeling is described in Sect.~\ref{depth}.
For modeling individual cores with more sophisticated density models
and using multiwavelength 3D RT, it will be worth re-evaluating
the mass estimates based on the 
original data and using a more sophisticated structure model with more than just a few
size parameters.
\begin{table*}
\caption{Core sample and their basic properties$^r$}.
\label{table:1}   
\centering                      
\begin{tabular}{l l c c c c c | c} 
\hline\hline             
Source        & Gal. long. & Gal. lat.  & Distance &  N$_c$                & Mass             & Size &  $\tau_{max}$ \\   
              & [$^\circ$] & [$^\circ$] & [pc]     & [10$^{26}$ m$^{-2}$] & [M$_\odot$]      & [kau]  & at 3.6 $\mu$m\\   
\hline                     
L260          & 008.65     & +22.17     & 160$^a$  &             1-4          & 1-3$^{a,b}$  & 36$^c$  & 1 at Ks$^m$ \\
ecc806$^l$    & 303.06     & -16.04     & 160$^s$            &    0.5-2$^q$                &  2-8$^q$                & 25$^c$ & 1      \\
L1262 (CB244) & 117.12     & +12.40     & 200$^d$            & 3-27$^p$             & 3-7$^d$      & 15$^d$& 4.5 \\
L1517A        & 172.50     & -07.98     & 140$^e$         & 1-4$^n$                & 1$^f$            & 20$^c$ & 1     \\
L1512 (CB27)  & 171.87     & -05.24     & 140$^e$            & 1-9$^p$              & 2.4$^g$          & 15$^c$ & 1.5\\
L1544         & 177.98     & -09.72     & 140$^e$             & 6-26$^k$             & 2.8$^g$          & 20$^g$ & 6 \\
L1506C        & 171.15     & -17.57     & 140$^e$            & 1-4                   & 4$^h$            & 25$^c$ & 0.5\\
L1439 (CB26)  & 156.06     & +06.00     & 140$^g$           & 0.7-6$^p$            & 1.6$^g$          & 15$^g$  & 0.5\\
L1498         & 170.14     & -19.11     & 140$^e$            & 1.6-6.4$^k$          & 0.51-0.83$^i$    & 30$^j$ & 0.5\\
\hline
\end{tabular}
\tablebib{
$^a$\citet{2002AJ....124.2756V},
$^b$\citet{2002ApJ...572..238C},
$^c$read from the {\it Spitzer} image using the distance,
$^d$\citet{2010A&A...518L..87S},
$^e$\citet{1994AJ....108.1872K},
$^f$\citet{2011A&A...533A..34H},
$^g$\citet{2009ApJ...707..137S},
$^h$\citet{2010A&A...512A...3P},
$^i$\citet{2005ApJ...632..982S},
$^j$\citet{2006A&A...455..577T},
$^k$\citet{2008A&A...492..703C},
$^l$official name is PLCKECC G303.09-16.04,
$^m$for L260, we used a constraint in the 2.2 $\mu$m band Ks,
$^n$derived from \citet{2011A&A...533A..34H},
$^p$\citet{2013A&A...560A..41L},
$^q$based on Herschel/SPIRE archive data of the Gould Belt Survey (PI: Andr{\`e})
$^r$Rho Oph 9 is not included.    
$^s$\citet{2011A&A...536A...7P}.    
}
\end{table*}
The meaning of $\tau_{max}$ is explained in Sect.~3.7.

To give an impression of the overall SFB levels, the
source geometry, sizes, and the influence of neighboring stellar sources,
we show in Fig.~\ref{all} the 3.6 $\mu$m
off-subtracted SFB images of all sources scaled to the same distance.
For the example of L260, Appendix A describes how we have determined for all cores this "off-core surface brightness" that is 
subtracted from the core SFB.
The linear scale in the plane-of-sky (PoSky) is indicated by the 10 kau bar in the lower lefthand corner.
To indicate the approximate location of the central part 
of the core, we have marked the minimum of the corresponding 8 $\mu$m maps (for L1498, L1262, L260, L1439, L1544, L1517A,
and L1512) where the cores are seen in extinction or the maximum of thermal emission maps in the FIR/mm (ecc806 and L1506C). Details are given in Sect.~4.

\section{Modeling approach}\label{sect3}

Coreshine modeling involves RT calculations in a
complex 3D core geometry. 
The scattering integral in the RT equation mixes radiation from all directions in all points.
Moreover, to describe a 3D density structure, about ten to hundreds of free parameters are needed
depending on the complexity of the structure. 
This makes 
coreshine modeling with an anisotropic illumination
of the dust grains
numerically difficult. A scan of the entire
parameter space of size distributions, chemical compositions, grain shapes, and
spatial density distributions is therefore too demanding for the current computer power.
In the following sections, we describe our basic assumptions for reducing the parameter
space and the complexity of the problem, the RT approximations, and the general modeling approach.

\subsection{Basic assumptions}\label{assumptions}
To reduce the complexity of the problem we make three basic assumptions.

\noindent
{\it Assumption 1:}
The grain size distribution $s(a)\ da$ is normalized to have the 
same gas-to-dust mass ratio across the core, and it has no
spatial change in shape or size limits.

This assumption is supported by the shallow gradients of the SFB
pattern observed across the cores with coreshine. 
In turn, 
\citet{2010A&A...511A...9S}
used a grain model with grain size increasing with
gas density, and the SFB model showed stronger gradients than 
observed.
Nevertheless, there are theoretical arguments for why the opacities might change
across the core. For example, grain coagulation and the thickness of the ice mantles are expected to increase toward the center of the cores,  depending, however, on how quickly turbulent mixing can smooth these
gradients again \citep[see also][]{2013A&A...559A..60A}. 
Relaxation of this approximation will be the subject of further publications.

\noindent
{\it Assumption 2:}
The size distribution is a power law,
and the sizes range from $a_1$ to $a_2$. 

We have varied the power-law index and
discuss its impact on the findings in Sect.~\ref{sectionL260}.
While there are studies suggesting that the spectral index of the distribution may be different from -3.5 as observed in the ISM
\citep{2001ApJ...548..296W}
or deviate from power laws
\citep{1994ApJ...430..713W,2009A&A...502..845O},
we use this distribution as a typical case where the total grain surface area is dominated
by the small grains while the mass predominantly rests in the large grains.
It needs to be verified whether currently available
data allow  constraining the detailed size-distribution beyond the power-law picture.

\noindent
{\it Assumption 3:}
The optical properties of the grains are derived from Mie theory
using spherical ice-coated silicate and carbonaceous grains.

Numerical calculations by
\citep{2009A&A...502..845O}
that are based on aggregate collisions show that grains in cores go through a phase of
compaction so that the spherical approximation might be better than 
initially expected from a growth picture that involves fluffy aggregates with
large geometrical filling factors. 
Nevertheless, future modeling with a more complex dust model should include non-spherical grains
\citep[as discussed, e.g., in][]{2014A&A...572A..20L}.
\citet{1983Natur.303..218W}
showed that grains in the Taurus cloud contain ice, and later studies have confirmed this finding for
other clouds.
\citet{2014A&A...568L...3A} find for the Lupus IV molecular cloud complex that the limits for the occurrence of ice mantles and coreshine are 
similar and possibly related.
We use a mass ratio of 1:4 for the carbonaceous and silicate component and assume the
same size distribution.
The details of the dust model are described in Appendix~\ref{Mie}.

\subsection{Optical depth}
The essential quantity for characterizing the impact of the dust
on the transport of radiation is the size-integrated optical depth between two points on a line
$\vec x=\vec x_0+\ell \vec n$ with the starting point $\vec x_0$ and the direction
$\vec n$. Based on the approximations listed in Sect.~\ref{assumptions} it has the form
\begin{equation}
\left< \tau \right>_s(\lambda,\ell_1,a_1,a_2)=
\int\limits_{0}^{\ell_1}d\ell\ t(\ell)
\int\limits_{a_1}^{a_2}da\ s(a)\ \sigma(\lambda,a)
\end{equation}
where $t$ is the spatial distribution of the H$_2$ number density, and $\lambda$  the 
wavelength. 
Integrating $t$ over space gives the total number of H$_2$ and, when multiplied with the H$_2$ mass, the assumed total core mass.
Since $\left< \tau \right>_s$ contains the line-of-sight (LoS) integral of the
gas density, it can be related to the H$_2$ number column density $N$ between the
two points.
The cross section $\sigma$ refers to absorption, scattering, or extinction, which defines
the optical depth 
$\left< \tau_{abs} \right>_s$,
$\left< \tau_{sca} \right>_s$, and
$\left< \tau_{ext} \right>_s$, respectively.

An important feature of the optical depth is that it depends not only on the core properties
but also on the dust properties, so that for the same core, for example, two dust models can give very
different central optical depths. This coupling of the core density structure and the dust properties
with the optical depth is not easy to disentangle: basic properties like total core mass, its size, 
or visual extinction are often derived by interpreting observational data with a specific dust model
and will change when the modeling requires a change in the dust properties.

In this paper, we use three approximations of
the general 3D RT equation for this application based on the assumed maximum optical depth
and give them simple names with the precise definitions provided in this section.

\subsection{"Full RT"}
For cores reaching optical depths for extinction beyond 1, shadowing effects and multiple scattering become important in the core.
This is the most complex of the three modeling variants, 
and the stationary 3D RT equation in this case has the form
\begin{eqnarray}
&&
\frac{dI(\vec x,\vec n,\lambda)}{d\ell}
=
- \left<\sigma_{ext}\right>_s(\lambda,a_1,a_2)\ t(\vec x)\ I(\vec x,\vec n,\lambda)
  + S(\vec x,\vec n,\lambda)+\cr
&&+
  t(\vec x)\ 
  \int\limits_{4\pi} d\Omega'\ 
    \left< \sigma_{sca}\ p\right>_s(\lambda,\vec n,\Omega',a_1,a_2)\  
    I(\vec x,\vec n',\lambda).
\end{eqnarray}
The equation describes the change in the intensity $I$ given at the location $\vec x$
in the direction $\vec n$ along the path $d\ell$. 
The source term $S$ contains the incident interstellar radiation field or nearby
radiation sources.
We have abbreviated the integrals over the grain sizes by
\begin{equation}
\left<\sigma_{ext}\right>_s=
\int_{a_1}^{a_2}da\ \sigma_{ext}(\lambda,a)\ s(a)
\end{equation}
and
\begin{equation}
\left<\sigma_{sca}\ p\right>_s=
\int_{a_1}^{a_2}da\ \sigma_{sca}(\lambda,a)\ p(\lambda,a,\Omega',\vec n)\ s(a).
\end{equation}
The probability of scattering radiation
from the solid angle $d\Omega'$ into the considered direction $\vec n$ is given by
the phase function $p$.
This non-local integro-differential equation requires  applying 
advanced numerical schemes like Monte Carlo
or ray-tracing in order to solve for the intensity emerging in the direction
of the observer \citep[for a review see][]{2013ARA&A..51...63S}.

\begin{figure*}[!ht]
\includegraphics[width=18cm]{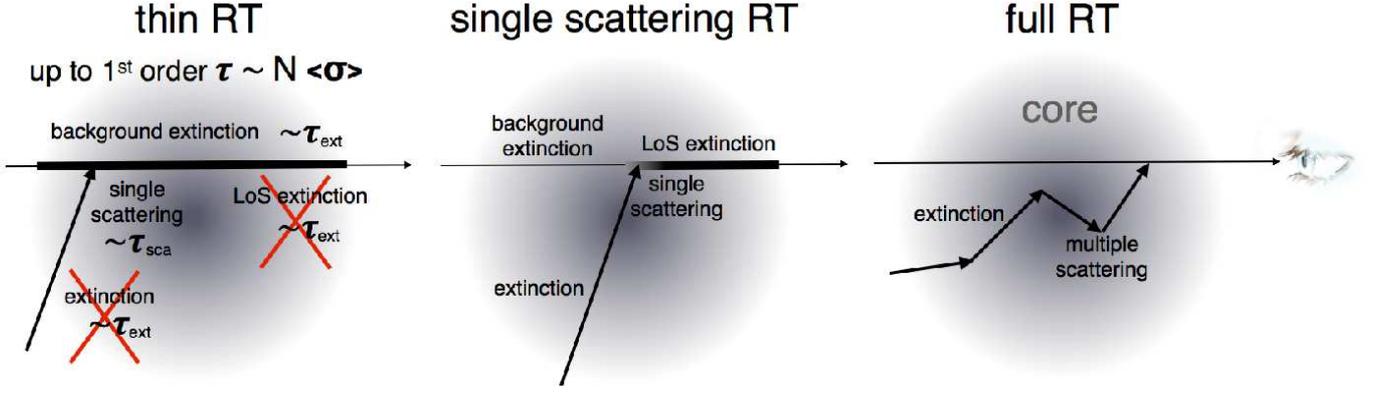}
\caption{
Illustration of the different RT methods used in the paper 
(the abbreviation "Thin RT" stands for "RT
in the optically thin limit").
        }
\label{illu}
\end{figure*}

An essential difficulty in analyzing radiation received from
a core with optical depths beyond 1 is that the structural information is mixed in at all 
LoS both by extinction effects and by multiple scattering. 
Since most cores show complex spatial structures, the error from assuming a
wrong density model will be propagated with every ray or photon propagation into the
derived SFB. It is important to note that this is not a problem of the
available solvers. It is a problem of the modeling, which tries to avoid using too many
free parameters 
that enter with the asymmetric core structure.
Shadowing effects also amplify the impact of anisotropic illumination.
The RT modeling of coreshine from the core L183 performed in
\citet{2010A&A...511A...9S}, for example, required the use of 100 3D Gaussian density clumps.
An illustration of the RT variants is given in Fig.~\ref{illu} with the full
RT case being 
described
in the righthand panel.

\subsection{"Single-scattering RT"}
When the optical depth for scattering is below 1 across the core, 
the impact of photons that scatter more than once is reduced.
In this case we can use the {\it \emph{single-scattering}}
approximation.
Instead of following rays or photons in arbitrary directions, we can separate
contributions along the LoS ($x$-direction) in each PoSky
position $(y,z)$.
Constructing the SFB from the 
background intensity $I_{bg}$ that undergoes the extinction and 
radiation that is scattered into the LoS from all directions, we find
\citep[see, e.g.,][]{2014A&A...563A.106S}
\begin{eqnarray}
&&I(\lambda,y,z,a_1,a_2)
=
e^{-\tau_{ext}(\lambda,y,z,a_1,a_2)}
 I_{bg}(\lambda)\ +I_{fg}(\lambda)+\cr
&&
 +\int\limits_{-\infty}^{+\infty}dx\ t(x)
  \int\limits_{a_1}^{a_2}da\ s(a)\ \sigma_{sca}(\lambda,a)\ 
  \left< IP \right>_\Omega(a,x,y,z)
\label{singleeq}
\end{eqnarray}
with the foreground SFB $I_{fg}$,
the direction-averaged product of incident radiation field, and phase function $P$
now giving the probability of scattering radiation from the solid angle $d\Omega$
into the LoS
\begin{eqnarray}
&&\left< I P\right>_\Omega(a,x,y,z)=e^{-\tau_{ext;cell-obs}(\lambda,x,y,z)}\ \times\cr
&&
\times 
\int\limits_{4\pi}d\Omega\ P(\lambda,\Omega,a,x,y,z)\ I_{ISRF}(\lambda,\Omega)\ 
e^{-\tau_{ext;sky-cell}(\lambda,\Omega,x,y,z)}.
\end{eqnarray}

Interstellar radiation on its way to the core position $(x,y,z)$
from a certain solid angle $\Omega$ undergoes extinction
described by $\tau_{ext;sky-cell}$.
The radiation that is scattered by the dust at $(x,y,z)$ and leaves towards the observer
again undergoes extinction with the optical depth $\tau_{ext;cell-obs}$.

The extinction of radiation still carries 3D density information from the entire
core into the intensity of radiation being scattered into the LoS, and with it
all errors of the density approximation. But the directions of the radiation transport
are known and can be analyzed in advance.

The advantage of this approach is that it allows precalculation of all number
column densities that are encountered by photons arriving from all directions
at the considered LoS and then propagate to the observer. This speeds up the calculations
when several wavelengths are considered and allows for the use of higher resolution
ISRF maps, for example. We have created a code version that makes use of this advantage 
\citep[used, e.g., in][]{2013A&A...559A..60A} and apply it also in this paper.

Owing to optical depth effects, the two bands at 3.6 and 4.5 $\mu$m can probe different
regions in the outer parts of cores with masses of a few M$_\odot$ when the
opacities change from 3.6 to 4.5 $\mu$m.
For a standard core profile (flat followed by power law with an index of -1.5),
a mass of 5 M$_\odot$, an outer radius of 10 kau, and an MRN-type distribution
of silicate grains, among others, the locations of the 
$\tau=1$ layers differ by
14 \% of the core radius, and the probed column number densities
differ by a factor of 1.8.  
For single-scattering RT like for full RT,
the SFB in both bands is a result of the combined
action of opacity and spatial density effects.

\subsection{RT in the optically thin limit ("Thin RT")}\label{RTot}
For some cores with low column densities and/or low-opacity grains, it is possible to
treat RT in the 
{\it \emph{single scattering}} and {\it \emph{optically thin}} 
approximation even for grains with sizes of several
$\mu$m in the wavelength range 3-5 $\mu$m.
The limiting $\tau$ values for the thin approximation can be related, for example, to the considered
order of the Taylor expansion of the exponential function
\citep{2014A&A...563A.106S}.

The SFB for an expansion to first order in $\tau_{ext}$ is
\begin{eqnarray}
&&I(\lambda,y,z,a_1,a_2) - I_{bg}(\lambda,y,z) - I_{fg}(\lambda,y,z) \approx\cr
&&-\left<\sigma_{ext}\right>_s(\lambda,a_1,a_2)\ N(y,z)\ I_{bg}(\lambda,y,z)\cr
&&+
N(y,z)\ 
\int\limits_{a_1}^{a_2}da\ s(a)\ 
\sigma_{sca}(\lambda,a)\ \int\limits_{4\pi}d\Omega\ P(\lambda,\Omega,a)\ I_{in}(\lambda,\Omega)
\label{Start1}
.\end{eqnarray}

The lefthand side of Eq. (\ref{Start1}) and the radiation
behind the core $I_{bg}$ can be determined observationally. Also for the incident field $I_{in}$, 
an approximation can be derived from observed allsky maps.
Then we can calculate the righthand side with a model for the density and dust opacities.
Unlike the cases of full or single scattering RT, the spatial and dust property problem
decouple for thin RT since both terms on the righthand side of Eq. (\ref{Start1}) contain the spatial variation 
in the form of $N$ for each pixel or LoS. 
Since for optically thin scattering, both bands see the same column of grains, the 
spatial density variation is separated from the opacity variation. 
This makes the low-density cores best-suited to 
studying the optical properties of the grains as performed for
translucent clouds and the diffuse ISM.

\subsection{Modeling surface brightness ratios instead of surface brightnesses}
For the modeling approach of the two images obtained at 3.6 and 4.5 $\mu$m 
for each source, we considered three methods partially based on the
ratio, $R$, of the off-subtracted images.

i) {\it Image modeling},
where we assume a dust model and a density structure, 
calculate model images in the two bands, and compare model and real images at each wavelength. 
The modeling process is to vary the density structure and opacity until the differences are minimized. 

ii) {\it "Full RT" $R$ and maximum surface brightness modeling},
for which we build the ratio map from the observed images, select a region where the optical depth is expected to be small, and determine the observed $R$ distribution. Then assuming a dust and density model, we calculate the model images in the two bands using "full RT", 
build the ratio map, and find the theoretical $R$ distribution in the selected region.
From the images we find the maximum SFBs.
The modeling process is to vary the opacity and the central column density until the differences are minimized. 

iii) {\it "Thin RT" $R$ and maximum surface brightness modeling},
by building a ratio map $R$ from the observed images in the two bands, selecting a region where the optical depth is expected to be small, and determining the observed $R$ distribution. Then assuming a dust model and a
maximum column density, we can calculate the maximum SFB.
The modeling process is to vary the opacity and the central column density value until we can reproduce the mean $R$ and the maximum SFB.

To avoid either very complex spatial modeling with many parameters or the introduction of large errors
from taking a density structure model that is too simple, we applied methods ii) and iii).
To build the ratio map
$R=I_{cs}(3.6\ \mu{\rm m})/I_{cs}(4.5\ \mu{\rm m})$ 
from the SFBs above the background caused by scattered light in the core $I_{cs}$,
we
divided Eq. (\ref{Start1}) for the two bands.
For each pixel in the two images, the ratio on the lefthand side can be determined from the observed
data
\begin{equation}
R_{obs}(y,z)=
\frac{I^{3.6}(y,z) - I^{3.6}_{bg}(y,z) - I^{3.6}_{fg}(y,z)}
     {I^{4.5}(y,z) - I^{4.5}_{bg}(y,z) - I^{4.5}_{fg}(y,z)}.
\label{Ro}
\end{equation}
Using the albedo
\begin{equation}
w^\lambda (a,a_1,a_2)= \frac{\sigma^\lambda_{sca}(a)}
             {\left<\sigma^\lambda_{ext}\right>_s(a_1,a_2)}
\label{w}
\end{equation}
and the mean product of albedo and phase function
\begin{equation}
\left< w^\lambda p^\lambda \right>_s(a_1,a_2,\Omega)= 
\int\limits_{a_1}^{a_2}da\ s(a)\ 
  w^\lambda(a,a_1,a_2)\ p^\lambda(\Omega,a),
\label{wp}
\end{equation}
we can express the righthand side ratio for optically thin and single scattering as
\begin{equation}
R_{theo}(a_1,a_2)=
   \frac{\left<\sigma_{ext}^{3.6}\right>_s}
        {\left<\sigma_{ext}^{4.5}\right>_s}\ 
 \frac{
       \int\limits_{4\pi}d\Omega\ \left< w^{3.6} p^{3.6} \right>_s(\Omega)\ I^{3.6}_{in}(\Omega)
- I^{3.6}_{bg}
        }
      {\int\limits_{4\pi}d\Omega\ \left< w^{4.5} p^{4.5} \right>_s(\Omega)\ I^{4.5}_{in}(\Omega)
-I^{4.5}_{bg}} .
\label{Rint}
\end{equation}
This equation contains only the dust properties (cross sections, phase functions, 
and size distributions) and the incident field 
(all-sky map $I_{in}$ and background $I_{bg}$ of the core).

Practically speaking, although some cores in the sample
can be modeled with optically thin RT, we used single-scattering 
RT since the numerical effort is only marginally greater.
The thin RT case is considered here because of its properties for separating dust opacity and structure,
an advantage that is kept to some extent for higher optical depths.

\begin{figure*}
\vbox{
\includegraphics[width=6cm]{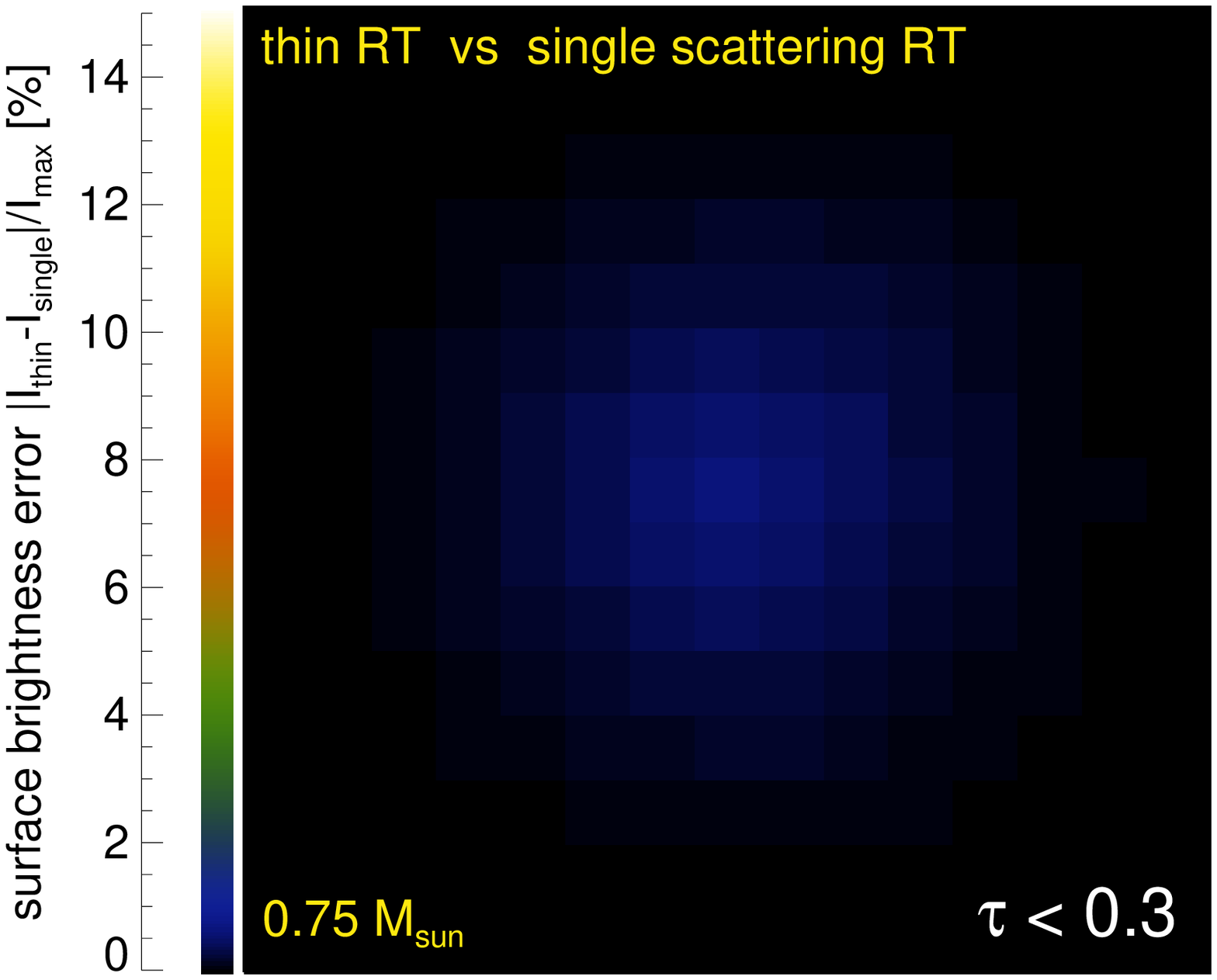}
\includegraphics[width=6cm]{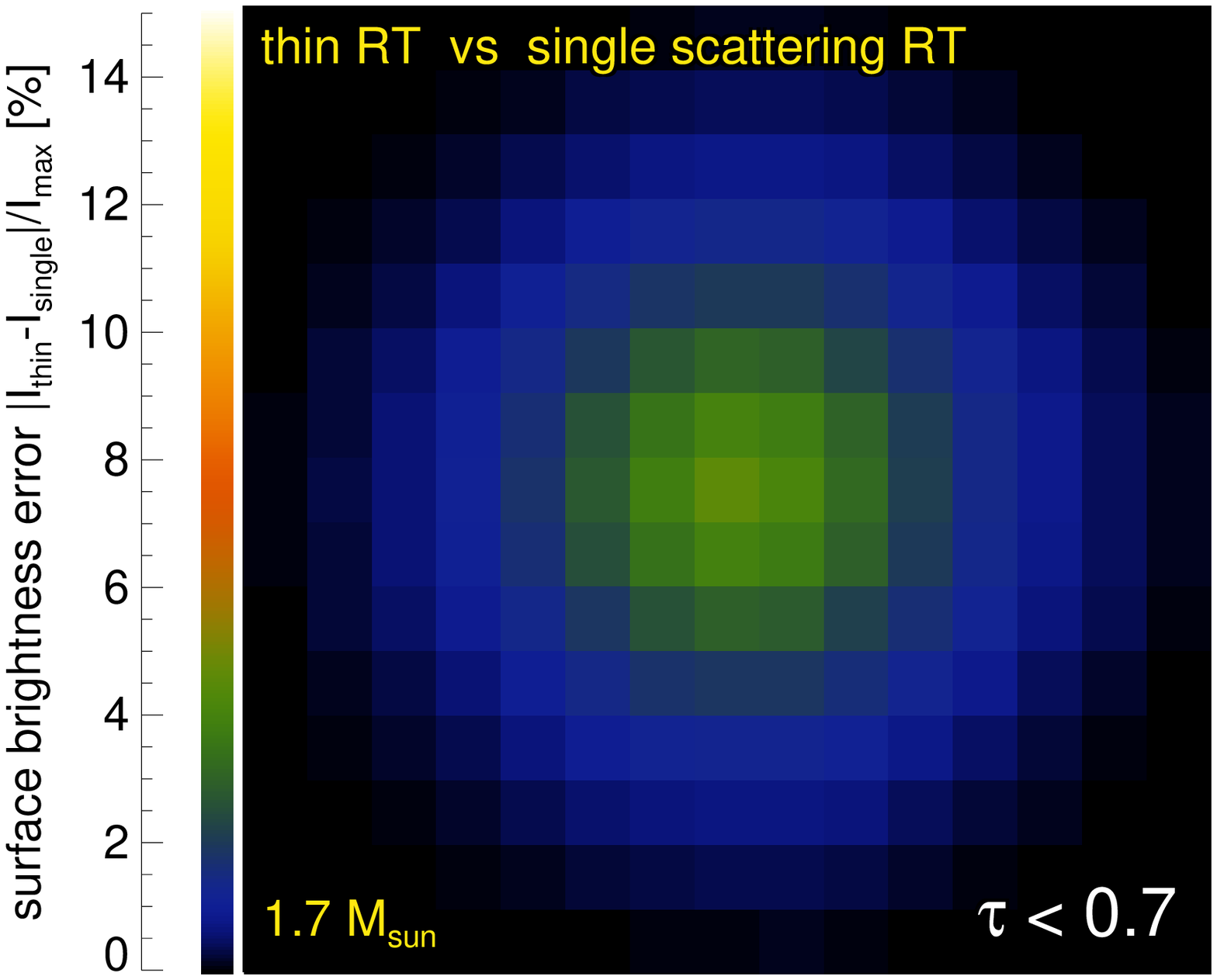}
\includegraphics[width=6cm]{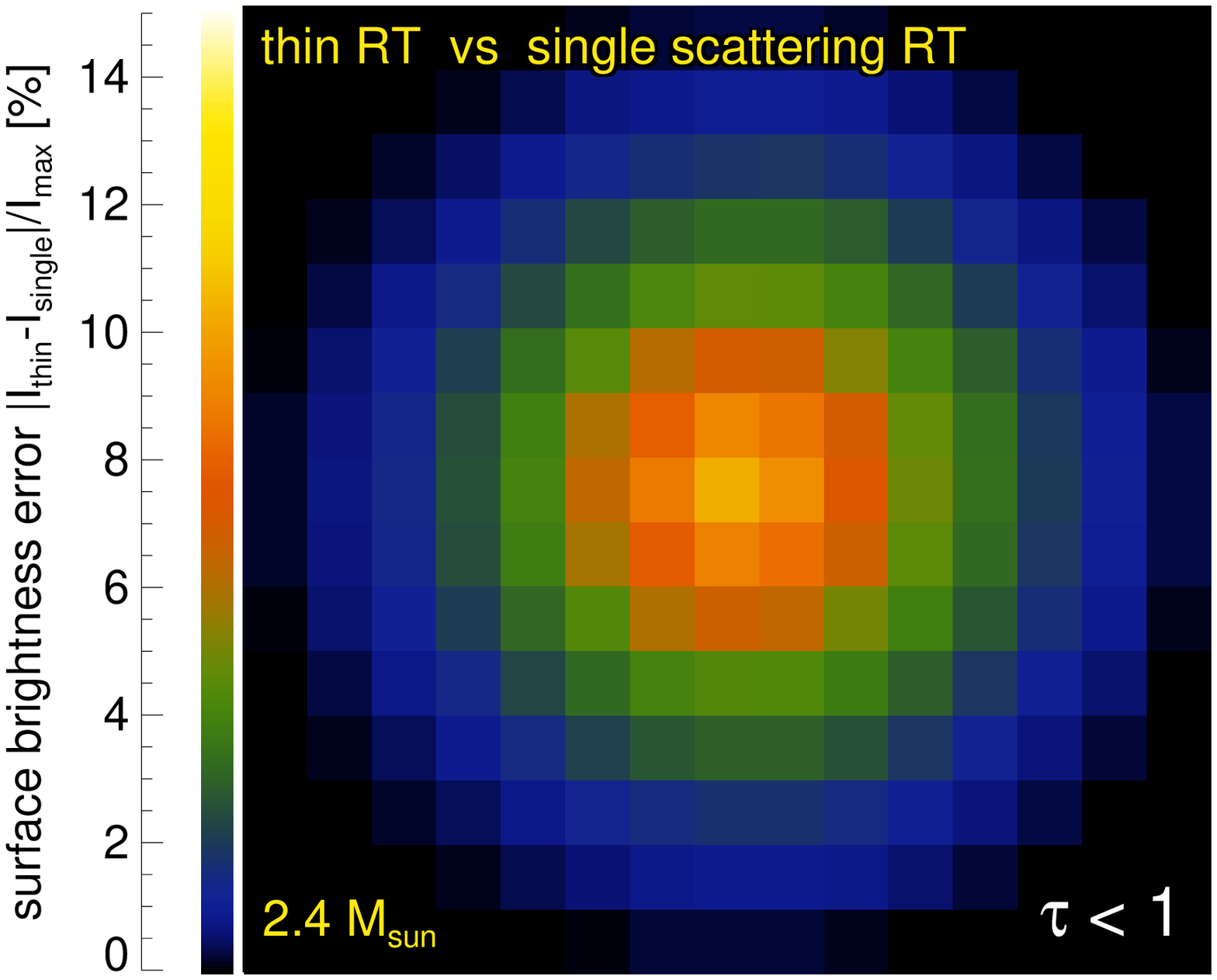}
}
\vskip 0.1cm
\vbox{
\includegraphics[width=6cm]{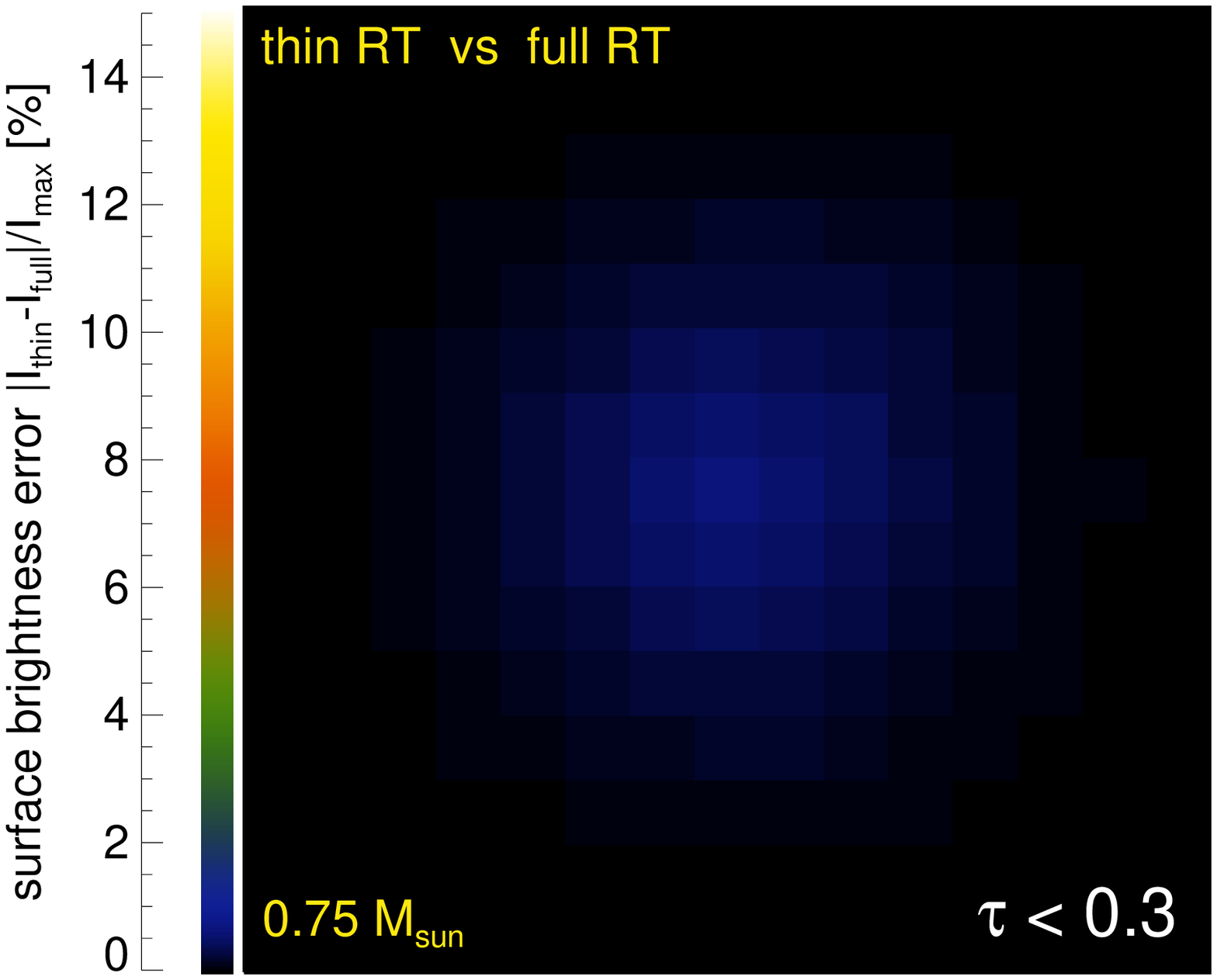}
\includegraphics[width=6cm]{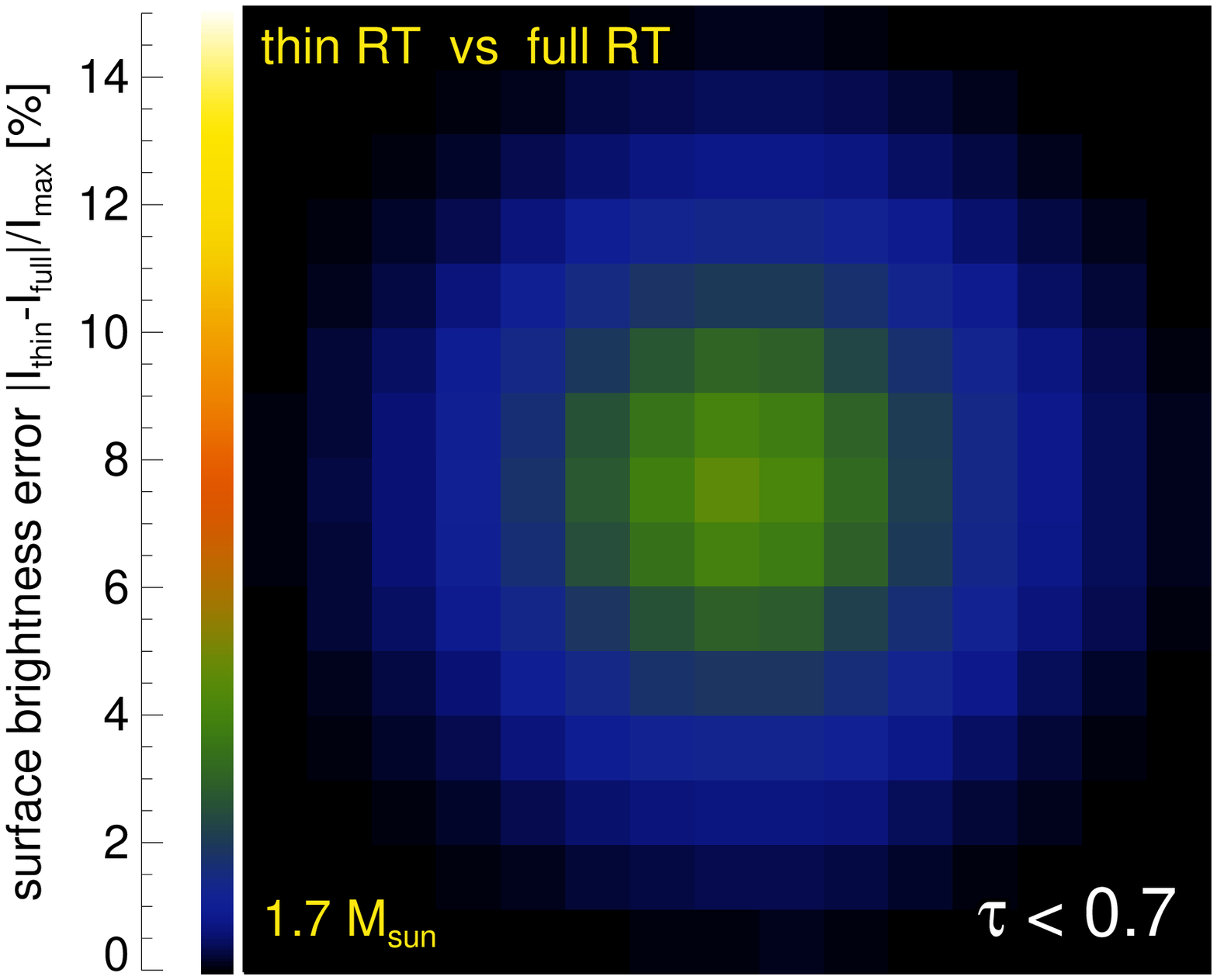}
\includegraphics[width=6cm]{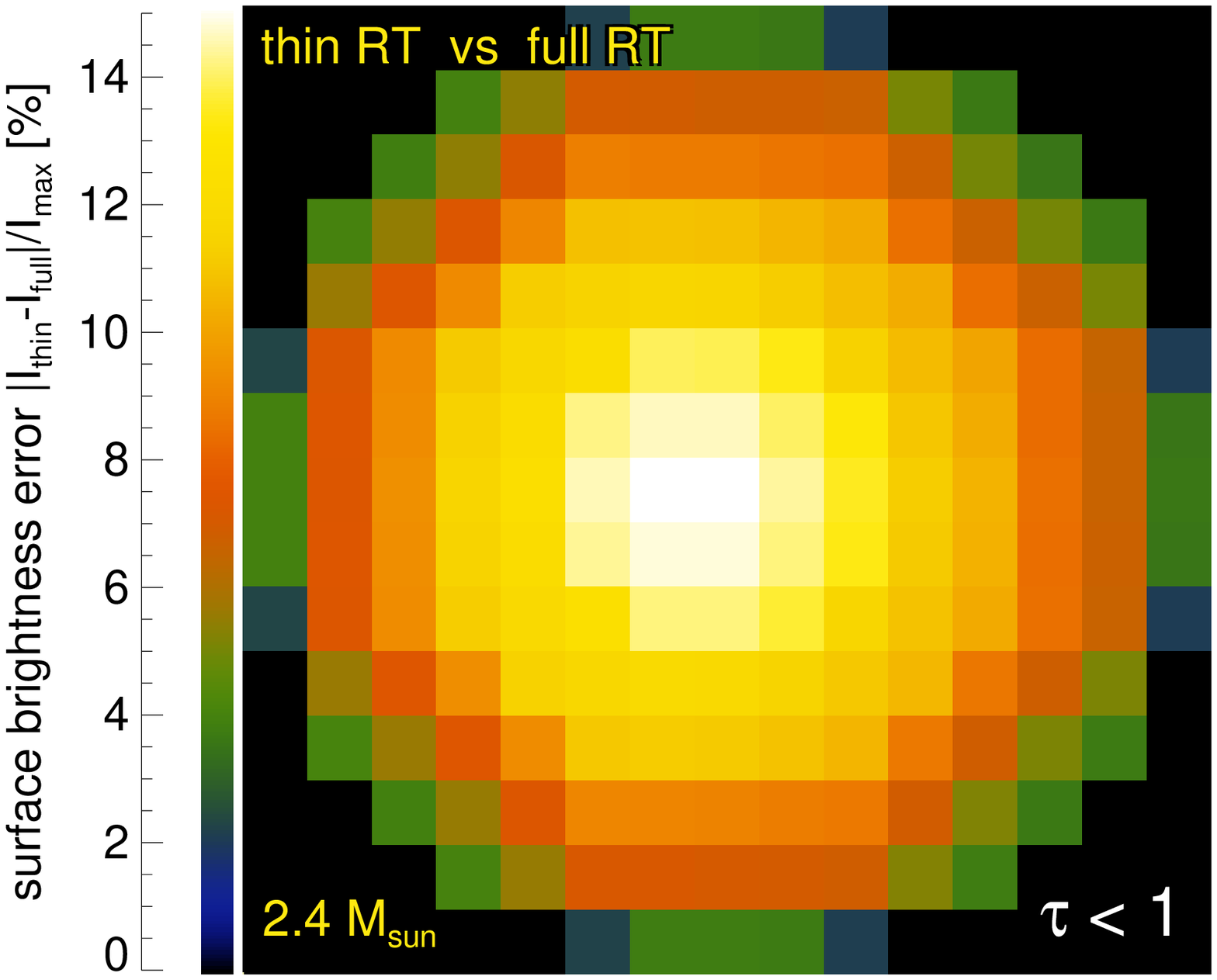}
}
\caption{
Relative error in using the RT approximations.
        }
\label{tauRT}
\end{figure*}

To investigate the errors from using the three approximate RT schemes, we performed RT calculations for model cores with the masses typical of the range of
low-mass cores 0.75,
1.7, and 2.4 M$_\odot$ and chose the core scale so that the optical depth for extinction through
the center is 0.3, 0.7, and 1, respectively.
Figure~\ref{tauRT} compares thin and single-scattering
RT in the upper panel. The images show the SFB error 
of the core at 3.6 $\mu$m $|I_{thin}-I_{single}|/I_{max}$ in percent
with the maximum SFB $I_{max}$ of $I_{full}$.
As expected, the maximum error for $\tau<0.3$ is small (about 2\%).
At $\tau<0.7$, the error is in the 5\% range, for $\tau<1$ it reaches
12 \%.
Comparing thin and full RT, the error up to $\tau<0.7$ is the same since
the extinction effects and multiple scattering can be neglected.
For $\tau<1$, both effects start to modify the resulting SFB,
and the error rises to 15\%. 

\subsection{The "limited optical depth" approach}\label{depth}
To decide which core allows for which RT scheme, we need to specify the optical depth
for the cores in the sample. However, neither the mass and density distribution in the core nor the optical properties of the
dust grains are well known (the targets of investigation in this paper). The ranges
in Table~\ref{table:1} give an indication of uncertainties estimated by authors and are often 
factors of a few.

Coreshine and scattered light from cores in general have a feature that can aid us in 
constraining the optical depth. When the central optical depth increases beyond one, the SFB profiles 
start to flatten and then show a central depression. 
The exact value when the depression is clearly visible depends on the properties of the core,
the incoming field, and the background field, but in our experience, $\tau$=2 is a reasonable 
value to expect the occurrence of depressions in the images.
Figure~\ref{all} contains two prominent cases (L1544 and Rho Oph 9) and a less pronounced one (L1262), but 
most of the cores have a pattern with a central maximum.
In the modeling, we can use this observational evidence to consider only models with
a central optical depth
below a few when no central depression is visible at the considered wavelength.
\citet{2013A&A...559A..60A} used this argument to exclude grains with sizes beyond 1.5 $\mu$m since with
the availability of NIR data, it is possible to find the wavelengths where depression sets in.

We therefore chose to apply a method which we call "limited optical depth" approach.
It adjusts density and opacities in a way that they
do not exceed a given $\tau$ limit. Fixing the density \citep[as done, e.g., in][]{2014A&A...572A..20L} and then
considering various dust models, $\tau$ may rise above a few and produce a central depression while the data show a 
maximum in SFB near the projected core center. 
Especially, models with steep density profiles and small kink radii (below which the profiles are assumed to be flat) accumulate most mass in the center leading to high central column densities.
Furthermore, central column densities are usually derived from thermal emission measurements in the
FIR/mm wavelength range based on a specific dust model, 
using the column densities with another dust model thus might be incorrect.
The limited optical depth allows us to avoid this specification
and the dust properties are consistently used only in the modeling of the coreshine.

In Table~\ref{table:1} we summarize the final maximum optical depth at 3.6 $\mu$m (2.2 $\mu$m in the case of
L260) that we have used for the core modeling. Cores without central depression have values below 2.
For the cores with $\tau_{max}>2$, we chose values that scale approximately with the average column density value 
given in the literature. 
The modeling is performed within the ranges of the mass and optical depth given
in Table~\ref{table:1}.

\subsection{Properties of the theoretical surface brightness ratio $R$}
\begin{figure}[!ht]
\includegraphics[width=9cm]{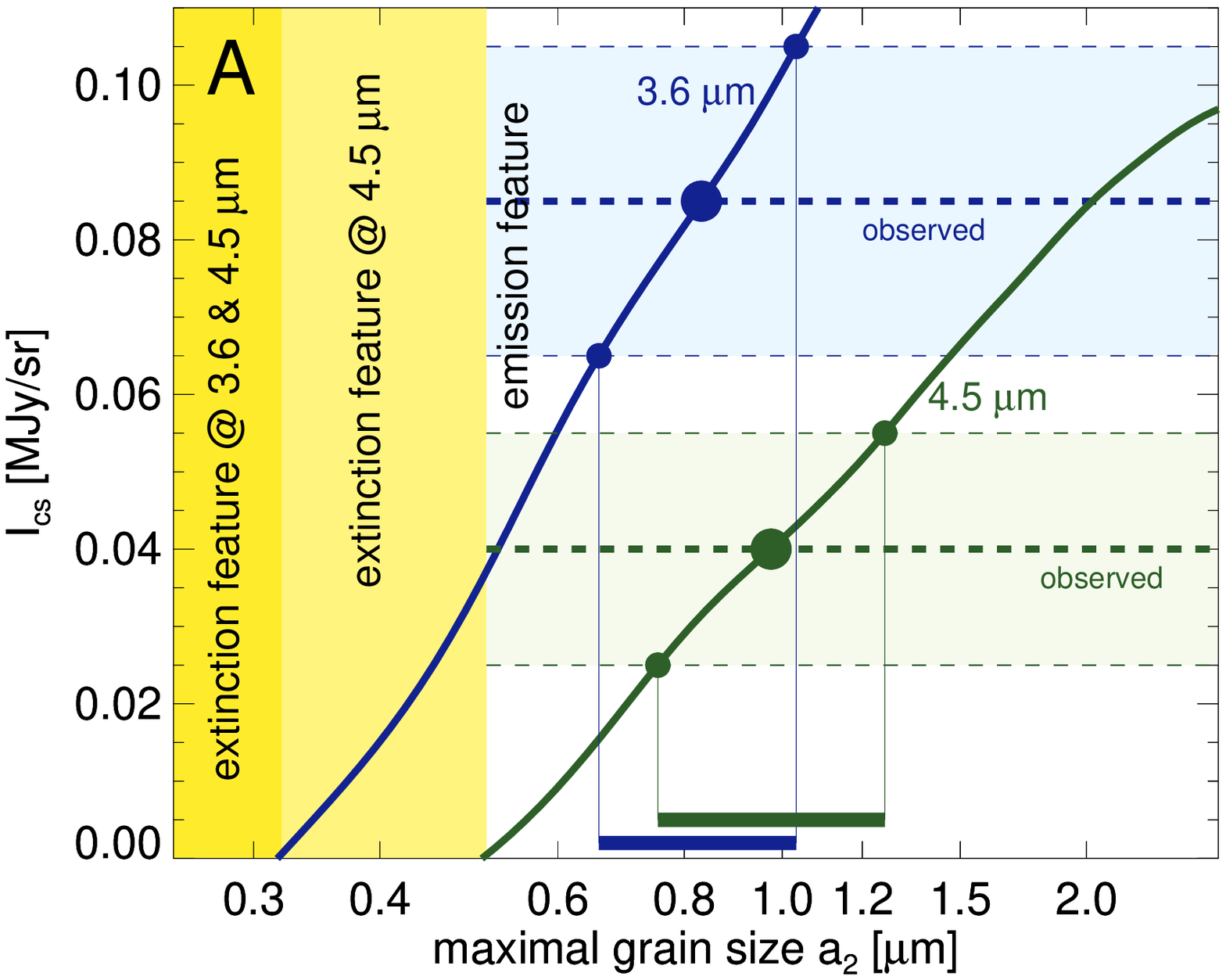}
\includegraphics[width=9cm]{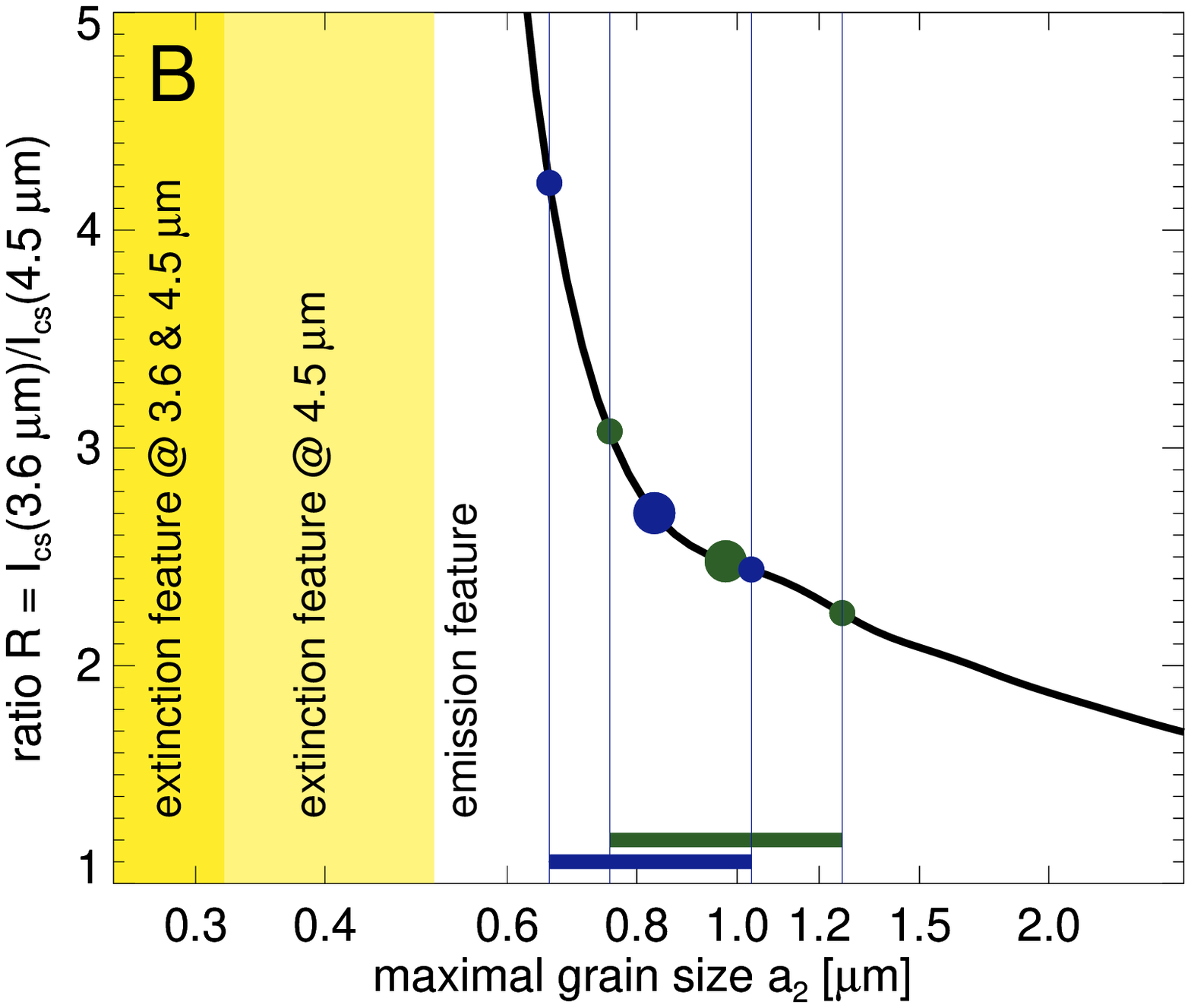}
\caption{
Panel A:
Background-subtracted theoretical surface brightness through the core center
at 3.6 (4.5) $\mu$m in blue (green),
respectively (see Eq.~\ref{singleeq}) as a function of the maximum grain
size of the dust size distribution for the core L260.
The roots of both functions indicate the regions where the core becomes visible
in excess of emission for each wavelength (region outside yellow regions).
The dashed lines give the observed maximum coreshine value with its error range, 
and the dots label the crossing points with the theoretical curve. 
Panel B:
Resulting theoretical ratio $R_{theo}$ of the surface brightnesses shown in Panel A
as a function of the maximum grain
size. The dots from Panel A mark the agreement of the central surface brightnesses with the
observed range for each wavelength. 
        }
\label{Ra2}
\end{figure}

In this section, we discuss how the cores can
appear in the two bands, which basic physical properties are responsible
for the appearance, and what the resulting range of theoretical 
SFB ratios is. 
Since the largest grains in the size distribution scatter the radiation more
efficiently,
our goal is to explore the maximum grain sizes when comparing the observed $R$ distribution 
across the core with the theoretical $R$ of a particular dust model.

Besides scattering the incident radiation toward the observer, the 
grains are also responsible for the extinction of background radiation along the LoS.
This background extinction introduces several complications.
The scattered light has to overcome the extinction of the background radiation
by the core to arrive at a SFB above the background and to 
appear in excess of emission.
The four possibilities are to see the core in extinction in both bands,
in just one of them, or in emission in both bands. 
Since for our core sample we have analyzed only parts of the cores with 
excess of emission in both bands, 
we concentrate on the last case. 

To keep the example simple, we consider the condition for excess of emission in the optically thin case.
From (\ref{Rint}), we find 
\begin{equation}
 I^{\lambda}_{bg} 
 < 
 \int\limits_{4\pi}
 d\Omega\ 
 \left< w^{\lambda} p^{\lambda} \right>_s(\Omega)\ 
 I^{\lambda}_{in}(\Omega)
\label{Bothemission}
\end{equation}
for both bands.
The most prominent effect on $R$ is the pole causing a singularity when the 4.5 $\mu$m denominator vanishes. Physically, this happens when 
the scattered light SFB is equal to the 
extincted background SFB. Increasing the maximum grain size,
the SFB in the two bands increases due to the higher scattering efficiency of larger grains.
In the picture of Rayleigh scattering, the condition $2\pi a \ll \lambda$ is less fulfilled for 
larger grains. The 4.5 $\mu$m band SFB benefits stronger from the increased maximum size due to
this condition and $R$ decreases.

Figure~\ref{Ra2} shows in Panel A the numerator and denominator of $R$
 for
the example of L260. 
We have used ice-coated silicate grains and an MRN-type size
distribution ranging from 0.01 $\mu$m to $a_2$. 
We follow the approach described in \citet{2013A&A...559A..60A} to calculate the incoming fields 
$I_{in}$ from zodiacal-subtracted 
DIRBE\footnote{Diffuse Infrared Background Experiment, see www.lambda.gsfc.nasa.gov/product/cobe/dirbe\_overview.cfm} maps.
The calculation is done in single-scattering approximation.

The left yellow regions indicate where the core is seen in extinction in both bands or just the 4.5
$\mu$m band. The observed coreshine SFB defines the limit of this region
when becoming zero.
The thick dashed lines give the maximum observed coreshine SFB, 
and we also show the uncertainty in the value due to the choice of background (see
Appendix \ref{off}) by thin dashed lines forming the light blue and green error ranges.
To indicate the resulting range of maximum grain sizes, we label the crossing points with the observed
values and their error range derived from the off-measurement uncertainty by circles. 
Bars near the x-axis aid in reading the $a_2$-range
corresponding to the agreement of observed and theoretical coreshine.
As is visible in the figure, the bars partially overlap, indicating that by extending 
the distribution up to sizes around 0.84 $\mu$m, the model reproduces the observed
data.

\citet{2013A&A...559A..60A}
have modeled the 2.2, 3.6, and 8 $\mu$m SFB along a cut through L260, and
found maximum grain sizes around 1 $\mu$m, which is close to what we find for our
ice-coated silicate and carbonaceous grains.
Their calculations required considering an enhanced incident DIRBE field boosted by a factor of 1.7, which might have been due to an error in the older RT code and is no longer needed in the new code version.

In Panel B, we show the ratio of the blue and green curves from panel A, which is
the theoretical band ratio $R$ with the same notations, and the bullets
indicate the crossing points of the model with the observed maximum coreshine SFBs. 
We show just the bullets and not the dashed lines from Panel A to be able to plot more than one R curve in the following figures without confusing the plots.
As expected, the curves show the increase
due to the pole in $R$ when the 4.5 $\mu$m SFB cannot overcome the background
extinction, and the core turns to an extinction appearance at smaller maximum 
sizes.

In the optically thin case, the curve in Panel B should be independent of assumptions
about the spatial density distribution within the core, while the SFB profiles
in Panel A assume a central column density or optical depth and therefore rely on spatial parameters.
For optical depths higher than unity both panels will be affected by the assumptions made
within the density distribution model.

We note that $R$ also depends on the position of the core
since the ISRF is anisotropic with a maximum at the GC, and
convolving it with the anisotropic phase function will lead to different
values for different core positions.
As a technical note we add that when referring to the theoretical quantities like the optical depth, 
we skip the $< >_s$ notation, indicating the integral over grain size to improve readability from
this point on.
\section{Coreshine modeling of the core sample} 
\begin{figure*}[!ht]
\vbox{
\includegraphics[width=6cm]{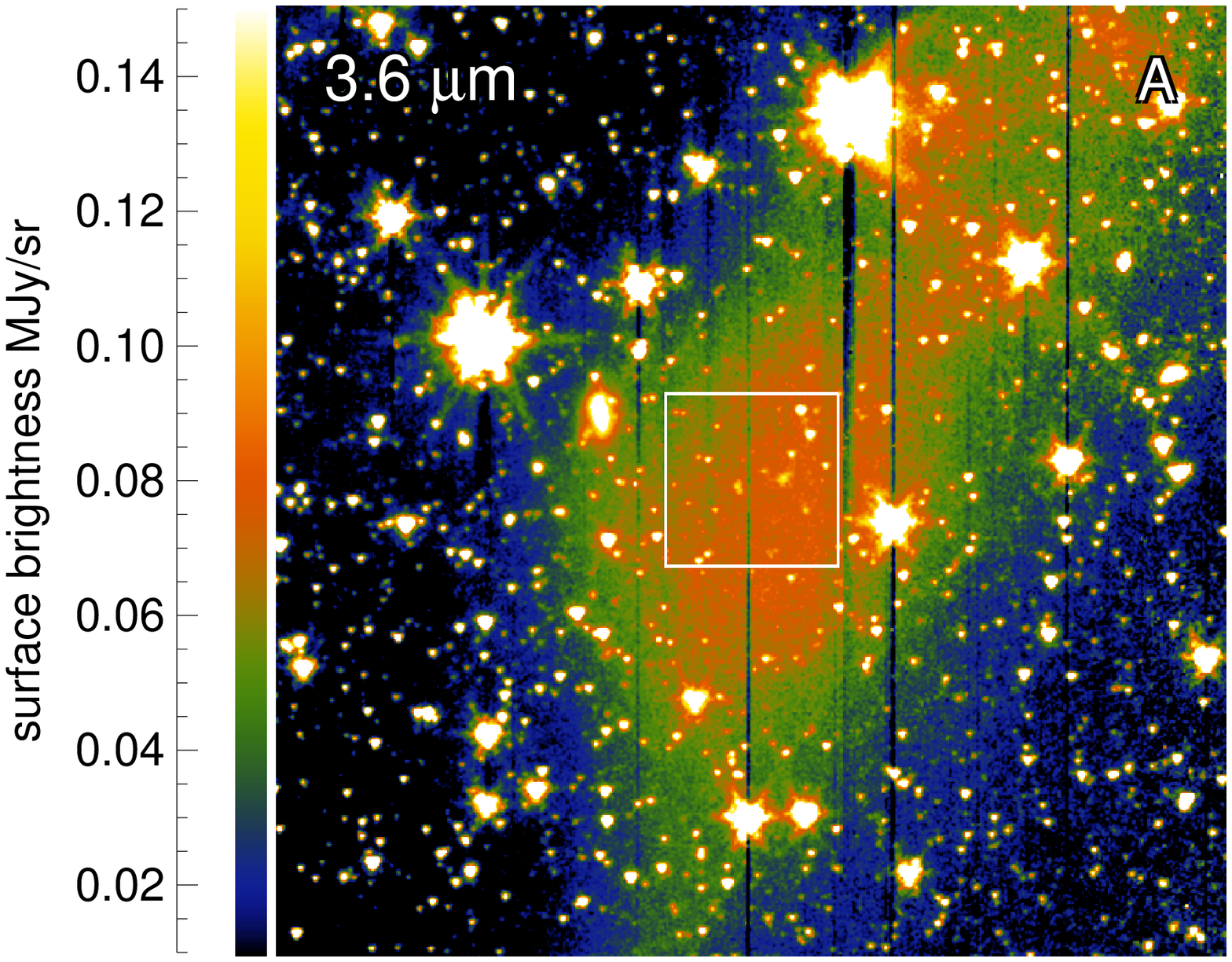}
\includegraphics[width=6cm]{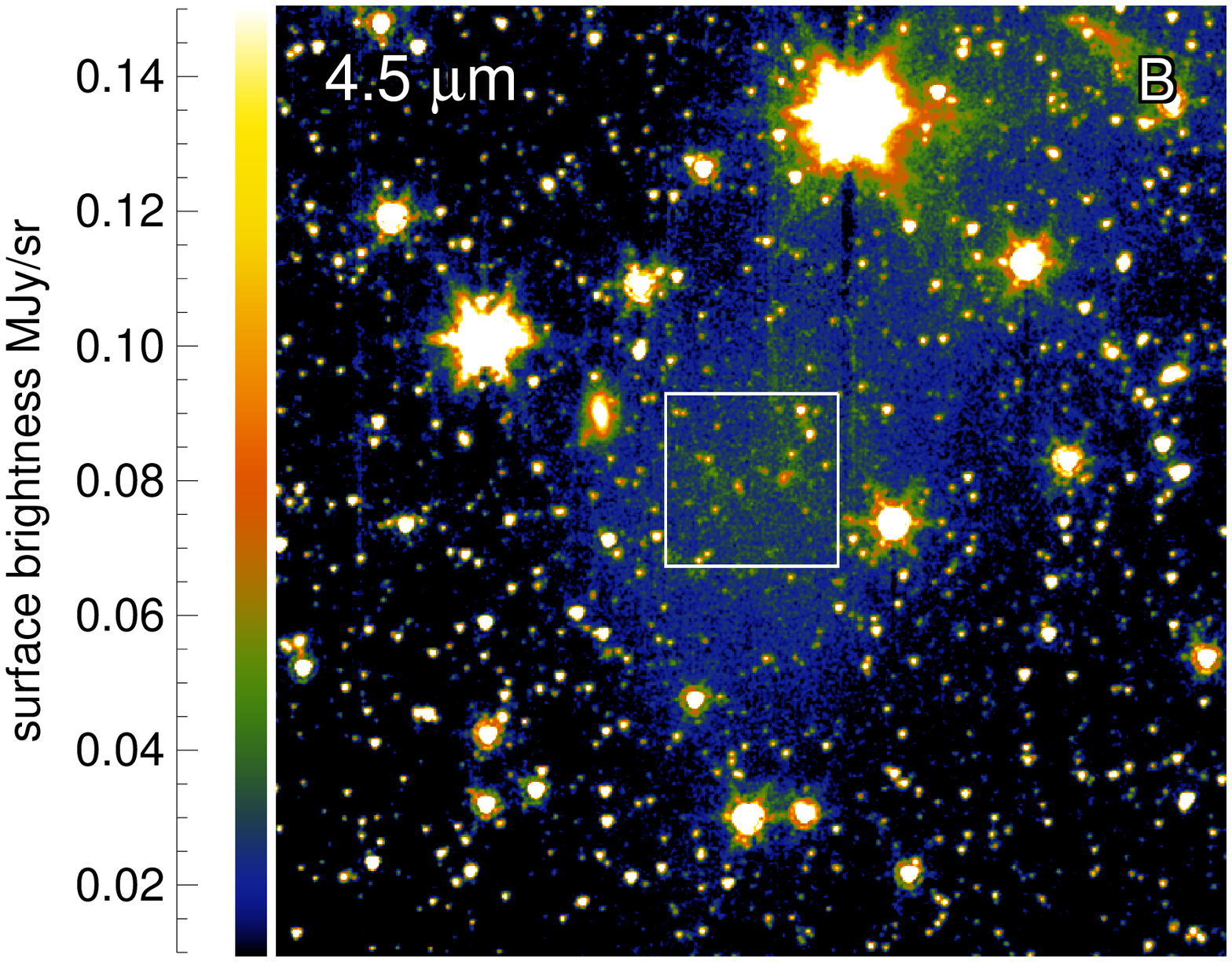}
\includegraphics[width=6cm]{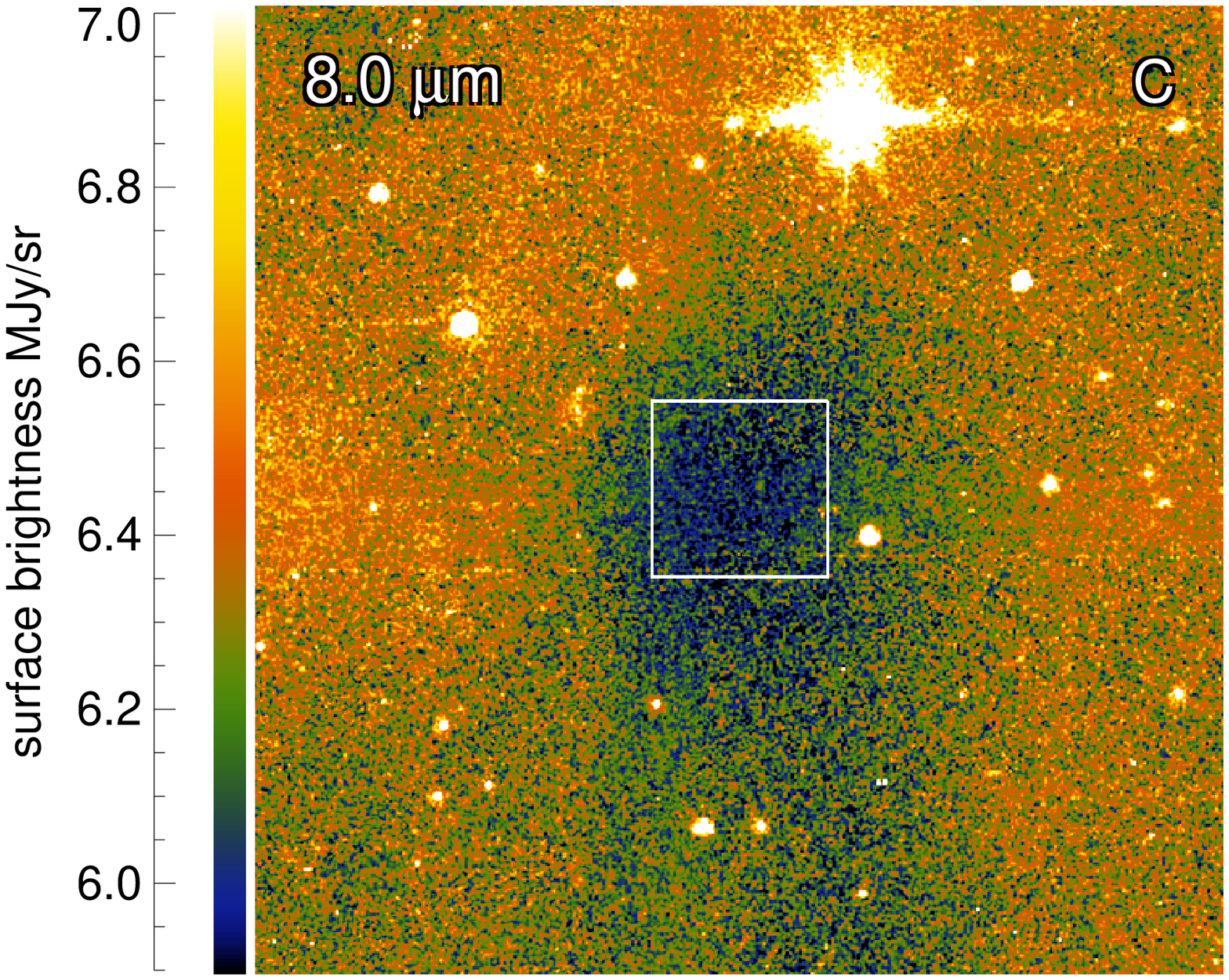}
}
\vskip 0.1cm
\vbox{
\includegraphics[width=6cm]{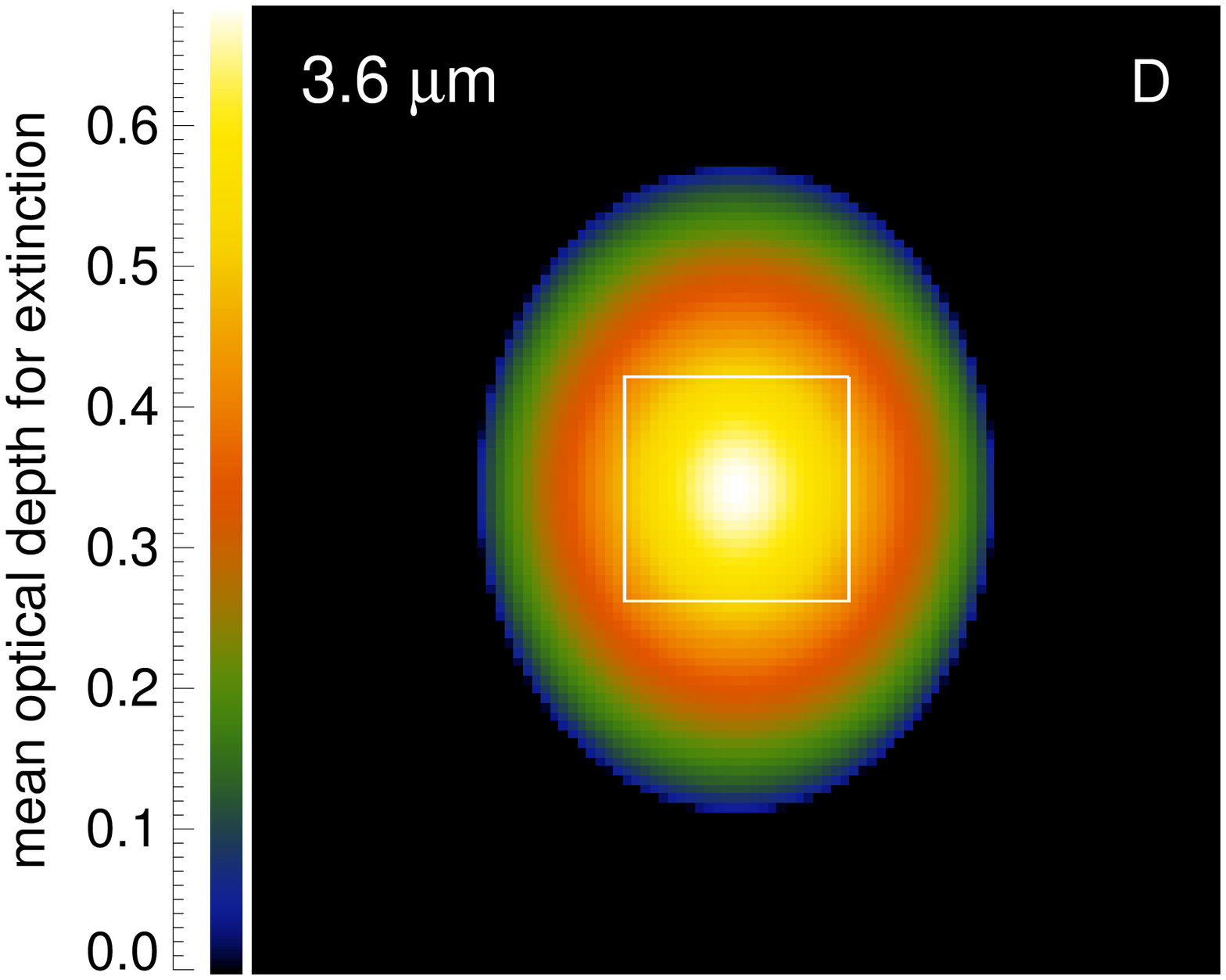}
\includegraphics[width=6cm]{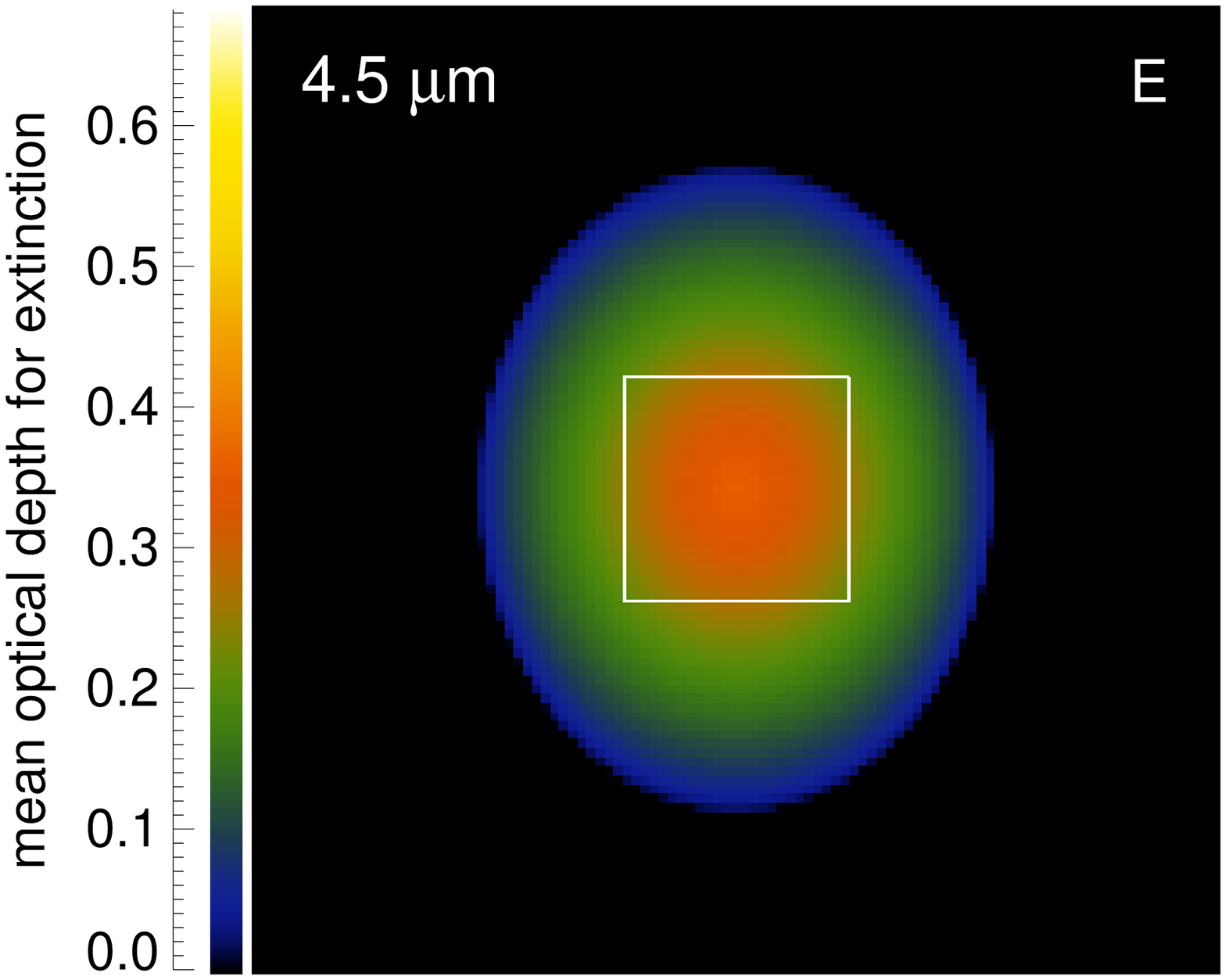}
\includegraphics[width=6cm]{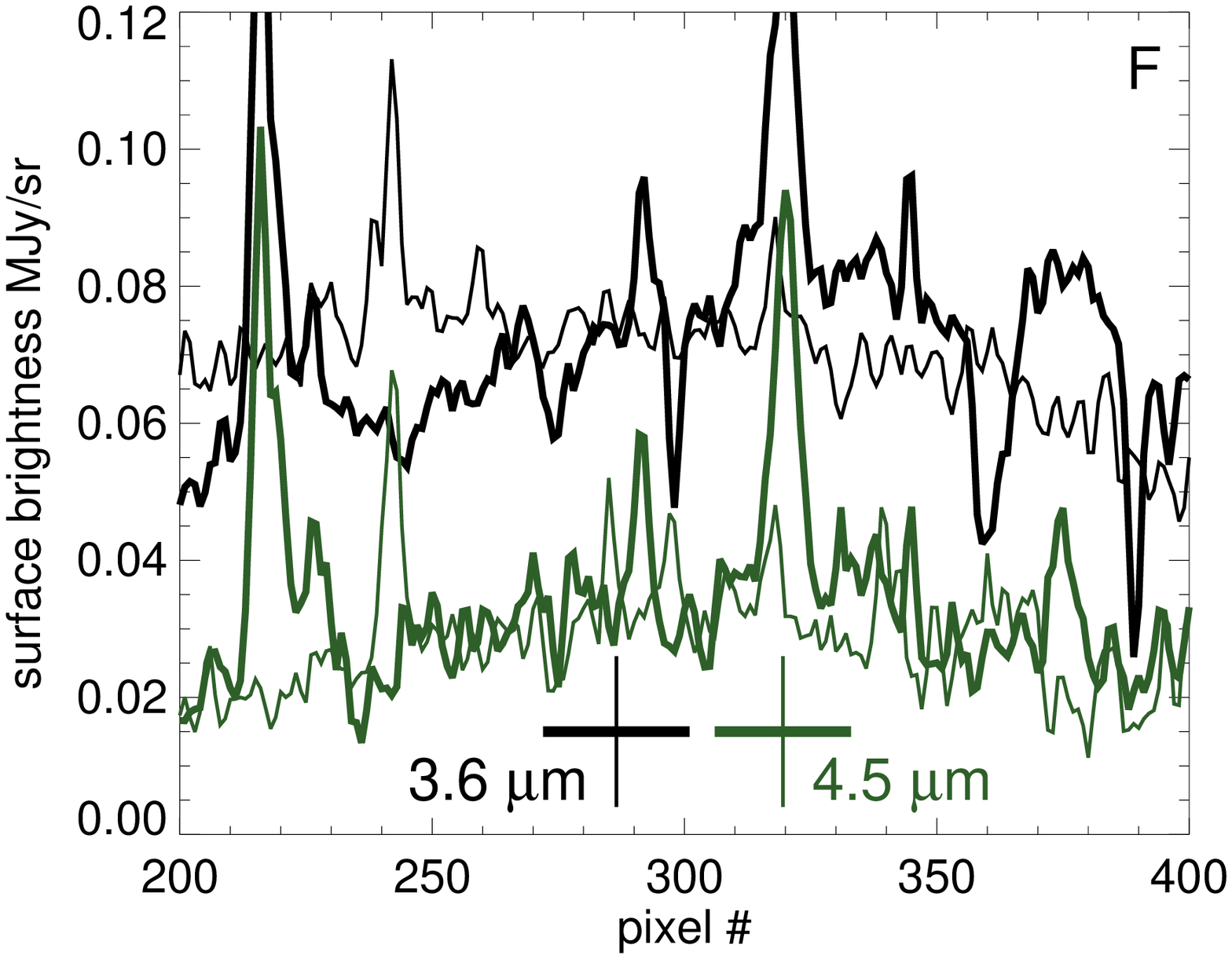}
}
\vskip 0.1cm
\vbox{
\includegraphics[width=6cm]{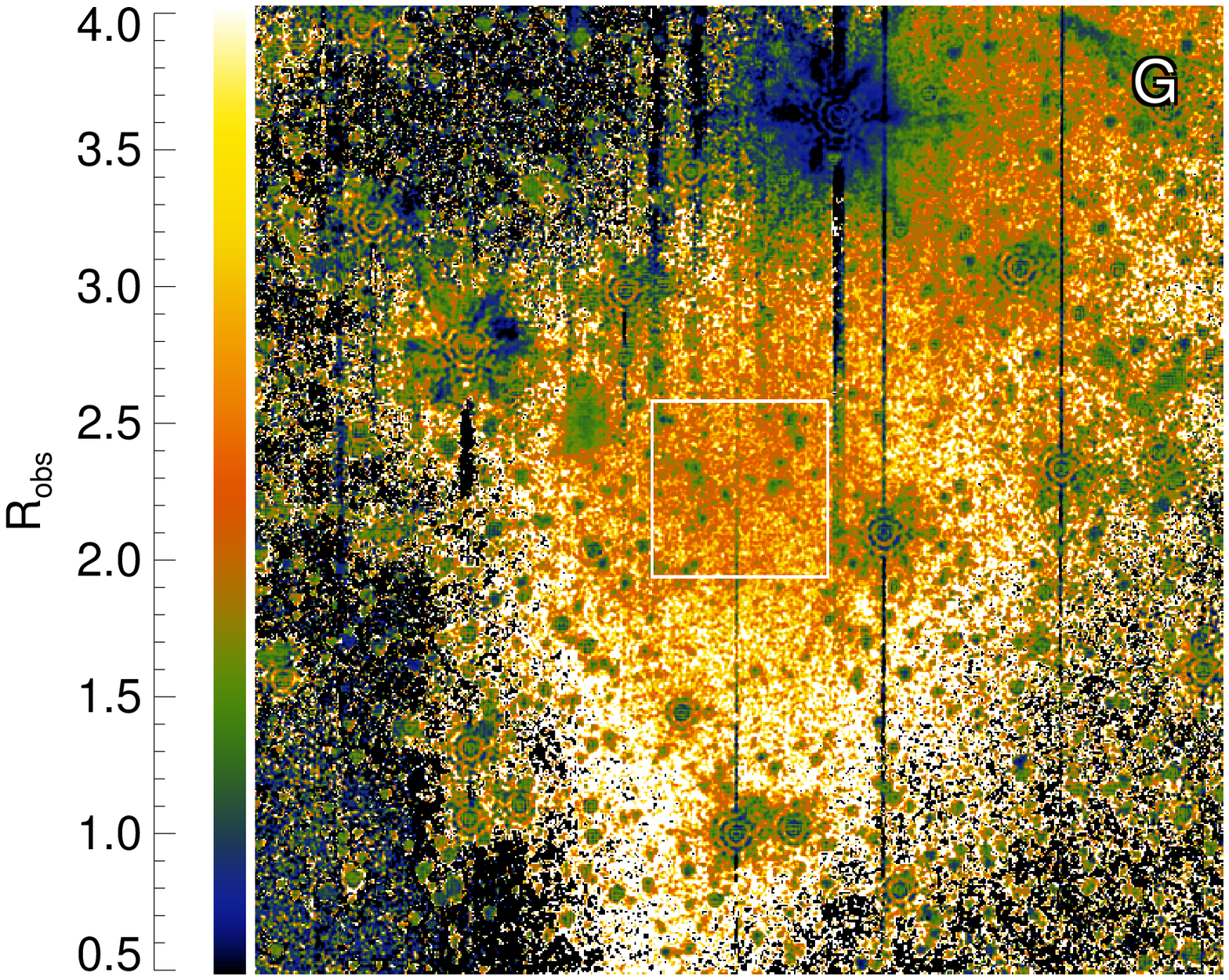}
\includegraphics[width=6cm]{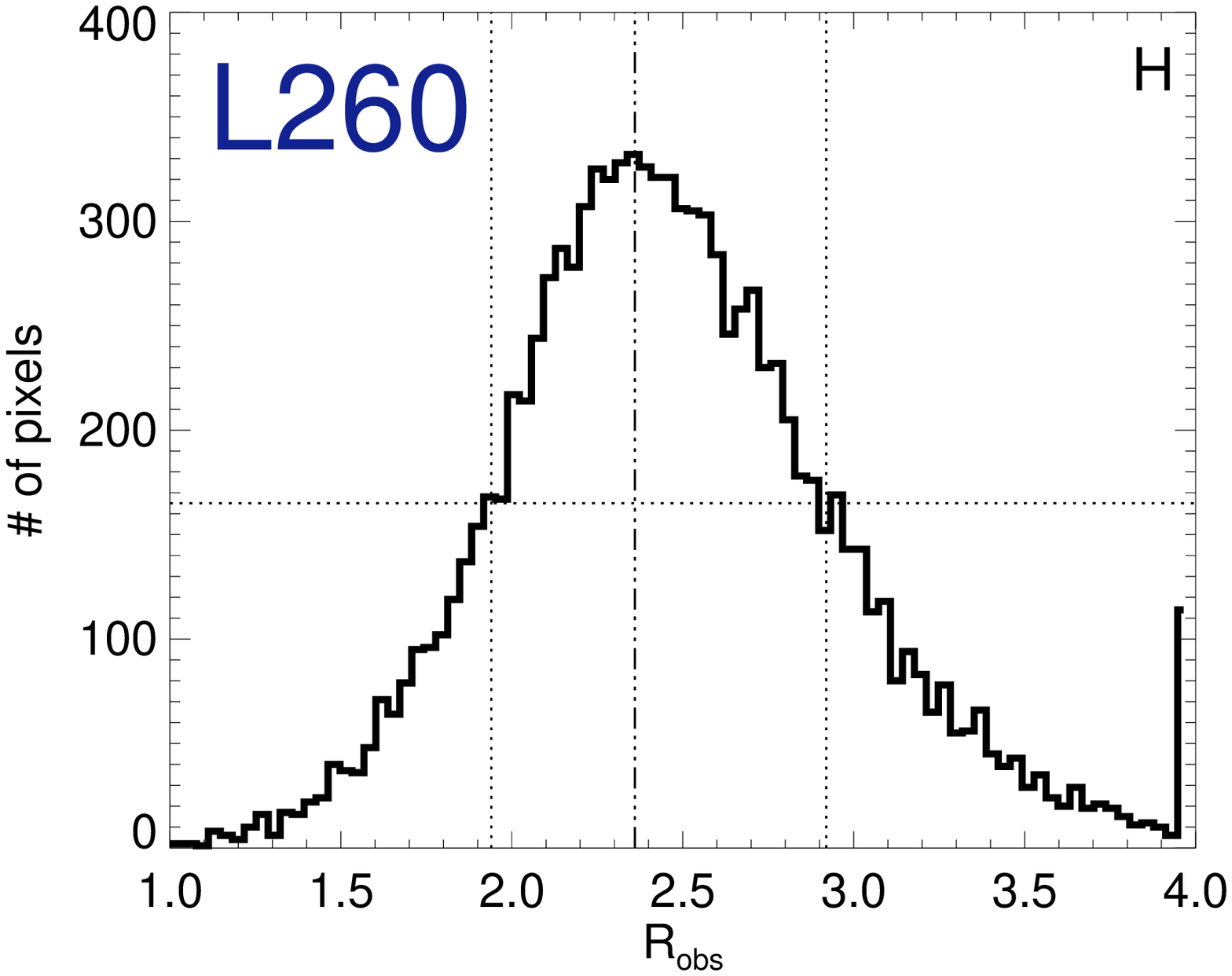}
\includegraphics[width=6cm]{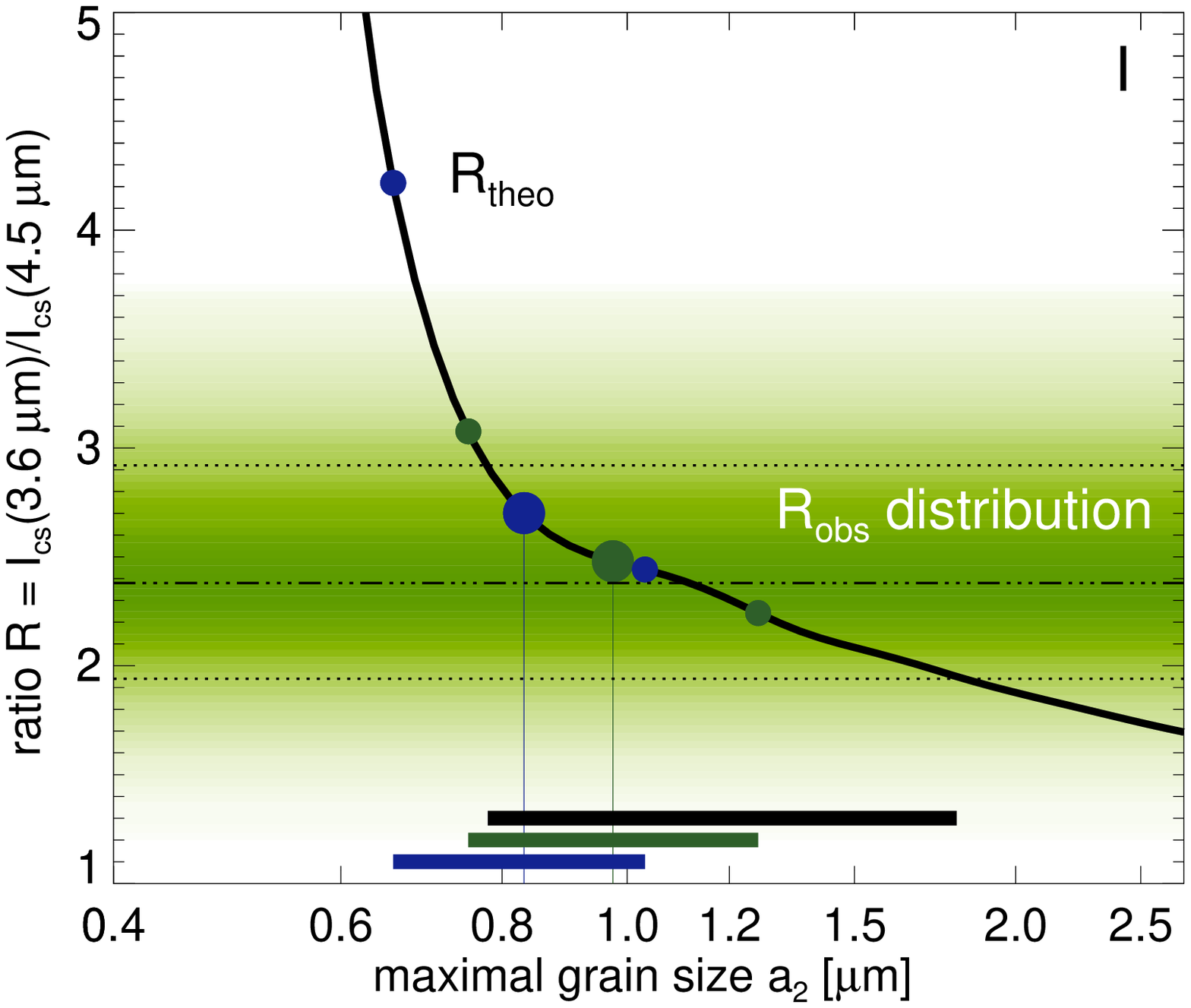}
}
\caption{
Data analysis for L260.
Warm {\it Spitzer} IRAC surface brightness maps (off-subtracted)
at 3.6 and 4.5 $\mu$m (Panels A,B)
and cold {\it Spitzer} map at 8 $\mu$m (Panel C) in a 5.5$^\prime\times$5.5$^\prime$ region around the core.
The white square indicates the 1.1$^\prime\times$1.1$^\prime$-region
where surface brightness ratios have been measured.
The optical depth for extinction of external radiation passing through the core and
to the observer (directionally averaged) for a core with the overall properties of L260 is shown in Panels D and E (3.6 and 4.5 $\mu$m).
Panel F: Horizontal (thick) and vertical (thin) cuts through the central 200 pixel in the two bands (black
for 3.6 $\mu$m and green for 4.5 $\mu$m).
Panel G: 
Map of the ratio of the off-subtracted surface brightnesses $R$ at 3.6 and 4.5 $\mu$m.
Panel H shows the $R$ pixel number distribution within the white square (pixels affected by point sources or by the column-pulldown effect have been masked).
Panel I: $R_{theo}$ as a function of the maximum grain size $a_2$ (thick black), 
with the notation from Fig.~\ref{Ra2}.
The observational $R$-distribution from Panel H is plotted and color-coded, and both the maximum value 
and the two half-maximum values are plotted as dashed lines. 
The horizontal black bar indicates the $a_2$ agreement range of $R_{theo}$ and $R_{obs}$.
       }
\label{L260}
\end{figure*}


The general approach taken in this section is to compare
the measured $R$ distribution in a region
of the core with the theoretical ratio $R_{theo}$ derived from a model core.
The core properties and the maximum grain size are varied without exceeding the limiting
optical depth (Sect.~\ref{depth}). 
Since two sets of 3.6 and 4.5 $\mu$m images can 
have the same ratio map but might differ strongly at each wavelength, we also
verify the agreement of the SFBs. 
For cores with $\tau<2,$ it is sufficient to consider a single SFB value.
We have chosen the maximum observed SFB and not some average for two
reasons. First averaging considers more regions with worse S/N, increasing the overall error, 
and second the averaging would bring in more assumptions about the way the averaging is done.
Even if the core has regions with $\tau>2$, we try to select a region in
the observed and model image that has optical depth below that limit. In this way we take advantage of the
optically thin approximation being valid, which removes spatial dependences from the problem.
For cores with $\tau>2$, we perform "full RT" modeling by calculating images, building ratio maps
and an $R$ distribution,
and fitting the observed $R$ distribution and the central column density within the optical depth limit.

Throughout the section, we make use of the approximate core properties summarized in 
Table~\ref{table:1}.

\subsection{Modeling details explained: the core L260}\label{sectionL260}
To describe the modeling approach in detail, we have chosen the core with the
brightest coreshine in the sample: L260. 
According to 
\citet{2014A&A...563A.106S}, the core is located in a region favorable to coreshine detection because
it is near the GC in the PoSky, but the latitude is high enough to prevent
a strong background. Still it has the second strongest 3.6 $\mu$m background of
the cores modeled in this paper.

Coreshine is a weak extended feature, and especially the 4.5 $\mu$m 
SFB is only factors 2-8 larger than the sum of instrumental and point source noises. 
Correspondingly, we have selected special regions in the images to 
minimize the impact of noise and stellar point spread function (PSF) on the analysis.
The two background SFBs $I^{3.6}_{bg}$ and $I^{4.5}_{bg}$ 
entering $R_{theo}$ are taken from the DIRBE maps by subtracting the 
stellar flux measured by WISE\footnote{www.nasa.gov/mission\_pages/WISE/}. 
We did not subtract a diffuse foreground component of the radiation field used as
background of the core since the current models
give only indications \citep[for a discussion see][]{2014A&A...572A..20L}. We 
briefly address its effect in Sect.~\ref{discussion}.
For 3.6 $\mu$m, the bands of IRAC and DIRBE are comparable, for 4.5
$\mu$m we interpolate from the other bands \citep[see also][]{2013A&A...559A..60A}.
The sum of the background and foreground contributions can be determined from the 
SFB near but off the core ("off" characterizes quantities derived in this way)
\begin{equation}
I^\lambda_{off}(y,z)= I^\lambda_{bg}(y,z) + I^\lambda_{fg}(y,z).
\label{off}
\end{equation}
The details of how this subtraction is carried out are explained in Appendix \ref{off}.
We note that the subtraction removes the possible zodiacal light contribution.

Figure~\ref{L260} shows the steps in the analysis for the core L260.
The off-subtracted SFB maps at 3.6 and 4.5 $\mu$m 
in a 5.5$^\prime\times$5.5$^\prime$ region around the core
are shown in the Panels A and B, respectively. 
The cold {\it Spitzer} 8 $\mu$m map is given in Panel C
to indicate the location and the shape of the core ("Coreplanets\_cores": Program ID 139, PI N. Evans).
L260 is seen in emission in 3.6 and 4.5 $\mu$m. 
The white square indicates the region where SFB ratios are measured. 
Since the core shows no sign of central depression in the two bands, the square is
placed in the central region of extinction in the 8 $\mu$m map.
The size and precise position of the square have been adopted to avoid
the impact from nearby stars and their PSF, to minimize the impact from
column pull-down in the image, and to recover the grain information in the part of the core
where the coreshine signal-to-noise is maximum in both bands. 
Alternative shapes of the measuring area would have been possible, but an investigation
of the ratio maps (Panel G, see farther down) showed that the results are not sensitive 
to the shape of the area.
All remaining pixels that
show impact from a stellar PSF or column pull-down are masked and not modeled.

To check the distribution of optical depths for extinction in the region where we measure $R,$
we show the direction--averaged optical
depth map of a core with the overall properties of L260 in Panels D and E.
The direction average takes into account that radiation is extincted while reaching the LoS from outside and on its way to the observer.
As simple core model, we use an ellipsoidal core with two of the three axes identical and 
a radial density profile that flattens in the inner part while following a power law with index -1.5 outside. 
Core mass and outer radius are taken from Table \ref{table:1}, and the kink radius is assumed
to be one-third of the outer radius. We note that both the density power-law index and the kink
radius will depend on the evolutionary stage of the core, but more evolved cores might have a
smaller kink radius, for example, and for L1544 we have used a different radial profile.
For L260, the mean optical depth values reach about 0.4 for 3.6 $\mu$m.
We there fore use single-scattering RT to model the core.
We note that the models shown in Panels D and E for the core sample do not aim to 
model the spatial density distributions but only serve to estimate the optical depth variation, with the exception 
of the cores with maximum optical depth above 2 where we create images in full RT based on the shown spatial model.

To read approximate values of the SFB band ratio, the noise level and
the contamination by stars,
we show in Panel F horizontal (thick) and vertical (thin) cuts through the center of the
white line frame for 3.6 (4.5) $\mu$m with black (green) lines.
While there are local gradients along the cuts that vary the surface brightness ratio,
the figure already allows reading an approximate ratio of about 2 to 4.
The cuts also indicate the level of stellar radiation impact and the overall noise of of about 0.005 MJy/sr.
Figure~\ref{Ra2} shows the two maximum surface brightness values and their error bars as dashed vertical lines. The error bar contains not only the error read from Panels A, B, and F in 
Fig.~\ref{L260}, but also the error in the off-measured surface brightness that is subtracted from the
measured surface brightness.
The entire map of off-subtracted surface brightness ratios $R$ is
shown in panel G. 
Most variations in the white square are due to stellar sources. 
We masked all pixels affected by
stellar PSFs and by column pull-down.
In panel H, we show the $R$ pixel number distribution across all
unmasked pixels in the white square. 
It has a maximum at $R=$2.37, and as a measure of its size, we use the approximate width of the
distribution at the half maximum value indicated by thin vertical lines (the full width at half maximum, FWHM, is about 1). 

In Panel I, the range of observational ratios $R_{obs}$
is compared to the theoretical ratio $R_{theo}$ as a function of $a_2$ as derived
in Sect.~\ref{sect3}, with the background values given in Table~\ref{table:2} and using the
core sky position for the phase function integration of the scattering integral.
The limited optical depth approach is applied, which prevents the maximum optical depth for
extinction to exceed 1 at $\lambda=$ 2.2 $\mu$m (see Table \ref{table:1}).
The green color-coding indicates the observed $R$-distribution from Panel H.

The crossing points of the theoretical $R$ curve with the observed coreshine range at 3.6 $\mu$m (blue)
and 4.5 $\mu$m (green) 
are shown as dots as in Fig.~\ref{Ra2} with small dots indicating the error range and large dots
giving the crossing point with the exact value.
There is an overlap of the $a_2$ ranges of the two bands indicated as blue and green
horizontal bars at the bottom of the figure and the black bar marking the $a_2$-range where $R_{obs}$ and $R_{theo}$ agree.
Thus the model is able to meet all three observational constraints for maximum grain sizes
between 0.78 and 1.02 $\mu$m.

We have performed extensive tests on varying the spectral slope of the
size distribution as suggested by
\citet{2001ApJ...548..296W}.
The lower grain size limit has no effect on the coreshine since the scattering is dominated 
by the largest grains as shown in \citet{2010A&A...511A...9S}.
The effect of the size distribution slope on the derived grain size limits
is as expected: flattening the distribution will decrease the limits, as
the number of bigger grains increases with a flatter slope. Since they are
responsible for most of the observed coreshine, less big grains are needed
to meet the observed flux. The impact on the limits is moderate since all
opacities are integrated over the size distribution. For a slope of -2 instead
of -3.5, the limits move by about 0.15 $\mu$m. More importantly, the overlap
between the $a_2$-ranges derived from the two observational constraints
remains the same. As a result, we argue that the derived grain size limits may
vary by several tens of percent when changing the slope of the size distribution
power law in a reasonable range of [-2 to -5], 
but this will not affect the conclusions. For the
models presented in this section, we have used a spectral slope of -3.5.

\subsection{ecc806 (PLCKECC G303.09-16.04)}
\begin{figure*}
\vbox{
\includegraphics[width=6cm]{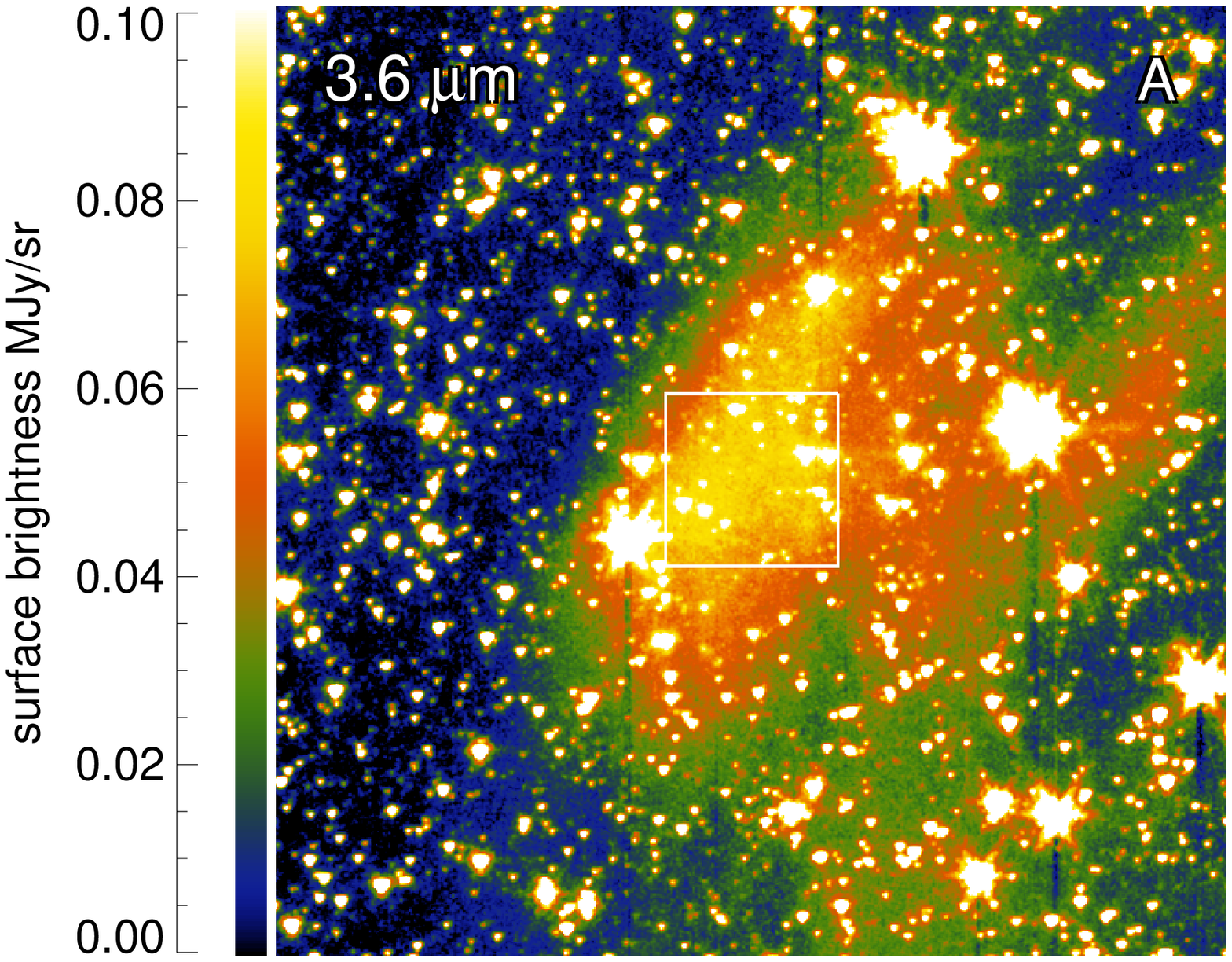}
\includegraphics[width=6cm]{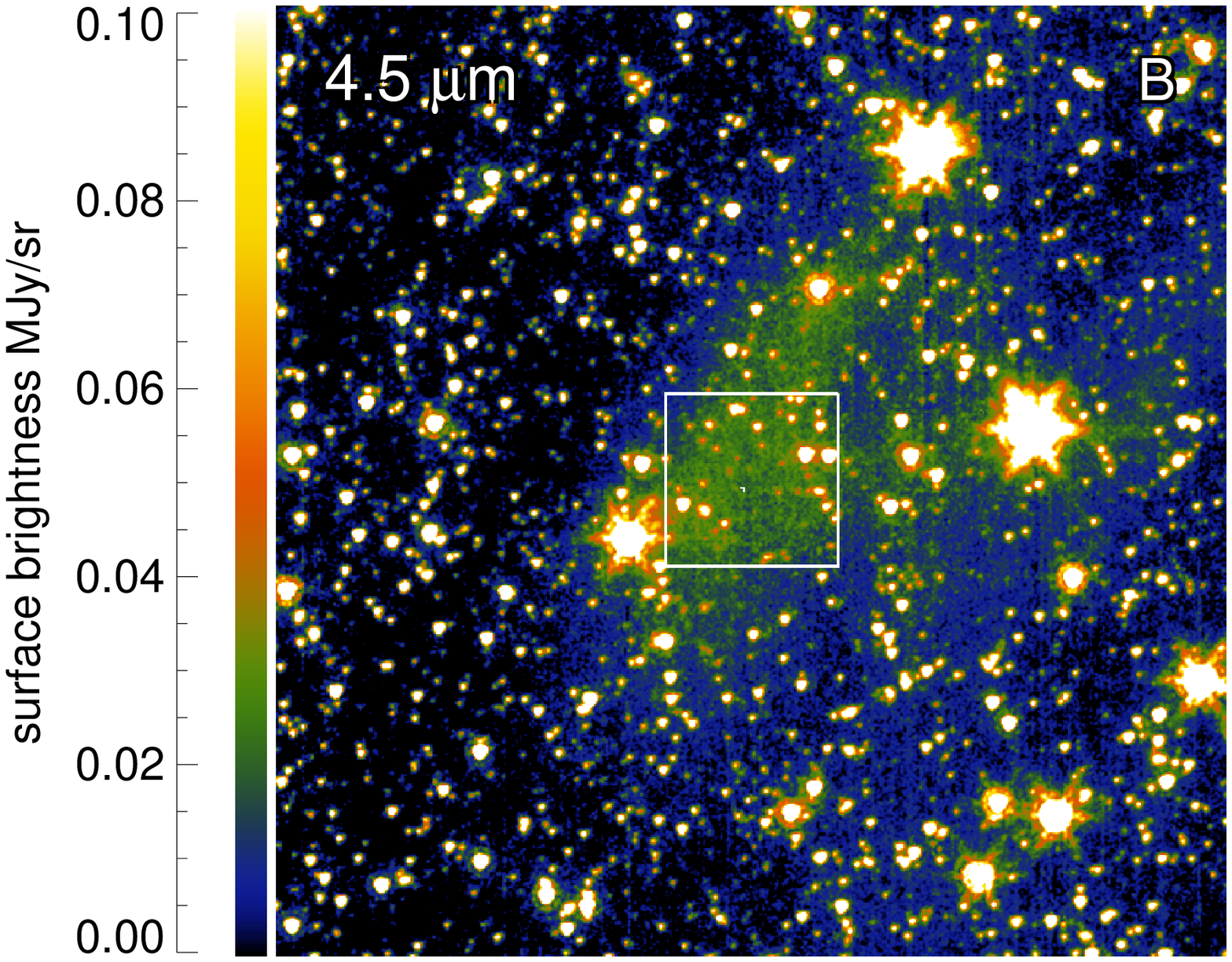}
\includegraphics[width=6cm]{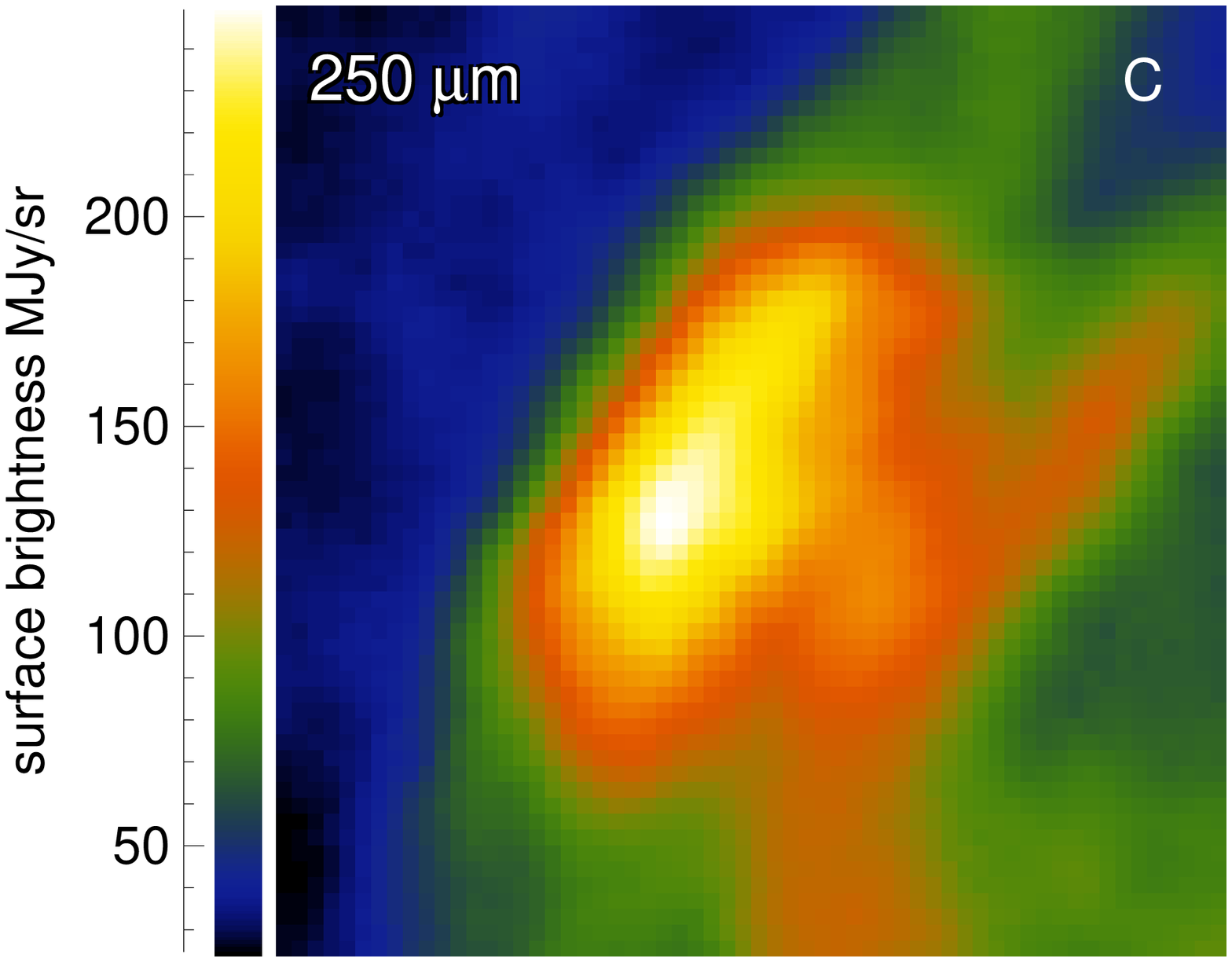}
}
\vskip 0.1cm
\vbox{
\includegraphics[width=6cm]{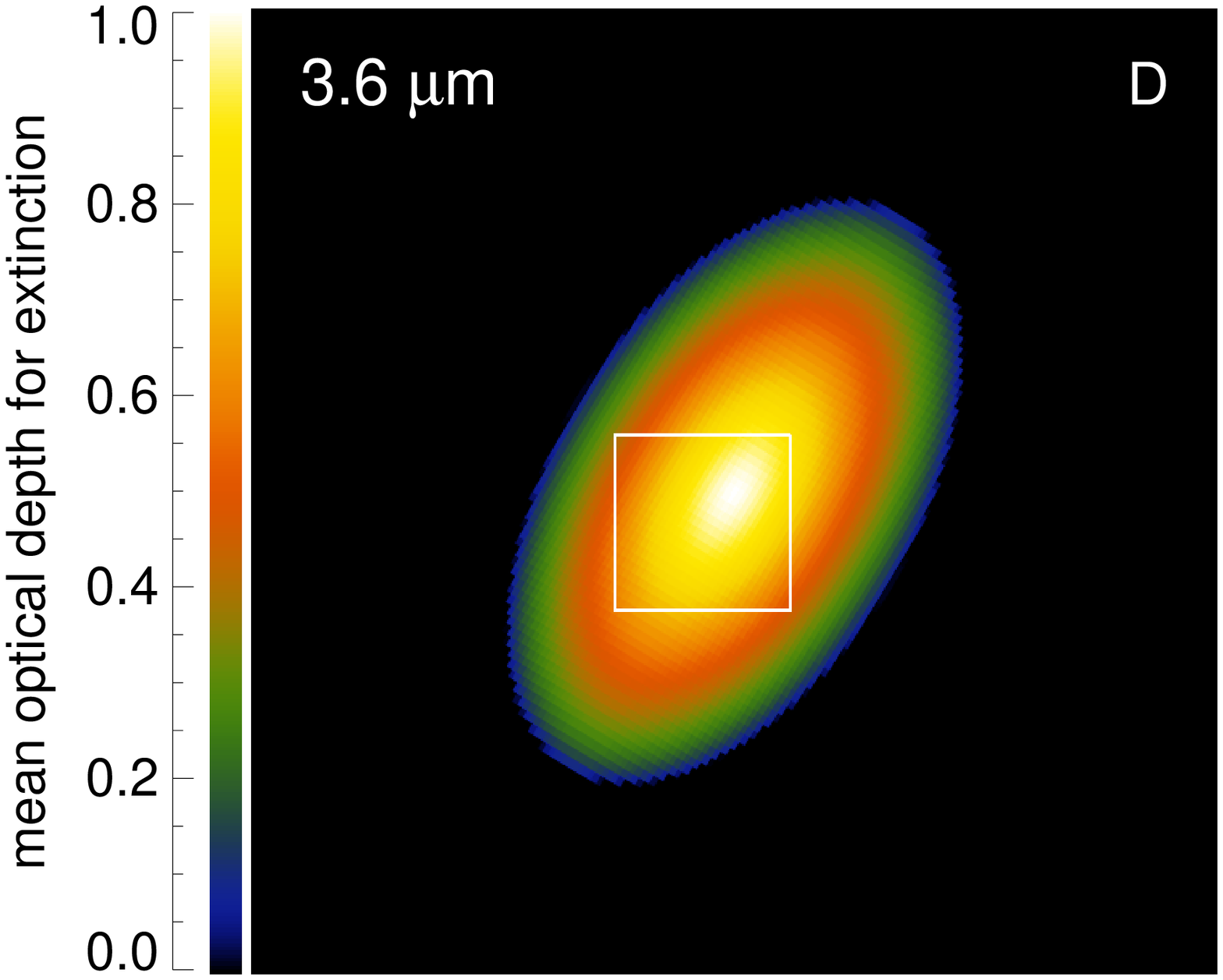}
\includegraphics[width=6cm]{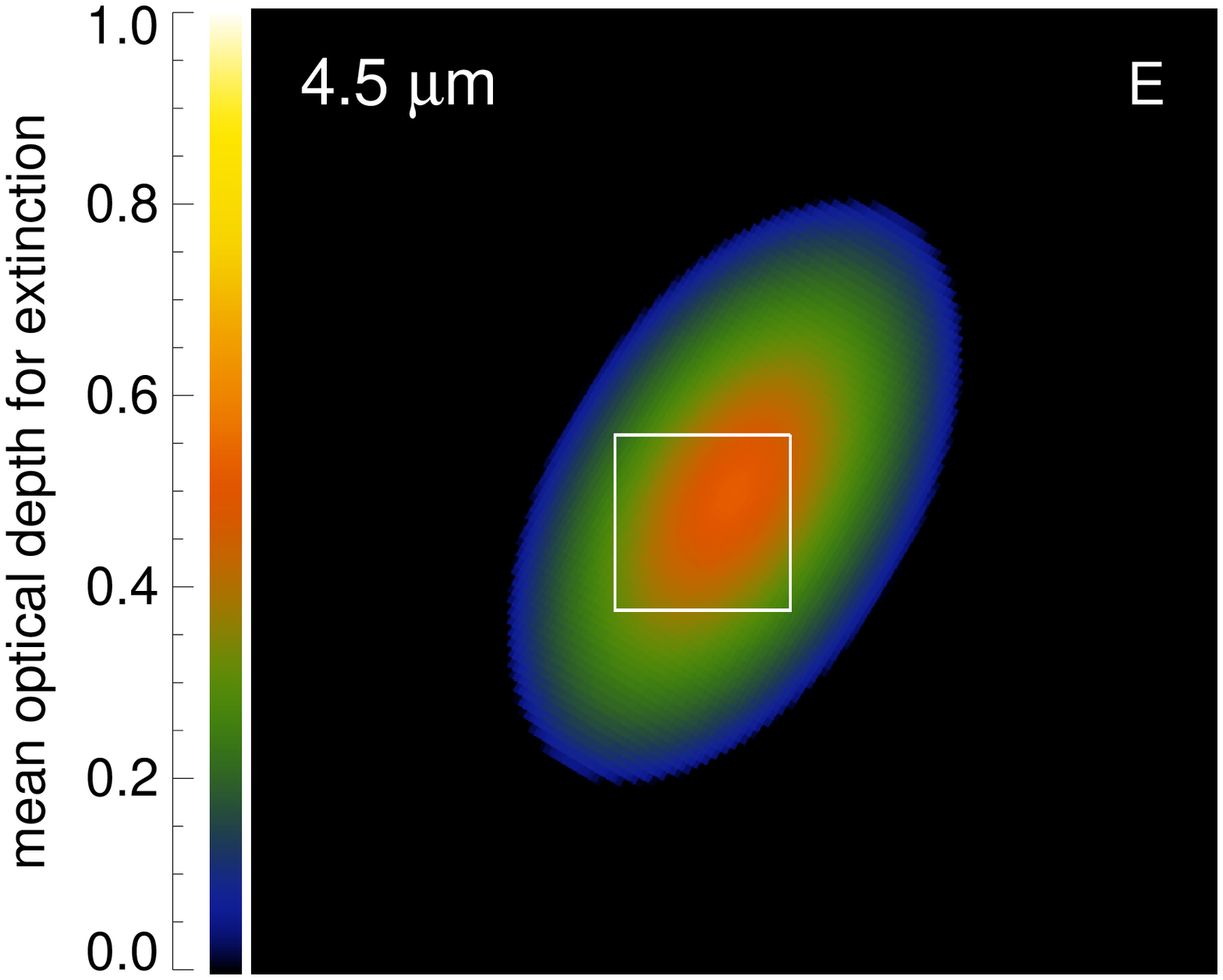}
\includegraphics[width=6cm]{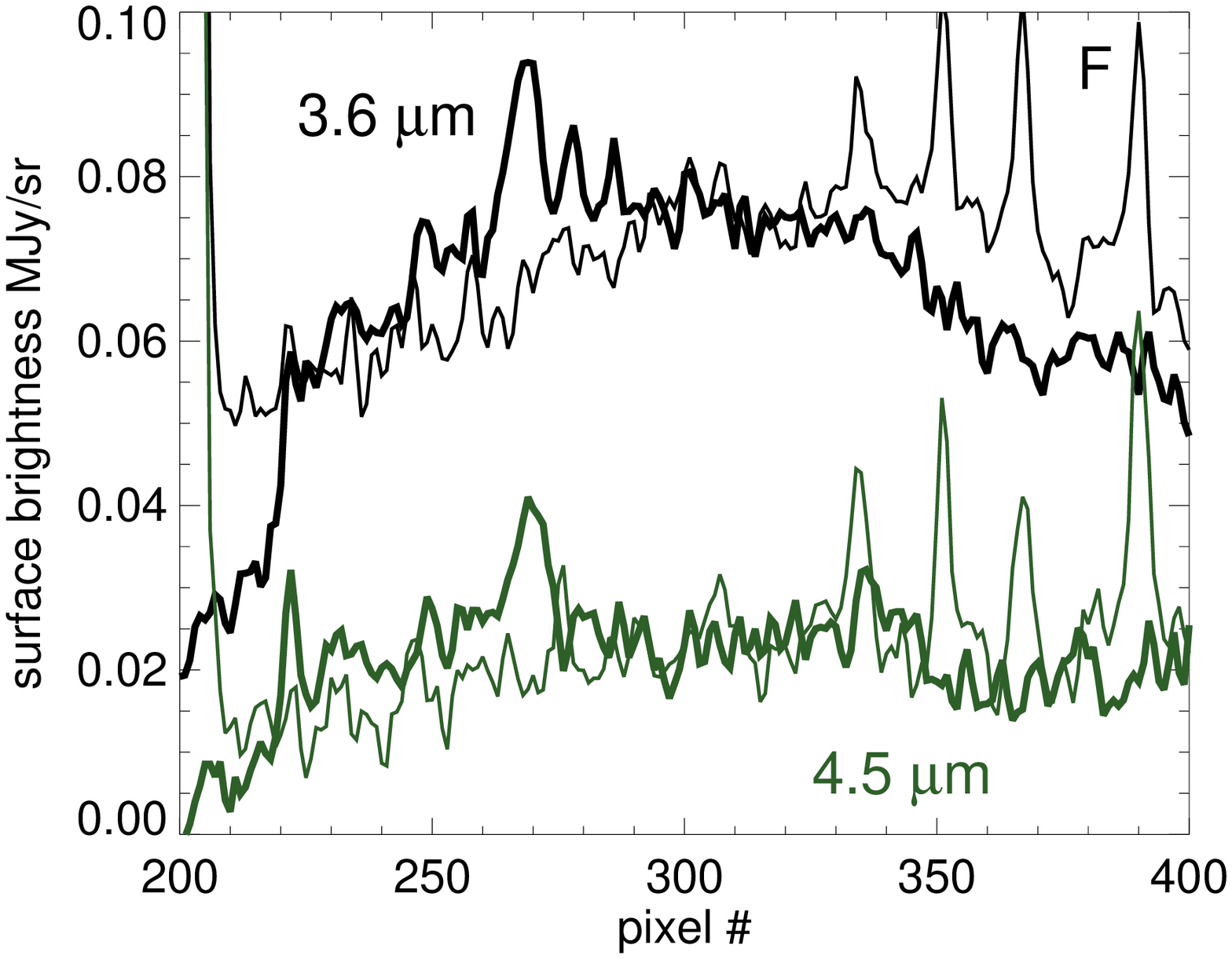}
}
\vskip 0.1cm
\vbox{
\includegraphics[width=6cm]{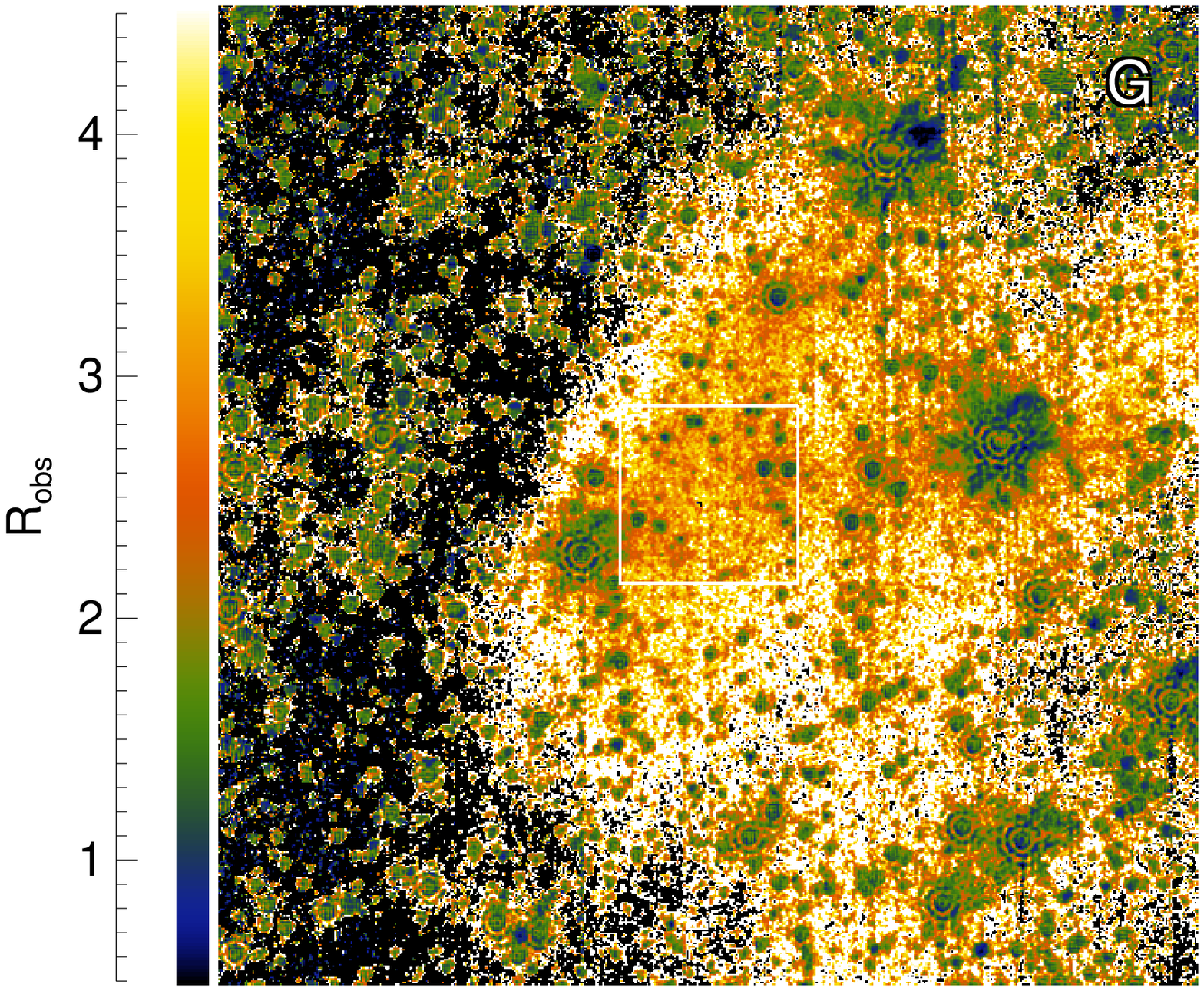}
\includegraphics[width=6cm]{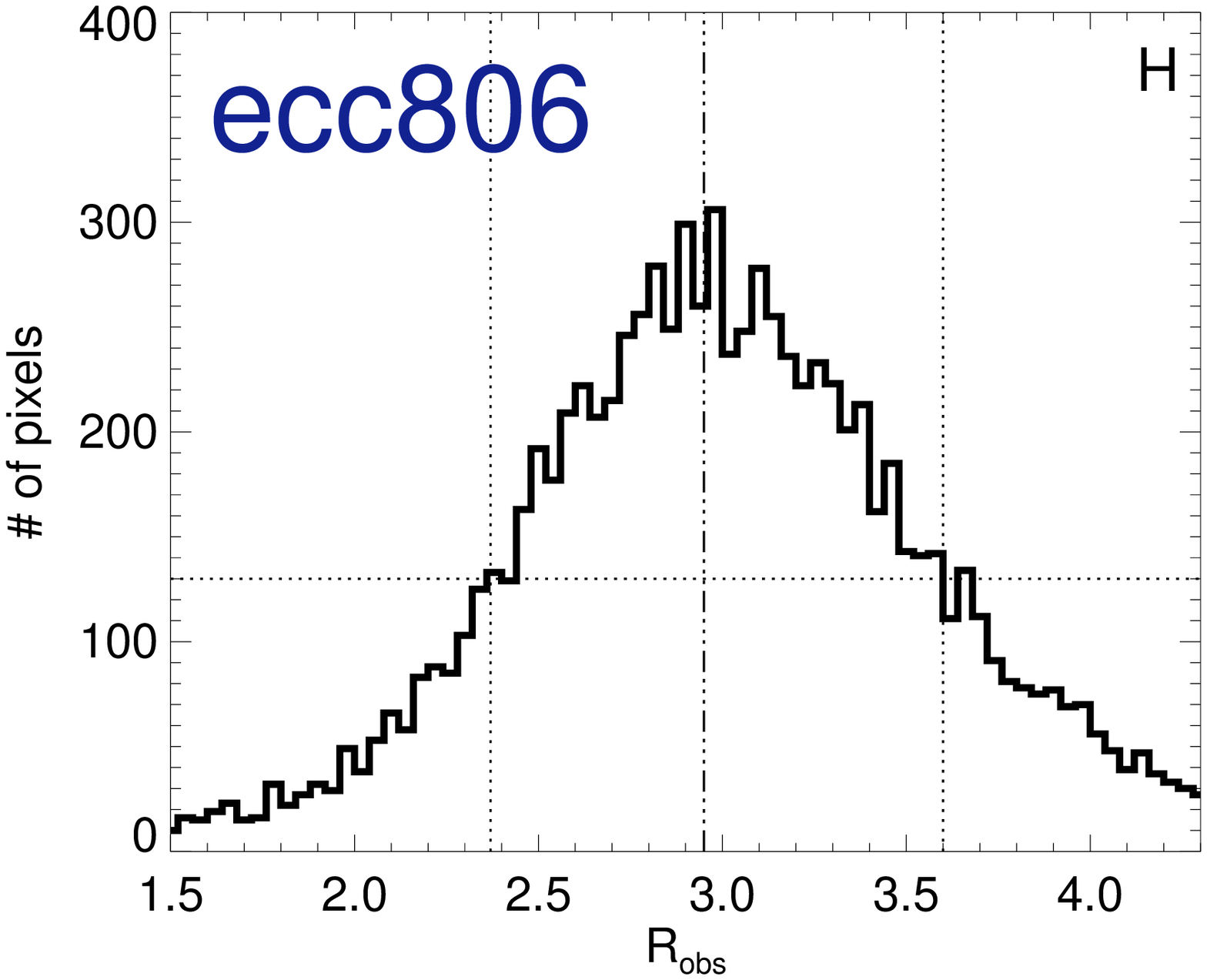}
\includegraphics[width=6cm]{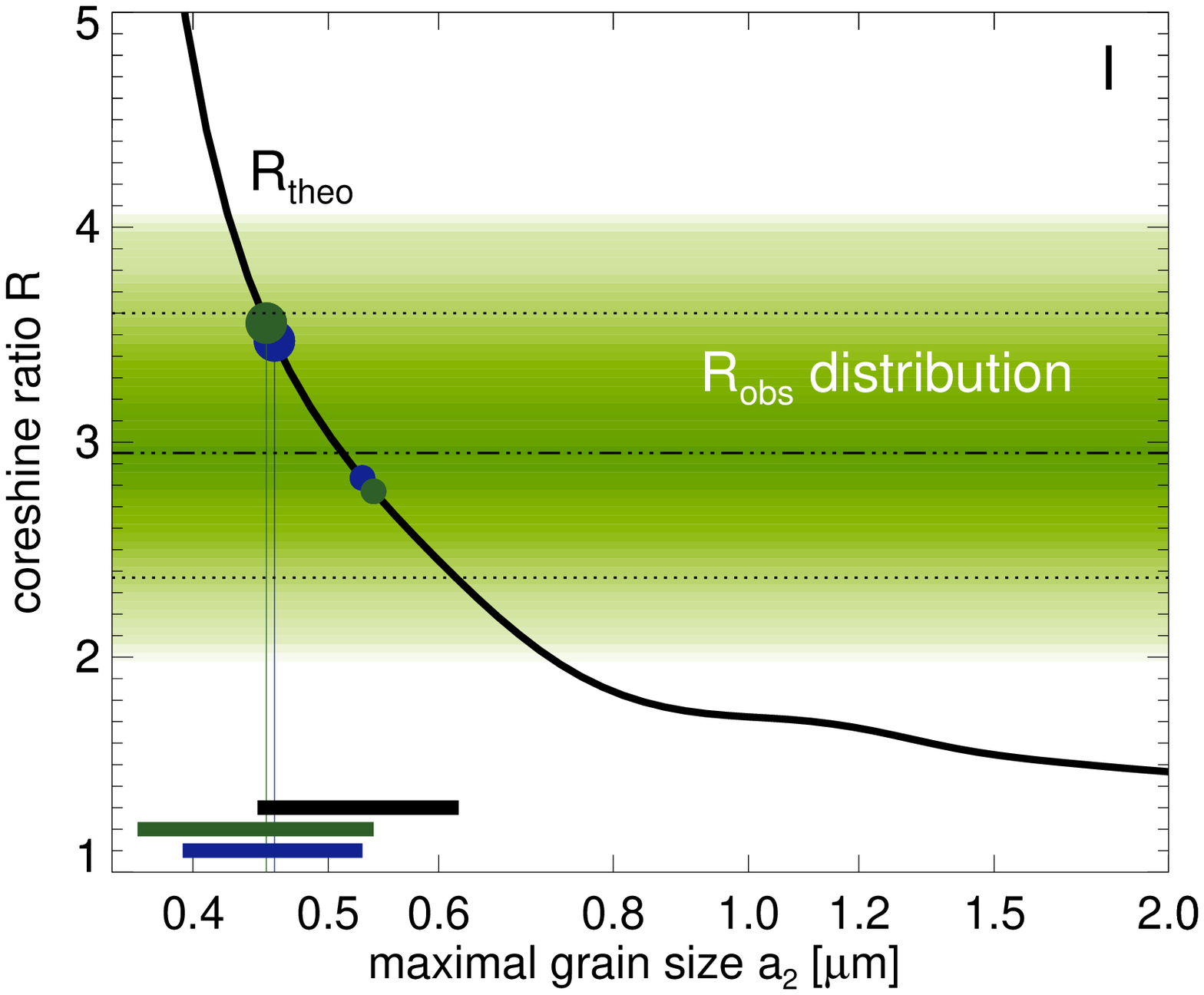}
}

\caption{
Data analysis for ecc806 (PLCKECC G303.09-16.04). 
For a detailed explanations of the panels see Fig.~\ref{L260}. Panel C
shows the SFB at 250 $\mu$m as observed by {\it Herschel}.
        }
\label{ecc806}
\end{figure*}
The second brightest core in the sample, PLCKECC G303.09-16.04 (we will use the shorter
name ecc806 here), is surrounded by 
more filaments with coreshine that are located outside the area
shown in Fig.~\ref{ecc806}.
As for L260, the position favors the detection of coreshine because it is close to the GC in longitude and
far below ($b$=-16.04$^\circ$) the Galactic plane. Moreover, it has the lowest background
in the sample in both bands.
There is no sign of depression in the images and in the cut in Panel F. We therefore use single-scattering RT. 
The chosen white frame encloses the core center with optical depths below the
$\tau$ limit.
In Panel C, the 250 $\mu$m map observed by {\it Herschel} (observation id 1342213209)
as part of the Gould Belt Survey (PI: Andr{\'e}) is shown for comparison.
As with the extinction pattern at 8 $\mu$m shown for the other sources, the thermal emission pattern also
matches nicely the coreshine pattern.
The SFB profiles are remarkably flat in the central part.
The $R$ distribution
is only slightly broader than that of L260 (FWHM about 1.2)
but with a maximum at larger $R$ near 3.
The fit indicates that the maximum
grain size in ecc806 is around from 0.4 to 0.52 $\mu$m.

\subsection{L1262}
\begin{figure*}
\vbox{
\includegraphics[width=6cm]{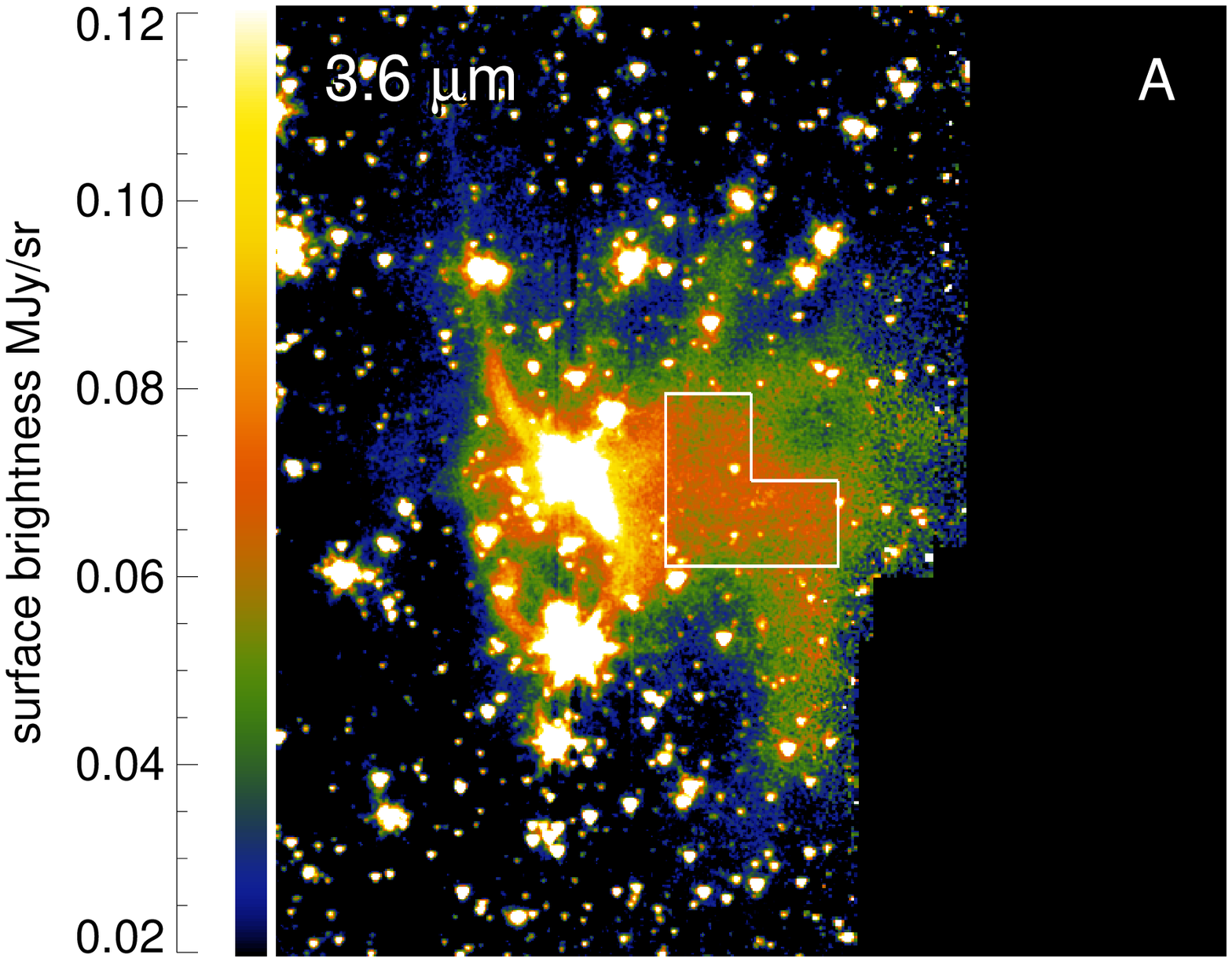}
\includegraphics[width=6cm]{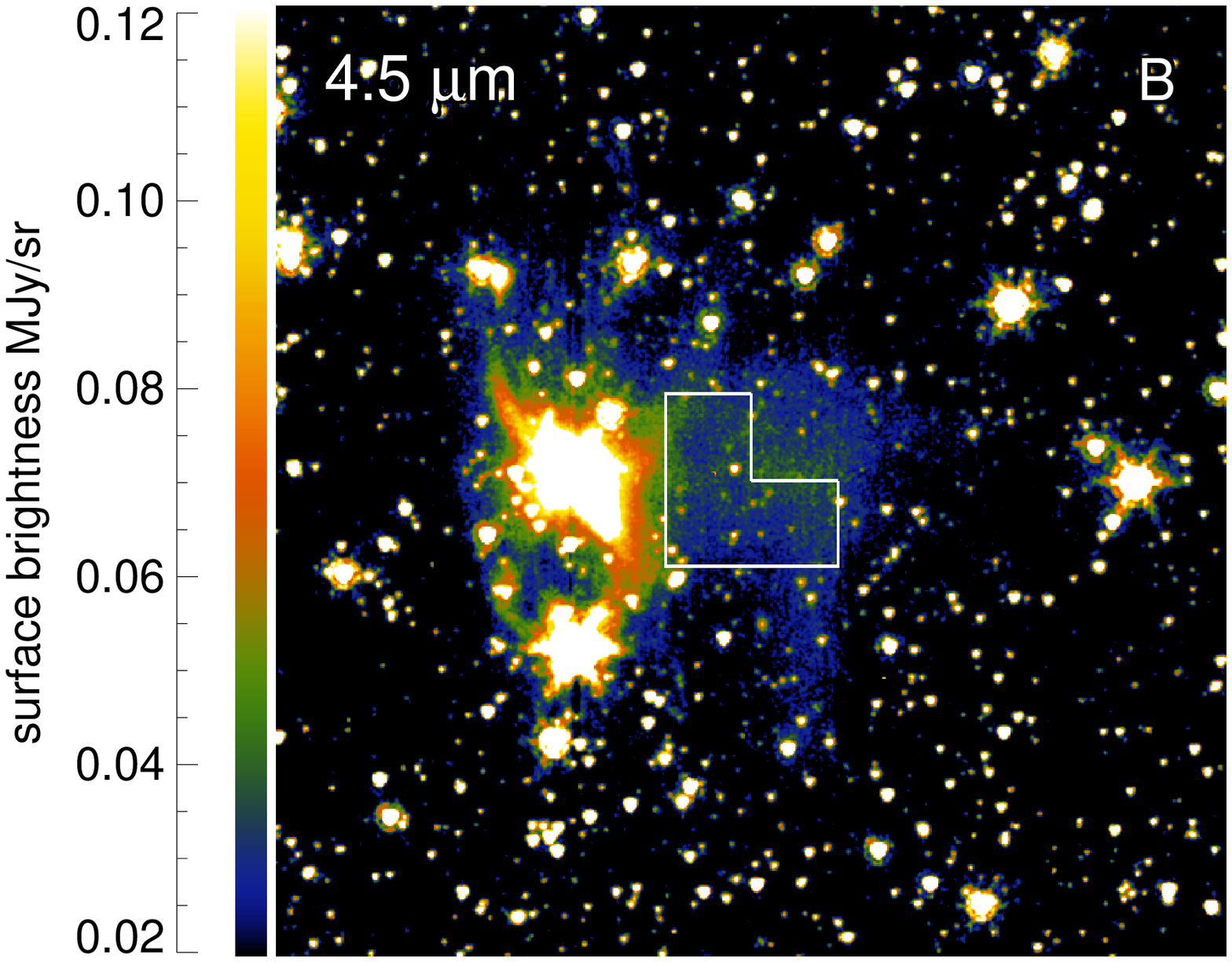}
\includegraphics[width=6cm]{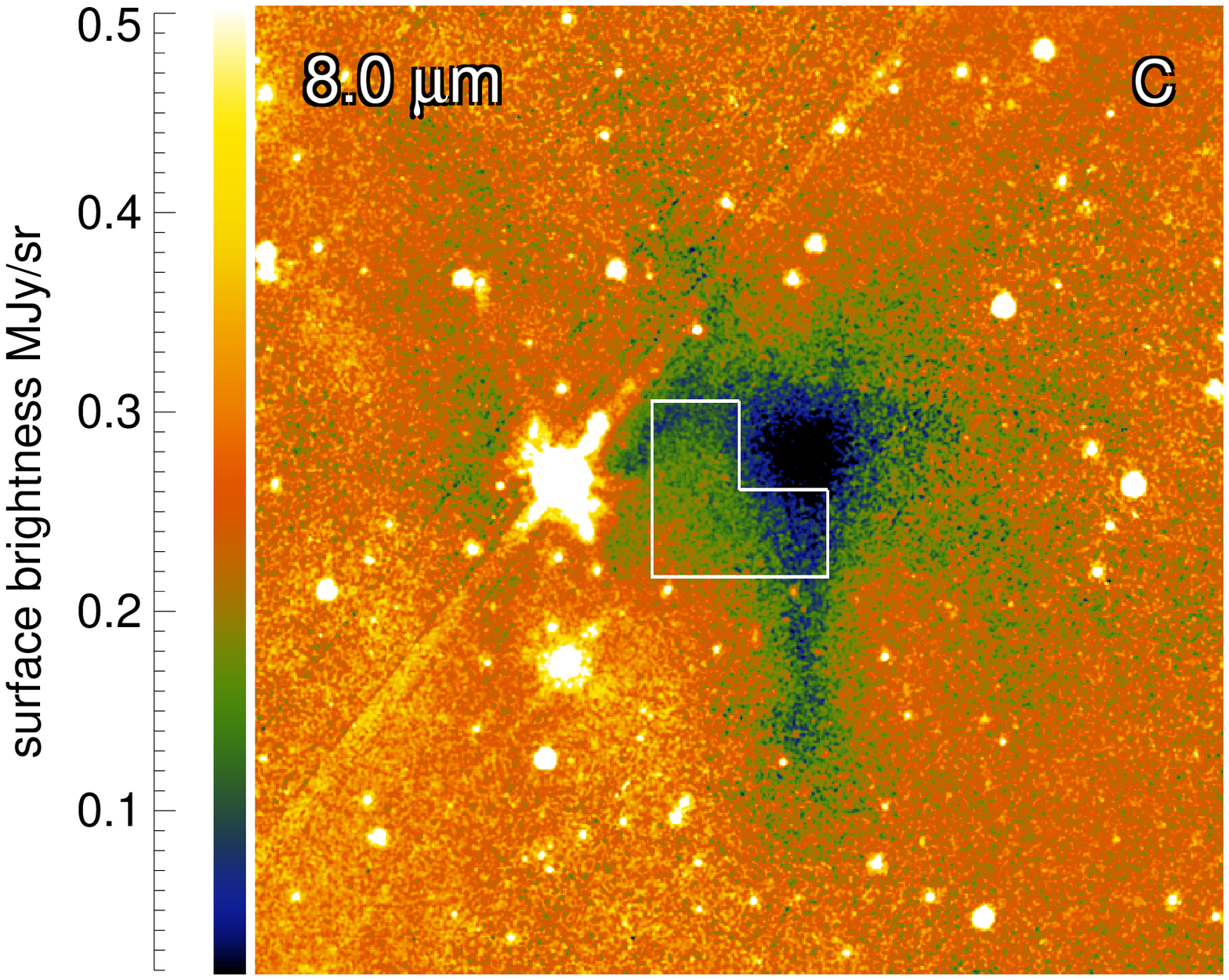}
}
\vskip 0.1cm
\vbox{
\includegraphics[width=6cm]{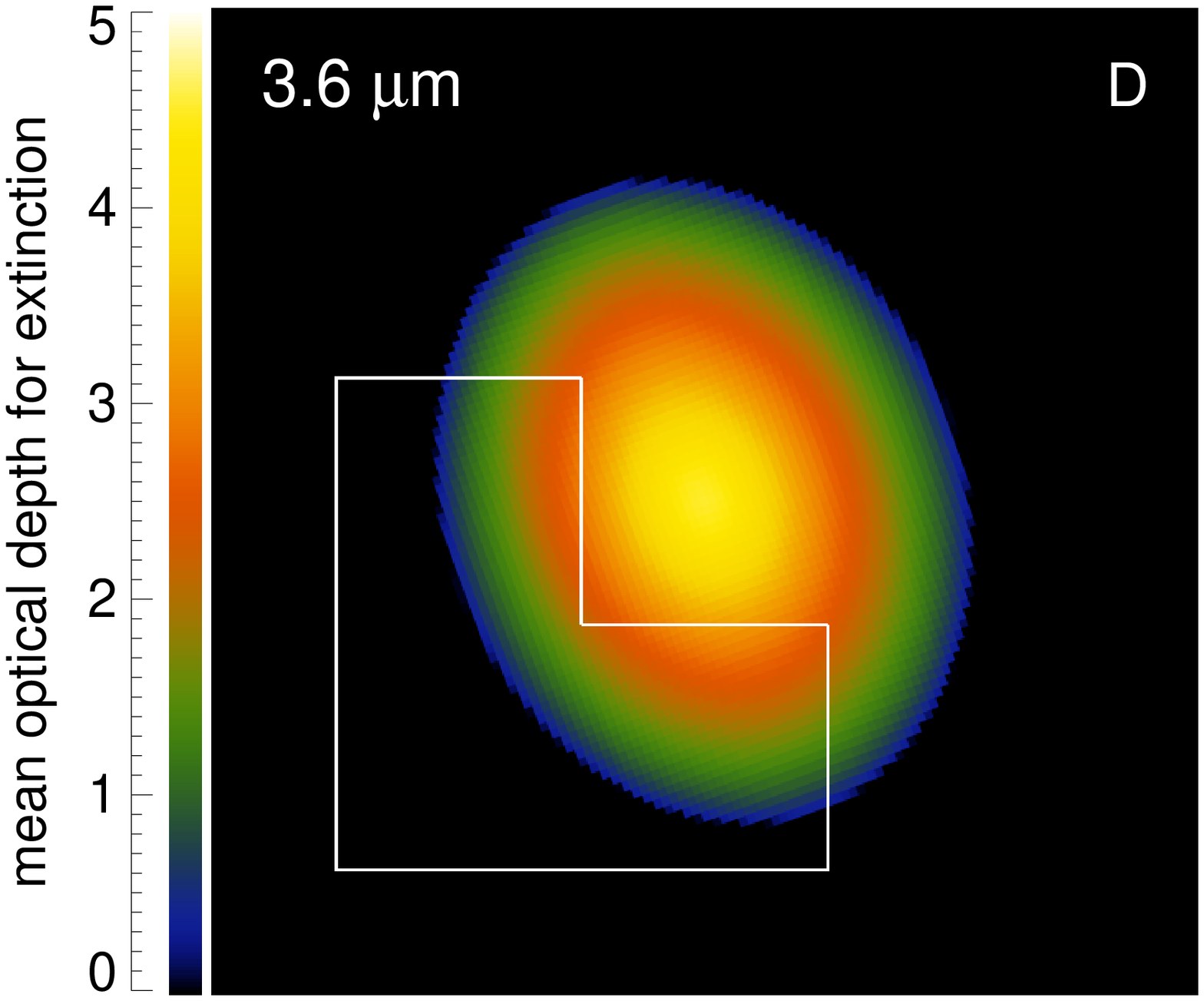}
\includegraphics[width=6cm]{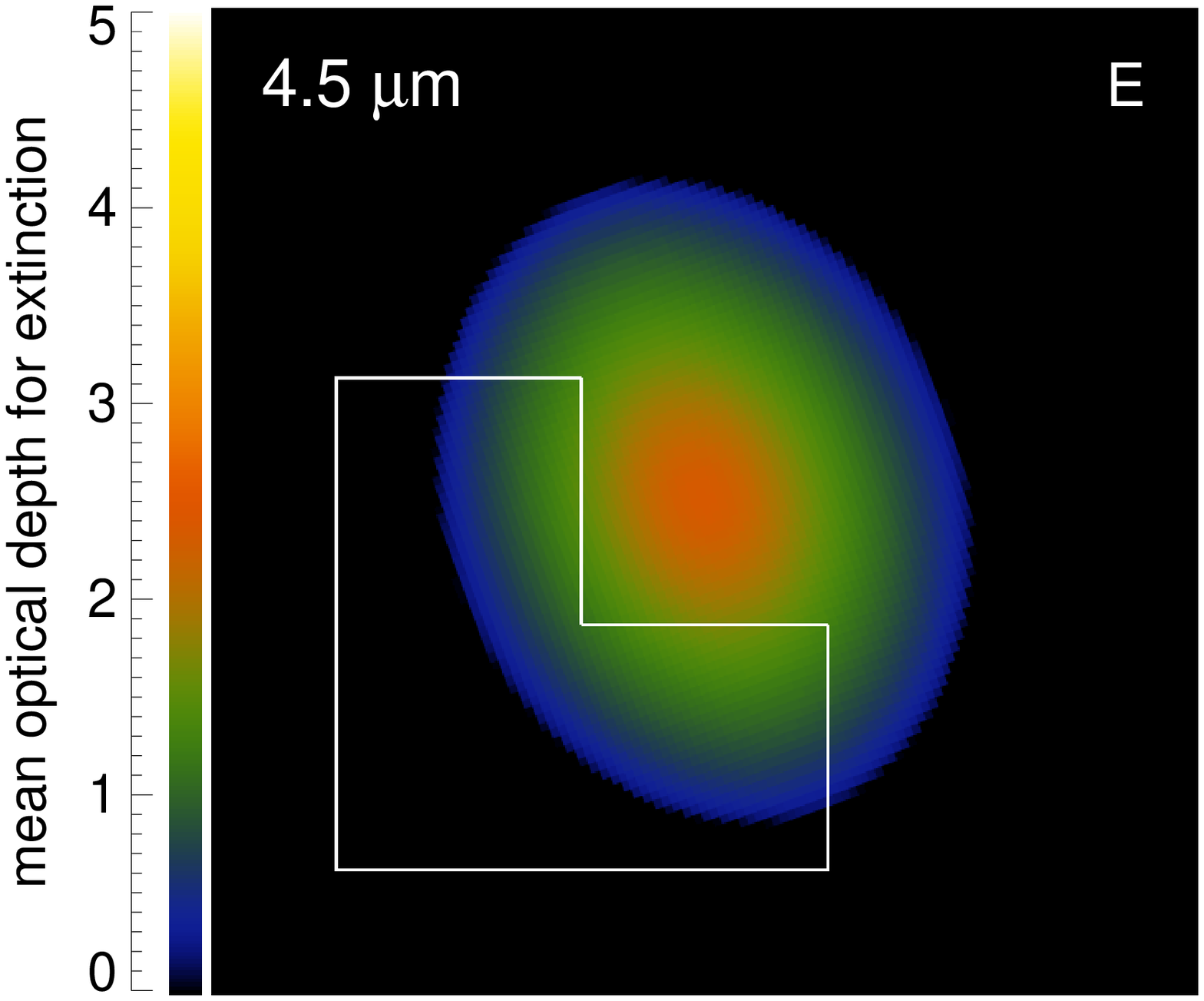}
\includegraphics[width=6cm]{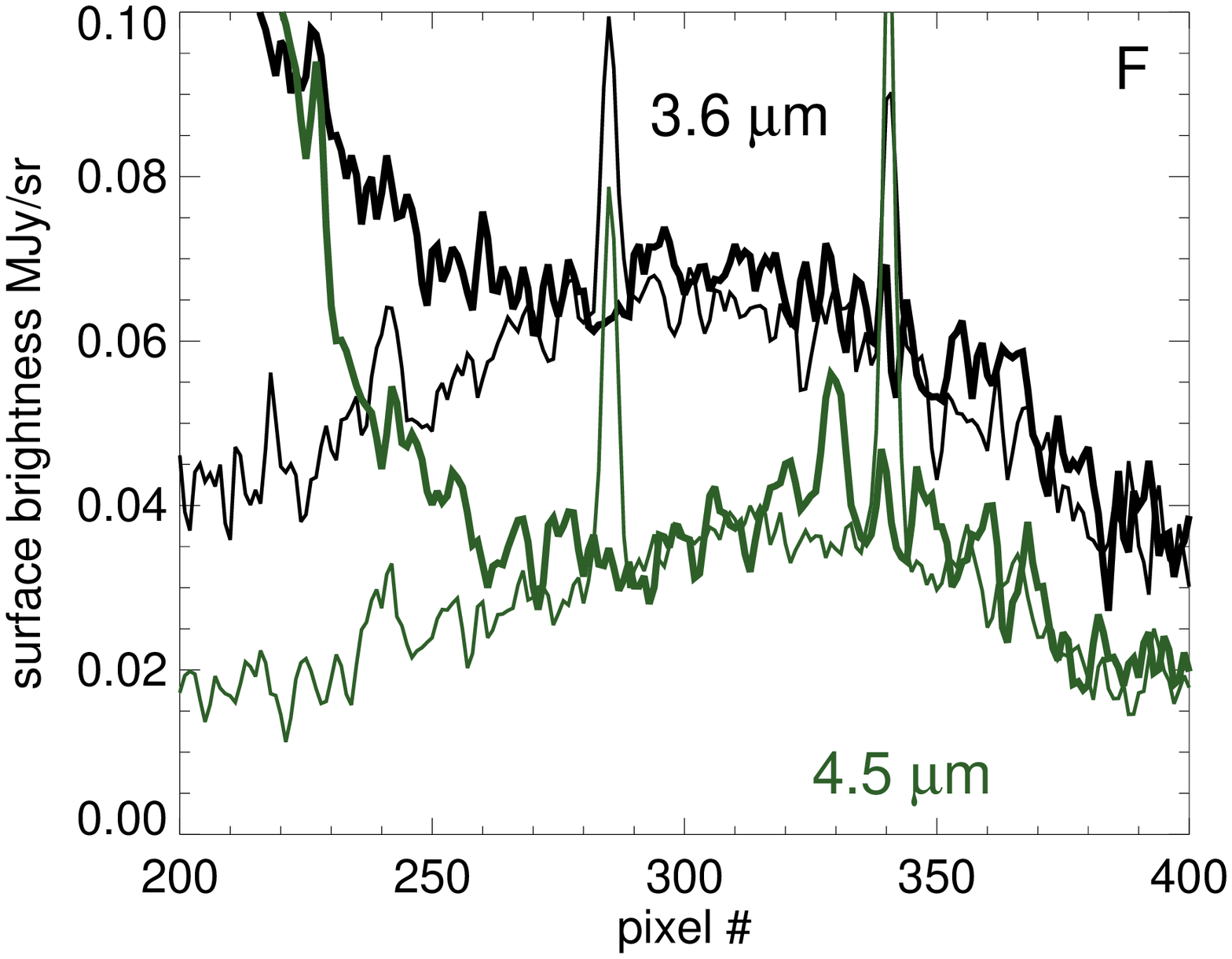}
}
\vskip 0.1cm
\vbox{
\includegraphics[width=6cm]{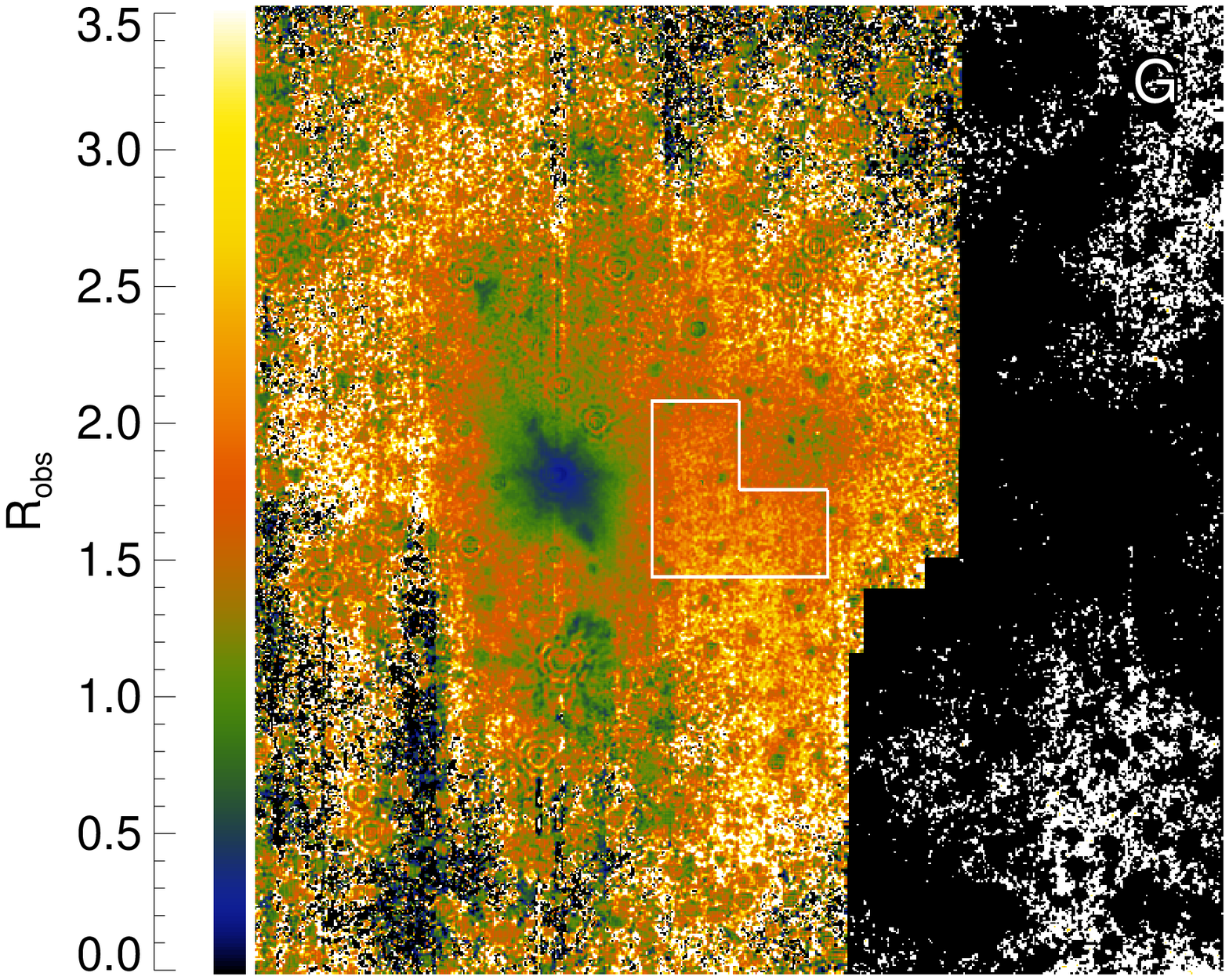}
\includegraphics[width=6cm]{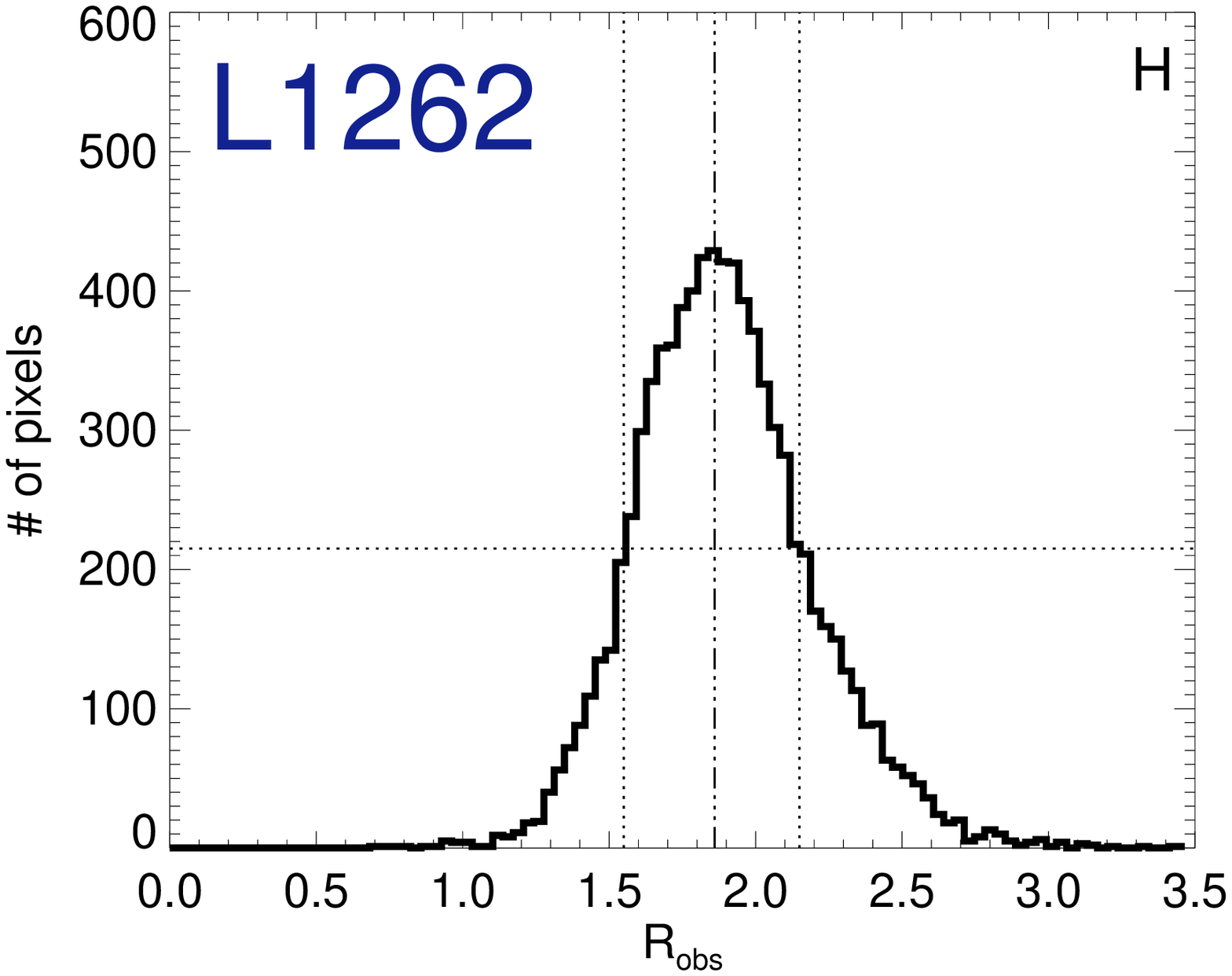}
\includegraphics[width=6cm]{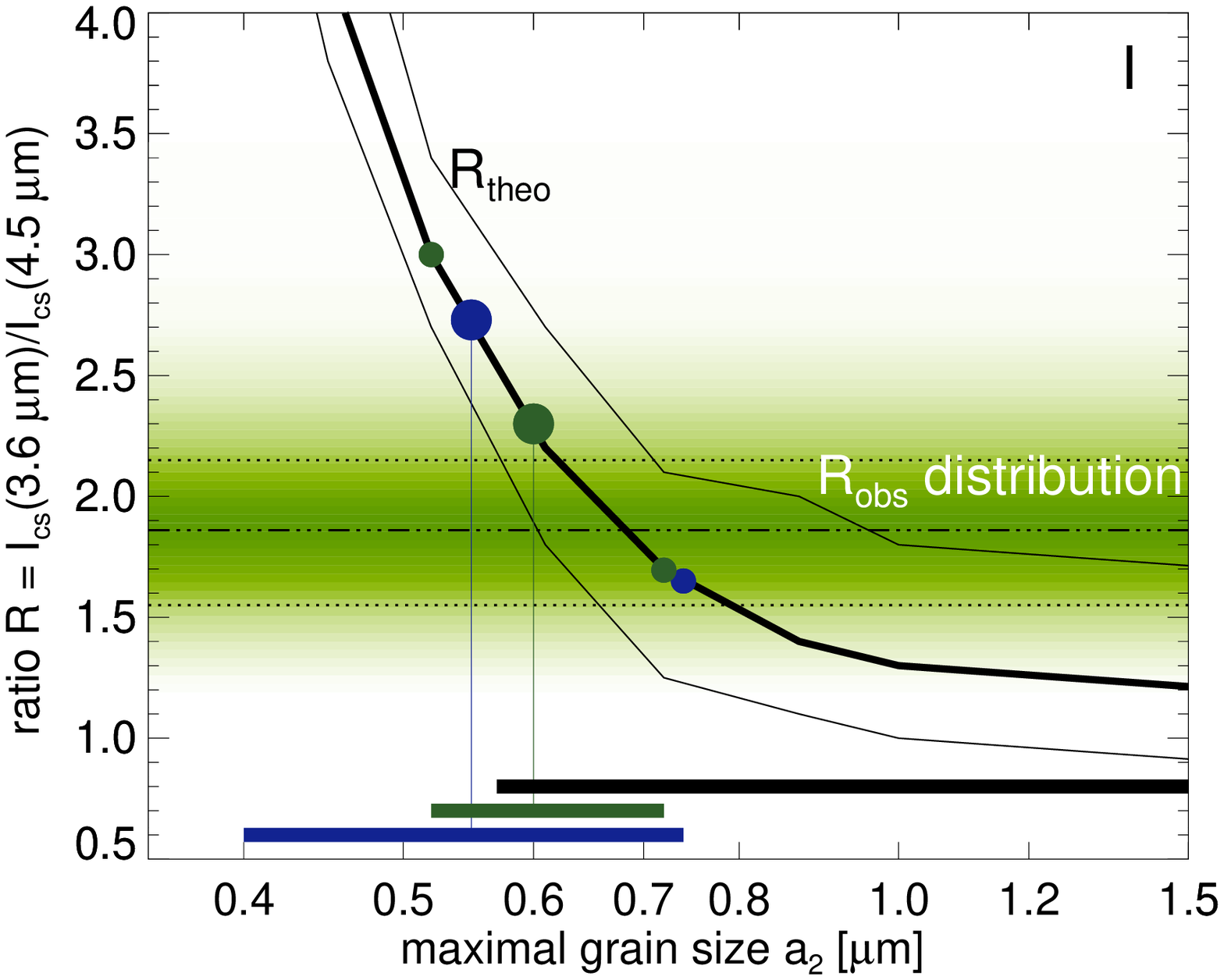}
}
\caption{
Data analysis for L1262. For s detailed legend see Fig.~\ref{L260}. The binary core harbors a
YSO to the left. The core was located at the IRAC 1 image border leading to a dark zone in Panels A and G. (The $R$ values in the dark right zone are not meaningful.)
Since the modeling is performed using full RT, Panel I compares the observed $R$ distribution as color-coded and the theoretical $R$-distribution with a
thick solid line for the mean value of the theoretical $R$ distribution and half maximum values as thin lines.
        }
\label{L1262}
\end{figure*}
Modeling of the starless core in the binary core L1262 (CB 244) is complex owing to the nearby YSO
with its outflows creating H2 emission in the shocked gas in both bands. 
L1262 also has a remarkably extended cloudshine that is visible at shorter
wavelength and is the most distant core in the sample (about 200 pc). 
The YSO and its warm-dust surroundings may give rise to an additional radiation field. Since the
density distribution is complex, it is hard to model this component in order to derive the intensity and directional dependency at the location of the starless core.
With 60$^\circ$ away from the GC, the phase function of the largest grains will not 
increase coreshine by beaming the strong GC radiationforward .

The core position in the PoSky is elevated enough ($b$=12.4$^\circ$) to have a low background, especially at 4.5 $\mu$m, which allows the detection of the weak coreshine signal in this band.
The mass of the core is rather uncertain
with a mean value of 5 M$_\odot$ and an error bar of 2~M$_\odot$.
However, the column density range given in Table \ref{table:1} also points to $\tau>2$, and the 3.6 $\mu$m image in Panel A reveals a depression right where the 
extinction against the background has a maximum in the cold 8 $\mu$m image in Panel C
("darkclouds\_IRAC": Program ID 94, PI C. Lawrence).
We therefore have chosen maximum optical depth values around 4.5 (scaled with the mean
column density values of the other cores) and use full RT.
We have chosen a white frame away from the YSO and its extended tail and excluded the part of the
square where the extinction seen in Panel C of Fig.~\ref{L1262} indicates the center of the core. As mentioned before, this improves the modeling because at lower optical depths, some of the spatial dependencies cancel out.
The observed $R$ distribution is about a factor of 2 narrower than for L260 with a lower mean value of 1.75.
Aside from the color-coded observational $R$-distribution,
Panel I shows the distribution of the theoretical $R$-values with a thick line for the mean
value and thin lines for the FWHM. 
The $a_2$ ranges overlap between 0.52 and 0.71 $\mu$m.

\subsection{L1517A}
\begin{figure*}
\vbox{
\includegraphics[width=6cm]{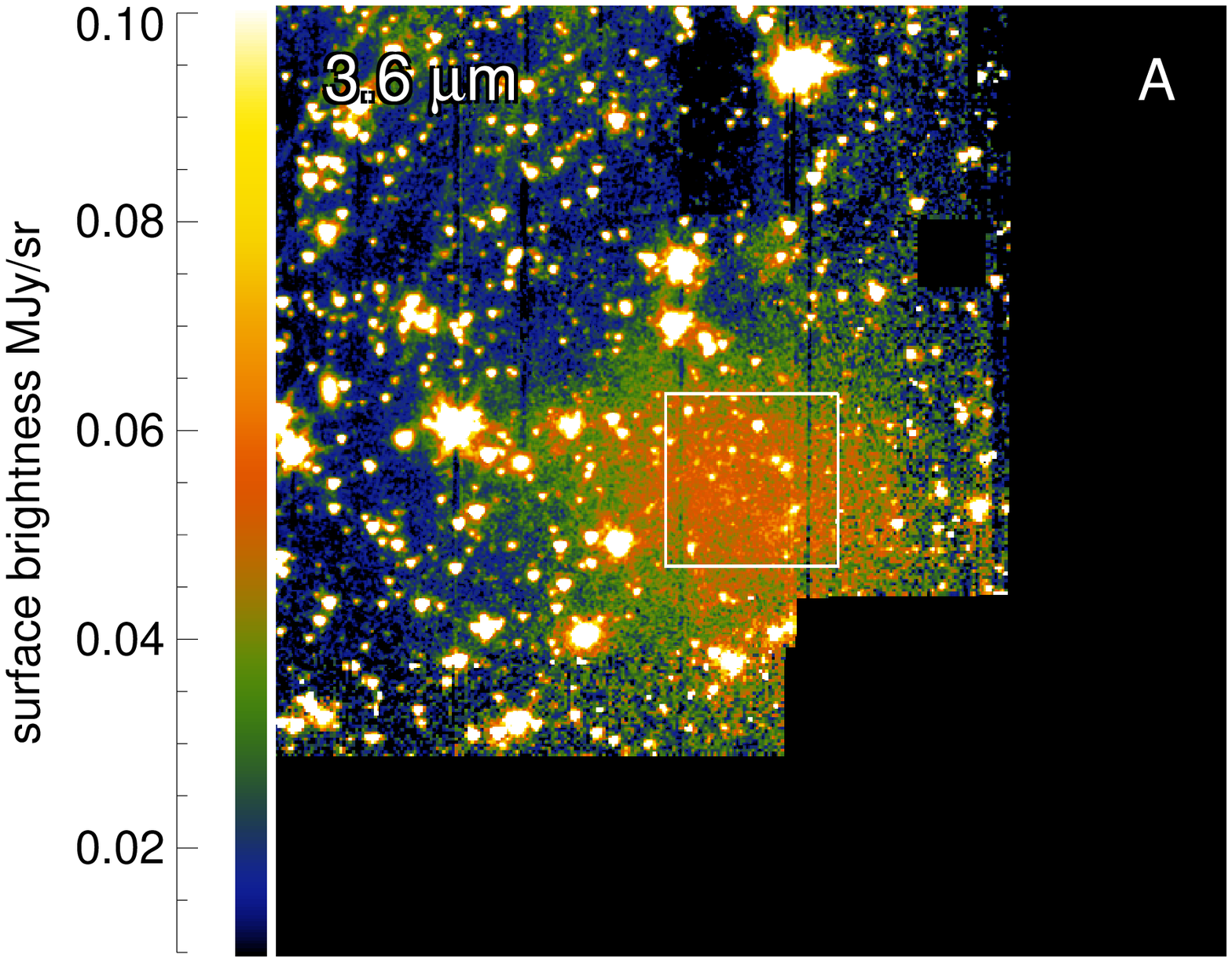}
\includegraphics[width=6cm]{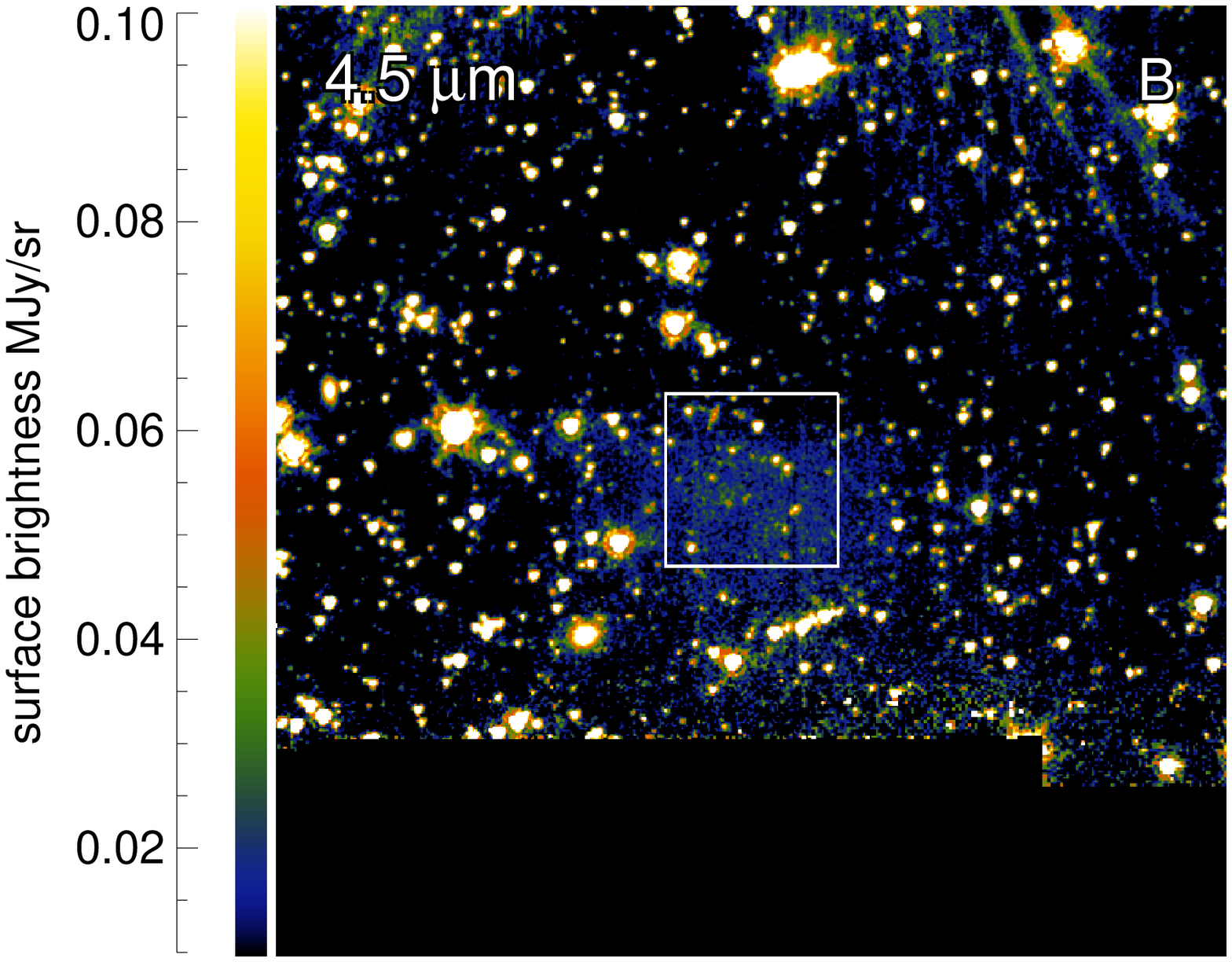}
\includegraphics[width=6cm]{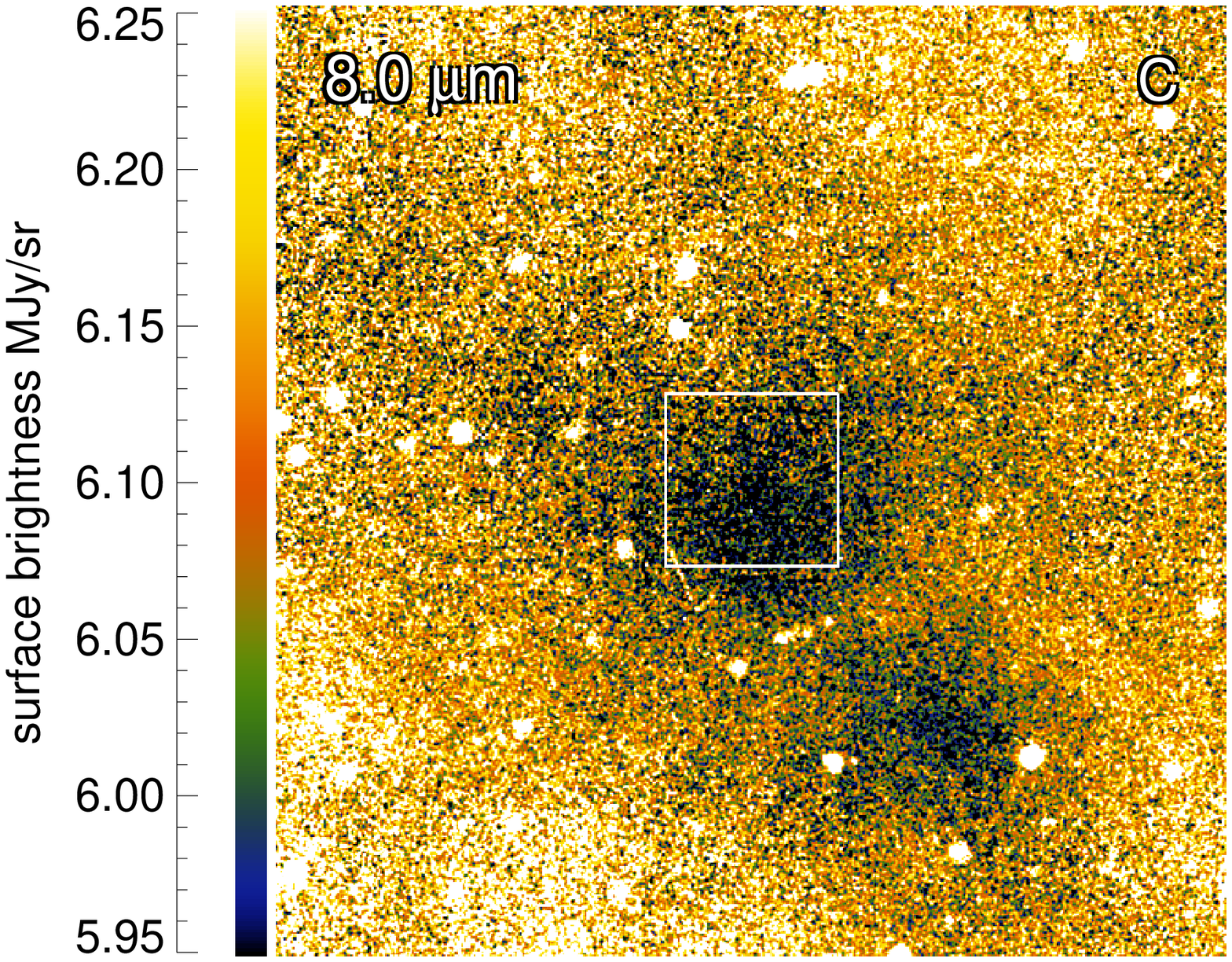}
}
\vskip 0.1cm
\vbox{
\includegraphics[width=6cm]{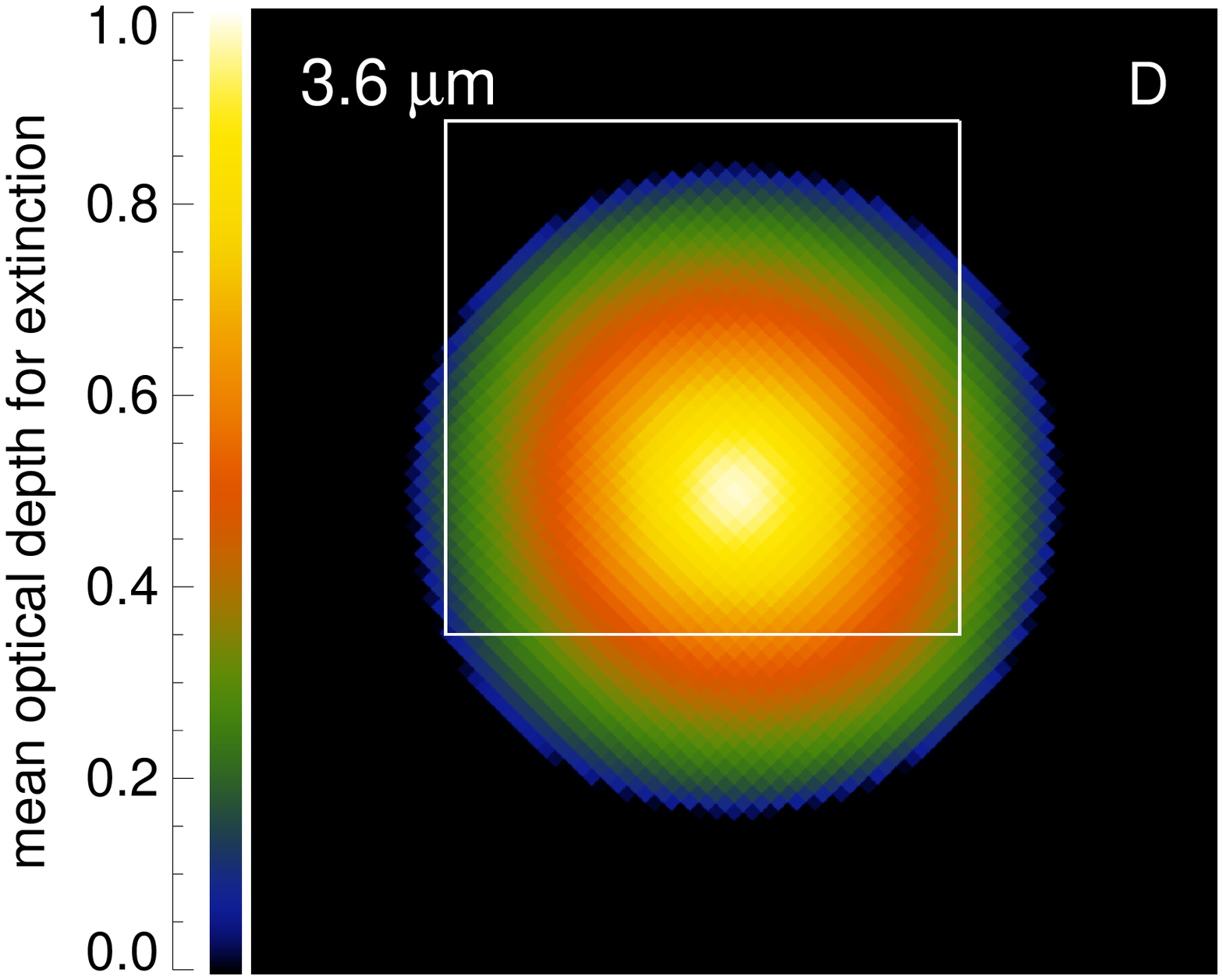}
\includegraphics[width=6cm]{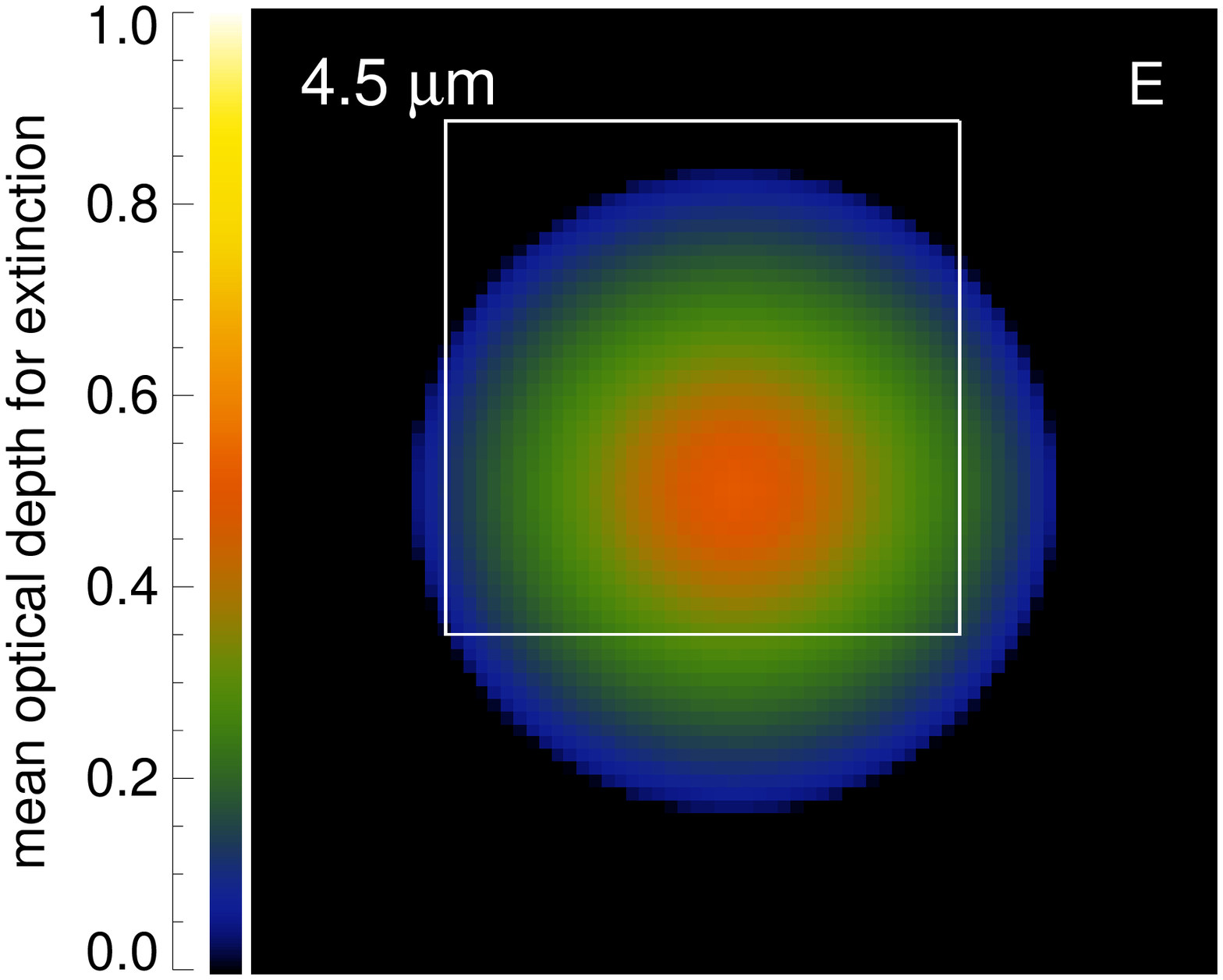}
\includegraphics[width=6cm]{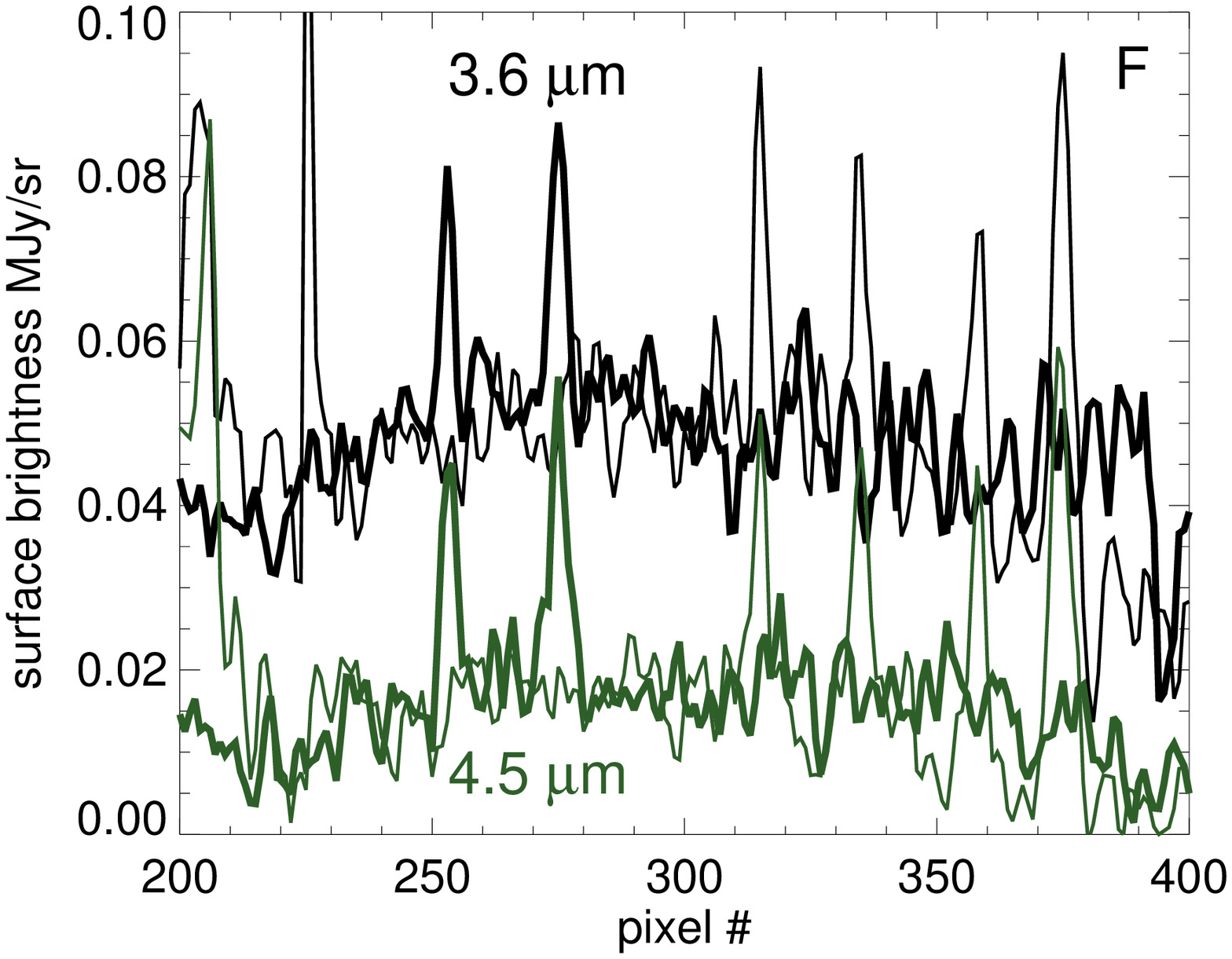}
}
\vskip 0.1cm
\vbox{
\includegraphics[width=6cm]{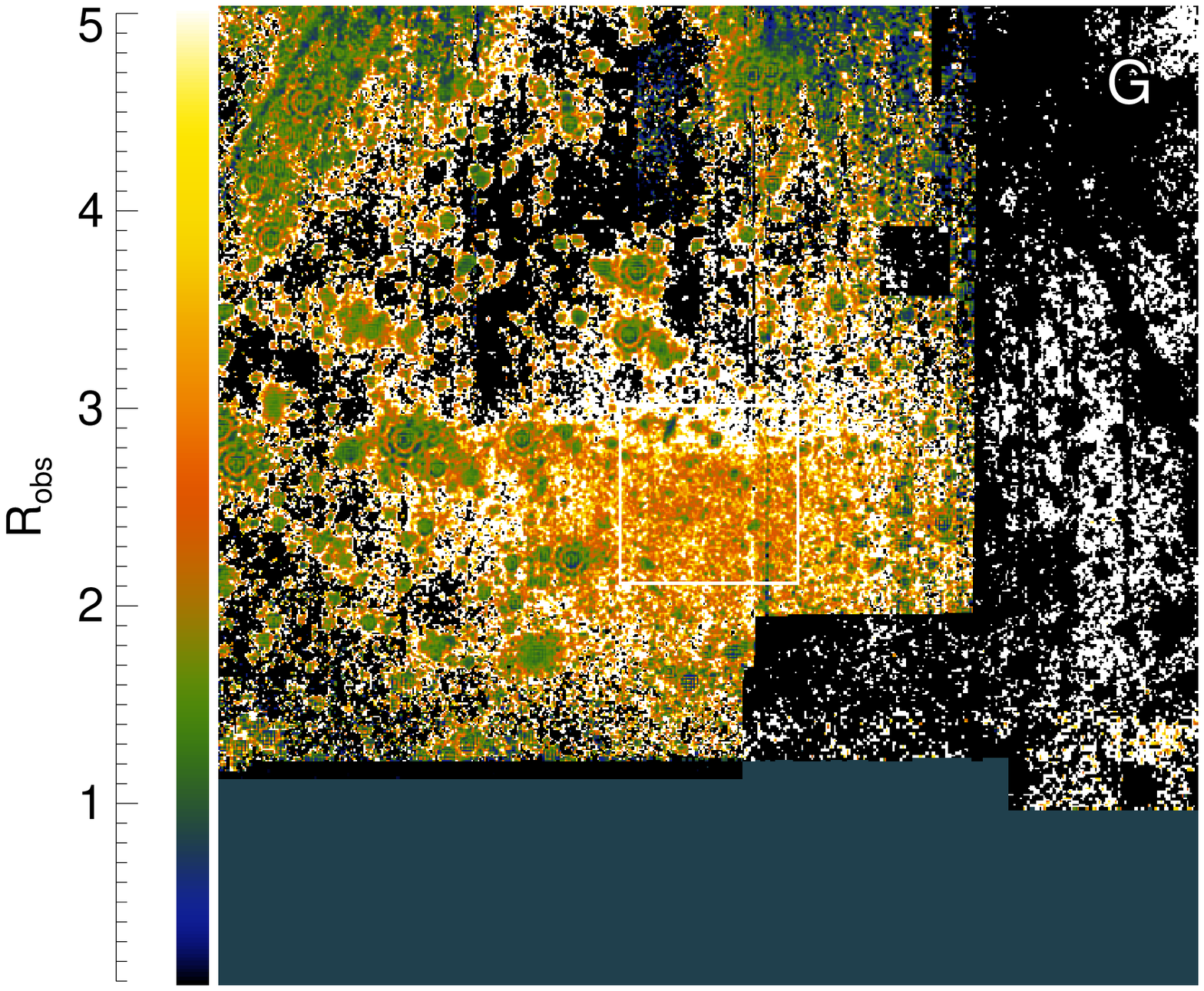}
\includegraphics[width=6cm]{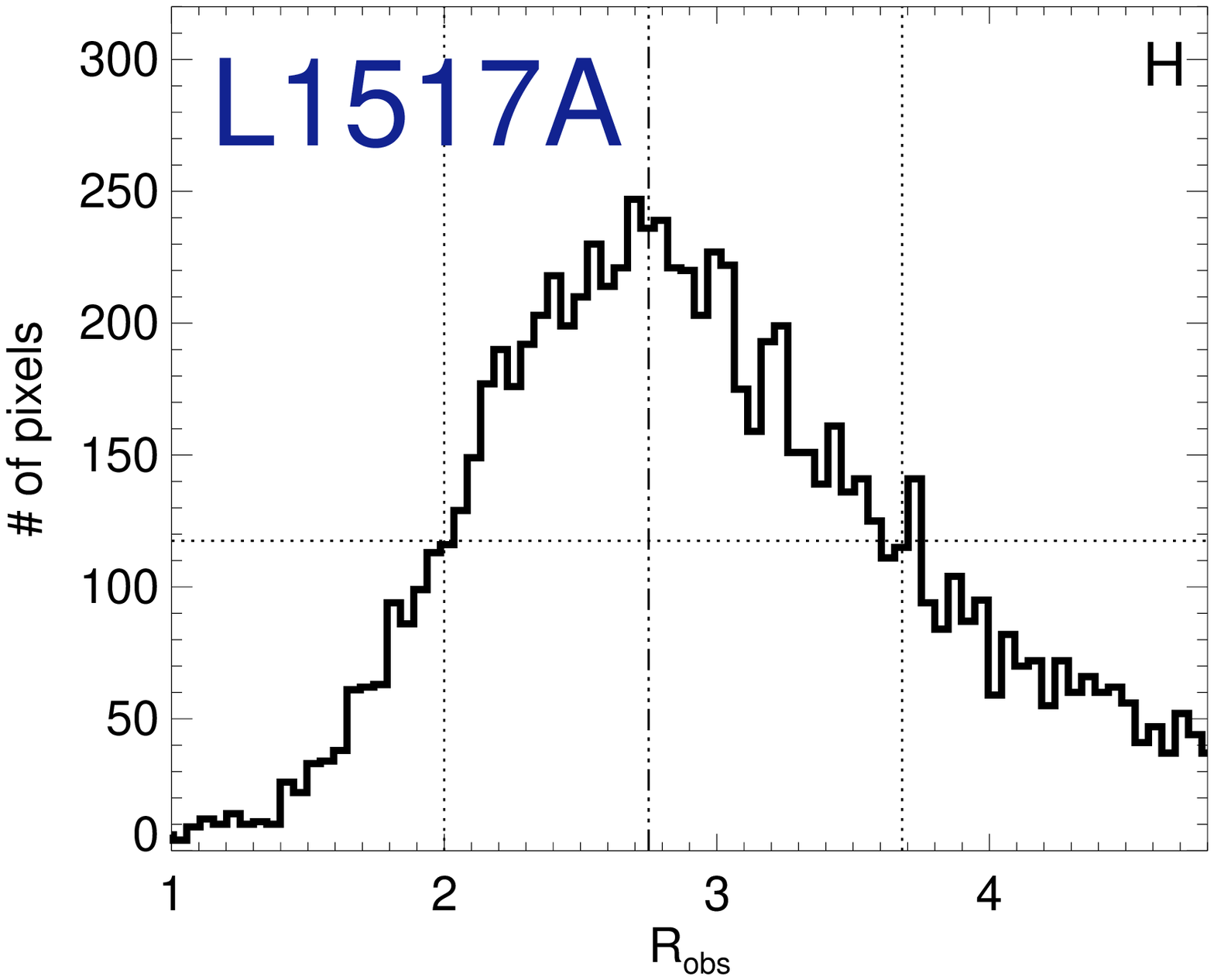}
\includegraphics[width=6cm]{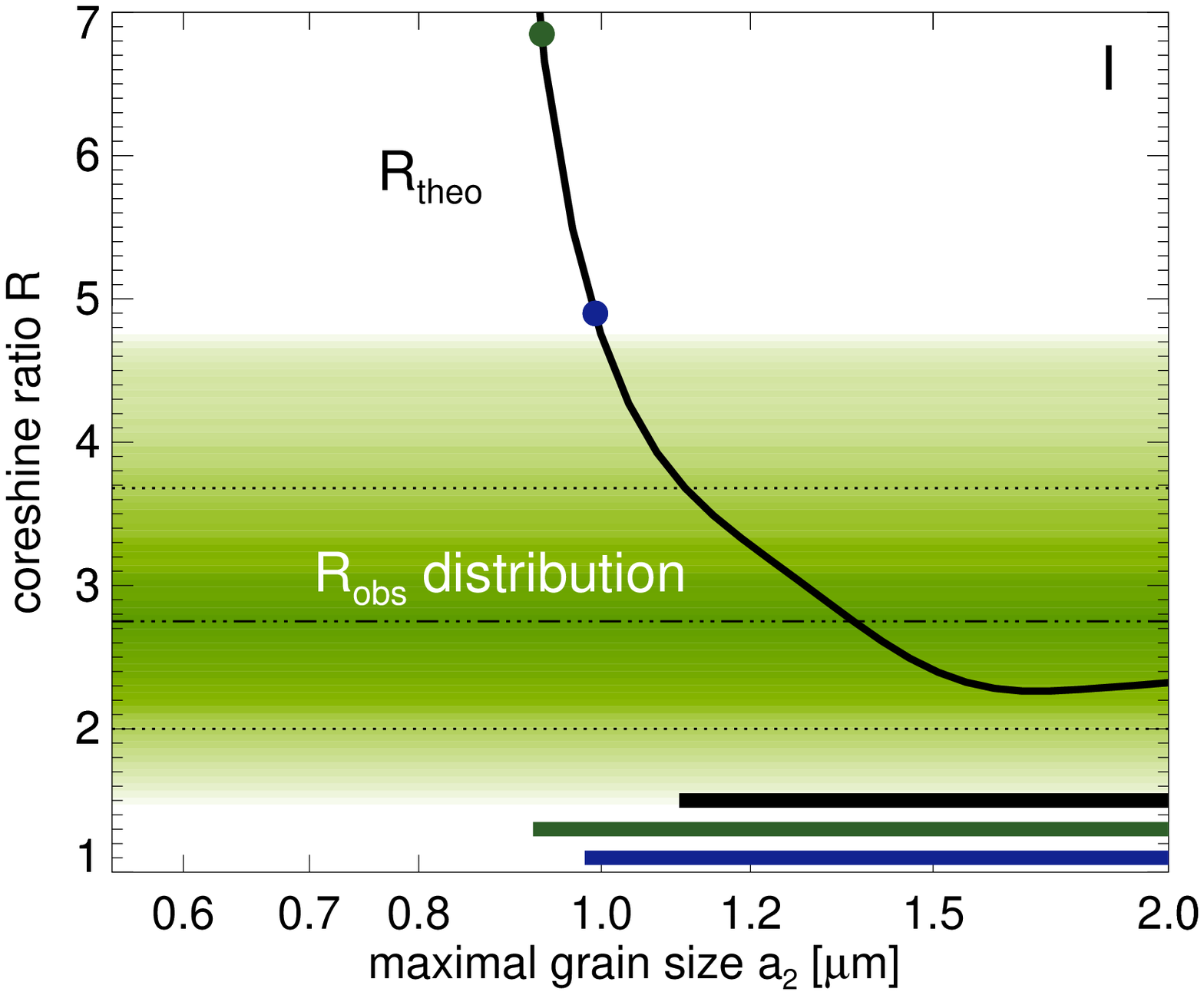}
}
\caption{
Data analysis for L1517A. For detailed legend see Fig.~\ref{L260}.
The core was located at the IRAC 1 and 2 image borders leading to a dark zones in Panels A,B, and G, (Here
the $R$ values in the dark lower and right zones are not meaningful.)
}
\label{L1517}
\end{figure*}
The L1517 cloud in Taurus contains four filaments with five embedded cores
\citep{2011A&A...533A..34H}.
For the modeling we have chosen core A
since it was visible in both bands in our new deep observation images.
The coreshine of all Taurus cores (in our sample L1517A and L1512, L1544, L1506C, and L1498 with $l$ around 180$^\circ$ discussed later) 
benefit from a side maximum in the dust- model phase function of large grains 
that predicts enhanced backward scattering of the strong 
GC radiation
\citep[see][]{2014A&A...563A.106S}.

The core is located at the border of the 3.6 $\mu$m image (Fig.~\ref{L1517}, 
see also Panel C, "darkclouds\_IRAC": Program ID 94, PI C. Lawrence),
but were able to select a region that includes the core center and most of the core. 
We considered using cold 3.6 $\mu$m data instead, but decided to stay with the warm data for the comparison to minimize the impact of systematic effects between the observing runs because the data show no strong increase in noise compared to the warm 4.5 $\mu$m data.
There is no indication of any depression, and the mass and column densities point toward
a maximum optical depth value for extinction below 1. We therefore use single-scattering RT.
The cuts show a clear constant factor between the two bands in both
cut directions, as one would expect from optically thin scattered light where 
the SFB is proportional to the column density. 
This should give rise to a narrow $R$ distribution. 
On the contrary, Panel H shows a broader $R$ distribution than L260 
(FWHM is about 1.5, peak at 2.75). We attribute this to the source having the strongest background in the sample.
The distribution also is asymmetric with a strong tail toward higher $R$. Panel G 
shows an increase in $R$ values in the upper part of the white frame
Probably because of the loss of the 4.5 $\mu$m coreshine signal, and we attribute the asymmetry
in Panel H to this signal loss.
The model is able to meet all observational constraints for  
$a_2$ values above 1.1  $\mu$m.

\subsection{L1512}
\begin{figure*}
\vbox{
\includegraphics[width=6cm]{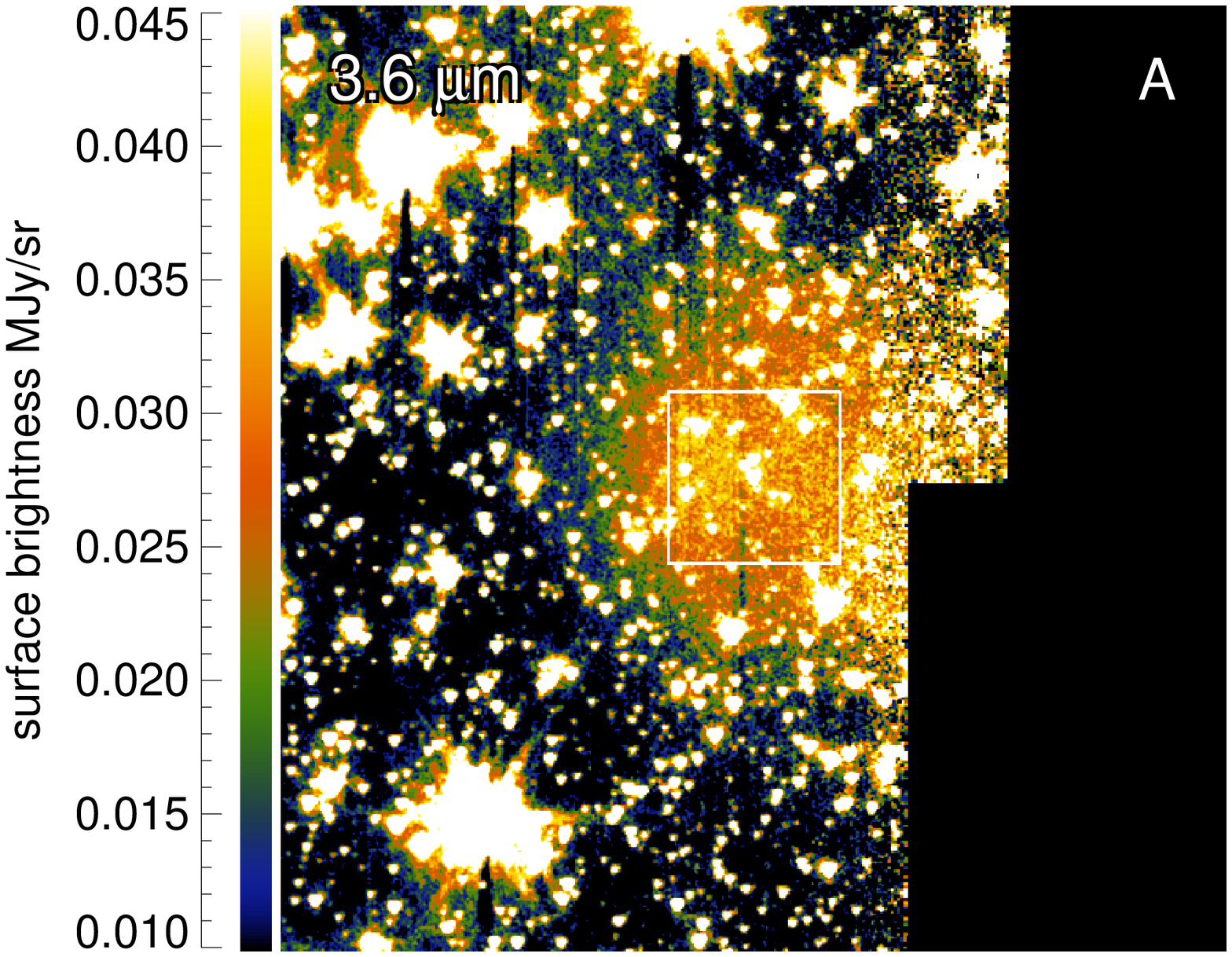}
\includegraphics[width=6cm]{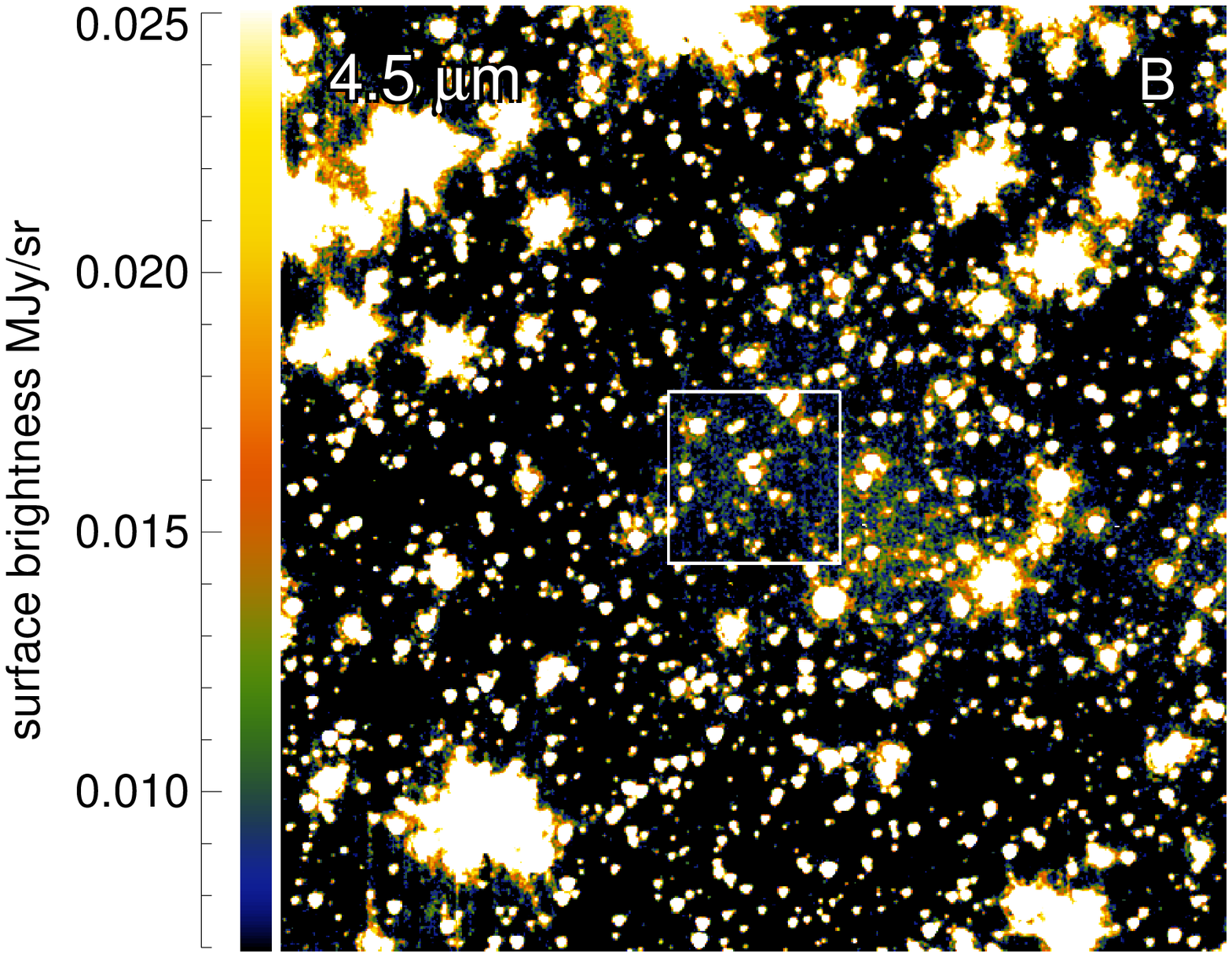}
\includegraphics[width=6cm]{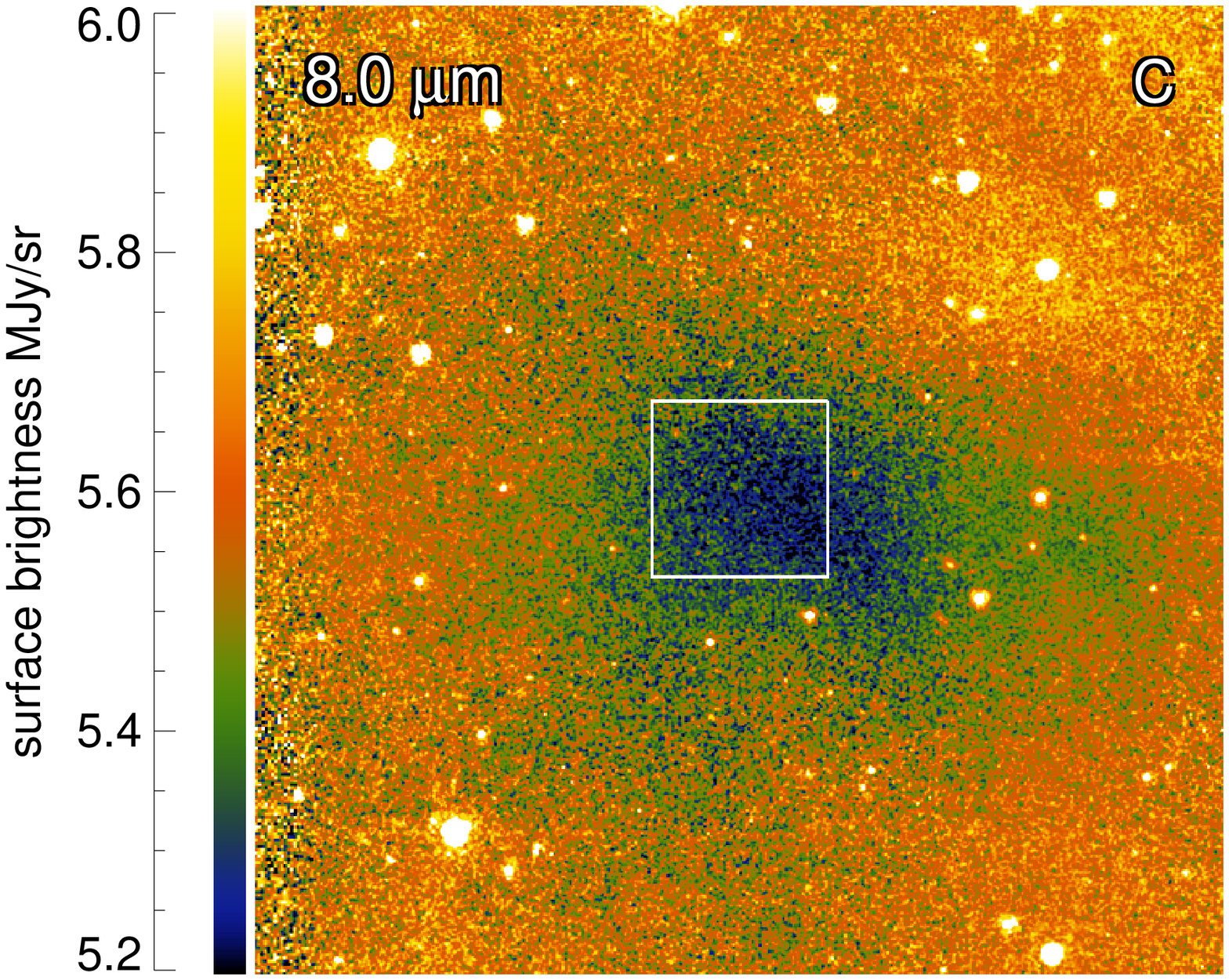}
}
\vskip 0.1cm
\vbox{
\includegraphics[width=6cm]{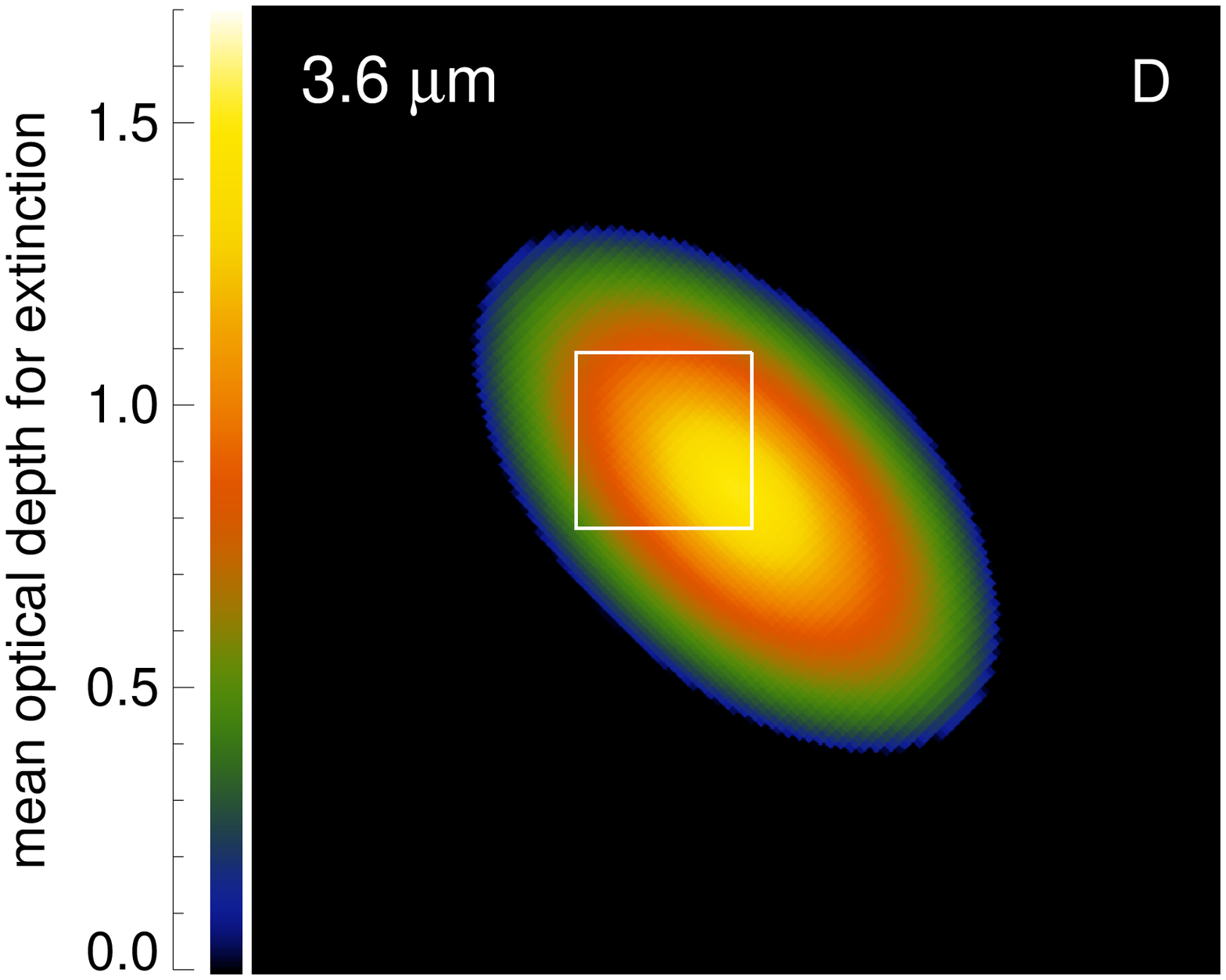}
\includegraphics[width=6cm]{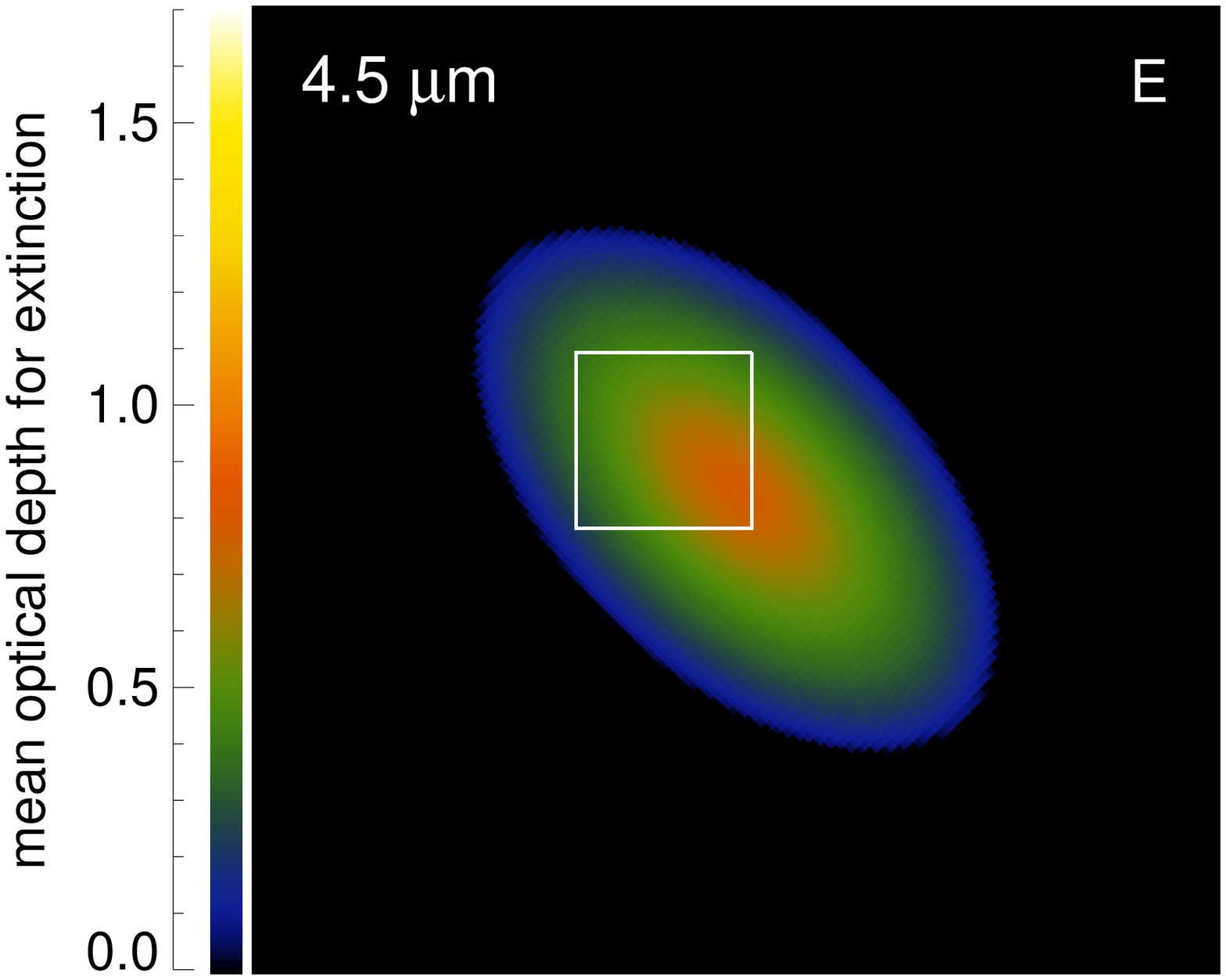}
\includegraphics[width=6cm]{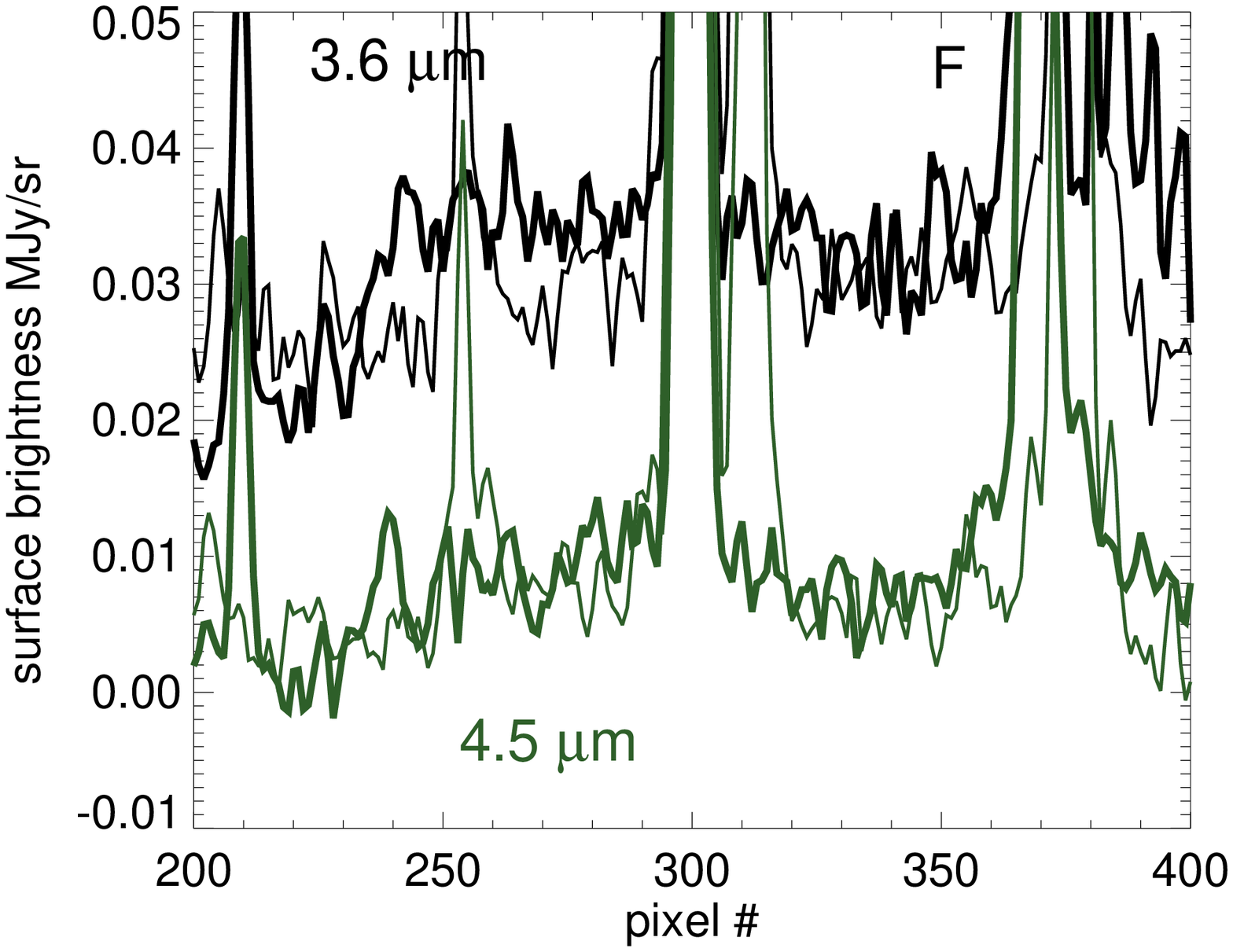}
}
\vskip 0.1cm
\vbox{
\includegraphics[width=6cm]{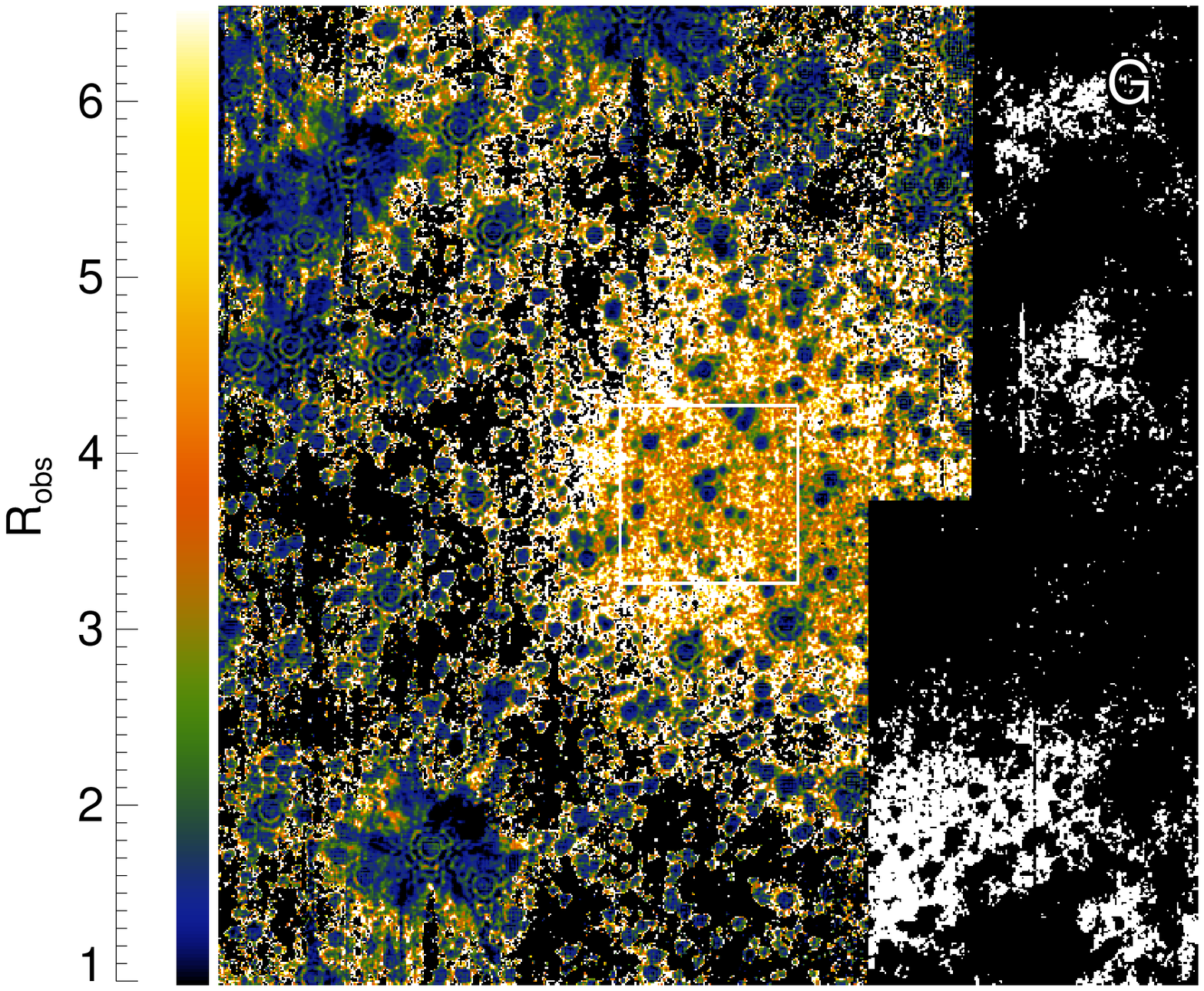}
\includegraphics[width=6cm]{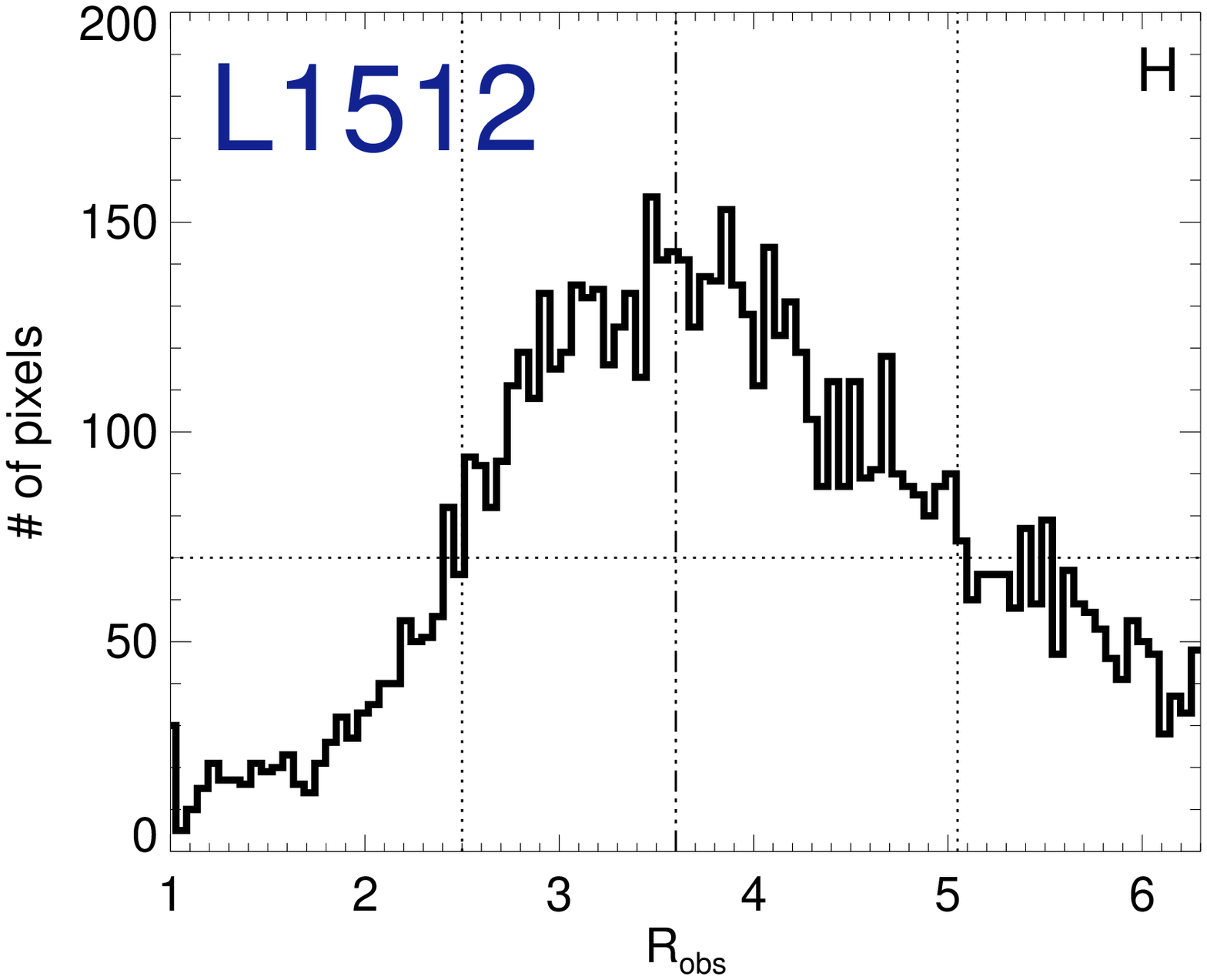}
\includegraphics[width=6cm]{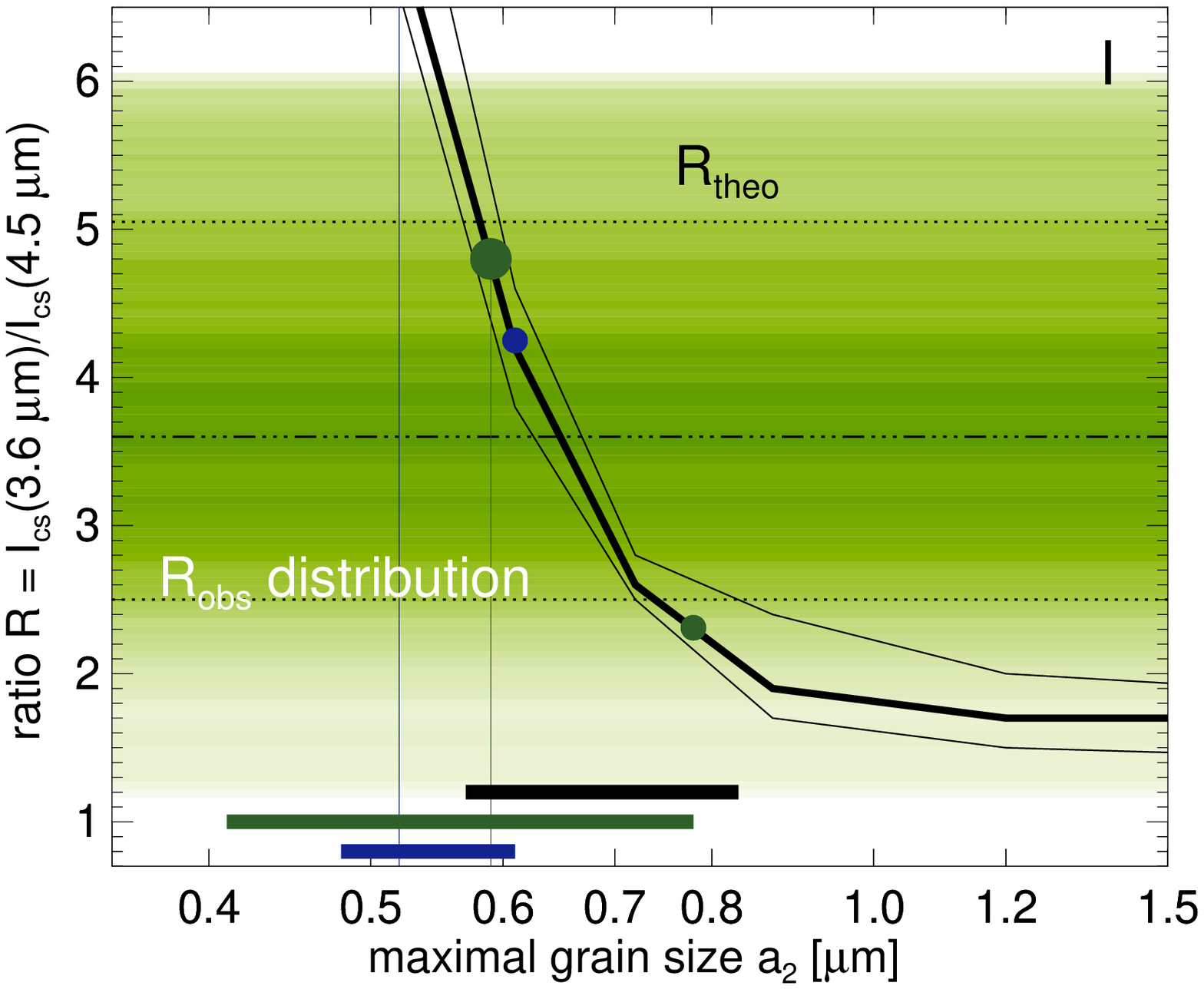}
}
\caption{
Data analysis for L1512 (CB27). For detailed legend see Fig.~\ref{L260}. Panel A contains
an image border that leads to meaningless ratios in Panel G in the dark righthand area. Panel I compares
the observed (dashed) and theoretical (solid) $R$ distribution characterized by the mean value (thick)
and the FWHM values (thin).
        }
\label{L1512}
\end{figure*}
This Taurus core is located close to the Galactic plane and has an enhanced background
(second strongest 3.6 $\mu$m background in the sample).
As for L1517A, we use the new warm data
showing the core at the image border instead of using center-image 3.6 $\mu$m cold {\it Spitzer} data in Fig.~\ref{L1512}.
The cold {\it Spitzer} map at 8 $\mu$m shown in Panel C is from the "darkclouds\_IRAC" program (ID 94, PI C. Lawrence).
Based on the core property estimates from Table \ref{table:1}, the optical depth analysis
indicates that the central part of an ellipsoidal representative of
L1512 would see optical depth above 1, and therefore extinction and second scattering
make full RT necessary. 
The image in Panel A and cuts like the one given in Panel F show little central depression though, but strong stellar PSFs make this analysis difficult.
We therefore performed full RT modeling.
The 4.5 $\mu$m signal is just about a factor 2 above the noise with the image showing
just hints of excess of emission.
The $R$ distributions are correspondingly broad (FWHM of about 2.6 around $R$=3.6)
and conclusions from the modeling must be made with caution.
The full RT distribution for $R$ overlaps with the observed $R$ values for maximum grains
between 0.48 and 0.61 $\mu$m.

\subsection{L1544}
\begin{figure*}
\vbox{
\includegraphics[width=6cm]{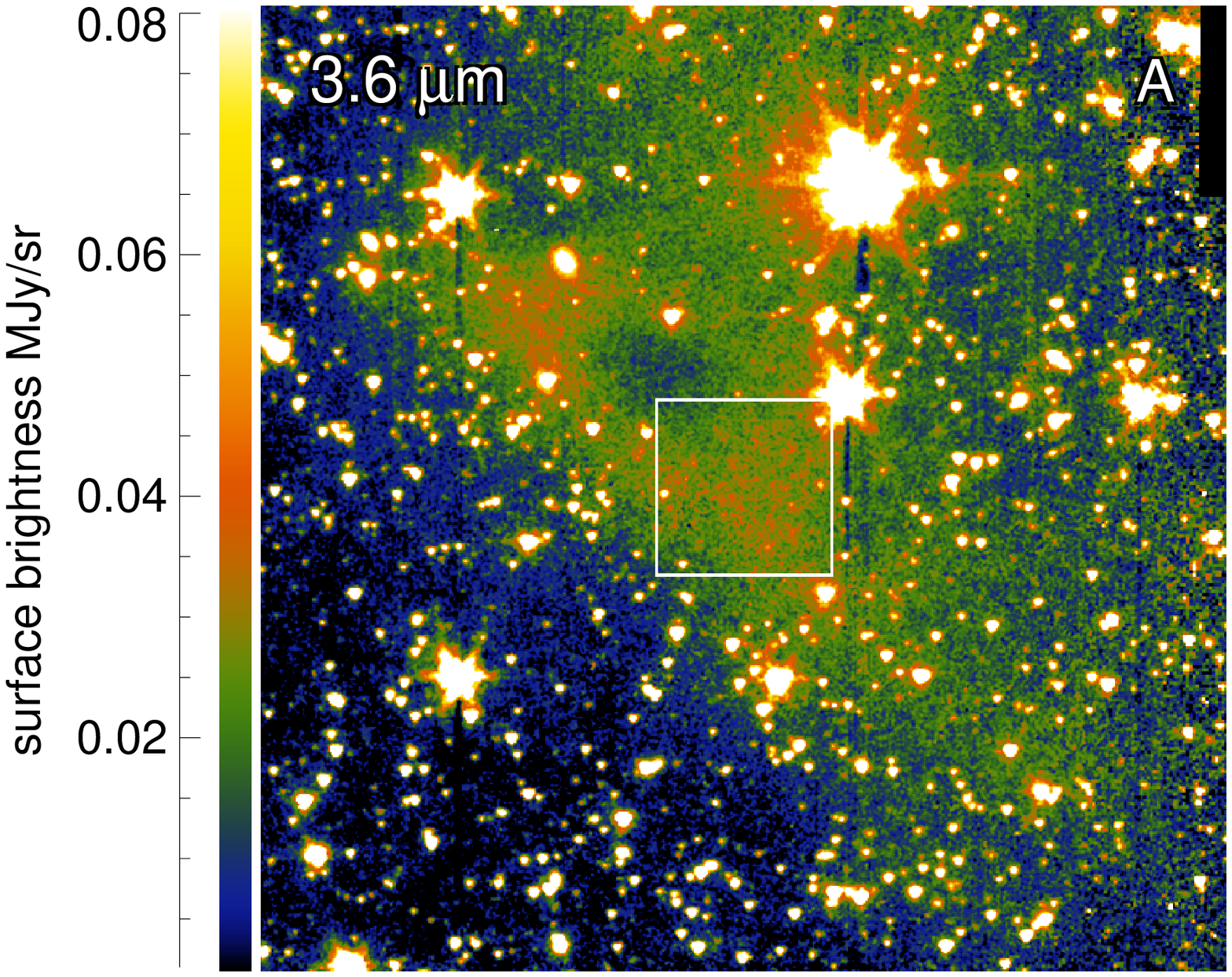}
\includegraphics[width=6cm]{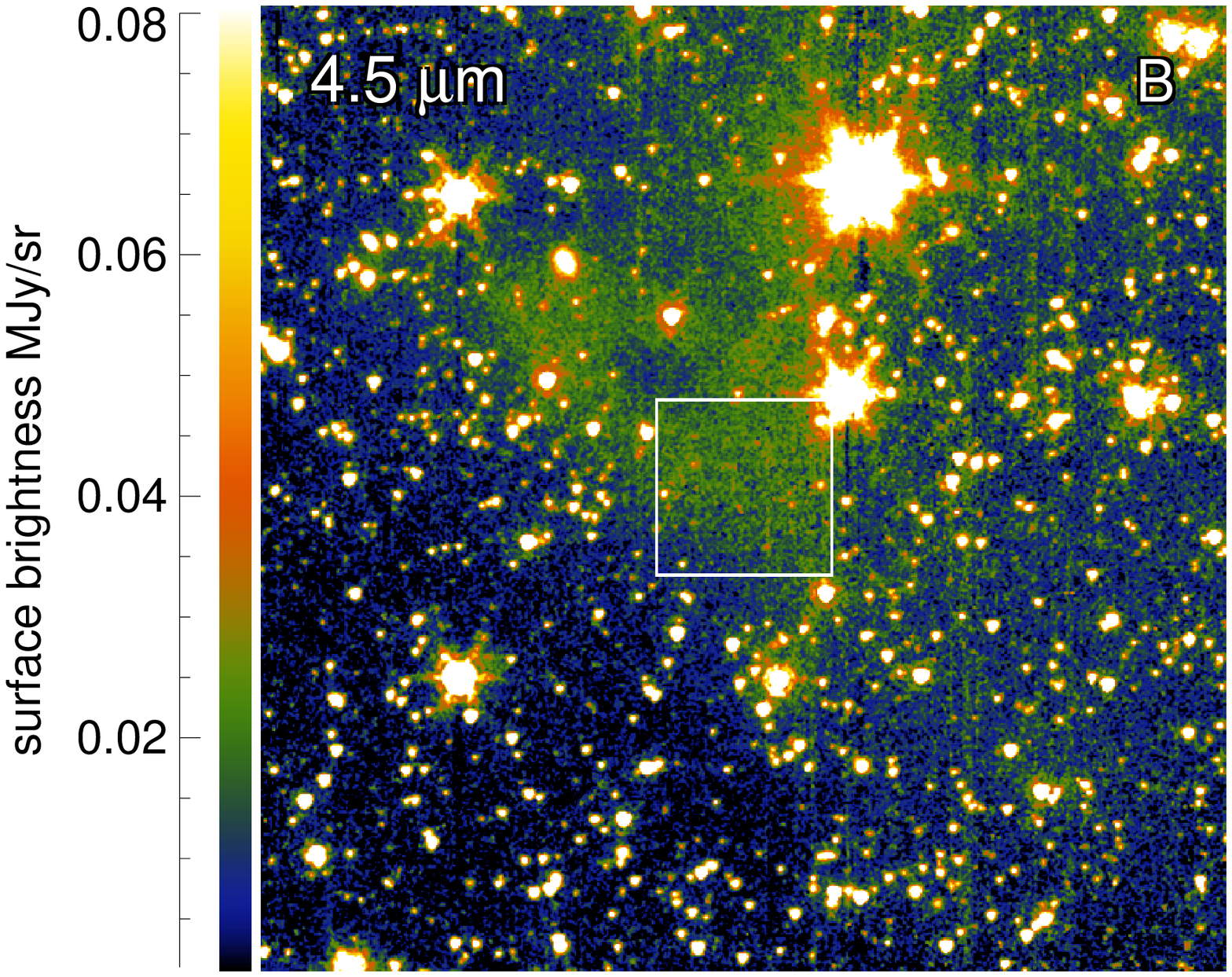}
\includegraphics[width=6cm]{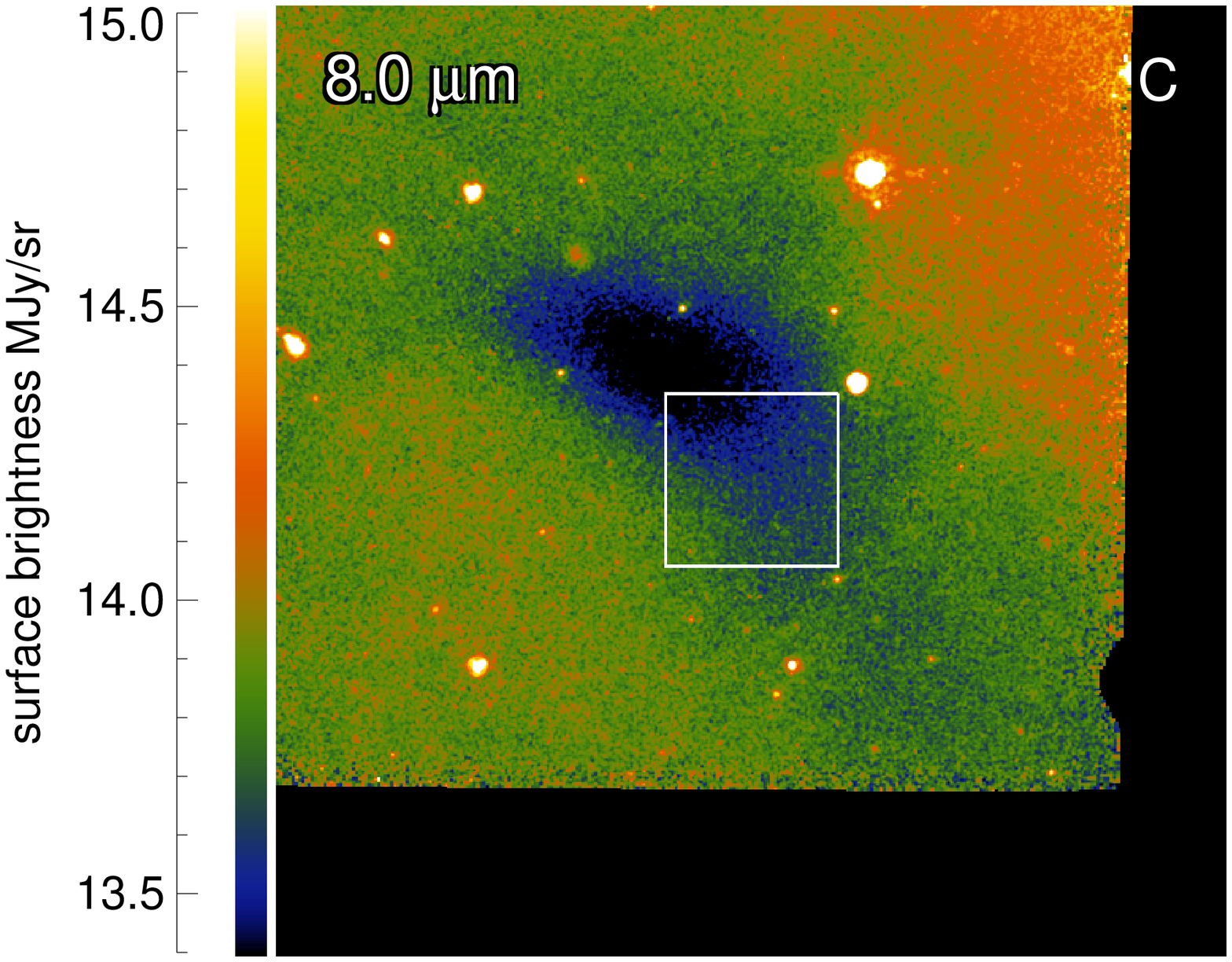}
}
\vskip 0.1cm
\vbox{
\includegraphics[width=6cm]{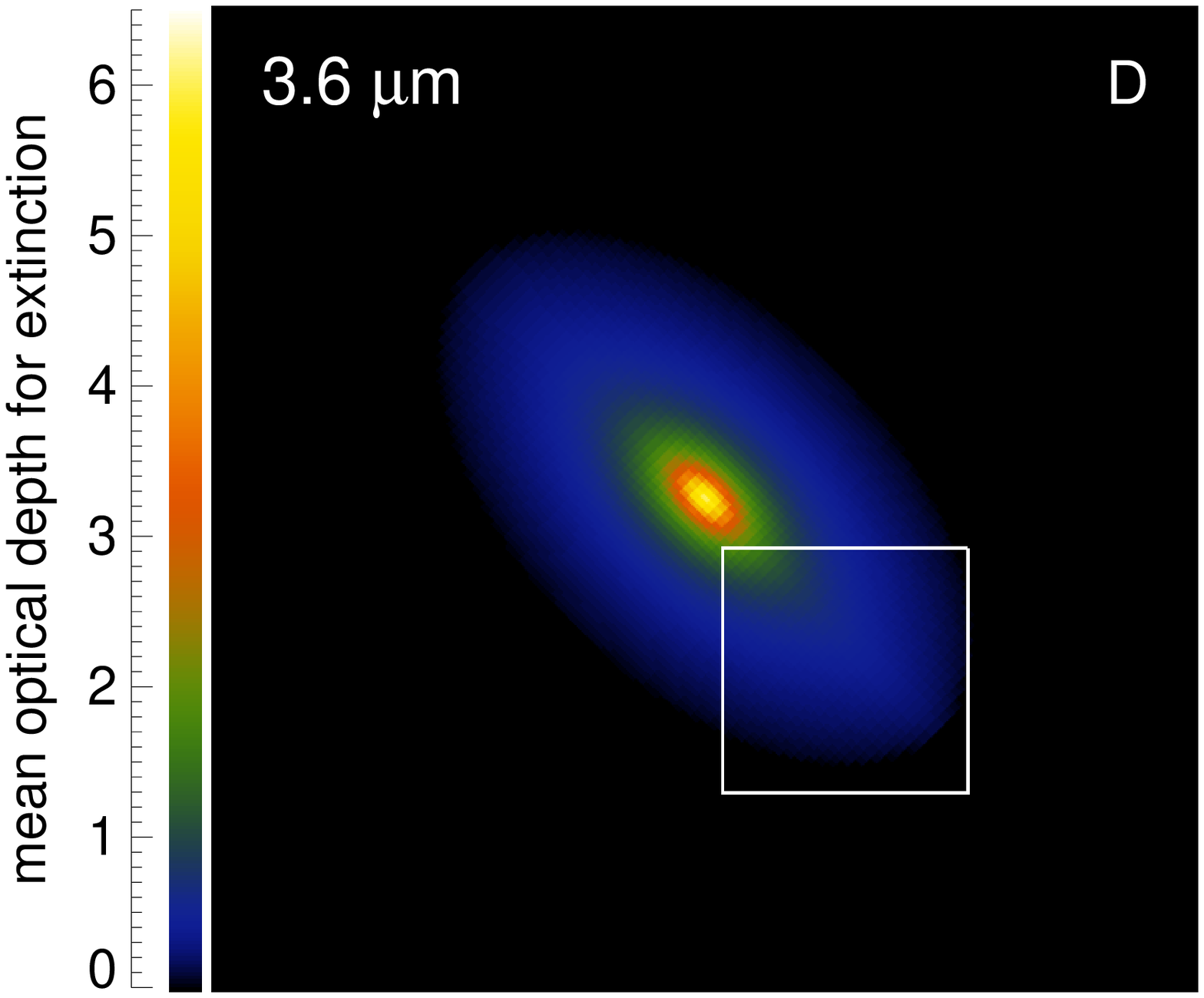}
\includegraphics[width=6cm]{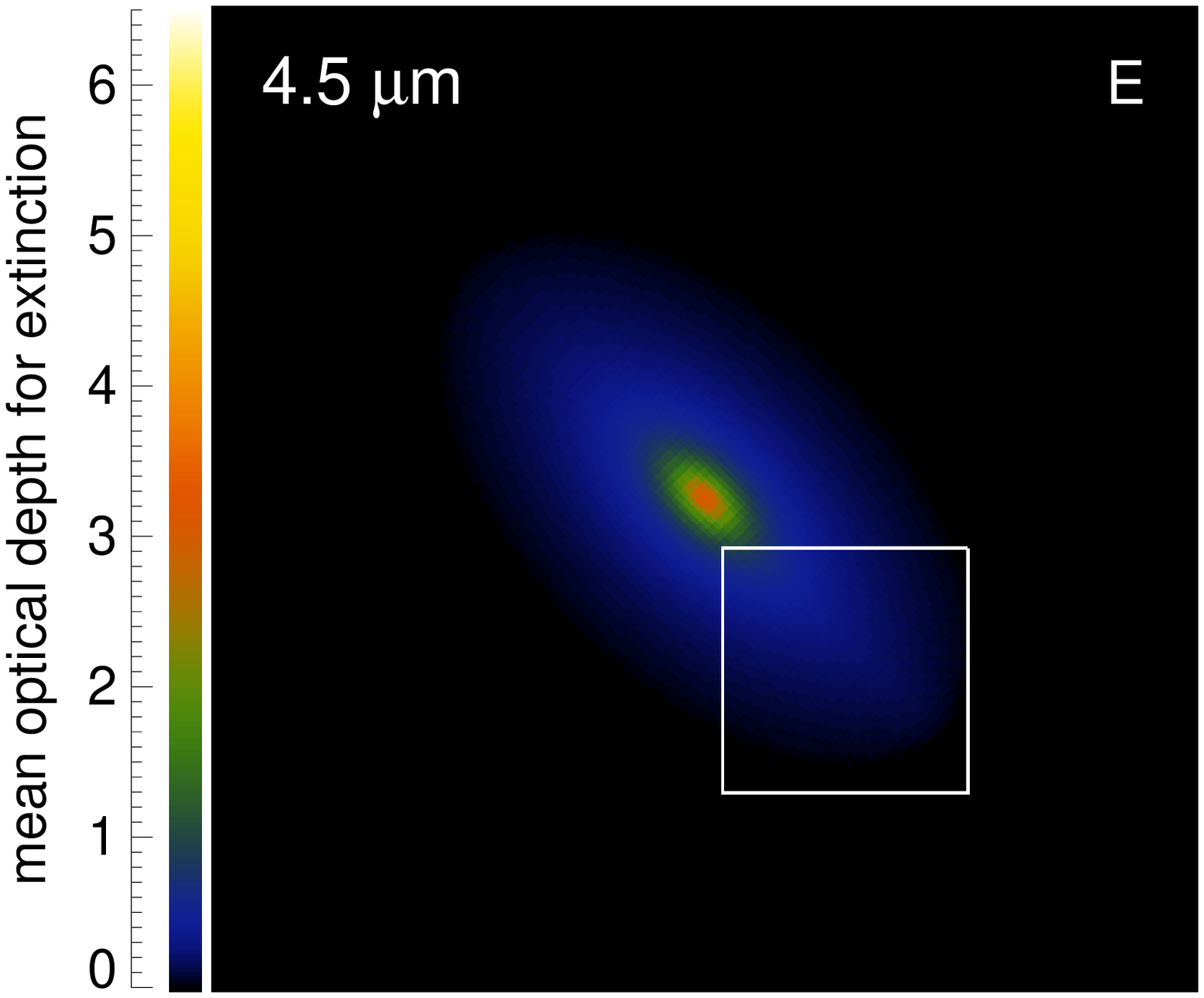}
\includegraphics[width=6cm]{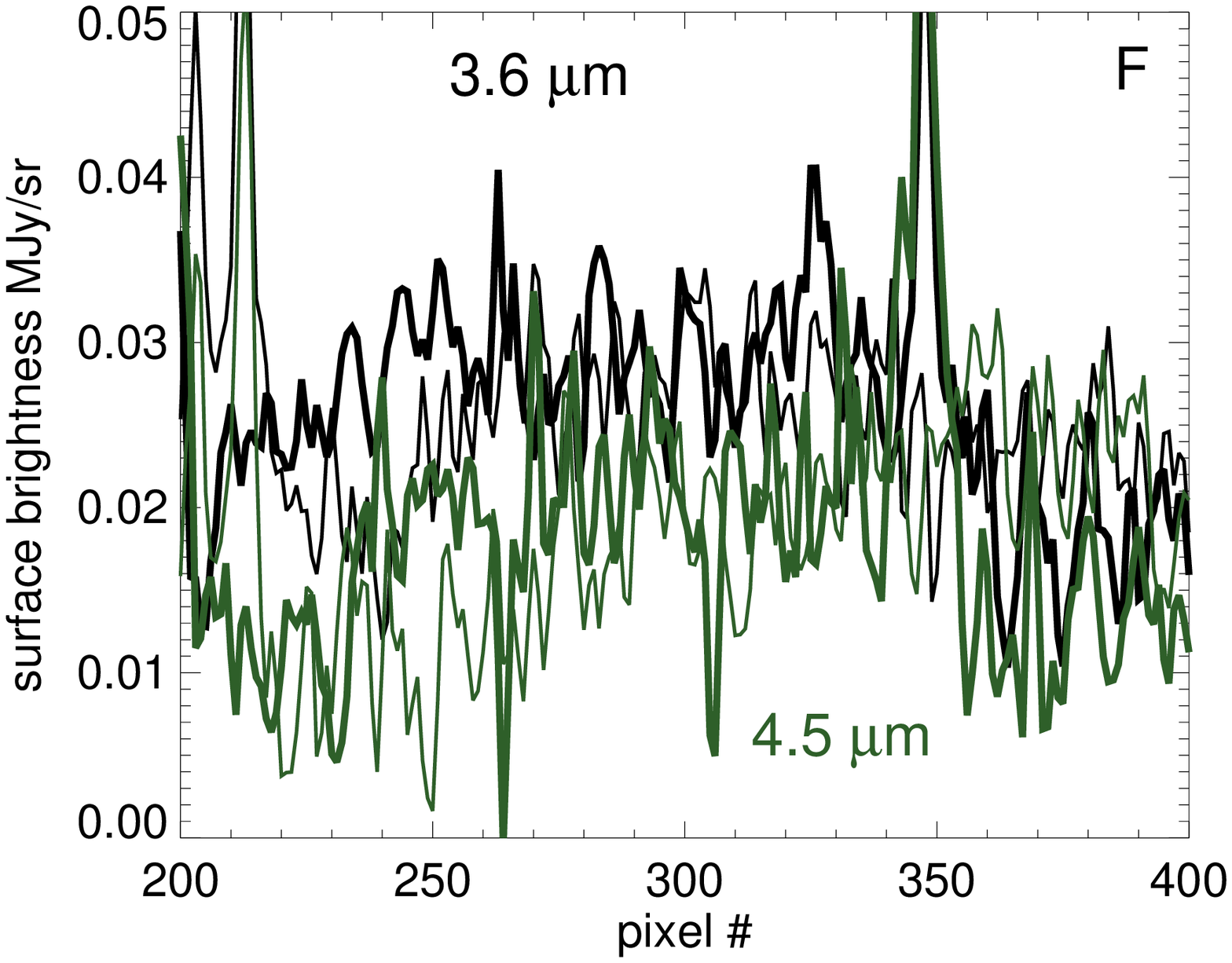}
}
\vskip 0.1cm
\vbox{
\includegraphics[width=6cm]{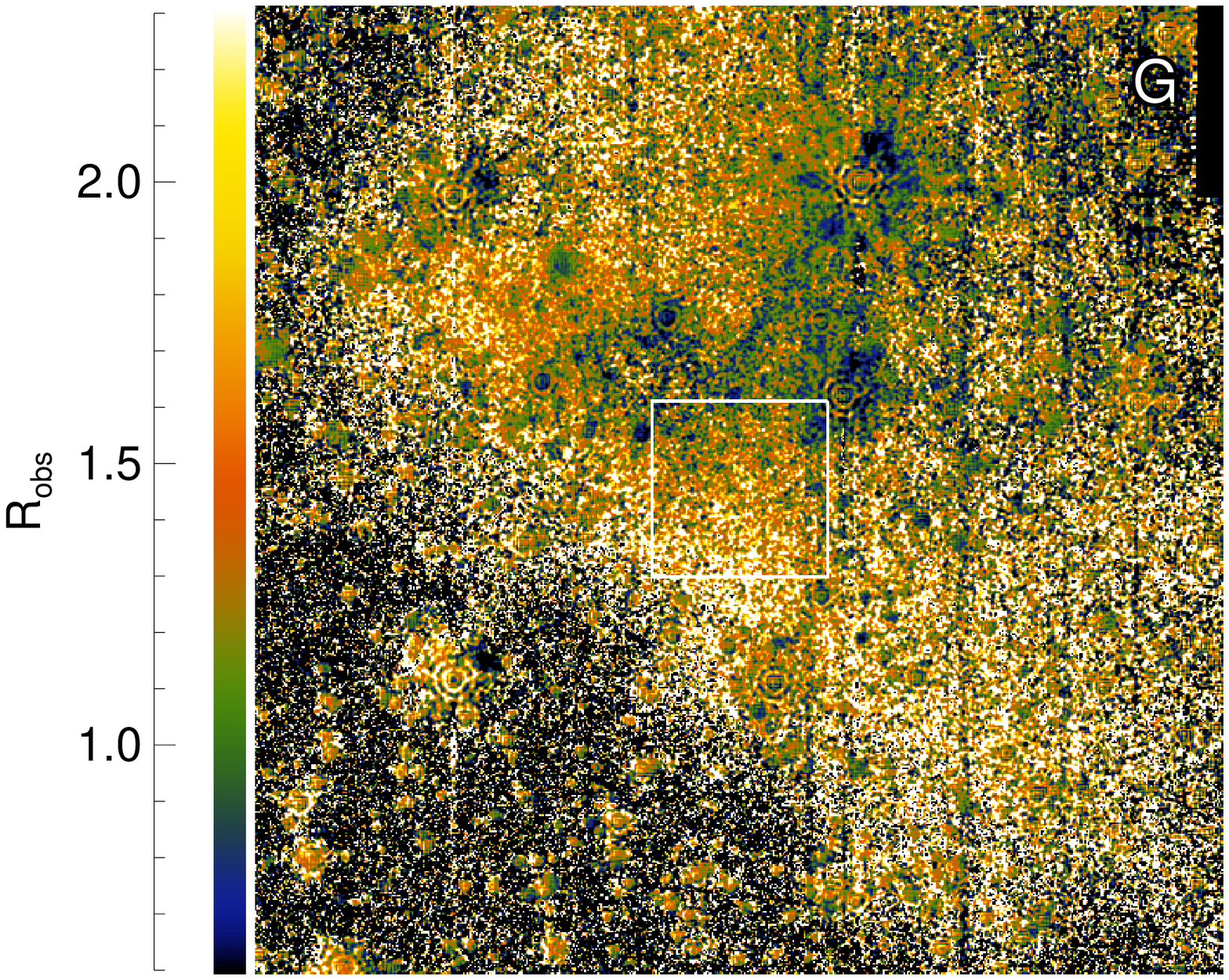}
\includegraphics[width=6cm]{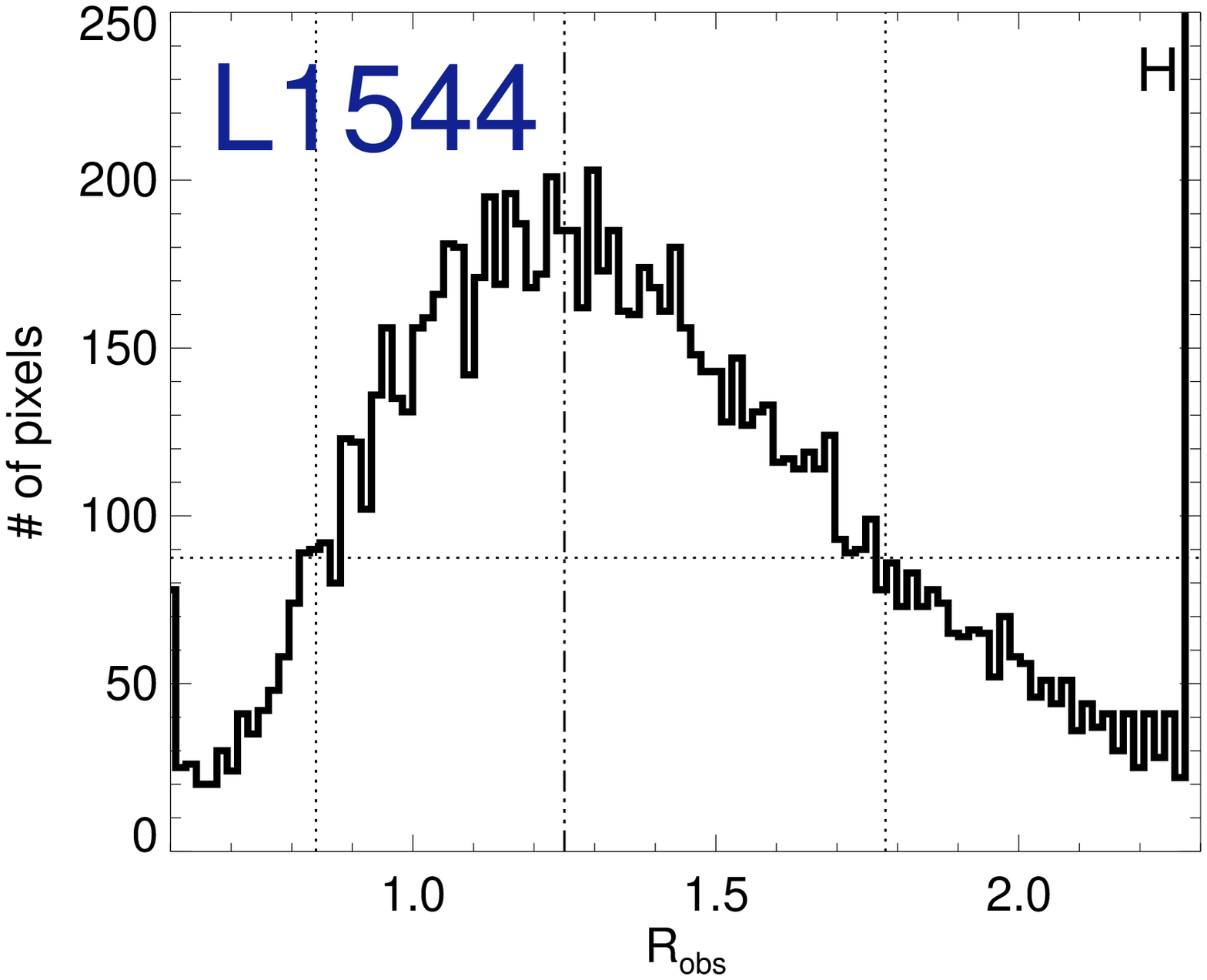}
\includegraphics[width=6cm]{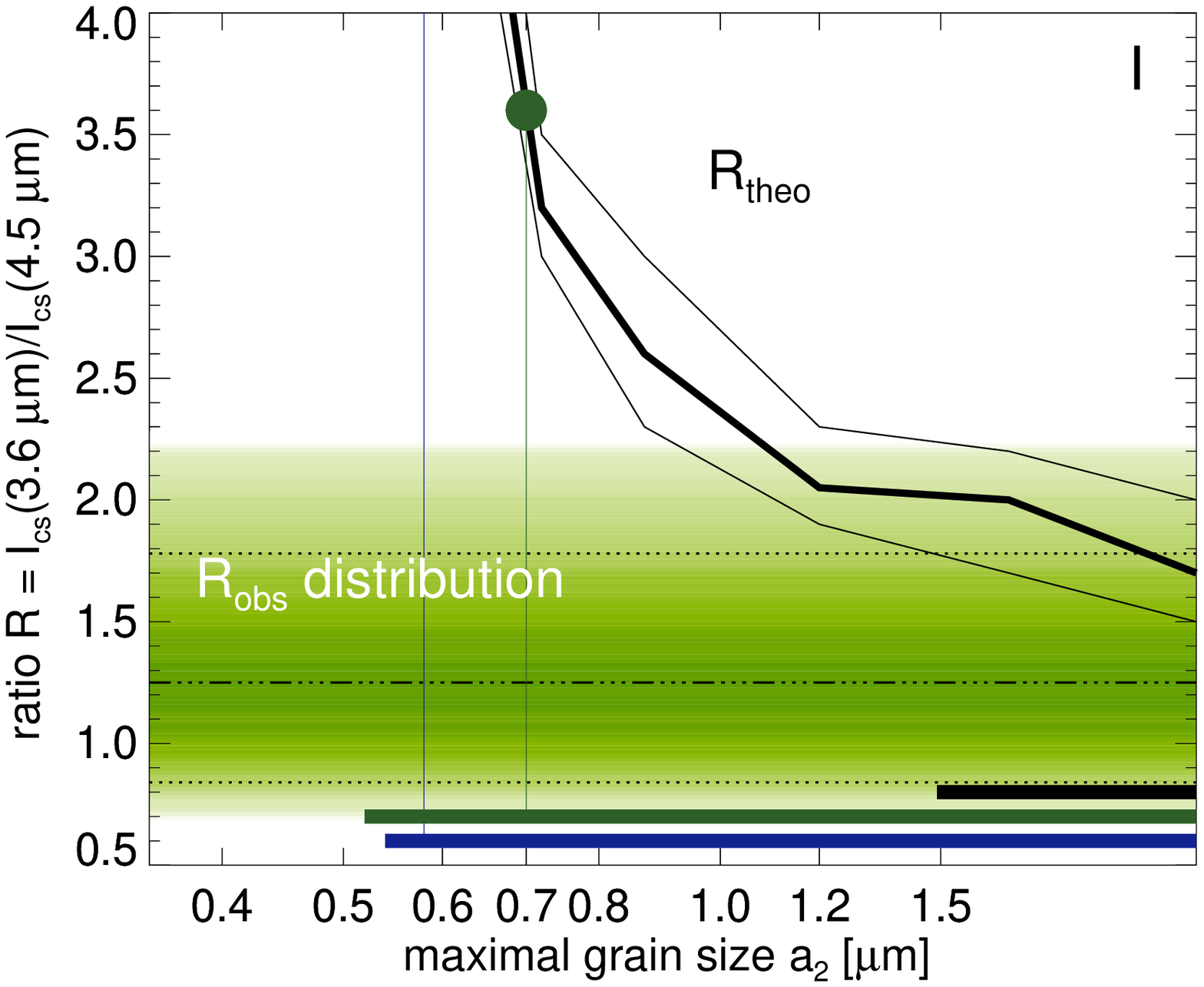}
}
\caption{
Data analysis for L1544. For detailed legend see Fig.~\ref{L260}. The core
shows an inner surface brightness depression, and the white frame was
chosen outside this region. The surface brightness signal is noisy owing to the many
stars in the field.
        }
\label{L1544}
\end{figure*}
The Taurus core L1544 is one of the few known cores across the sky that show
a prominent inner depression of SFB in both
bands. 
While the mass seems to be comparable to that of L1512 showing no clear depression,
the central SFB of L1544 very likely drops because of optical depth effects of the
centrally condensed radial profile. The 8 $\mu$m map shows a strong extinction pattern at the core center ("cores2deeper": Program ID 20386, PI P. Myers).
This would mean that we can expect moderate
optical depth in the outer parts of the elliptically shaped core and a strong rise in
$\tau$ near the center. Using a profile with a power-law index of -2.5 in the outer parts
and a flattening starting at the kink radius of 1 kau, the mean optical depth maps shown
in Panels D and E in Fig.~\ref{L1544} provide such a pattern. Scaling $\tau$ with the estimated central column 
density of L1544, we use a $\tau$ limit of 6 at $\lambda$=3.6 $\mu$m.
The white square region was placed outside the depression, and we use full RT.
The derived ratios are low with a mean value of 1.25, and the FWHM is comparable to
that of L260 (about 0.95).
But as is visible in Panel F, the data are extremely noisy with a dense stellar 
background field interfering and with a loss of the
4.5 $\mu$m signal in the lower white frame part that causes a
wing in the $R$ distribution at high $R$.
The model suggests large grains beyond $a_2$>1.5 $\mu$m to reproduce the SFBs and ratios.

\subsection{L1506C}
\begin{figure*}
\vbox{
\includegraphics[width=6cm]{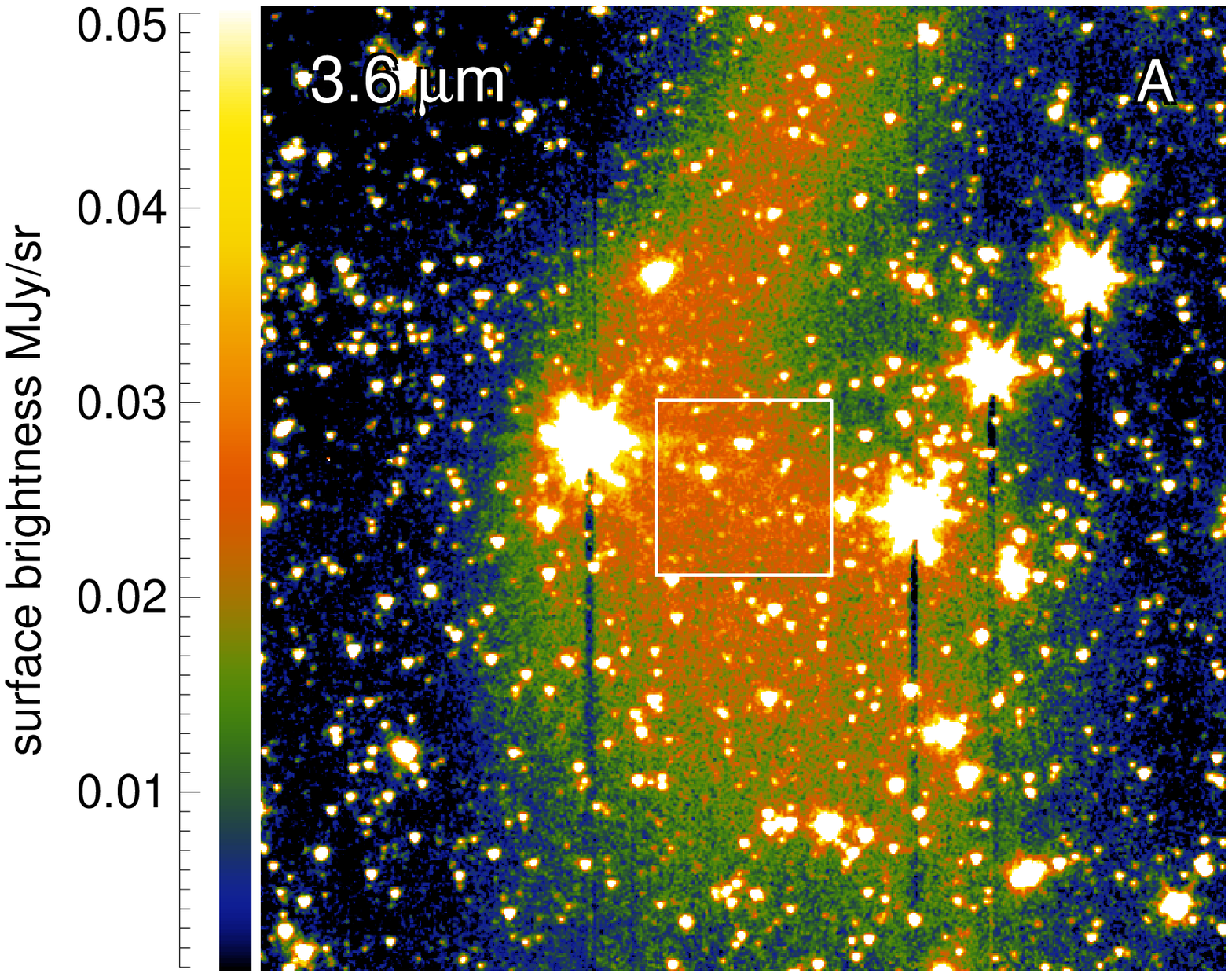}
\includegraphics[width=6cm]{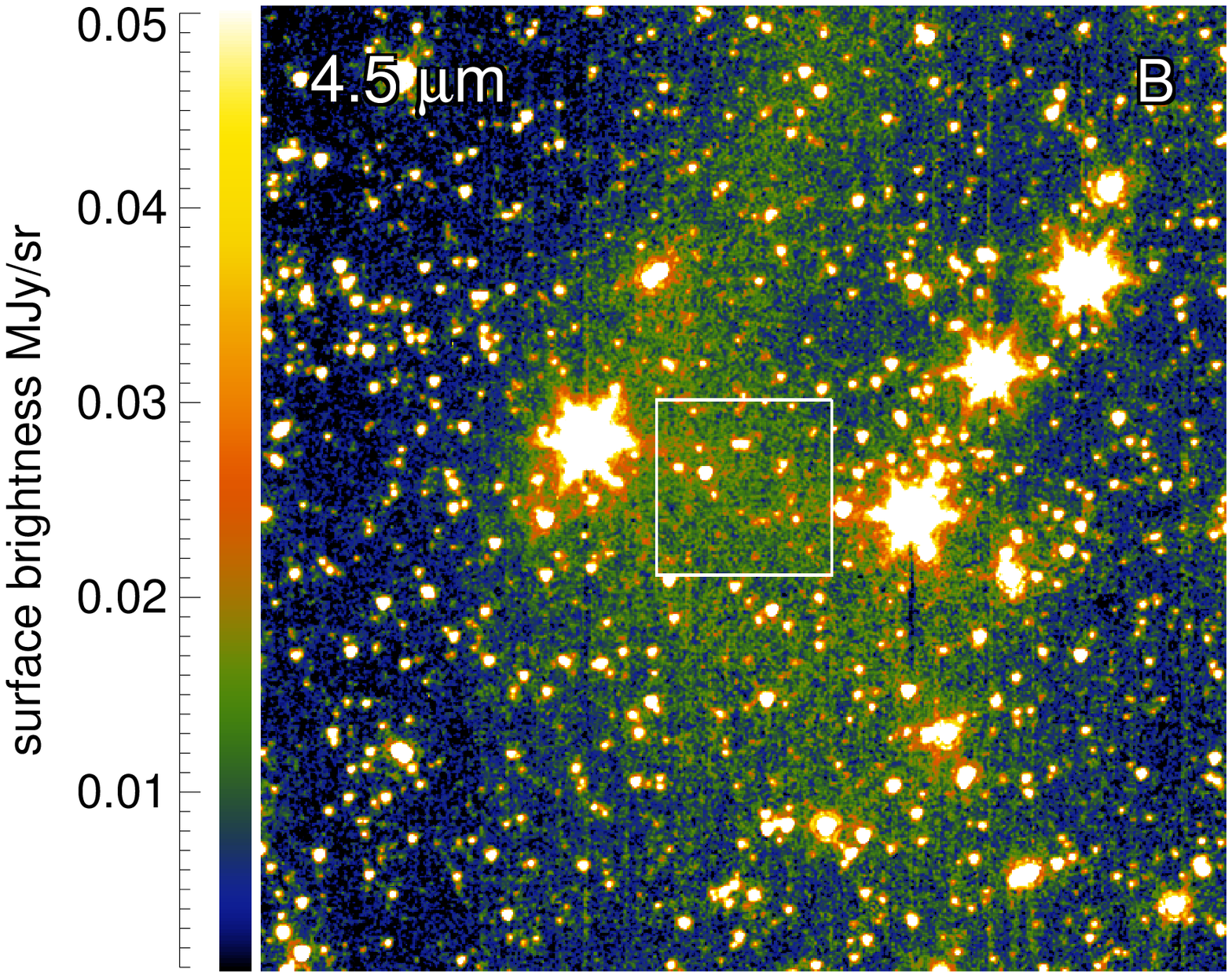}
\includegraphics[width=6cm]{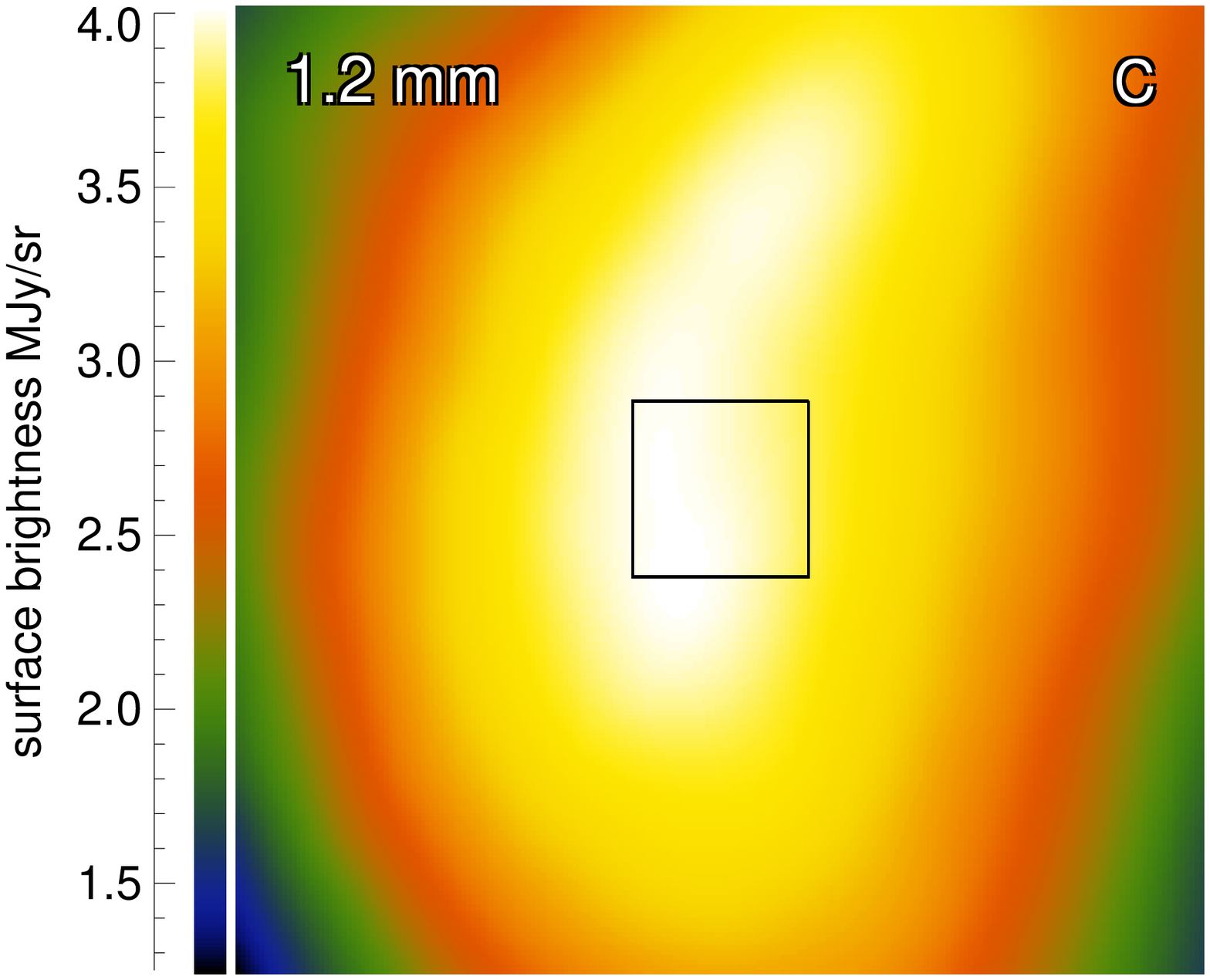}
}
\vskip 0.1cm
\vbox{
\includegraphics[width=6cm]{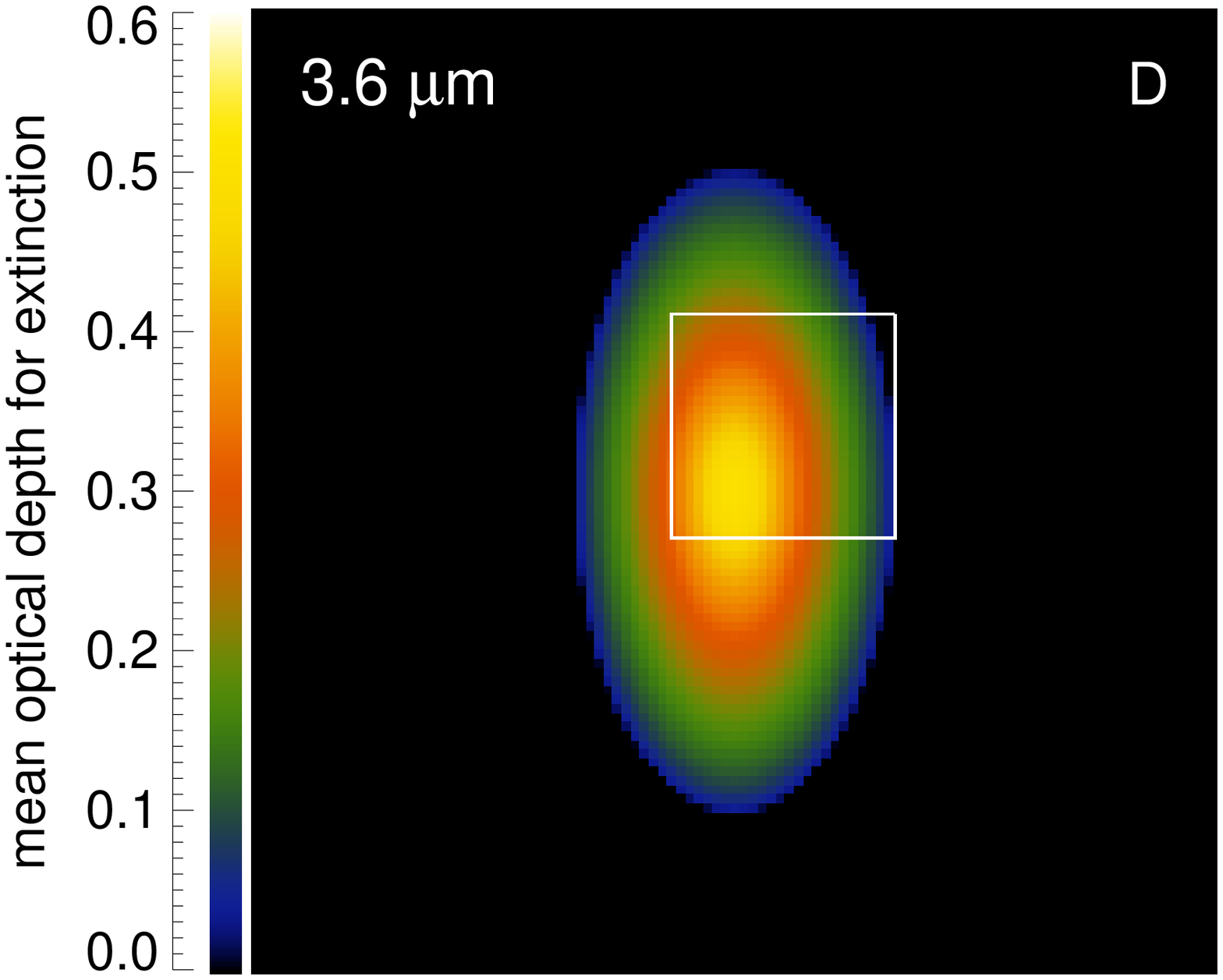}
\includegraphics[width=6cm]{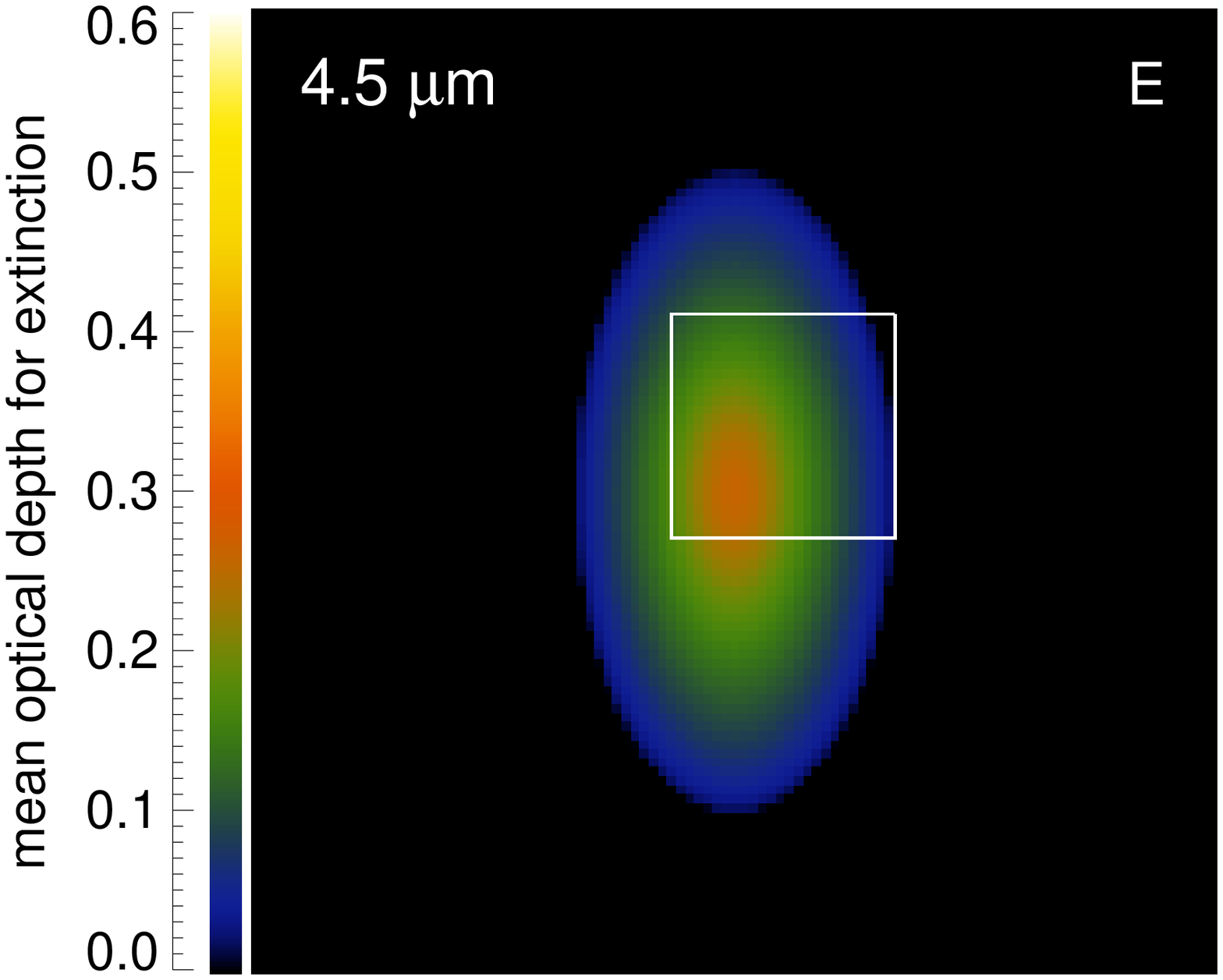}
\includegraphics[width=6cm]{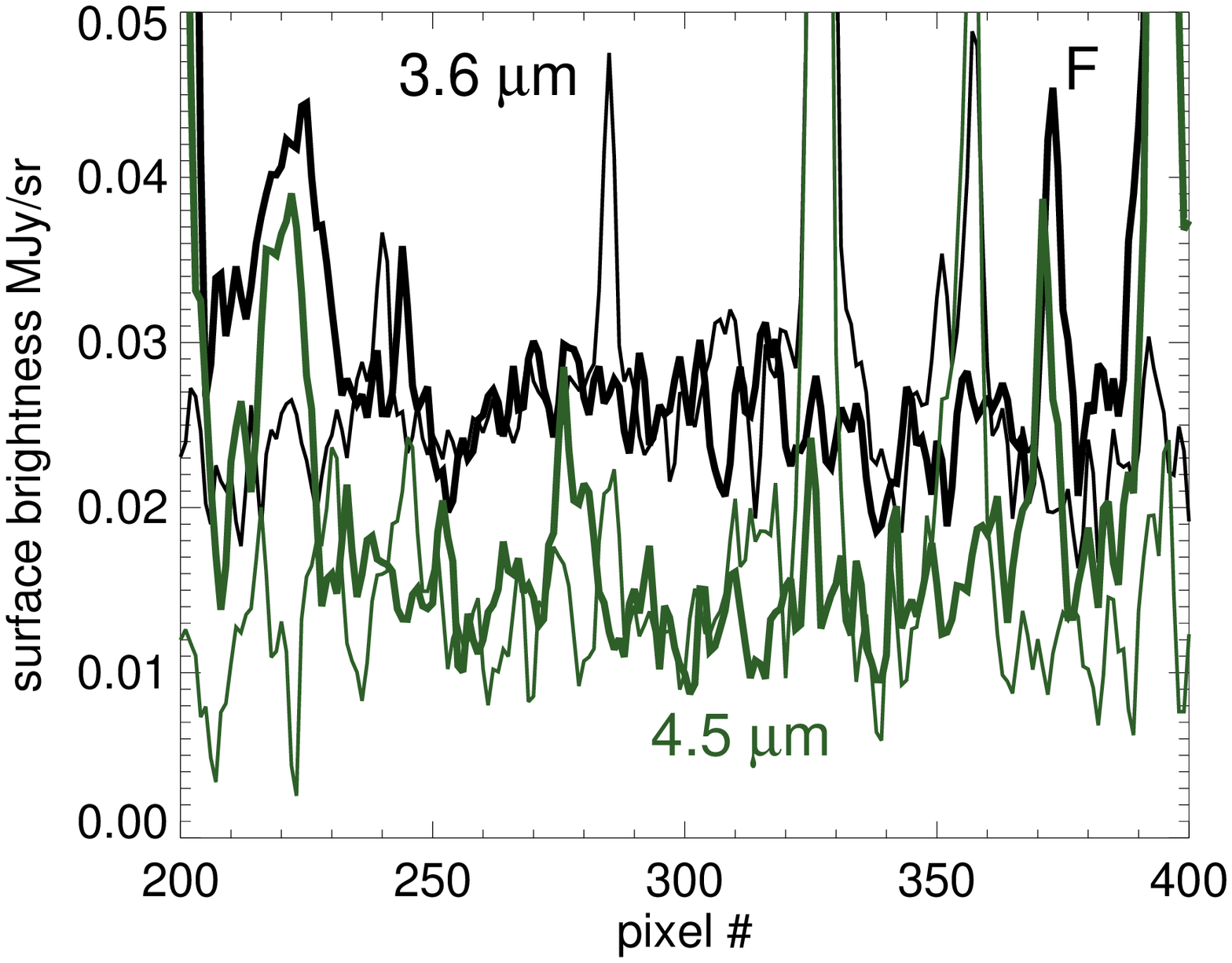}
}
\vskip 0.1cm
\vbox{
\includegraphics[width=6cm]{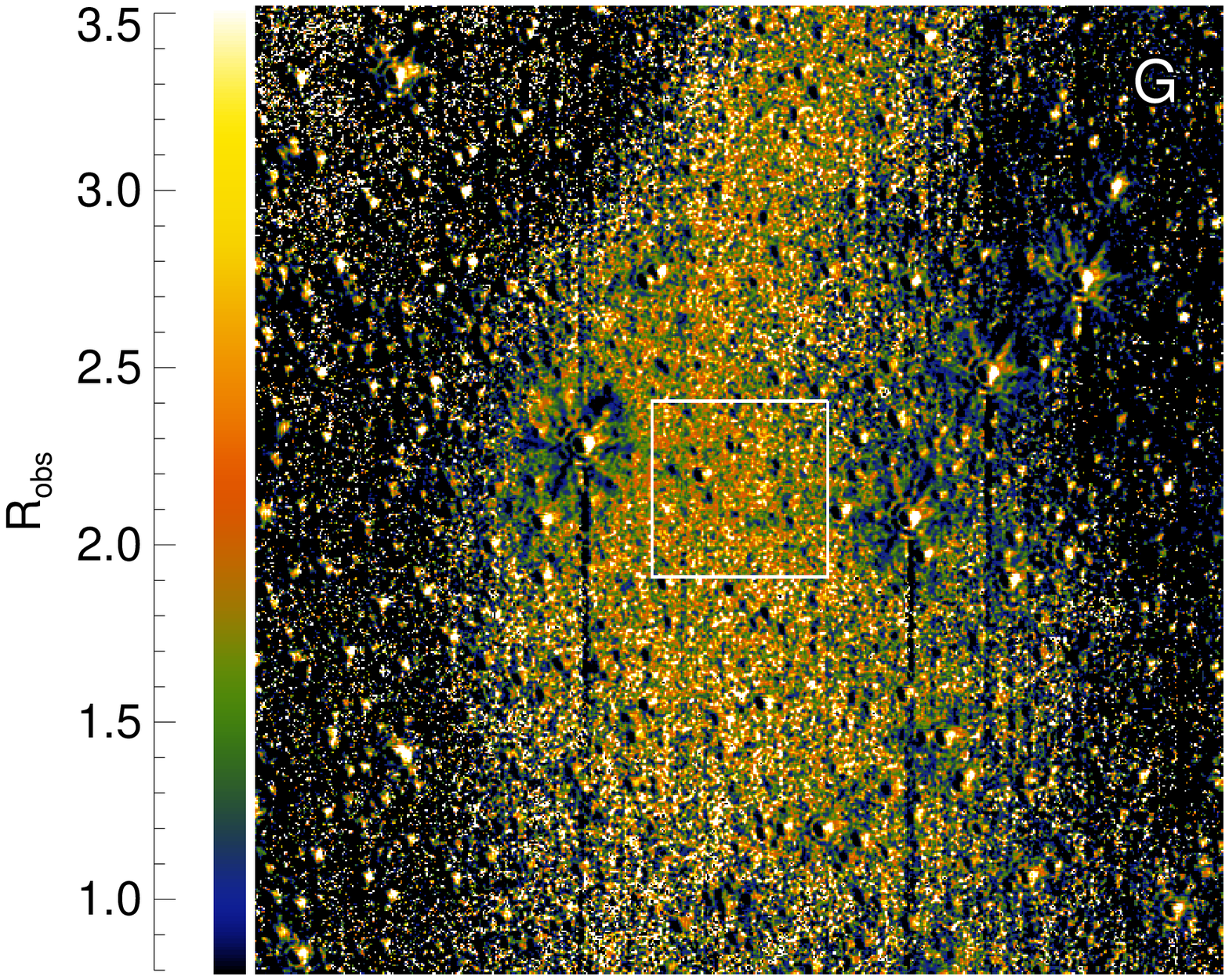}
\includegraphics[width=6cm]{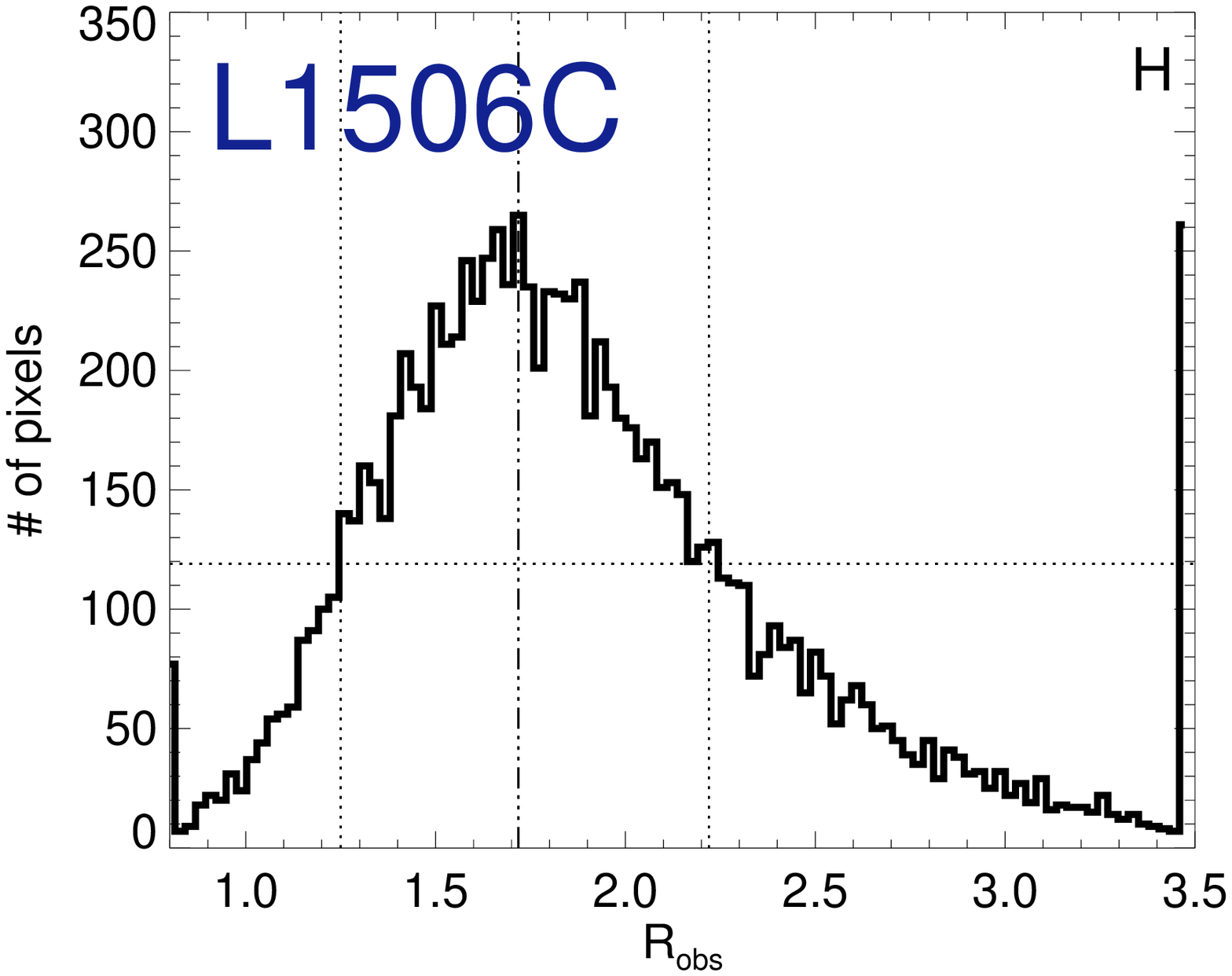}
\includegraphics[width=6cm]{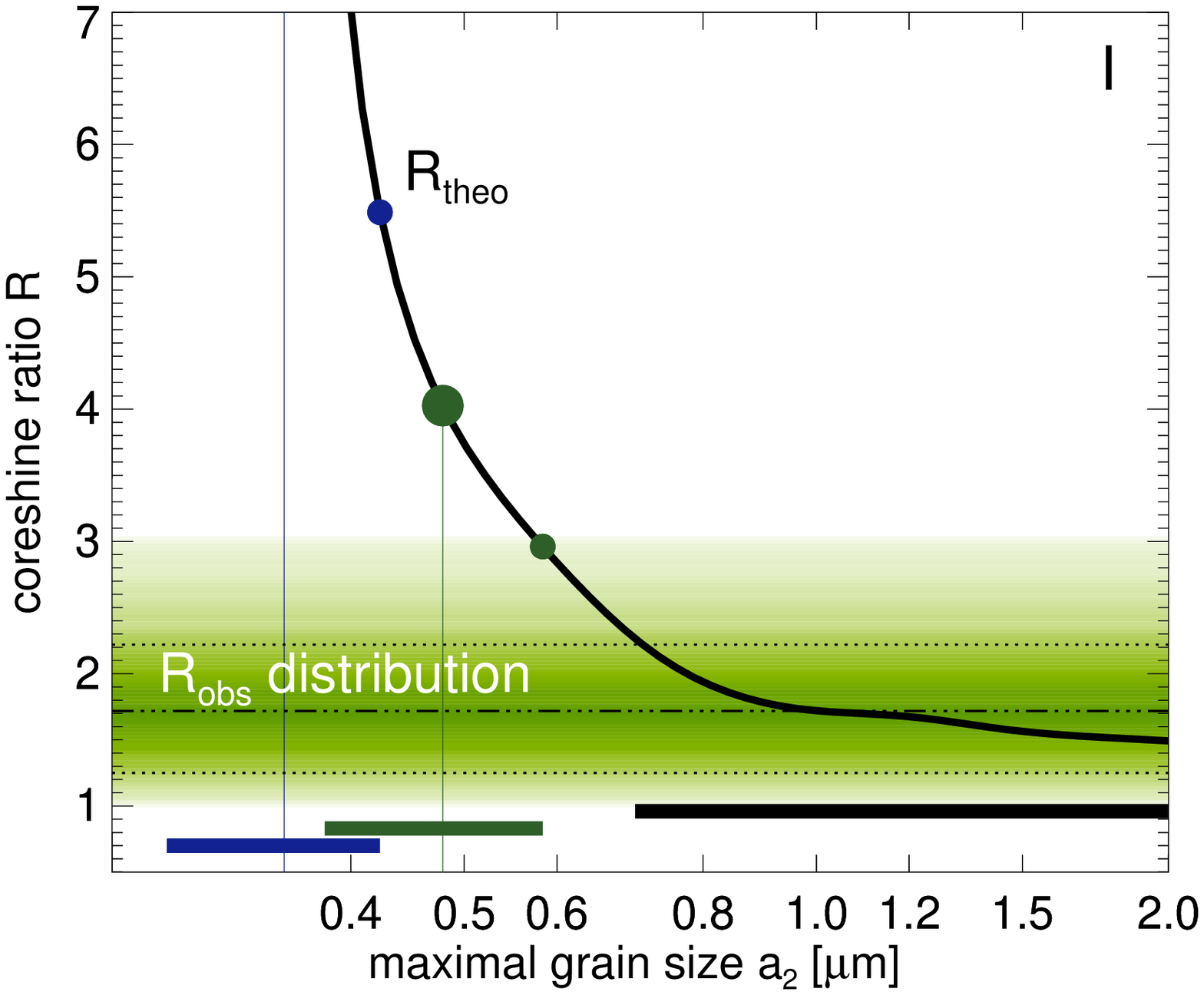}
}
\caption{
Data analysis for L1506C. Panel C shows the MAMBO II map reproduced from
\citet{2010A&A...512A...3P}.
For detailed legend see Fig.~\ref{L260}.
        }
\label{L1506C}
\end{figure*}
This extended Taurus core with low density, low turbulence, but strong depletion
\citep{2010A&A...512A...3P}
was discussed in 
\citet{2014A&A...564A..96S}
as a core that either is still contracting to become a prestellar core
\citep[as suggested by][]{2010A&A...512A...3P} 
or  has passed
through a core phase and is dissolving again. 
Since the core is invisible at 8 $\mu$m,
we show in Fig.~\ref{L1506C} instead a MAMBO II map
\citep[reproduced from][]{2010A&A...512A...3P} at 1.2 mm
revealing
the same pattern (in thermal emission) as the coreshine. 
The large Galactic latitude of -17.57$^\circ$ helps to make the coreshine visible.
The 4.5 $\mu$m signal is just
a factor 2 over the noise but the entire core pattern is nicely visible
and exhibits no strong impact of potential noise confusion. 
From the noisy cuts displayed in Panel F, it is difficult to judge on potential central depressions, since the images in Panels A and B do not reveal a darkening, and the basic
parameters suggest optical depths below 1 in the center. As a result, we placed
the white frame near the core center, thereby avoiding the two large stellar PSFs, and use 
single-scattering RT. The $R$ map in Panel G shows the interesting feature that
the entire clump visible in the image seems to have $R$ values comparable to the mean
value measured in the white frame as 1.7 (FWHM about 1).

The model is able to account for the SFBes based on
grains with maximum sizes around 0.4 $\mu$m, while the ratios produced by such
grains of about 7 would be too high to match the observed low values between $R$=1.2 and 2.2.
This steep ratio points toward grains with sizes larger than 0.7 $\mu$m.
The analysis of L1506C presented in \citet{2014A&A...564A..96S} using
a grain growth model and simple spatial mixing of the size distribution
pointed toward large grains beyond 1 $\mu$m in size. However, including
the spatial mixing in the growth process would likely reduce the efficiency of
the growth process, leading to smaller maximal sizes.

\subsection{L1439}
\begin{figure*}
\vbox{
\includegraphics[width=6cm]{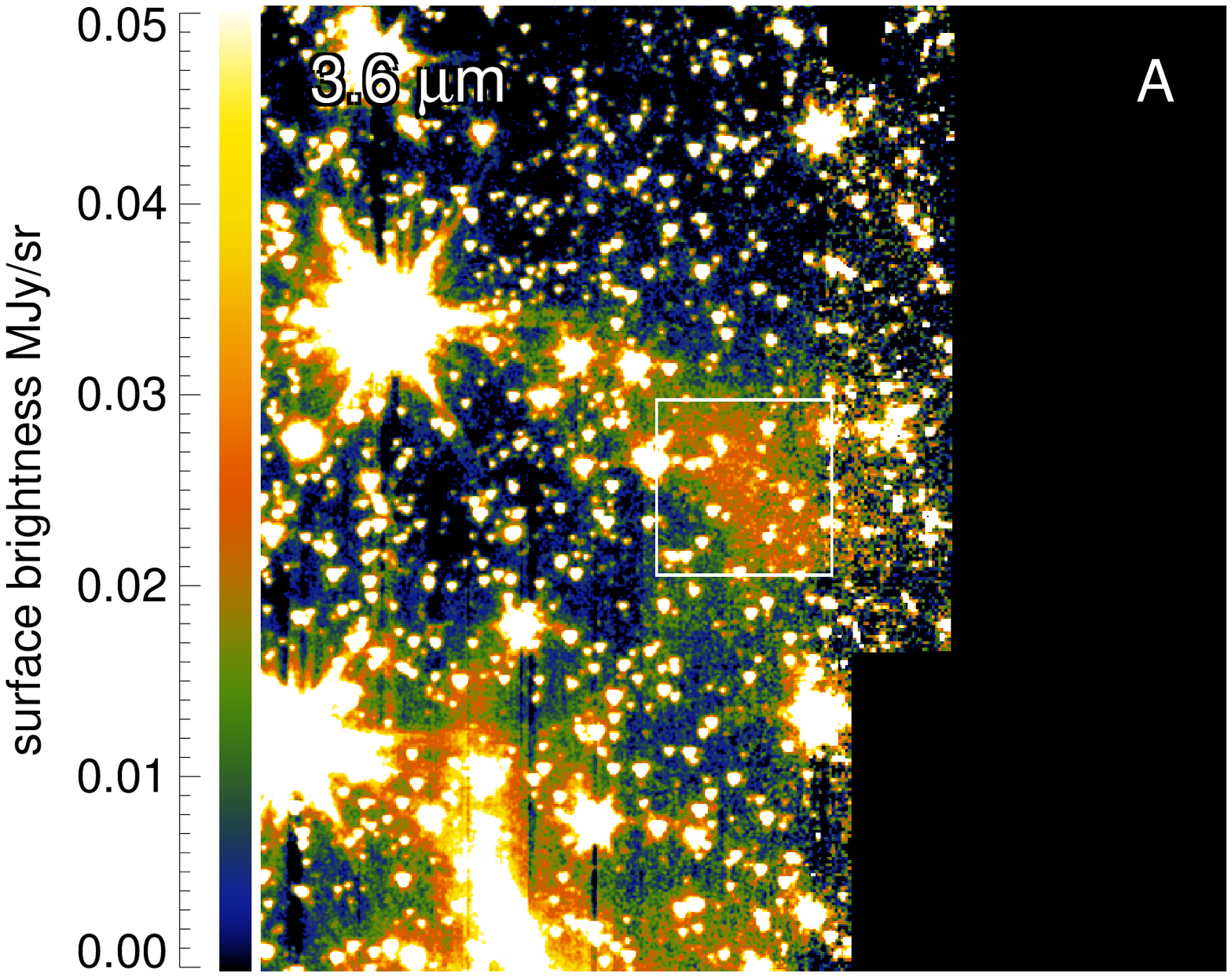}
\includegraphics[width=6cm]{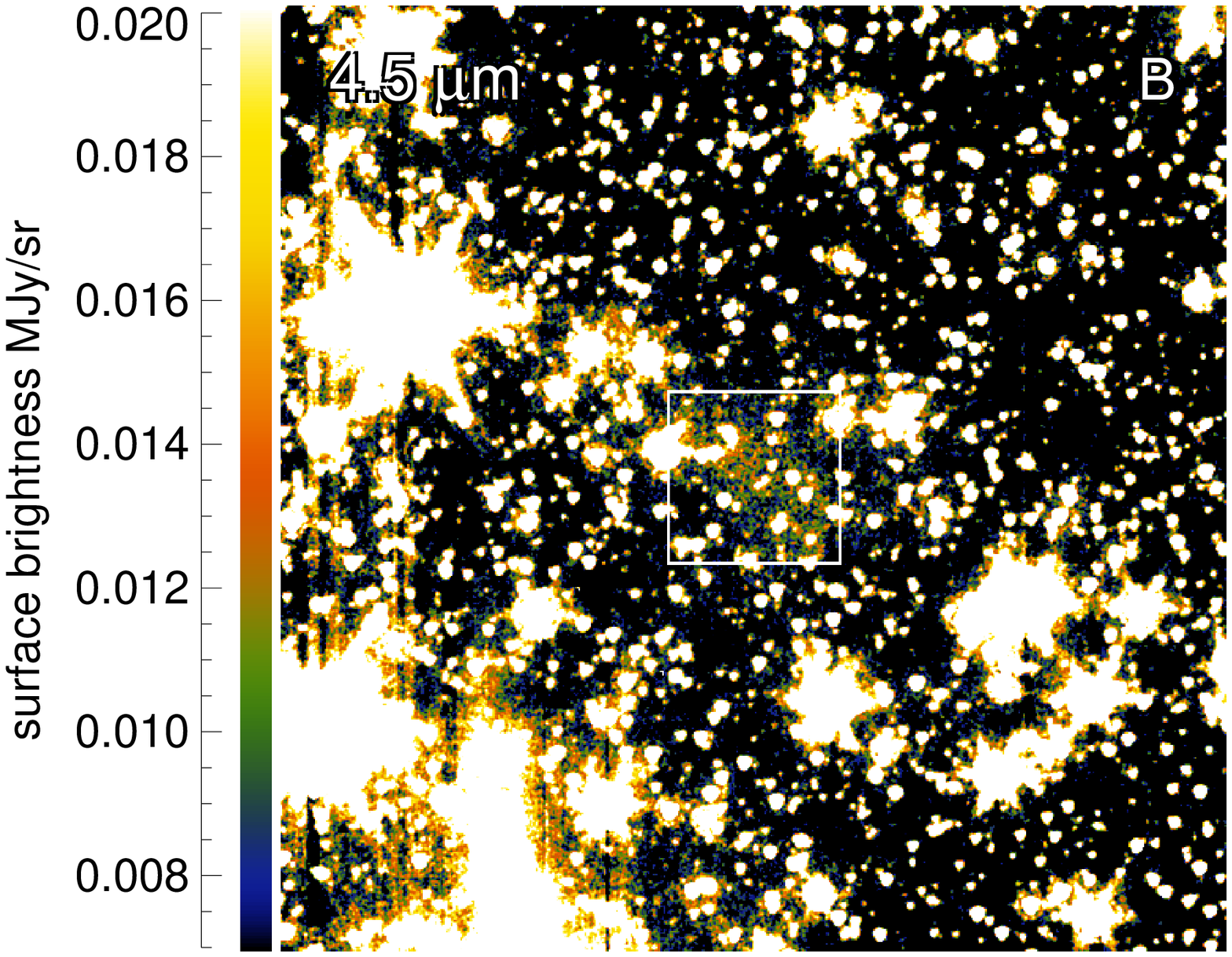}
\includegraphics[width=6cm]{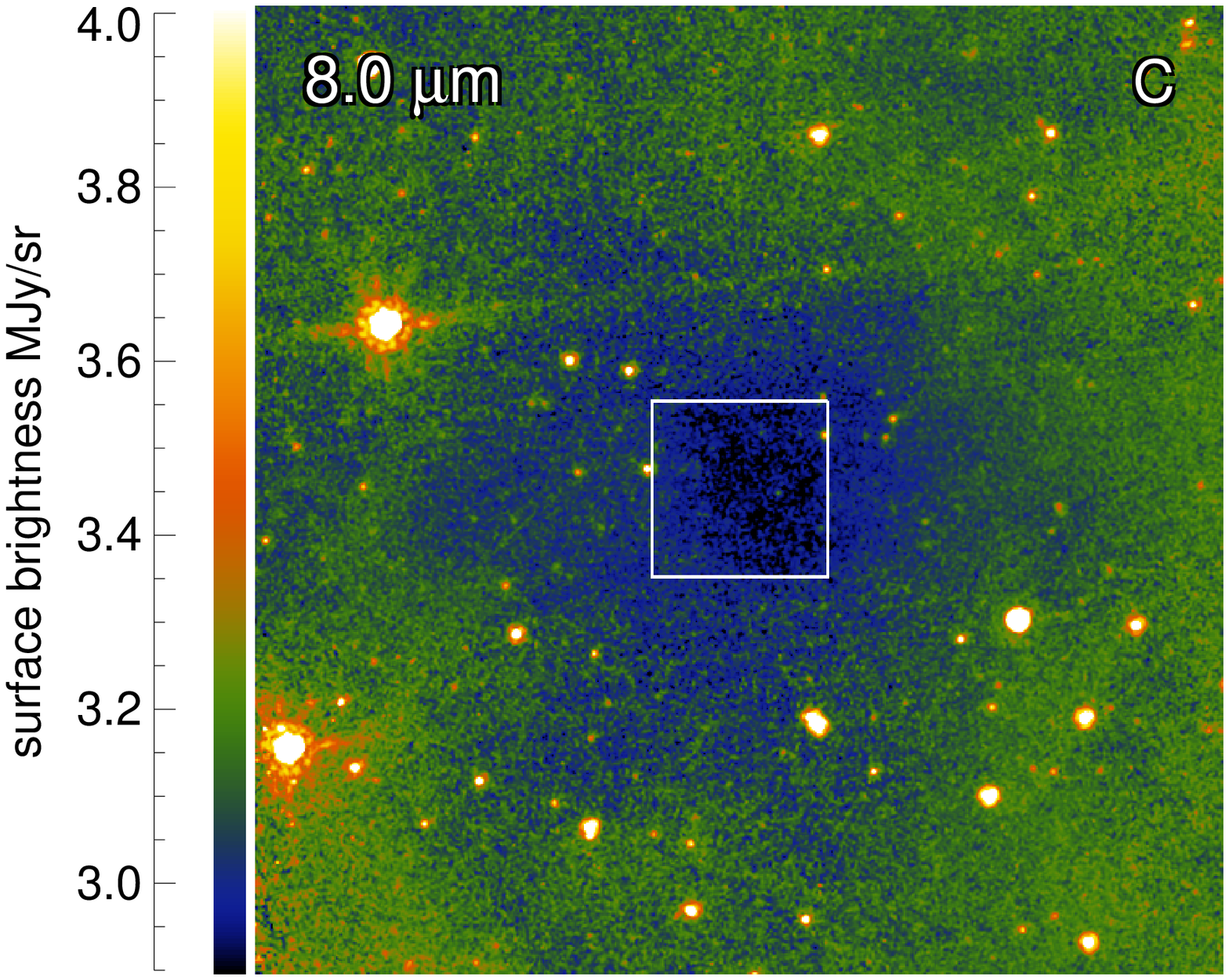}
}
\vskip 0.1cm
\vbox{
\includegraphics[width=6cm]{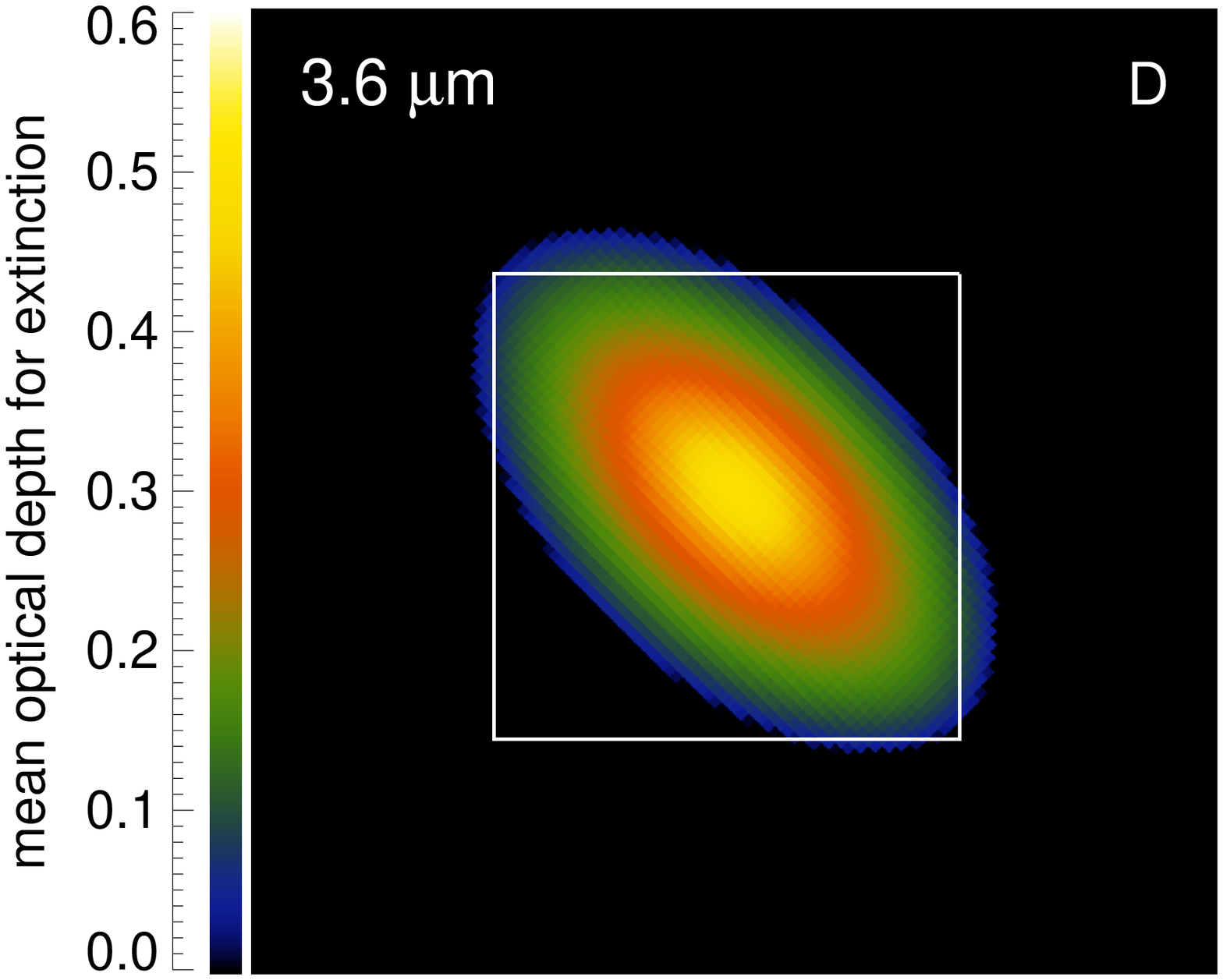}
\includegraphics[width=6cm]{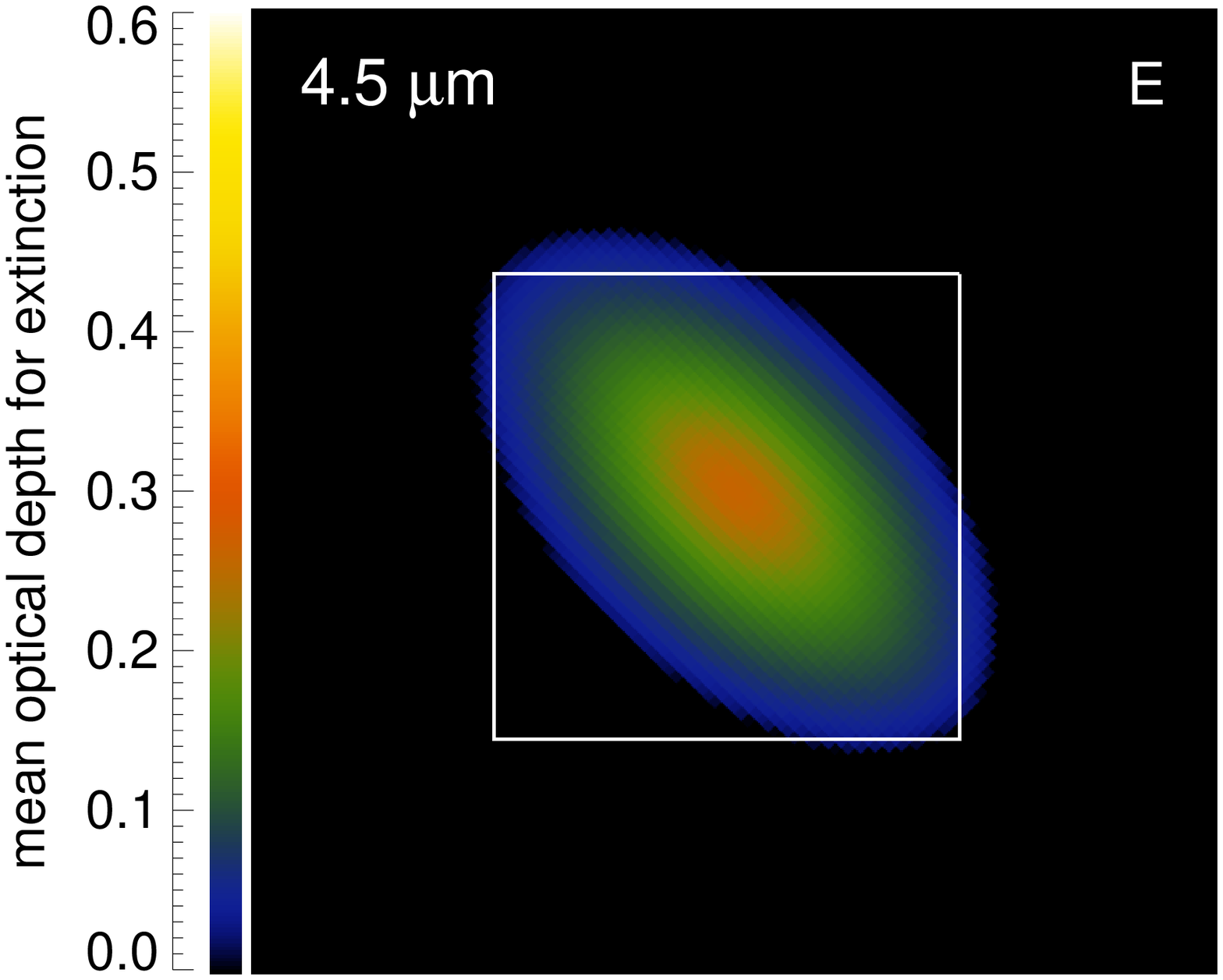}
\includegraphics[width=6cm]{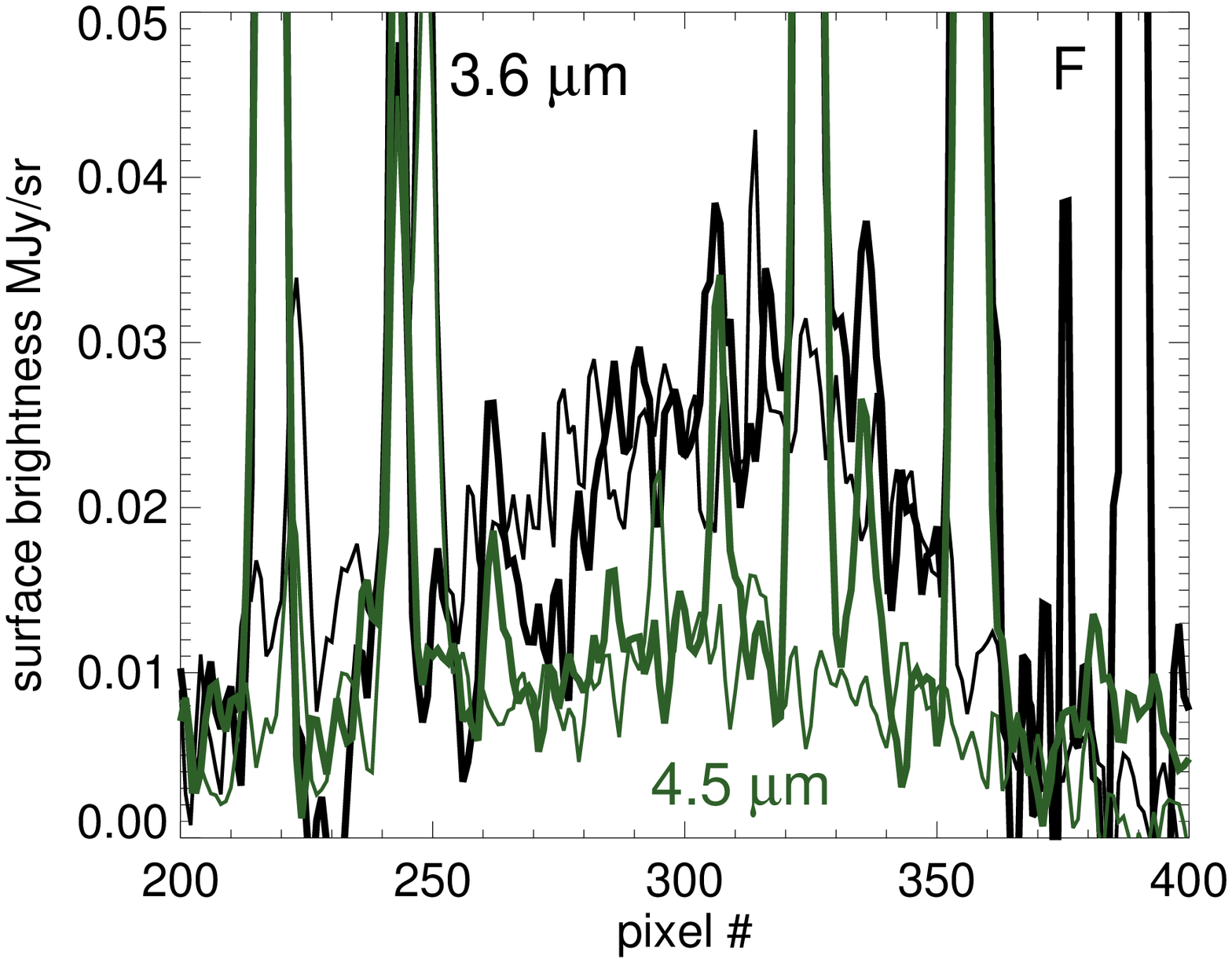}
}
\vskip 0.1cm
\vbox{
\includegraphics[width=6cm]{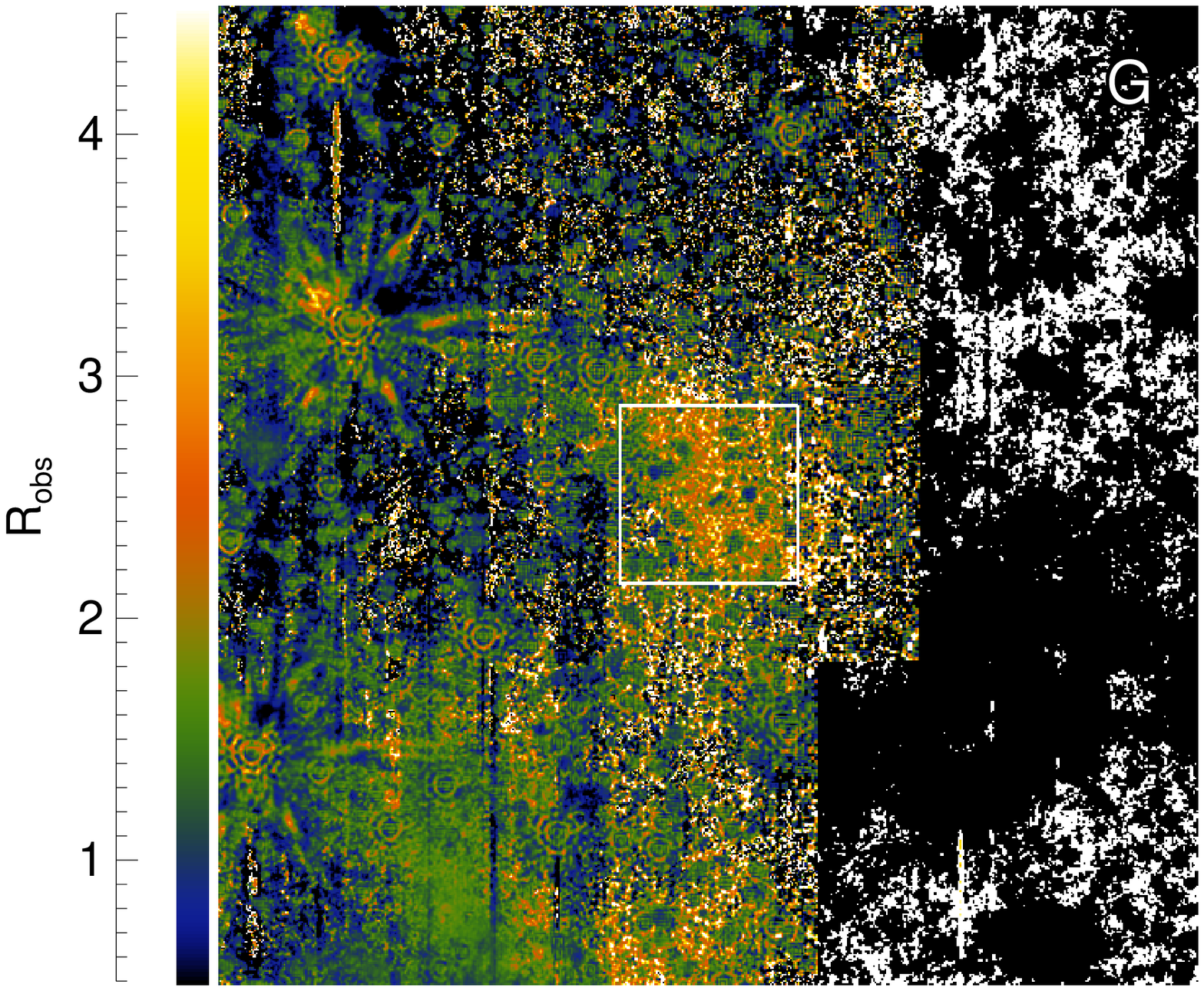}
\includegraphics[width=6cm]{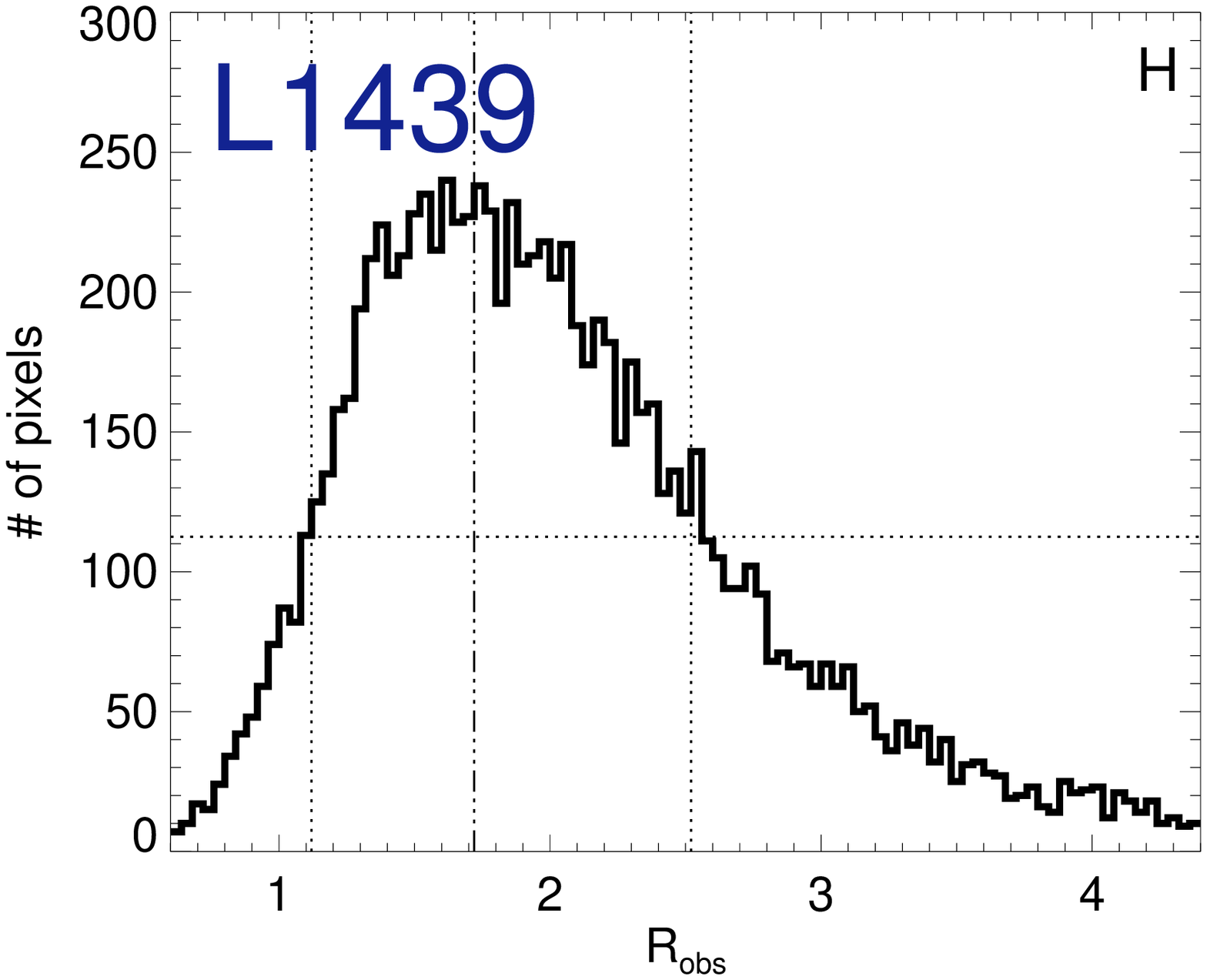}
\includegraphics[width=6cm]{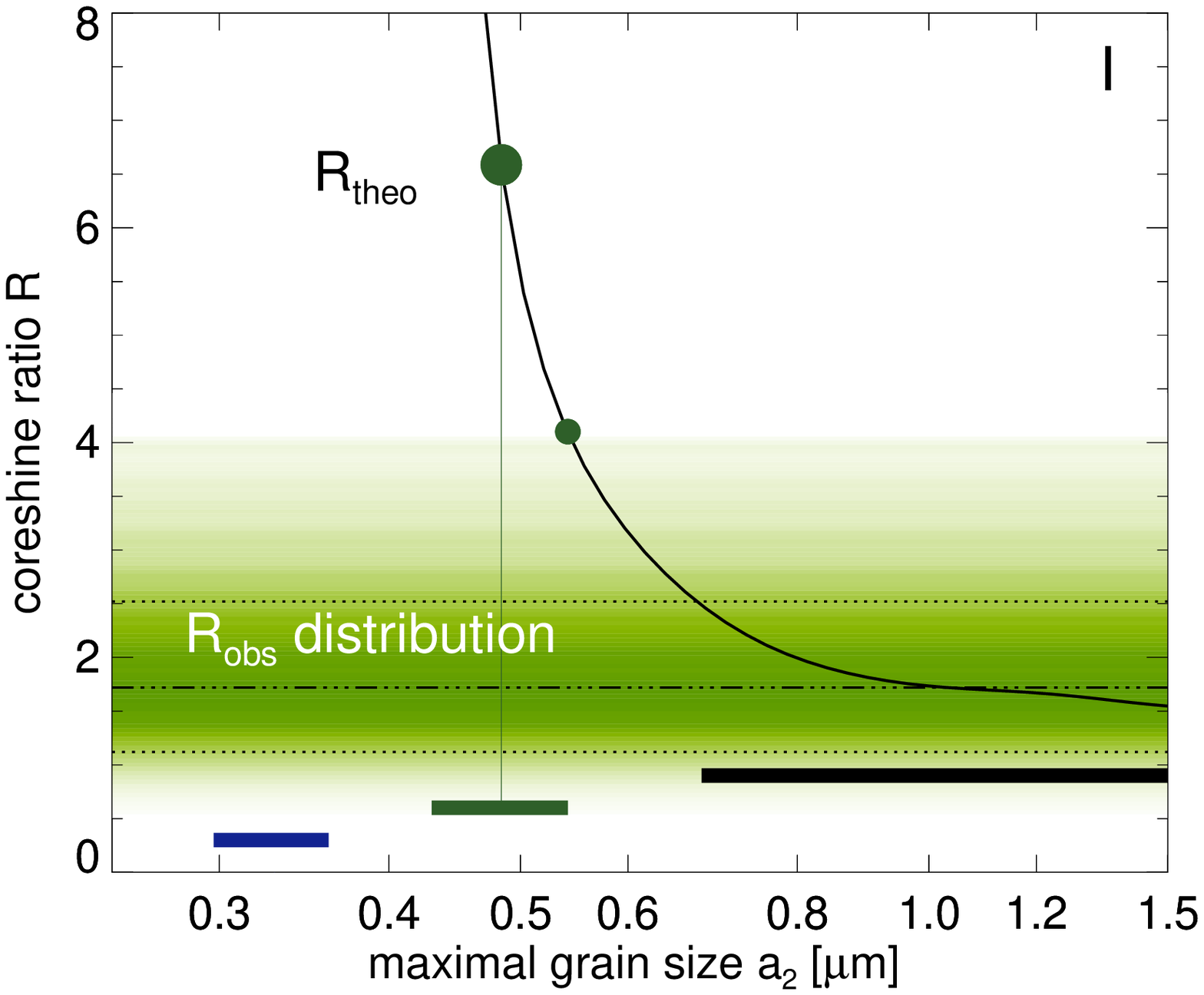}
}
\caption{
Data analysis for L1439 (CB26). The 3.6 $\mu$m image (panel A) contains an edge which creates
a dark region in the ratio map in Panel G where the ratio is not defined.
For detailed legend see Fig.~\ref{L260}.
        }
\label{L1439}
\end{figure*}
The core L1439 (CB26) is located only 6$^o$ above the Galactic plane in the PoSky and is therefore
seen against a crowded star field. The 4.5 $\mu$m signal is only a factor 2 above noise
(Fig.~\ref{L1439}).
At its rim, a YSO creates a jet, and it is not clear whether the stellar component
or the warm dust adds local radiation to the ISRF.
The cold {\it Spitzer} map at 8 $\mu$m is shown in Panel C ("darkclouds\_IRAC": Program ID 94, PI C. Lawrence).
The core parameters listed in Table \ref{table:1} make central optical depths below 1 likely.
We therefore placed the white frame right on the core and used single-scattering RT.
The modeling with a variable $\tau$ limit led to values around 0.5 for 3.6 $\mu$m.
The core is located at the border of the 3.6 $\mu$m frame so that the dark righthand
part
of the $R$ map displayed in Panel G is meaningless.
The $R$ pixel number distribution
shown in Panel H of Fig.~\ref{L1439} contains an asymmetric wing from the map parts with a weak 4.5 $\mu$m
signal. It has a maximum around 1.7 and a FWHM of 1.4.
Panel I reveals that the model
produces no common size range to meet the observed SFBs 
and the observed low $R_{obs}$ ratios.

\subsection{L1498}
\begin{figure*}
\vbox{
\includegraphics[width=6cm]{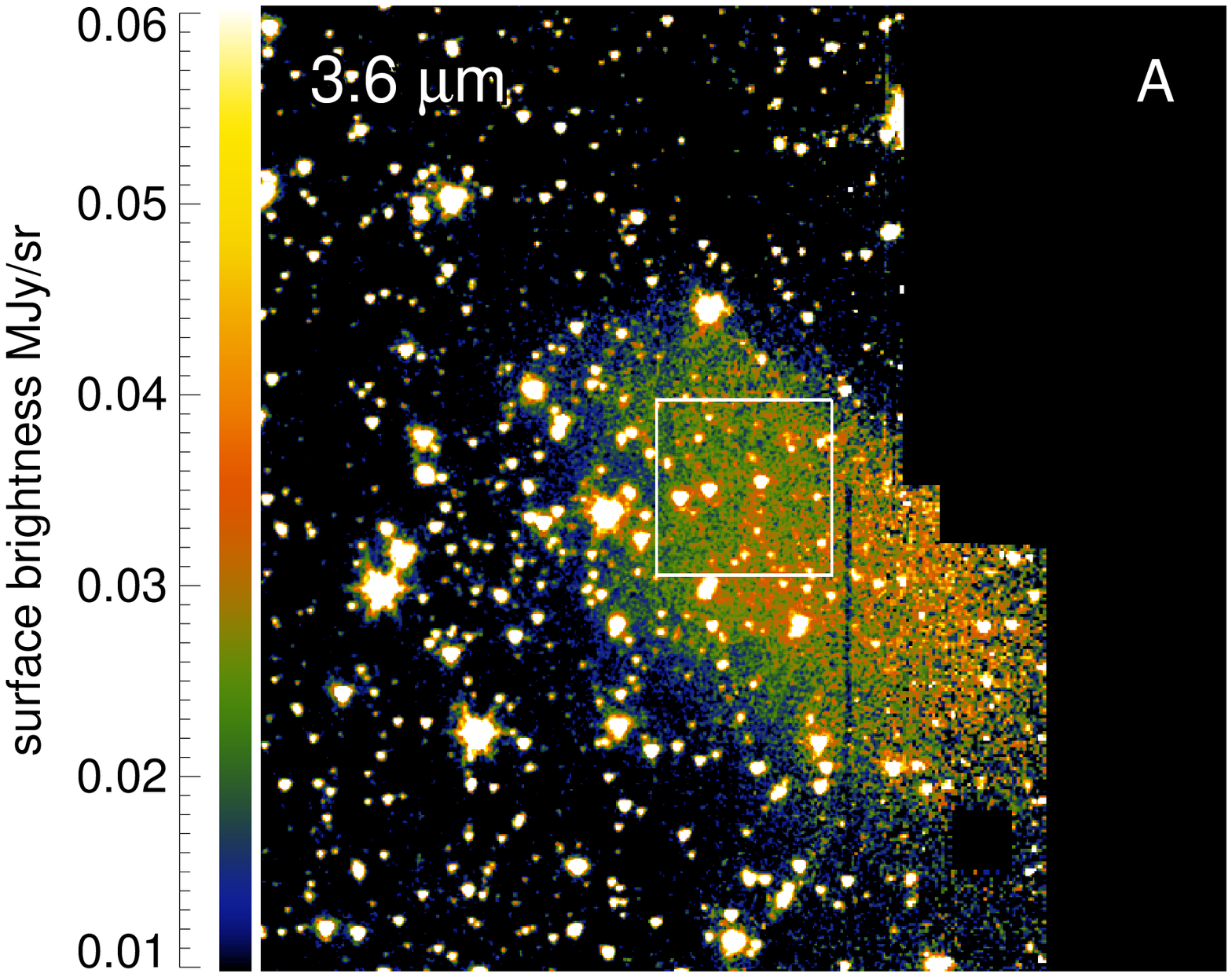}
\includegraphics[width=6cm]{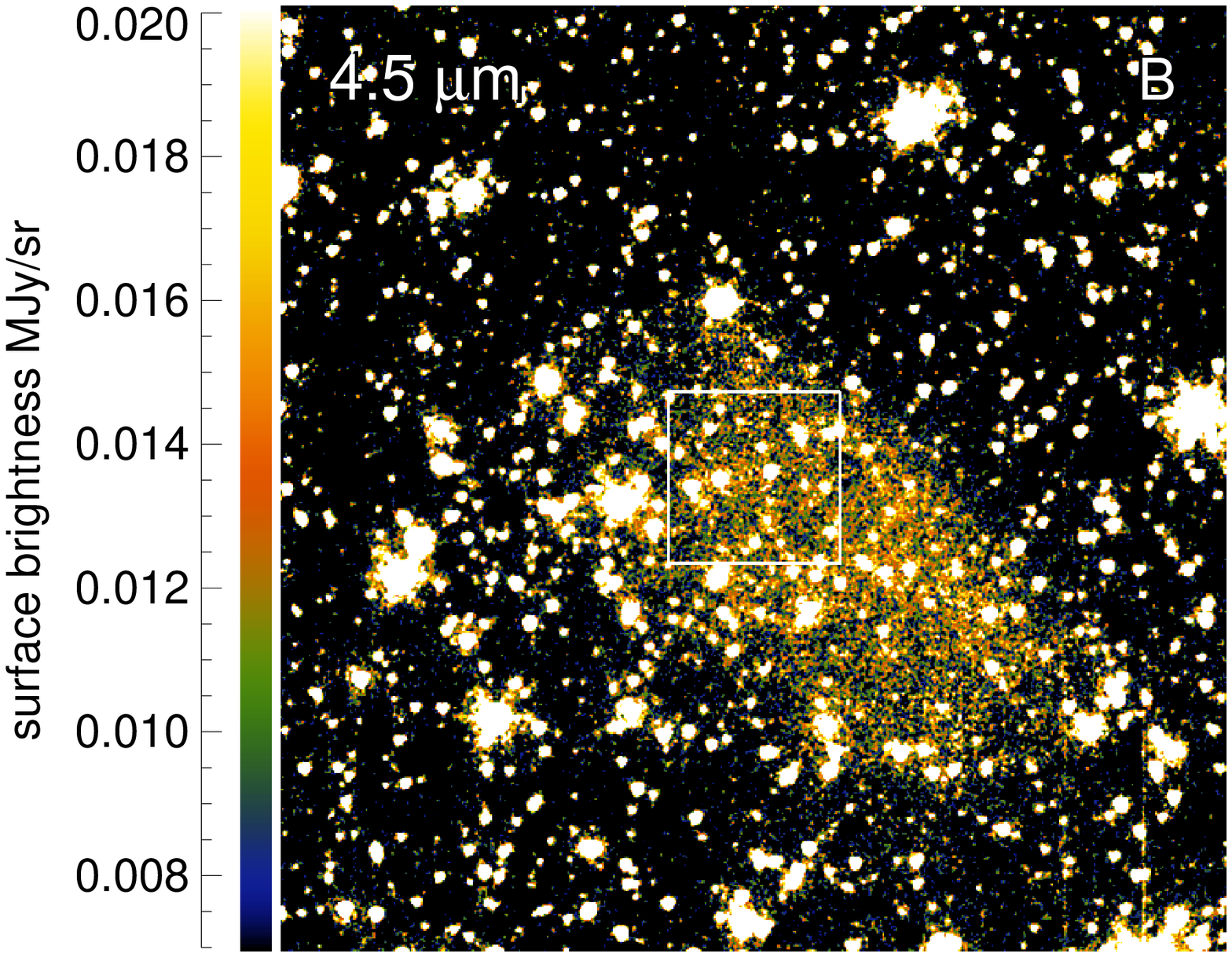}
\includegraphics[width=6cm]{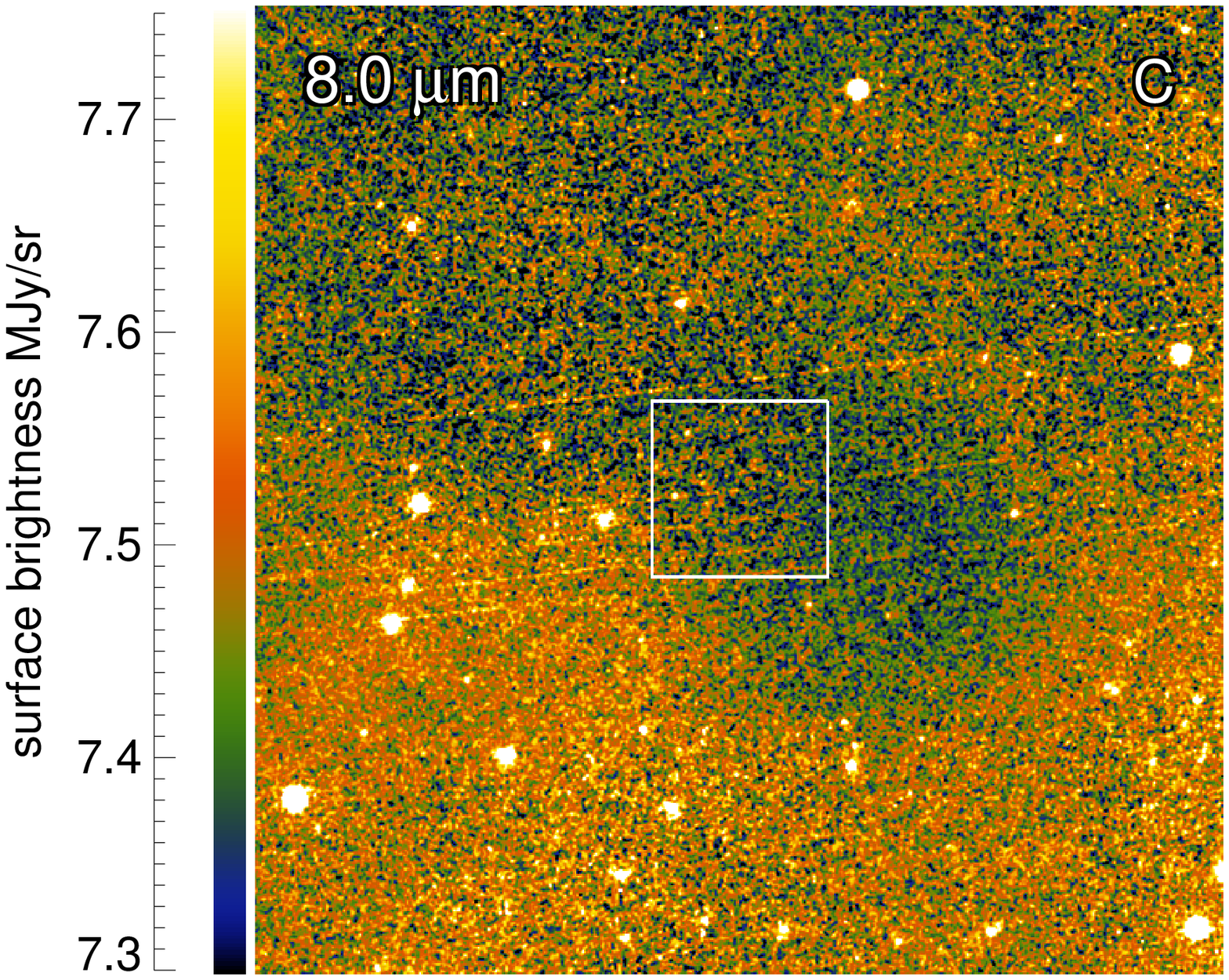}
}
\vskip 0.1cm
\vbox{
\includegraphics[width=6cm]{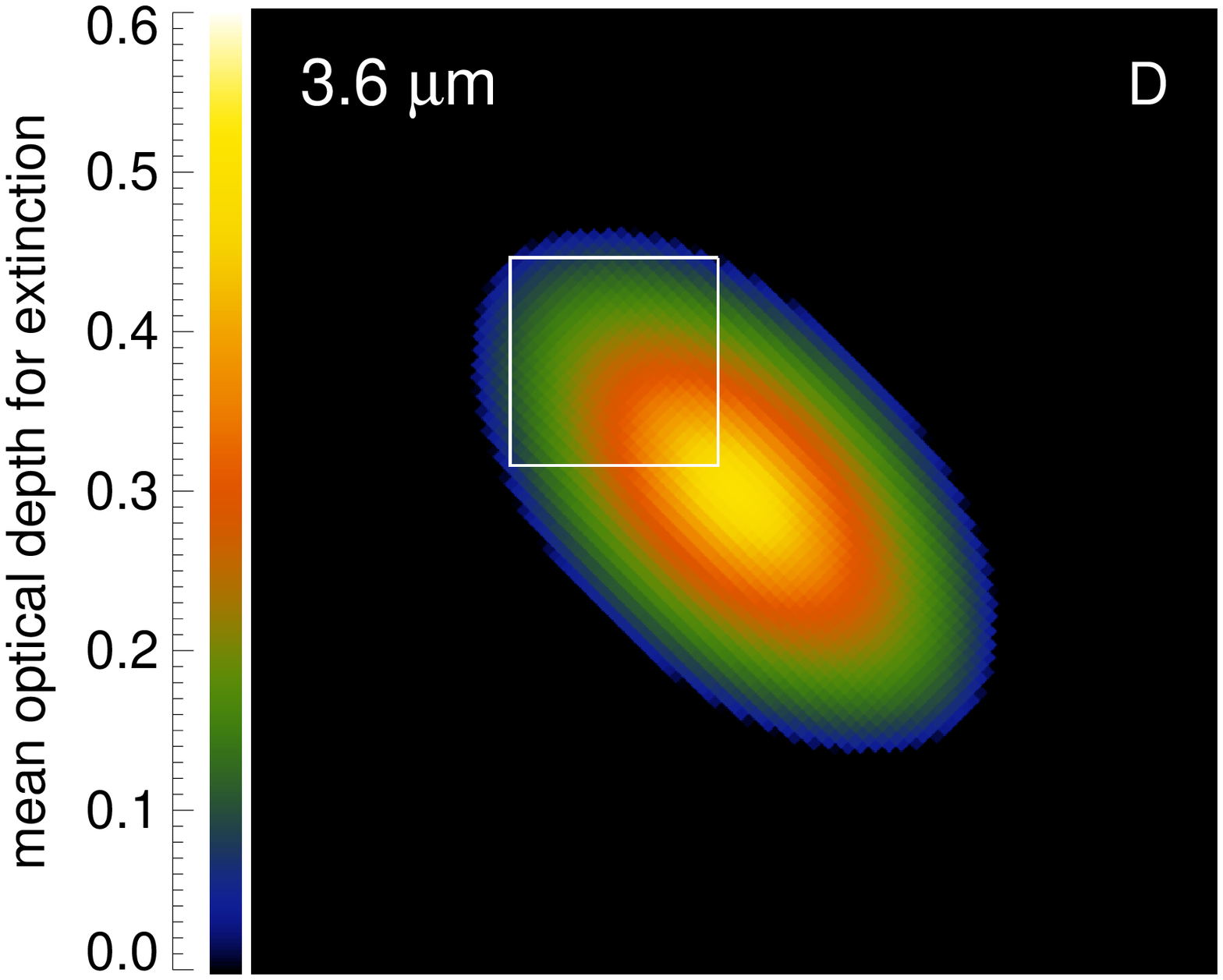}
\includegraphics[width=6cm]{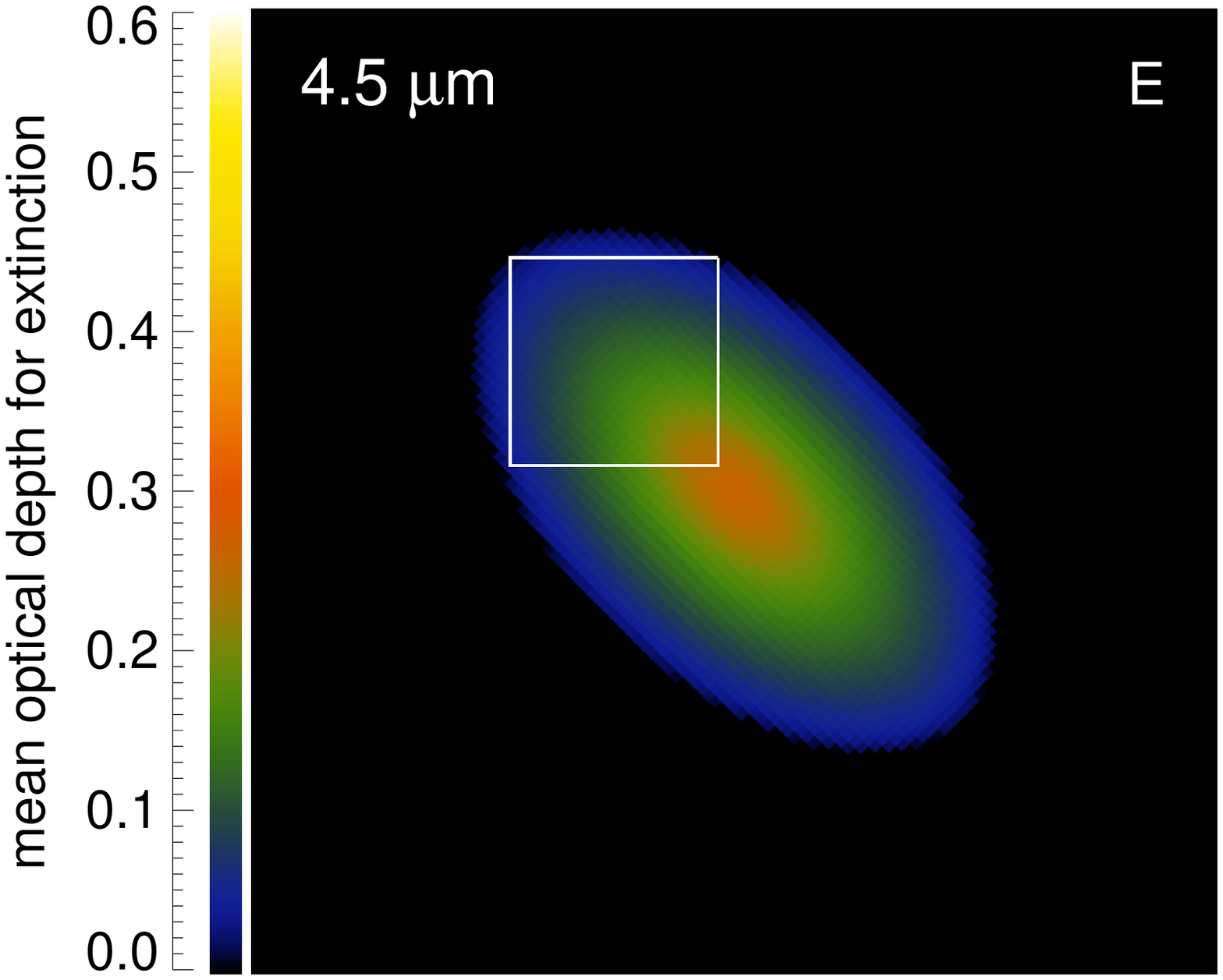}
\includegraphics[width=6cm]{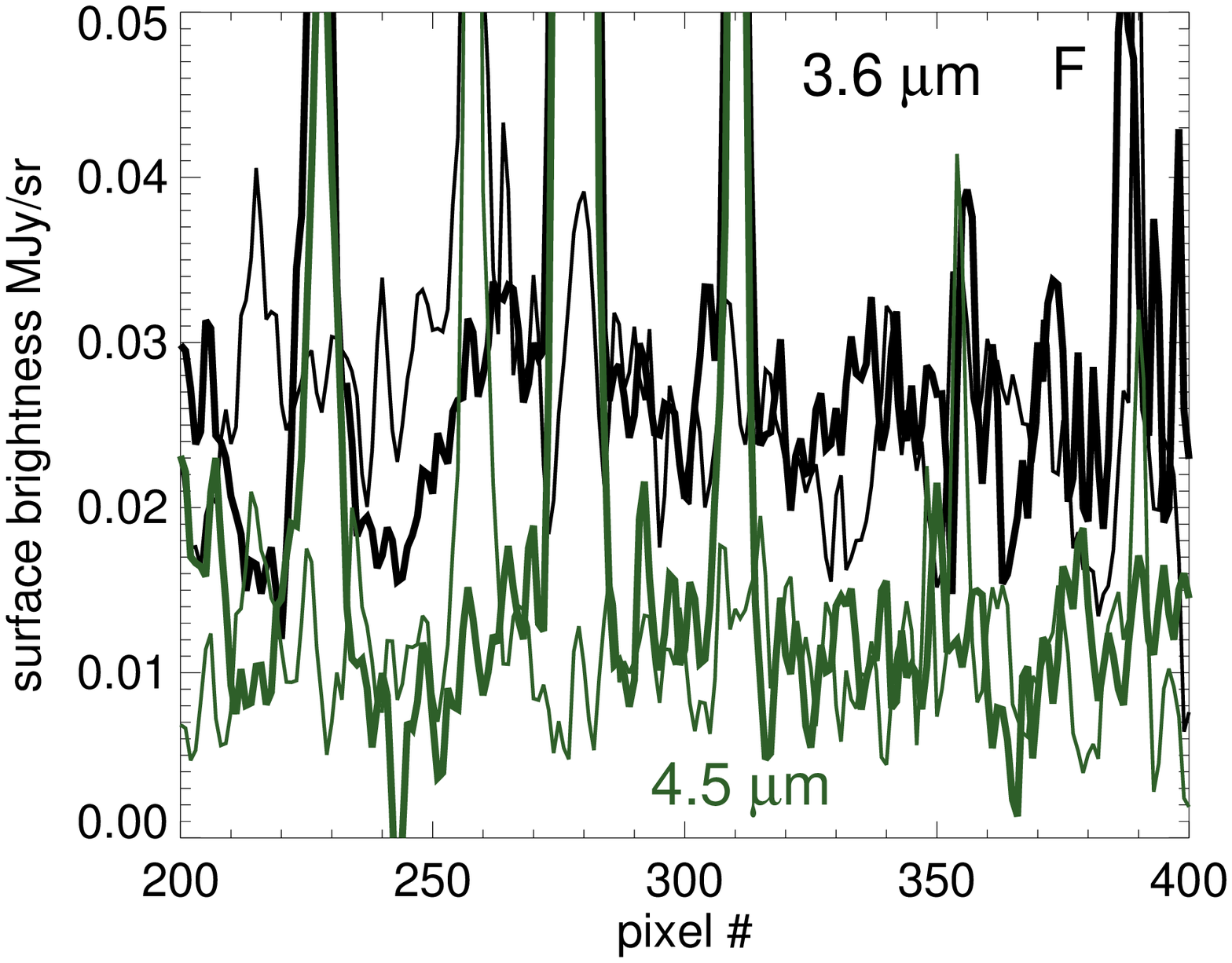}
}
\vskip 0.1cm
\vbox{
\includegraphics[width=6cm]{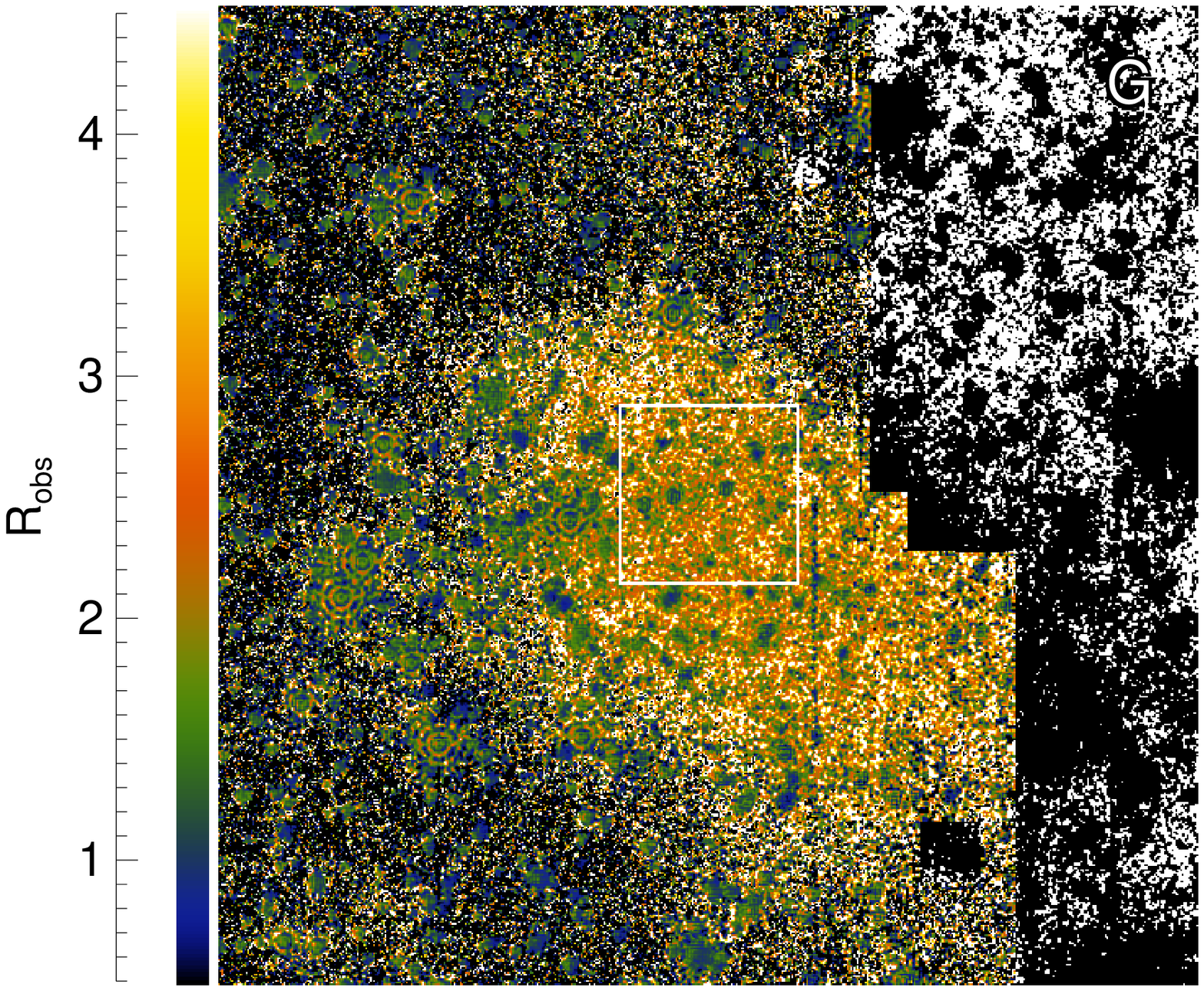}
\includegraphics[width=6cm]{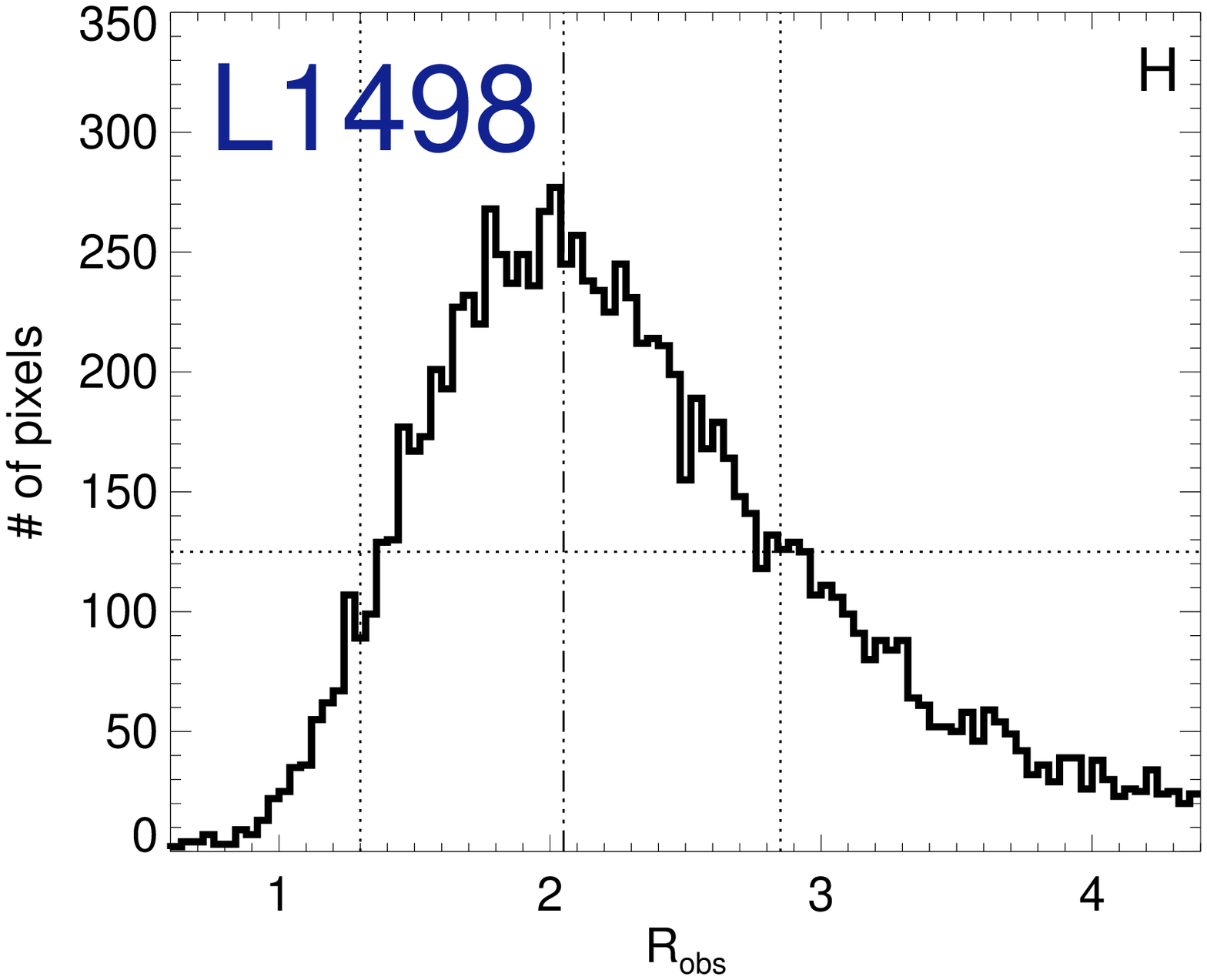}
\includegraphics[width=6cm]{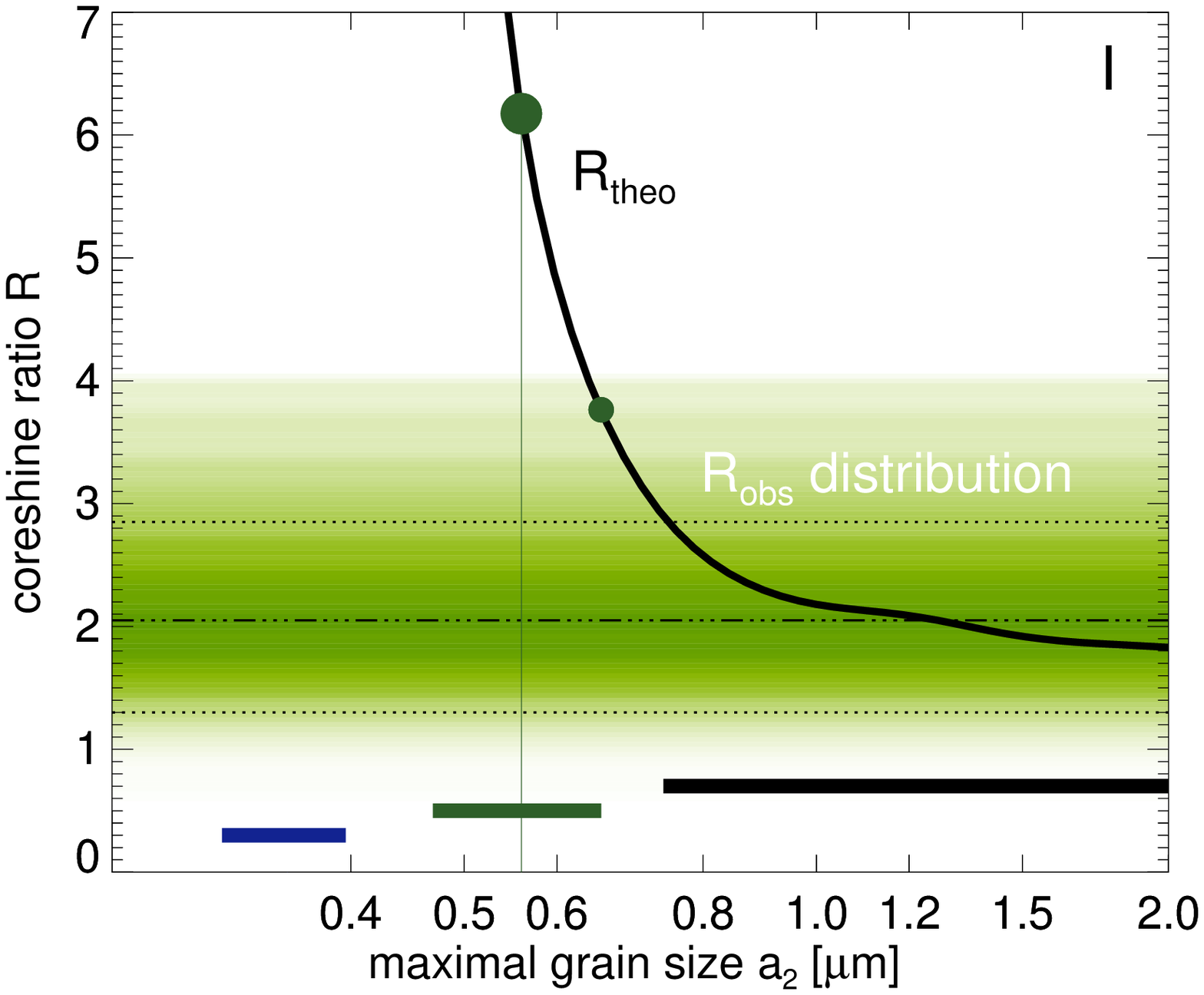}
}
\caption{
Data analysis for L1498. For detailed legend see Fig.~\ref{L260}.
        }
\label{L1498}
\end{figure*}
L1498 in Taurus
is discussed to be on the verge of becoming a core
\citep{2004ApJ...610..303L}, and only
a faint extinction pattern is visible at 8 $\mu$m in Panel C of Fig.~\ref{L1498} 
("darkclouds\_IRAC": Program ID 94, PI C. Lawrence). 
Positioning of the white frame was guided by the 3.6 $\mu$m image border so
that it was placed in one half of the ellipsoidal shape of the core. The $R$ maps
in Panel G show that this choice does not largely affect the results since the other
parts of the core seem to have similar values to those of the core L1506C.
The core parameters suggest 
central optical depths in the range of L1506C, so we use the same
optical depth limit and single-scattering RT.
Given the low
3.6 $\mu$m signal, L1498 shows a surprisingly strong 4.5 $\mu$m 
SFB yielding to low ratios around 2 (FWHM is 1.55).
The model grains produce too much coreshine when using the grains with size above
0.77 $\mu$m as suggested by the model to fit the $R_{obs}$ distribution so that the 
$a_2$ ranges do not overlap.

\subsection{Rho Oph 9}
\begin{figure*}
\vbox{
\includegraphics[width=6cm]{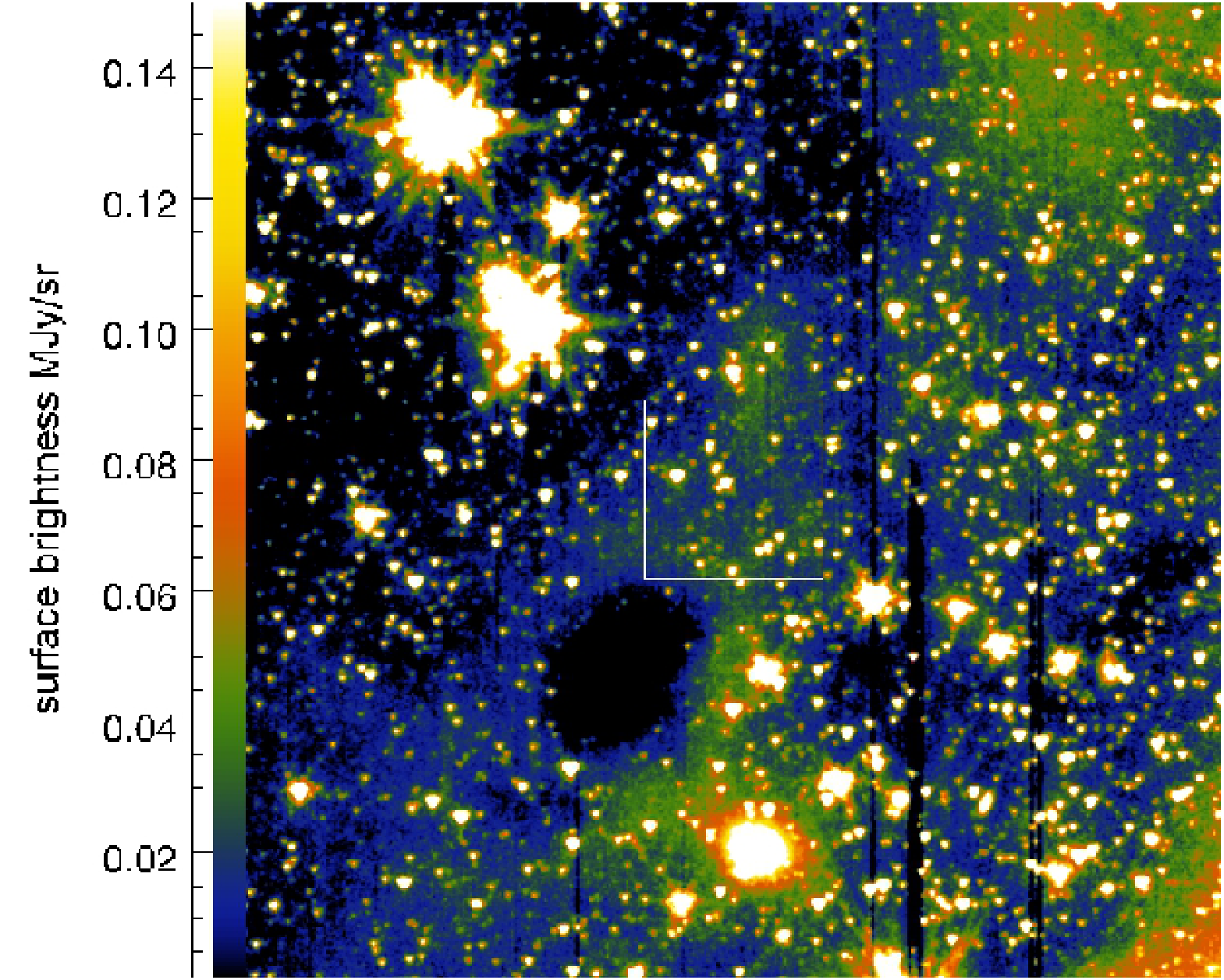}
\includegraphics[width=6cm]{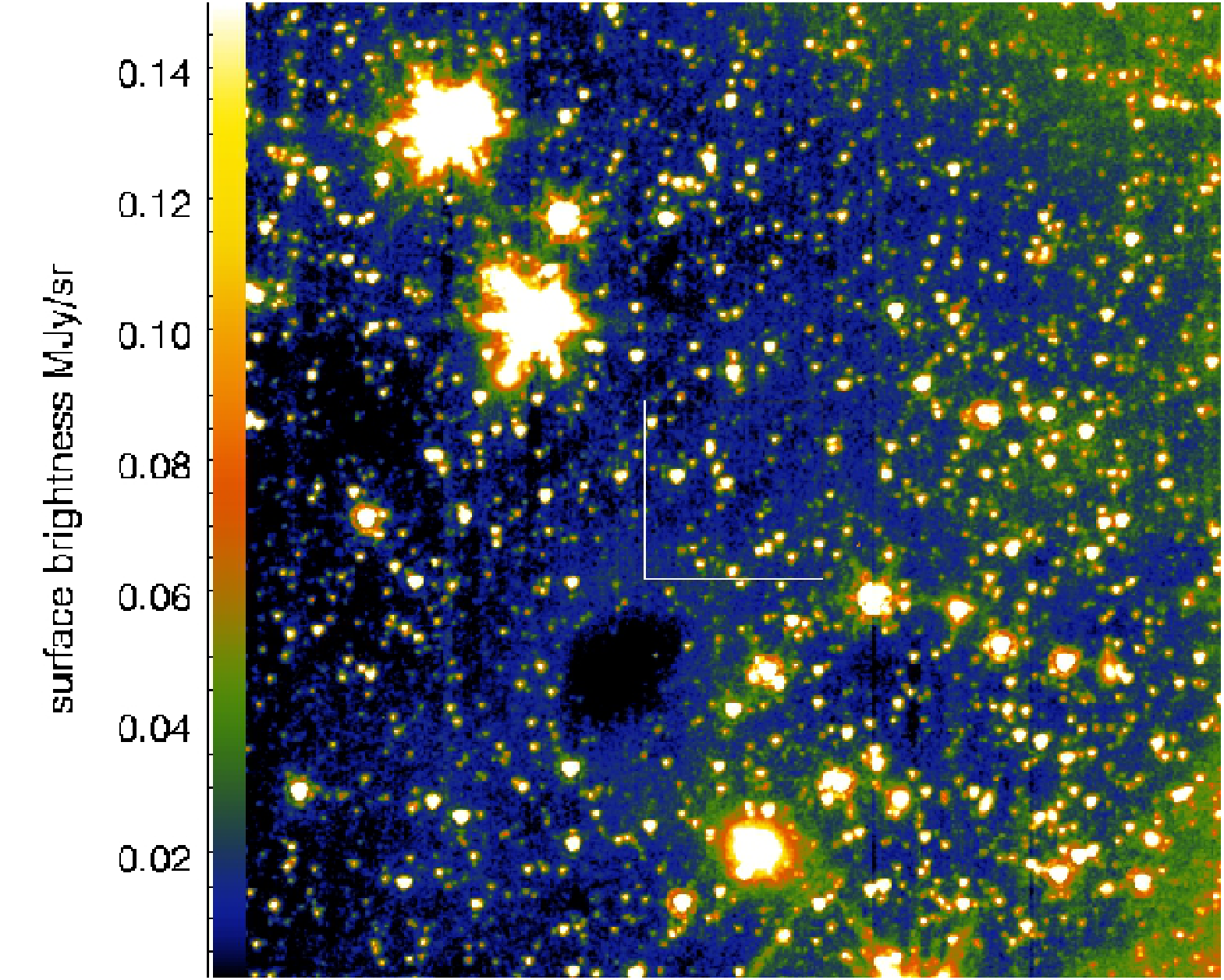}
\includegraphics[width=6cm]{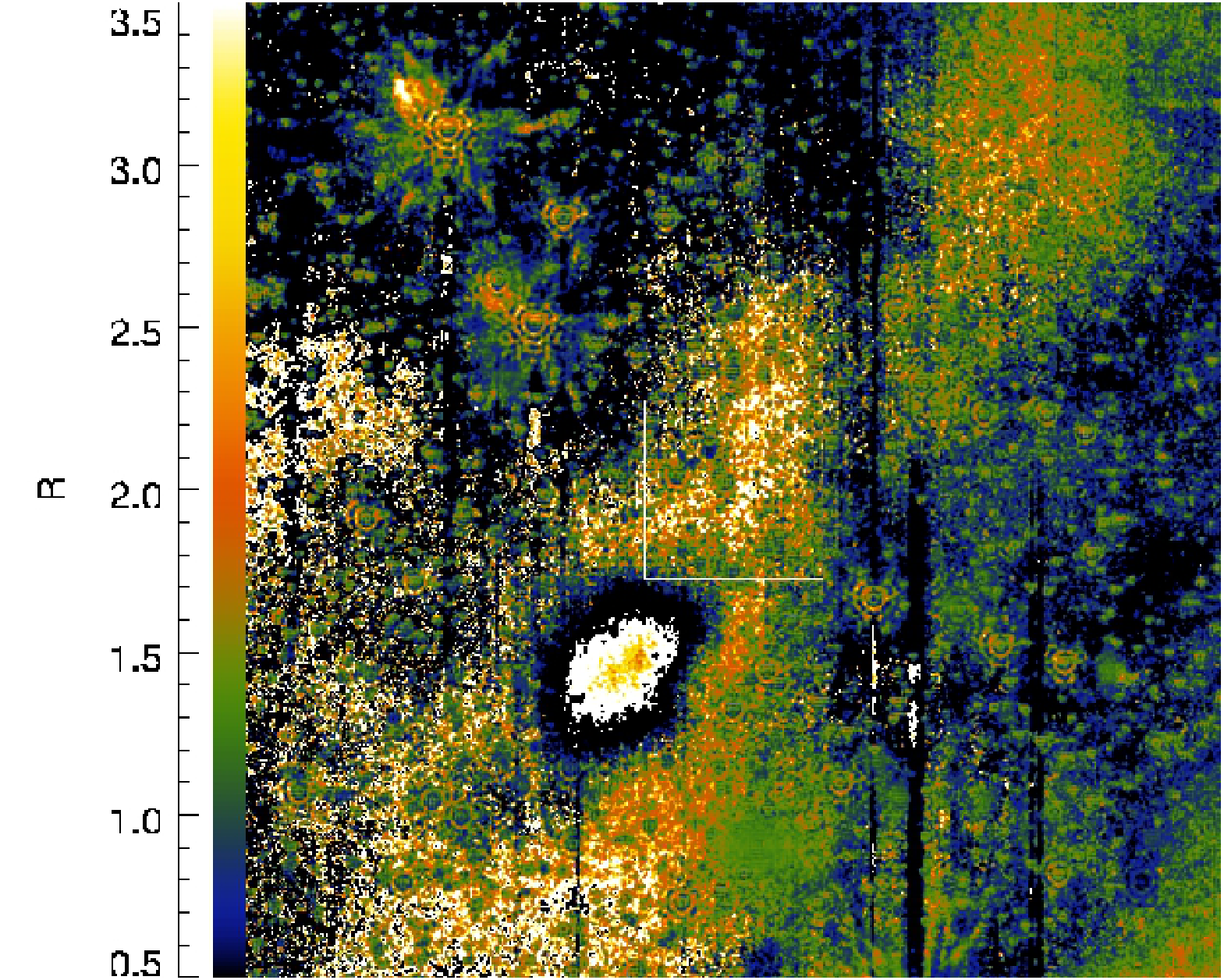}
}
\caption{
3.6 and 4.5 $\mu$m maps, and ratio map for Rho Oph 9. 
        }
\label{rho_oph9}
\end{figure*}
The core Rho Oph 9 is located in the most complex environment of all cores
in the sample (Gal. coord. 354.37 +16.17). The cold {\it Spitzer} data shown in
\cite{2010Sci...329.1622P}
show nearby PAH regions in all four bands with reduced SFB
at 4.5 $\mu$m owing to the lack of a main PAH feature in this band.

The excess of emission that is seen near the strong extinction pattern is
also present at 4.5 $\mu$m similar to the pattern seen in L1544. 
Unlike the PAH emission, it disappears
for larger wavelength in the cold {\it Spitzer} data 
as expected from scattered light. 
Figure~\ref{rho_oph9} shows the 3.6 and 4.5 $\mu$m warm {\it Spitzer} images.
The overall pattern is similar to L1544 with a strong central SFB
depression and an ellipsoidal emission region around. But disentangling 
the coreshine emission from variations in the background and PAH emission
turned out to be difficult.
We have not modeled the core since we could not determine an
off-region that was clearly free of PAH emission and near the core,
and hope to return to the source when more data are available.
We show the maps and the ratio map for completeness.

The peak and background SFBes of the modeled sources
are summarized in Table \ref{table:2} along with the derived
observed SFB ratios $R_{obs}$.

\begin{table*}
\caption{Peak and background surface brightnesses at 3.6 and 4.5 $\mu$m
and derived observed coreshine ratios $R$.}
\label{table:2}
\centering
\begin{tabular}{l c c c c c c c}
\hline\hline
Source       & I$^{3.6}_{peak}$ & I$^{4.5}_{peak}$ & I$^{3.6}_{bg}$ & I$^{4.5}_{bg}$ & R$_{obs}$\\
             & [MJy/sr] & [MJy/sr] & [MJy/sr] & [MJy/sr] & \\
\hline
L260         & 0.085   & 0.04   & 0.098 & 0.114 & 1.9-2.9 \\
Ecc806       & 0.075   & 0.022   & 0.026 & 0.036 & 2.3-3.6 \\
L1262 (CB244)& 0.070   & 0.035   & 0.031 & 0.042 & 1.4-2.2 \\
L1517A       & 0.055   & 0.02    & 0.150 & 0.171 & 2.0-3.7 \\
L1512 (CB28) & 0.035   & 0.01    & 0.052 & 0.082 & 2.5-5.0 \\
L1544        & 0.030   & 0.02    & 0.040 & 0.090 & 0.8-1.8 \\
L1506C       & 0.027   & 0.015   & 0.032 & 0.086 & 1.2-2.2 \\
L1439 (CB26) & 0.025   & 0.01    & 0.041 & 0.066 & 1.1-2.4 \\
L1498        & 0.025   & 0.01    & 0.046 & 0.090 & 1.3-2.9 \\
\hline 
\end{tabular}
\end{table*}
%

\section{Discussion}\label{discussion}
\begin{figure}
\includegraphics[width=9cm]{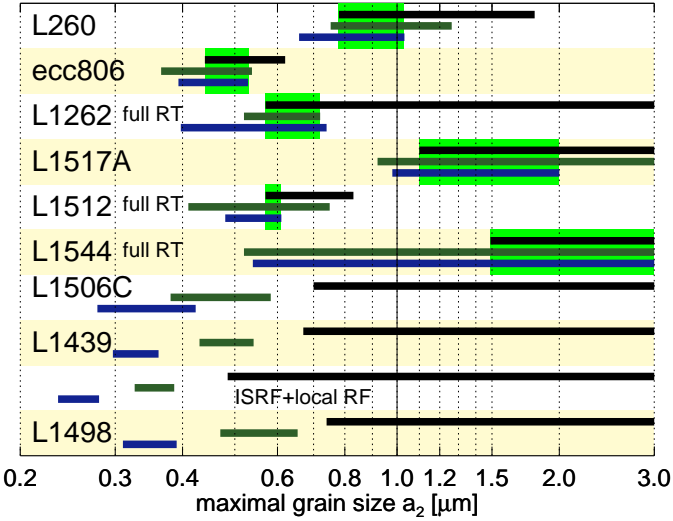}
\caption{
Summary of the derived maximum grain size ranges for the modeled sources.
The black bar gives the range where $R_{theo}$ is in the FWHM range of
$R_{obs}$, and the blue (green) bar shows the range for the model maximum coreshine
to be within the observed maximum coreshine range at 3.6 (4.5) $\mu$m, respectively.
For L1439, we also give size ranges for a model that includes  
a local radiation field.
Light green boxes indicate ranges of agreement.
        }
\label{sum}
\end{figure}
\begin{figure}
\includegraphics[width=9cm]{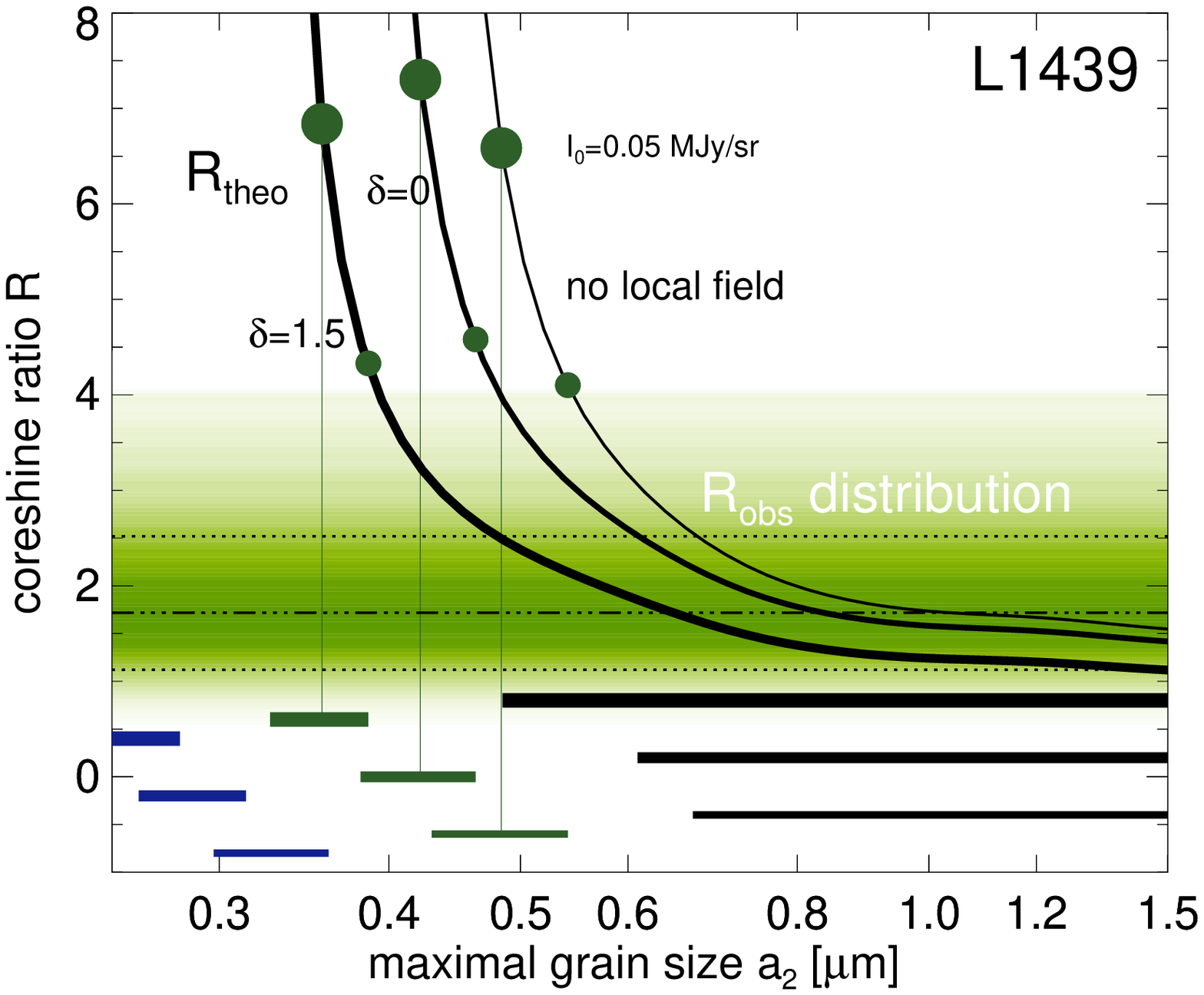}
\caption{
Impact of a local radiation field with no directional variation
and a wavelength dependency following $(\lambda/2.2 \mu m)^\delta$ on the
coreshine ratio of L1439 as a function of the maximum grain size.
        }
\label{local}
\end{figure}
In Fig.~\ref{sum} we summarize the $a_2$ ranges derived from the modeling
with the same notation as in Fig.~\ref{L260}: blue, green, and black bars for agreement
with the 3.6 $\mu$m maximum SFB, with the 4.5 $\mu$m maximum SFB, and with the observed SFB ratio $R_{obs}$, respectively.

For L260, ecc806, L1262, L1517A, L1512, and L1544, the bars overlap so that the model is able to reproduce the observations with a single grain size distribution. 
The derived maximum grain sizes for these sources are around 0.9, 0.5, 0.65, 1.5, 0.6, and > 1.5 $\mu$m, respectively.
The value derived for L260 of about 0.9 $\mu$m is consistent with the results derived in
\citet{2013A&A...559A..60A}.

The grain size limit of about 0.5 $\mu$m in ecc806 is surprisingly low given that it is bright in 
coreshine (about 0.075 MJy/sr) and that large grains dominate the scattering. 
At a longitude of about 303$^\circ$, radiation from the GC already needs large
scattering angles to reach us. 
Because grains larger than 0.5 $\mu$m scatter mainly in the forward direction, their contribution  to the observed signal is reduced. 
However, the low background due to its high elevation (b about -16$^o$) aids for seeing a 
strong coreshine.

The structure around the starless core in the binary core system L1262 
is complex and not represented well
by an ellipsoid. Correspondingly, the derived images carry the errors from the
spatial density because of both shading and multiple scattering. 
Nevertheless, maximum grain sizes of 0.65 $\mu$m agree with the
measured ratios and maximum SFBs in the chosen region outside
the core center. They may actually be smaller since the nearby YSO could contribute
to the local radiation field both by
direct irradiation and by causing hot dust and PAH emission in its vicinity. Further exploration
with a better spatial model may answer that question.

Strong coreshine (about 0.055 MJy/sr) also comes from L1517A but on top of a much
stronger background. Correspondingly the grain size limit $a_2$> 1.5 $\mu$m
derived by the model is larger than, for example, that of ecc806 with similar maximum SFB values to enable a stronger scattering. Like all cores in
Taurus, L1517A benefits from side maxima of the phase function supporting
back scattering of the strong GC signal.

For the cores with the lowest coreshine SFB L1506C, L1439, and L1498, the three different $a_2$ ranges do not overlap in one coherent way. 
The maximum grain sizes to explain 
the ratios are in all cases larger with just a lower limit.
To investigate whether this systematic trend arises from a simplification
of the model,
we consider a local radiation field in addition to the ISRF. Since the cores
are located in star-forming regions where warm dust is present near the YSOs, such a local radiation field
is not unlikely. Moreover, the DIRBE map reveals radiation on top of the
stellar component from the regions where cores are located, with a spectrum
increasing with wavelength in the considered bands. 
While the radiation sources near the core will also contribute to
the DIRBE flux from that region, if it is
not shielded along the LoS, sources very close to the core will make an increased contribution
to the local field due to the squared distance dependency of the flux.

The directional dependence of the local field is unknown and is likely to be different
from that of the ISRF. To explore the impact of a local field, we assume a simple
field with no directional variation and an intensity of 0.05 MJy/sr (about 1/6 of the mean
DIRBE field) at 3.6 $\mu$m. We modify the spectral shape by a factor $(\lambda/2.2\ \mu$m$)^\delta$
so that a variation in $\delta$
takes the rise in the local field intensity with wavelength into account.
Figure~\ref{local} shows three cases for the core L1439. The thin line gives the theoretical
$R$ without a local field. When adding a constant field without spectral variation 
($\delta=0$, medium thick line), more radiation is scattered toward the observer. The grains
can thus be smaller to explain the observed SFB, but the overlap between the ranges for each band, and the ratio is not improved.
Assuming a local field with an increase described by $\delta=$1.5, this trend 
continues to lead to no improvement in the fit.
For cores with an YSO in the vicinity or even embedded, a local anisotropic radiation field is added
to the incident field, and its impact on the coreshine ratios was discussed in 
\citet{2014A&A...572A..20L}. They find higher coreshine SFBes and lower observed coreshine SFB ratios $I_{3.6}/I_{4.5}$ for embedded cores than for starless cores. This would make it more
difficult in our model to converge to a common maximum grain size.
Alternatively, the size distributions explored in that paper and in \citet{2014A&A...564A..96S} 
mark the beginning of a more systematic search of the large size distribution parameter space for
a dust model that can explain the presence or absence of coreshine in the various observational
bands. 

For L260,
\citet{2013A&A...559A..60A}
also used the observed Ks band to constrain the size distribution. At shorter wavelengths,
the central depression is more likely. It needs to be 
verified that peaks in the grain size distribution at sizes around 1 $\mu$m will 
be compatible with the coreshine and extinction observed at shorter wavelengths.
This emphasizes the
importance of deep Ks band and NIR observations in general of cores with coreshine.
Moreover, \citet{2014A&A...572A..20L} argue that both the NIR/MIR ratio for the core outer layer and the absence of emission at 5.8 $\mu$m for any layers eliminate a mix of silicates and carbonates that both include grains above about 2 $\mu$m in meaningful quantity.

A word of caution is needed for the uncertainties in the background estimates. 
A better estimate of the foreground diffuse radiation field would improve the current uncertainty
that enters coreshine modeling. Weaker background would increase the coreshine SFB in both
bands so that smaller grains would be sufficient. 
The cores where the model fails would require
the opposite.
Moreover, the background values derived by \citet{2014A&A...572A..20L} using a different method 
can vary from one region to the other by a factor
of up to two. This can modify the results for the cores with weak coreshine or strong background.
For future 3D modeling of a single source, it could be important to put more effort into modeling
the foreground and background contributions.

One particular finding about L1506C and L1498 is interesting. Both cores show
little variation in $R_{obs}$ across the entire core as far as 4.5 $\mu$m coreshine is detectable. For both cores this ratio is small, as expected from the presence of larger grains. Both
are discussed as being on the verge of becoming low-mass cores with typical properties,
and also show central gas depletion that requires a density that might not be reached in
these cores.

For both cores the existence of larger grains would be surprising if the grains are to be formed
by coagulation since their densities are not high enough to coagulate them, as indicated by
studies of \citet{2014A&A...564A..96S} for L1506C. 
One of the hypotheses presented in that work was that L1506C went through a phase of higher
density and stronger turbulence to explain the existence of larger grains and strong gas depletion. With the similarities between the two cores also emerging from the work presented here, this could also
be the case for L1498.

\section{Summary}
This work is based on deep 3.6 and 4.5 $\mu$m IRAC {\it Spitzer}
warm mission data taken during Cycle 9.
The sample is composed of ten cores, 
nine of which show coreshine at both 3.6 and 4.5 $\mu$m, 
and one where coreshine is unconfirmed owing to strong PAH emission.
The observations and the work in this paper were motivated by the expectation that
coreshine observed at MIR wavelengths might be dominated by dust grains in the micron
size range and that 
opacity models predict a stronger impact of the largest grains at longer wavelengths.

Given the vast parameter space of the general grain scattering problem in cores caused mainly by the
unknown density structure, the presented model approach 
makes use of the fact that in the optically thin limit, both the 
observed coreshine intensity and the intensity losses by extinction through the core are 
proportional to the column density of grains.
By considering ratio maps of 3.6 and 4.5 $\mu$m intensities, the impact of the spatial density distribution
is expected to be reduced for the cores with moderate optical depth $<$ 1.
Moreover, we use the constraints from the observed SFBes based on
the expected column densities. 
Since the core properties are known only within factors of a few, and the opacity changes caused
by increasing the grain size can be substantial, it is essential for the modeling not
to exceed the observed optical depth limits, which are provided in the form of depression
of the central SFB for cores with optical depth for extinction above about 2.
When the column density is varied within its uncertainties, we make sure that this optical depth limit is never exceeded.

We applied a 
model based on the opacities (cross sections and phase functions)
of ice-coated silicate and carbonaceous spheres and a size-distribution
following a power law (index -3.5) ranging from 0.01 to a variable maximum grain size
for both species.
The incident field is approximated by DIRBE maps, and the radiation right behind the core
is constructed from the absolutely calibrated DIRBE map with point source contribution
subtracted using the WISE catalogue.

We outlined three ways to perform RT and describe the gain and
potential errors by comparing them. We illustrated that the spatial and opacity information is blended into the derived SFB by the action of multiple scattering
and extinction for increasing optical depth. 
For an example core, we found that the error of the optically thin approximation is about 14\% compared to full RT as long as the optical depth for extinction stays below 1.
The theoretical coreshine ratios are described and applied for a model core with the properties of
L260. They have a singularity for the limiting grain size where the 
net effect of core extinction and scattering is zero and then decline toward $R_{theo}$-values around 1 for grains beyond 2 $\mu$m in size.

We described the modeling procedure in detail for L260, also using longer wavelength
maps like 
cold Spitzer data at 8 $\mu$m to construct a simple spatial model. The model was used to  estimate the optical depth variation.
Selecting a region in the ratio map of the two bands that primarily shows optical depth smaller 1, the pixel distribution as a function
of the ratio serves to characterize the observed ratios.
Then we applied either single-scattering or full RT depending on the assumed maximum
optical depth to derive theoretical $R$ values or distributions to compare them to $R_{theo}$
along with the constraints from the observed maximum coreshine SFBes.

For the cores L260, ecc806, L1262, L1517A, L1512, and L1544, the model is able to account for the observed central SFBes and the ratio maps with maximum grains size around 0.9, 0.5, 0.65, 1.5, 0.6, and > 1.5 $\mu$m, respectively. 
The low grain size limit for ecc806 
is remarkable given the bright coreshine from this core, and might be related to the low background.

The maximum grain sizes found agree with the findings of earlier studies
listed in the introduction, suggesting that coreshine indicates the presence of grains larger than found in the
diffuse ISM. No obvious correlation with basic core properties is evident for our
sample of nine modeled cores.

The model fails to simultaneously reproduce the SFBes and ratio maps of
L1506C, L1439, and L1498. The grain size limits derived from the ratio maps 
are larger than the limits from the coreshine intensity. 
Since the coupling to the spatial structure is moderate ($\tau_{ext}<1$), this possibly
points to assumptions that are too simplified or to the action of an additional component.

Assuming a constant local field with a rising spectral shape on top of the ISRF
described by the DIRBE all-sky map does not improve the modeling for L1439.
The flat $R$ distribution of the cores L1506C and L1498 gives rise to speculations
that both cores host larger grains across the core, either as a primordial component
or created in a former more dense and more turbulent phase as discussed in
\citet{2014A&A...564A..96S} for L1506C.

Rho Oph 9 is in a complex environment with strong PAH emission. We did not model it in this paper and showed the maps and the ratio map for completeness.

The presented results encourage further exploration of the size distribution
parameter space. Observational efforts to gain information at shorter wavelengths
will be important for supplying further constraints. Finally,
full 3D modeling based on more realistic density structures is required with the option
of also fitting thermal emission maps.

\begin{acknowledgements}
JS, MA, and WFT acknowledge support from the ANR (SEED ANR-11-CHEX-0007-01).
MJ and V-MP acknowledge support from the Academy of Finland grant 250741.
This work is based on observations made with the {\it Spitzer} Space Telescope, 
which is operated by the Jet Propulsion Laboratory, 
California Institute of Technology under a contract with NASA.
\end{acknowledgements}

\bibliographystyle{aa}
\bibliography{4dot5}

\begin{appendix}

\section{Determination of the off-core surface brightness}\label{off}
\begin{figure}
\includegraphics[width=9cm]{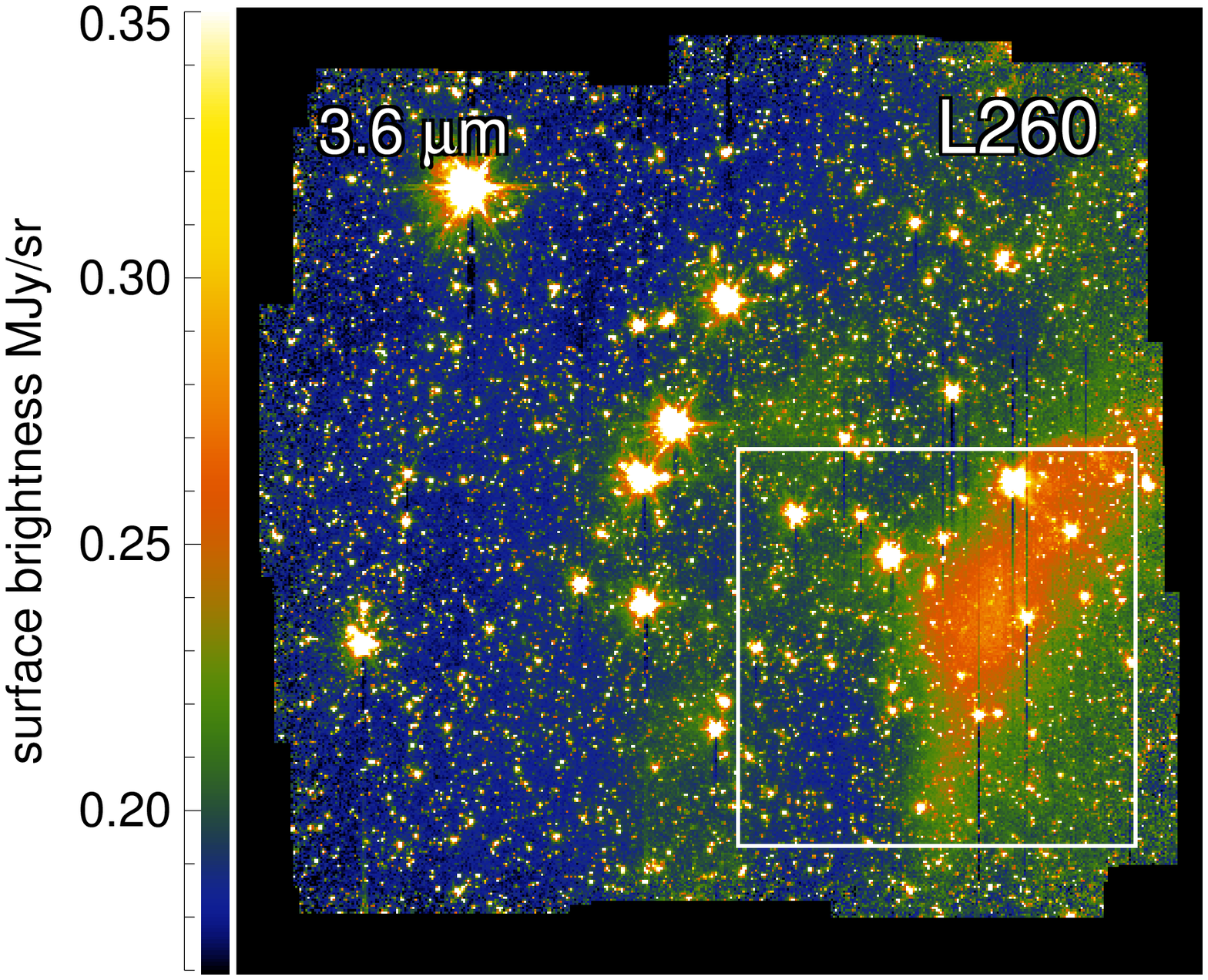}
\includegraphics[width=9cm]{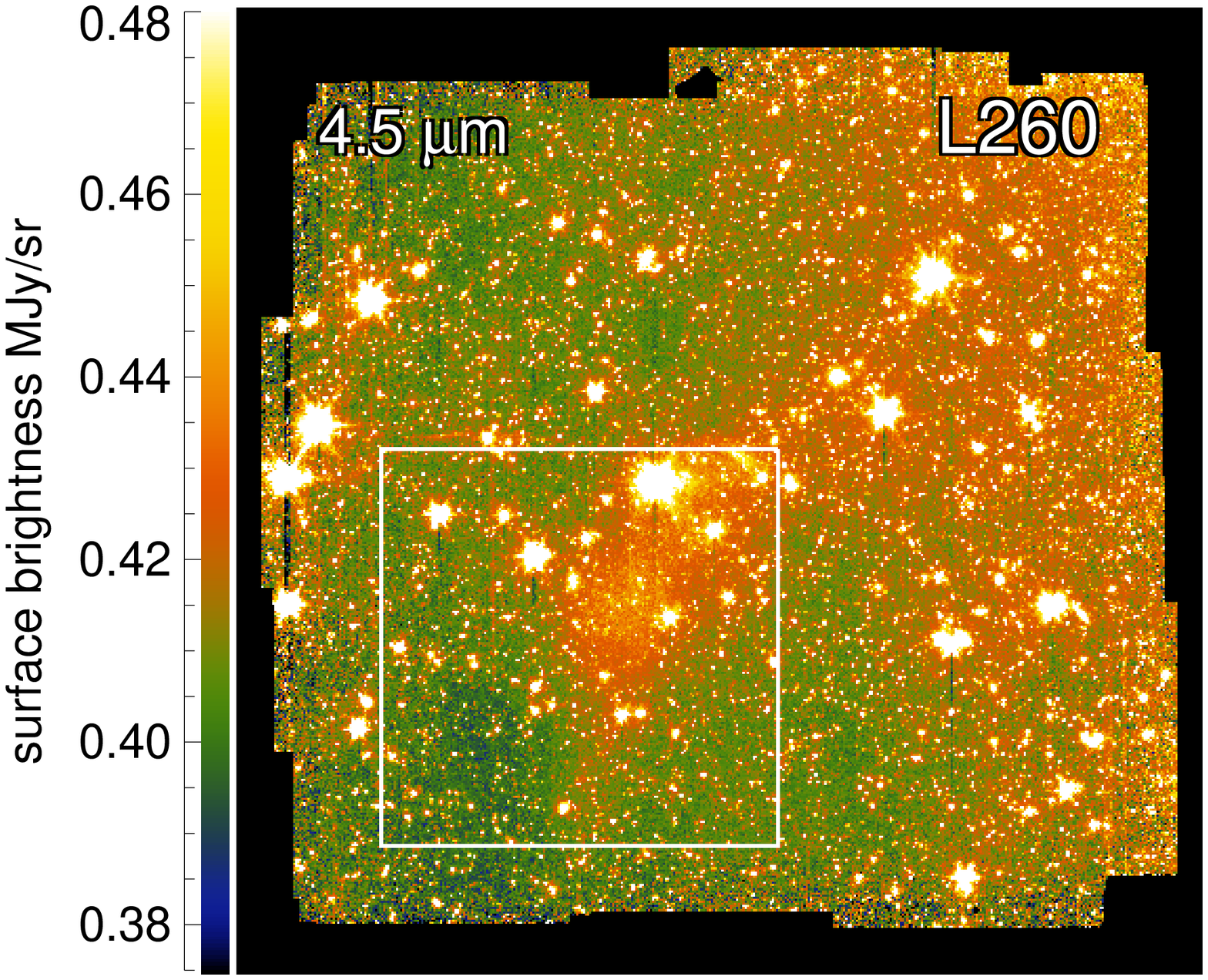}
\caption{
Entire IRAC images containing the core L260 in the white frame (see
also Fig.~2) for 3.6 $\mu$m (top) and 4.5 $\mu$m (bottom), respectively.
        }
\label{IRAC}
\end{figure}
In this appendix, we describe how we have estimated the surface brightness
near to but off the core and which variation in the determined ratios can be
expected
when this off-core measurement is performed differently.

The off-core surface brightness needs to subtracted in Eq. (\ref{Ro})
\begin{equation}
R_{obs}(y,z)=
\frac{I^{3.6}(y,z) - I^{3.6}_{off}(y,z)}
     {I^{4.5}(y,z) - I^{4.5}_{off}(y,z)}
\label{ARo}
,\end{equation}
since the IRAC measurements are not absolute and contain instrumental, as well
as back- and foreground, contributions. Since $I_{off}(x,y)$ is needed at
any PoSky location
of the core, but where it cannot be measured, an approximate value needs to be
determined
for each point (x,y).
Ideally, the region where $I_{off}$ is measured is chosen
(i) to be near the core (to represent the surface brightnesses at core
location as closest as possible),
(ii) to avoid outer core parts (which emission should not be subtracted),
(iii) to contain no stellar contributions, and
(iv) to enable interpolation of surface brightness variations across the
core.

In former work, constant $I_{off}$ were determined by circular averages
around the core
\citep[e.g.,][]{2012A&A...547A..11N} or
choosing the local region near the core with the lowest surface brightness
\citep[e.g.,][]{2009ApJ...707..137S}.
For cuts through the image, variations have also been used, e.g., by
linearly interpolating $I_{off}$ from locations
left and right from the core
\citep[e.g.,][]{2010A&A...511A...9S,2013A&A...559A..60A}.

As is visible from the extinction features of cores at 8 $\mu$m
\citep[see, e.g.,][]{2010Sci...329.1622P}, almost all cores
have a shape deviating from simple spherical symmetry.
Judging where outer parts of the core gas are located and where
we are looking at variations in the foreground or background
gas is therefore difficult, especially for cores near the
Galactic plane when the LoS likely crosses other regions.

In the following, we use the core L260 with the strongest coreshine
surface brightness in
our sample to illustrate how we have performed the off-core
measurement and how the results depend on the choice of the background.

In Fig.~\ref{IRAC}, the two IRAC images are shown for L260 in the bands
3.6 $\mu$m (top) and 4.5 $\mu$m (bottom), respectively. The white frame shows the region
around the core that is used for the top panels in Fig.~\ref{L260}.
As is visible from both images, there is a large scale horizontal gradient 
across the background of the core.

\begin{figure*}
\vbox{
\includegraphics[width=9cm]{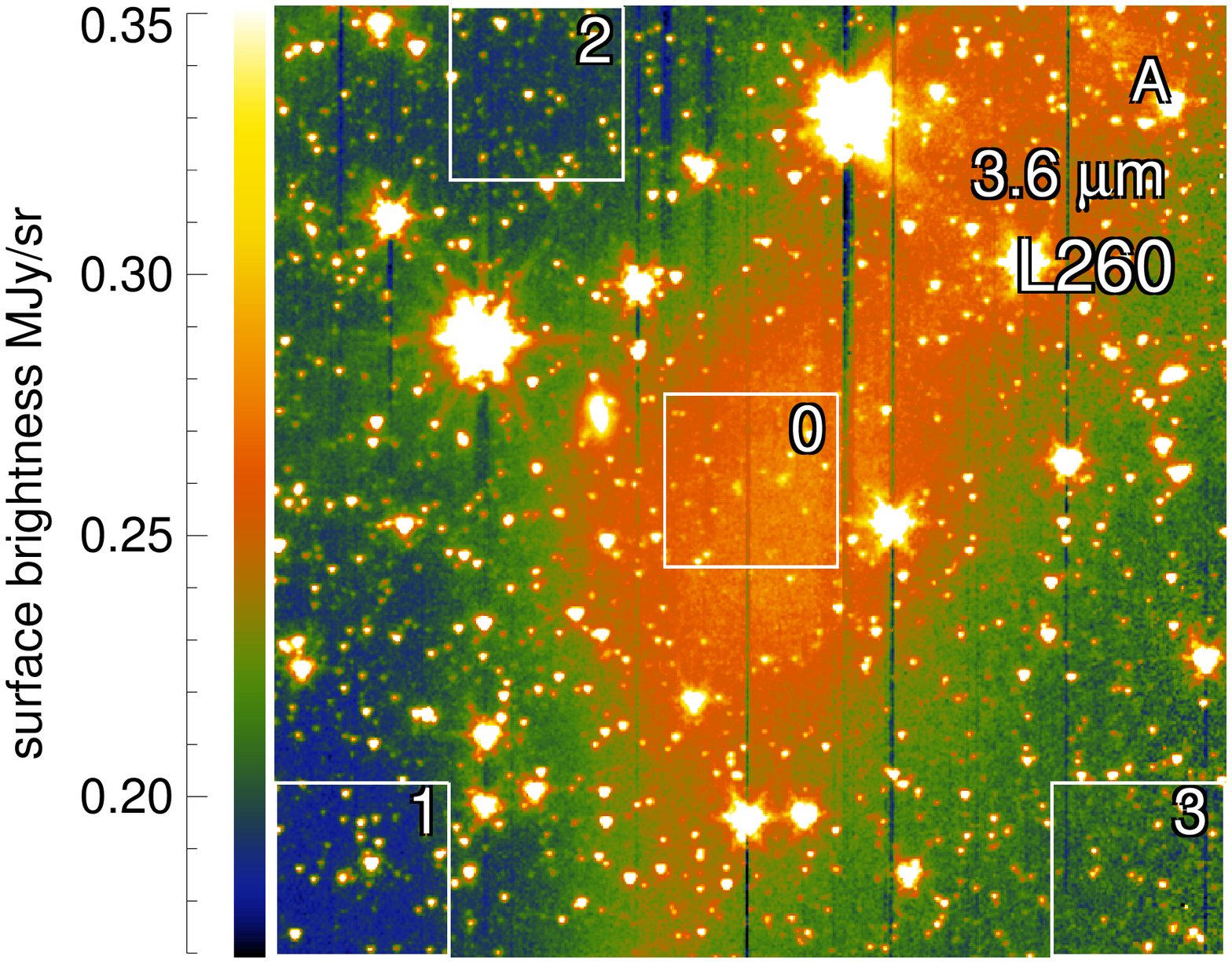}
\includegraphics[width=9cm]{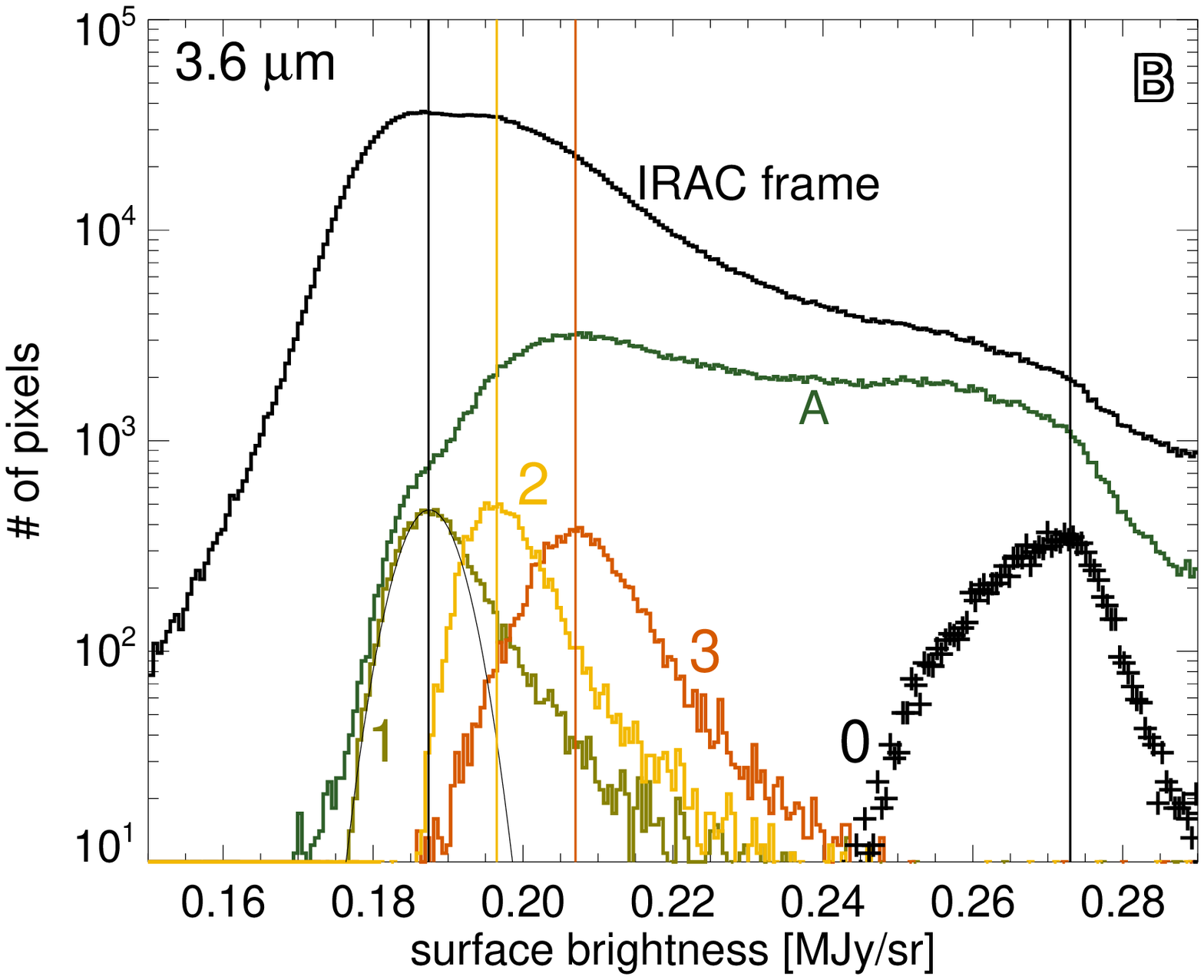}
}
\vskip 0.1cm
\vbox{
\includegraphics[width=9cm]{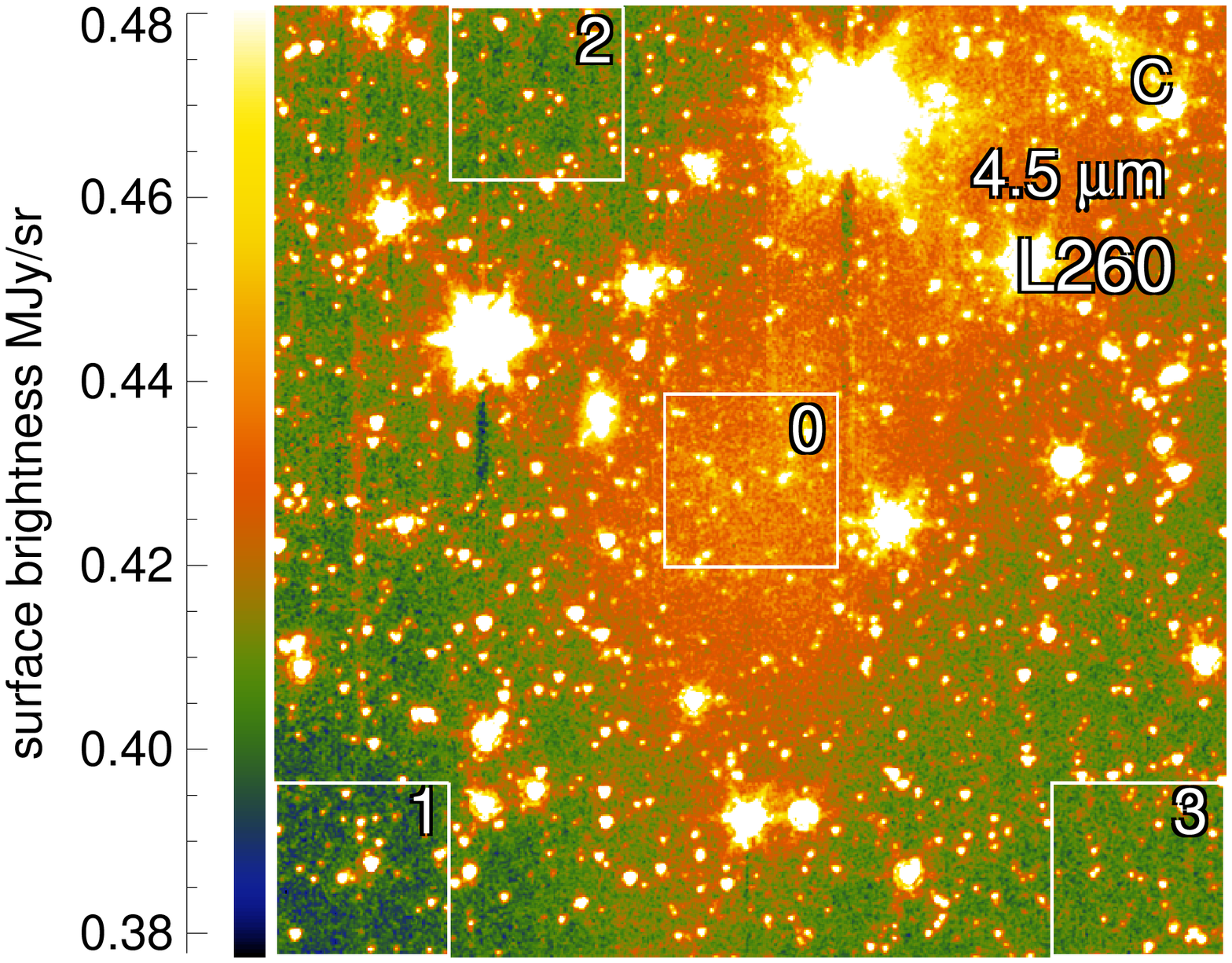}
\includegraphics[width=9cm]{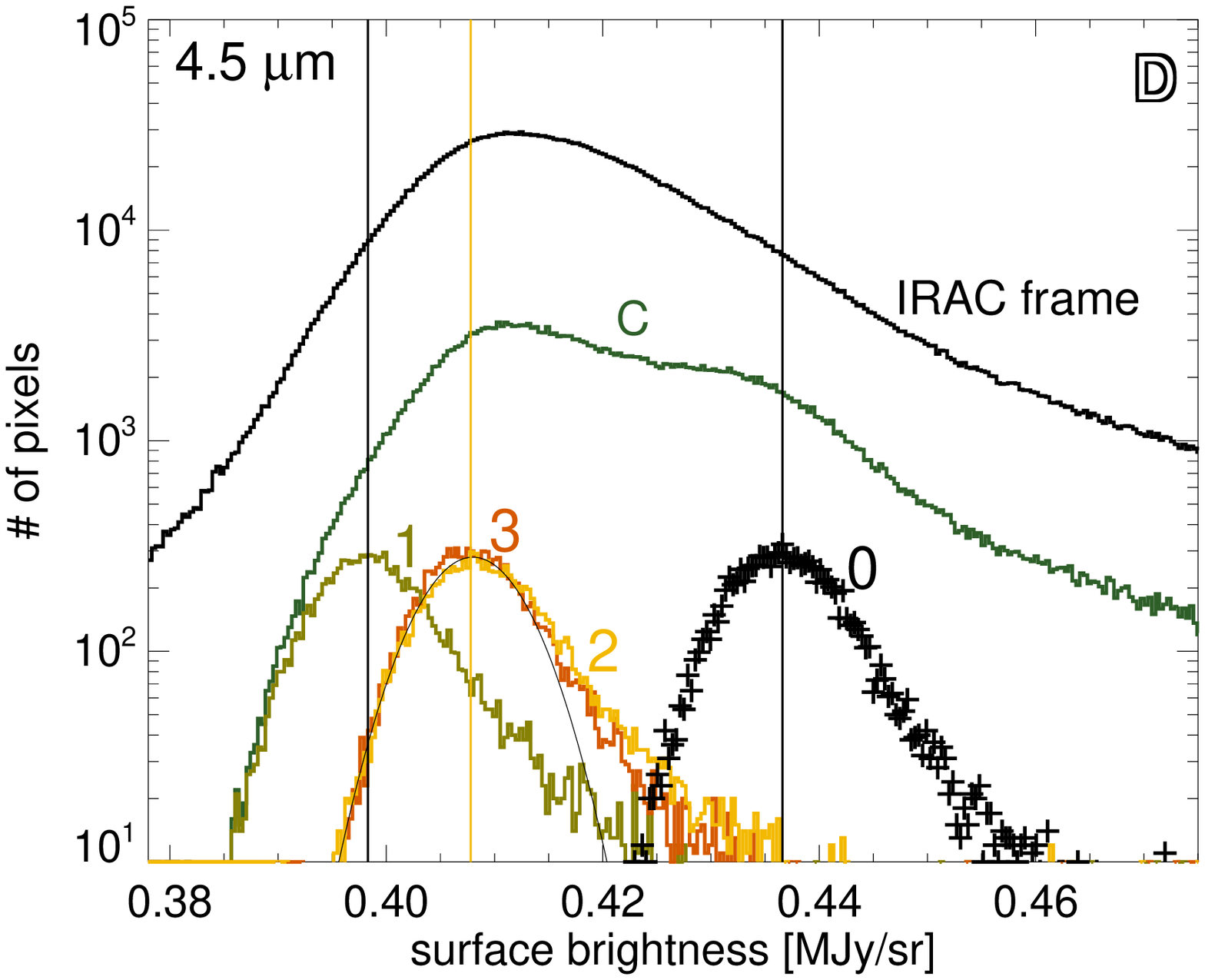}
}
\caption{
Background choice for L260
        }
\label{bg4frames}
\end{figure*}

We have chosen three spots near the core to measure $I_{off}$ and indicate
their location in Panel A of Fig.~\ref{bg4frames} as numbered white frames. We also give
for comparison the frame 0 in which we measured $R$.
In Panel B we show the number distributions of pixels as a
function of their surface brightness for the entire IRAC image.
The distributions contain the stellar sources that are visible in
the high SFB wing of the distribution as they are above the mean SFB of each frame: the frame around Core (A), the off-frames 1-3, and the $R$-measurement frame 0.
The gradients visible in the IRAC image results in a mean surface brightness 
shift from 1 to 3. We therefore interpolate $I_{off}$ from 1 and 3 which leads
to a value close to the mean surface brightness seen in Frame 2.

The situation changes for 4.5 $\mu$m. Using the same frames as indicated
in Panel C, the surface brightness distributions in Frames 2 and
3 are almost identical. Nevertheless, the gradient between 1 and 3 remains,
and we also interpolate $I_{off}$ at 4.5 $\mu$m from both frames.

To estimate the uncertainty in the derived range of observable $R$, we compare
the mean $R$ from interpolating between 1 and 3 and between 2 and 3.
Using the mean surface brightnesses of the four frames, we get R=2.3 for
Frame1 to 3 off measurement, and R=2.41 for Frames 2 to 3.
We performed this procedure of testing the variation in various frames
for all sources discussed here.

\section{Mie calculations\label{Mie}}

The absorption and scattering coefficients of large grains can be
exactly computed for spherical particules only using Mie theory
\citep{Bohren1983asls.book.....B}. Other grain shapes rely on
approximate numerical models, or their validity is restricted to small
grain sizes.

In this study the water optical constants are extracted from the
database of \cite{Hudgins1993ApJS...86..713H}. The silicate data are
taken from \cite{Draine1984ApJ...285...89D}, and the amorphous carbon
data are provided by \cite{Zubko1996MNRAS.282.1321Z}.

To simulate porous grains, we use an effective medium formulation
where the inclusions are made of
vacuum. Amorphous carbon is also considered as inclusions in the
silicate matrix.

Consider a particulate composite consisting of a matrix and including various sizes and shapes made of material other than the matrix.
Under certain conditions, the composite can be homogenized; i.e., the
composite can be replaced by a homogeneous dielectric medium with the
same macroscopic electromagnetic response and a certain effective
permittivity.  \citet{Landau1960ecm..book.....L} and independently
\citet{Looyenga1965Phy....31..401L} proposed an effective medium
formulation that take inclusion connections for all
volume fractionss into account (hereafter LLL model).  The effective dielectic
function is a volume-fraction weighted-average of the spherical
constituents for the composite and is correct to the second order in
the differences in permittivities, although dipole-dipole interaction is
still not taken into account. For $N$ constituents, the effective
dielectric function is
\begin{equation}
\epsilon_{\rm eff}^{\rm LLL}=< \epsilon_i^{1/3}>^N_{i=1}=\left(\sum_{i=1}^Nf_i\epsilon_i^{1/3}\right)^3,
\end{equation}
where $\epsilon_{\rm i}$ is the dielectric function of the material
$i$, and $f_i $ are the volume fractions of arbitrary shape.  Here, $N$ is the
total number of inclusions of a given composition. The sum of all
volume fractions has to be lower than 1. For two constituents, the
formulation is extremely simple:
\begin{equation}
\epsilon_{\rm eff}^{\rm LLL}=\left[ \left( \epsilon_{\rm inc}^{1/3}-\epsilon_{\rm mat}^{1/3}\right) f_{\rm inc} + \epsilon_{\rm mat}^{1/3}\right]^3,
\end{equation}
where $\epsilon_{\rm mat}$ is the dielectric function of the matrix.
It is clear that the LLL model is symmetric with respect to the
constituents. Other formulations of the effective medium dielectric
function exist \citep{1904RSPTA.203..385M, 1935AnP...416..636B}, each of them with their
own strengths and weaknesses. We chose to use the LLL model for its
validity for all volume fractions.

Water ice is assumed to form a mantle on the top of the porous
silicate + amorphous carbon core. Absorption and scattering
cross-sections, as well as the phase-functions for ice-coated porous
spherical grains, are computed using the latest version of the Mie
routine for coated spheres provided by \cite{Toon1981ApOpt..20.3657T}.
\end{appendix}

\end{document}